\documentclass[final,5p,times,twocolumn]{elsarticle}

\usepackage{amssymb}
\usepackage{amsthm}
\usepackage{amsmath}
\usepackage{amstext}
\usepackage{fixmath}
\usepackage{import}
\usepackage{background}
\usepackage{afterpage}
\usepackage{xcolor}
\usepackage{txfonts} 
\usepackage{hyperref}
\usepackage{txfonts} 
\usepackage{dcolumn}
\usepackage{enumitem}
\usepackage{multirow}
\usepackage{colortbl}
\usepackage{bm}% bold math
\definecolor{myblue}{rgb}{0.70,0.80,0.83}
\definecolor{myyellow}{rgb}{0.9,0.93,0.65}
\definecolor{darkgreen}{rgb}{0.0,0.4,0.0}
\newcolumntype{b}{>{\columncolor{myblue}}c}
\newcolumntype{y}{>{\columncolor{myyellow}}c}
 
\journal{Physics Reports}

\newlength\mylen
\SetBgScale{1}
\SetBgAngle{0}
\SetBgColor{violet!0}
\SetBgContents{\tikz{\draw[step=\mylen] (-.1\paperwidth,-.1\paperheight) grid (1.2\paperwidth,1.1\paperheight);}}

\journal{Physics Reports}

\newcommand{\sh}{H$_3$S}

\newcommand{\lah}{LaH$_{10}$}
\newcommand{\mg}{MgB$_2$}

\newcommand{\mus}{$\mu_c^*$}
\newcommand{\tc}{T$_{\text{c}}$}
\newcommand{\ef}{E$_{\rm F}$}
\newcommand{\omlog}{$\omega_{\text log}$}
\newcommand{\ep}{{\it e-ph}}

\newcommand{\Tr}{{\rm Tr}}
\newcommand{\vect}[1]{{\boldsymbol{#1}}}
\newcommand{\ii}{{\sl{i}}}

\newcommand*{\chichi}{{\chi\mkern-12mu\chi}}

\begin{document}

\begin{frontmatter}

\newcommand{\UniRoma}{Department of Physics, Sapienza Universita' di Roma, Italy}
\newcommand{\MPIHalle}{Max-Planck Institute of Microstructure Physics, Weinberg 2, 06120 Halle, Germany}
\newcommand{\UniAquila}{Dipartimento di Fisica Universit\`{a} degli Studi di L'Aquila and SPIN-CNR, I-67100 L'Aquila, Italy} 
\newcommand{\UTokyo}{Department of Applied Physics, Hongo Bunkyo-ku, 113-8656, Japan}
\newcommand{\RIKEN}{RIKEN Center for Emergent Matter Science, 2-1 Hirosawa, Wako, 351-0198, Japan}
\newcommand{\MPIMainz}{Max-Planck Institute for Chemistry, Hahn-Meitner-Weg 1 55128 Mainz, Germany}

\title{A Perspective on Conventional High-Temperature Superconductors at High Pressure: Methods and Materials}

\author[label1]{Jos\'e~A. Flores-Livas} % -> working now
\author[label1]{Lilia Boeri}            \address[label1]{\UniRoma}
\author[label2]{Antonio Sanna}          \address[label2]{\MPIHalle}
\author[label3]{Gianni Profeta}         \address[label3]{\UniAquila}
\author[label4,label5]{Ryotaro Arita}   \address[label4]{\UTokyo}\address[label5]{\RIKEN}
\author[label6]{Mikhail Eremets}        \address[label6]{\MPIMainz}

\begin{abstract}  
Two hydrogen-rich materials, H$_3$S and LaH$_{10}$, 
synthesized at megabar pressures, have revolutionized the field of condensed matter physics 
providing the first glimpse to the solution of the hundred-year-old problem of room temperature superconductivity. 
The mechanism underlying superconductivity in these exceptional compounds is the conventional electron-phonon coupling. 
Here we describe recent advances in experimental techniques, superconductivity theory and first-principles 
computational methods which have made possible these discoveries. 
This work aims to provide an up-to-date compendium of the available results on superconducting hydrides 
and explain how the synergy of different methodologies led to extraordinary discoveries in the field. 
Besides, in an attempt to evidence empirical rules governing superconductivity in binary hydrides under pressure, 
we discuss general trends in the electronic structure and chemical bonding. 
The last part of the Review introduces possible strategies to optimize pressure and transition temperatures 
in conventional superconducting materials as well as future directions in 
theoretical, computational and experimental research.
\end{abstract}

\begin{keyword}
High-pressure chemistry \sep Hydrides \sep Conventional superconductivity 
\sep Density-functional theory \sep Structure prediction 
\end{keyword}

\end{frontmatter}
\tableofcontents

  \newpage 

\section{Introduction}

\begin{figure*}[t]
  \centering          
  \includegraphics[width=1.55\columnwidth]{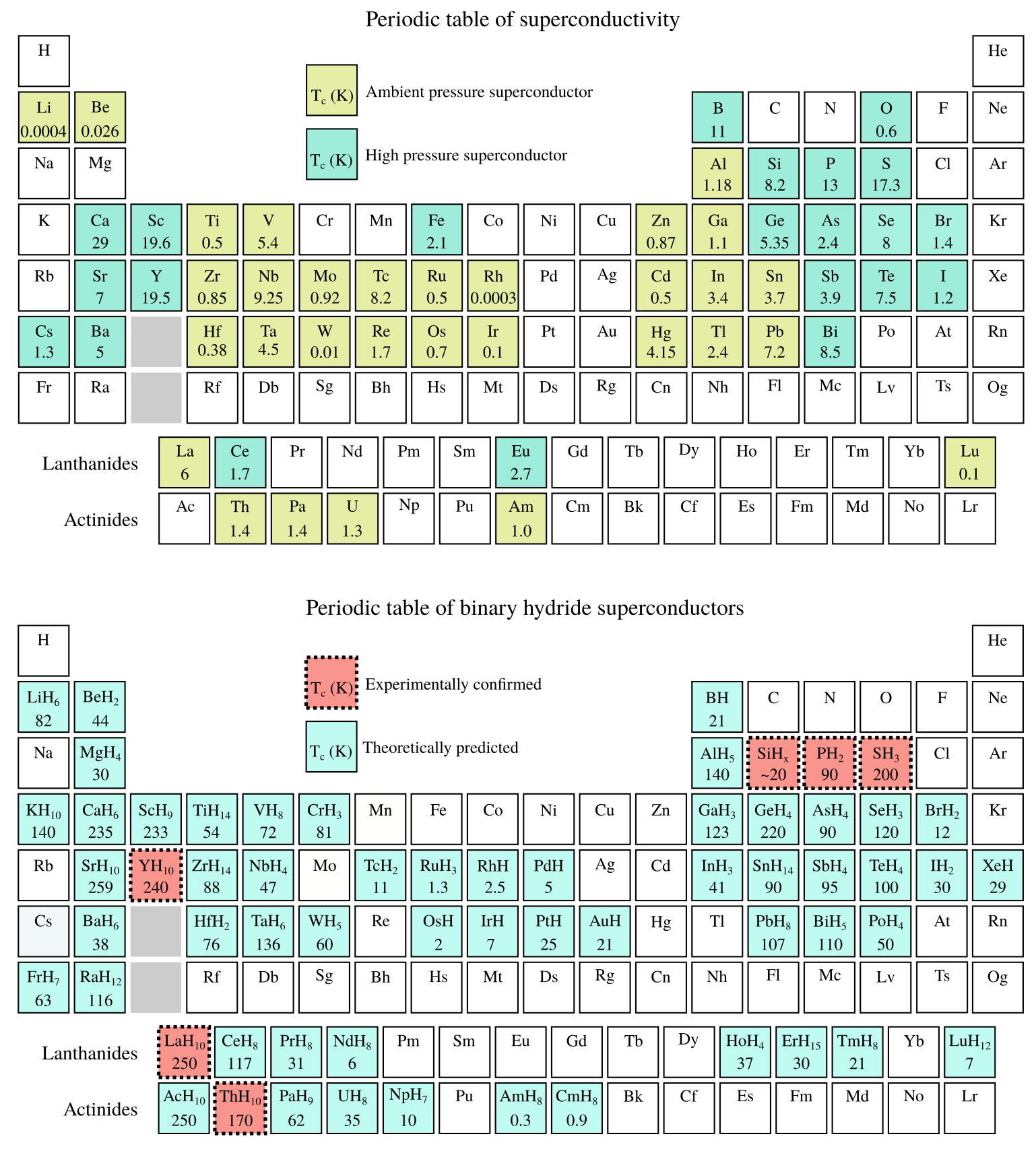}
  \caption{~Top: Periodic table of superconducting elemental solids 
  and their experimental critical temperature (\tc). 
  Bottom: Periodic table of superconducting binary hydrides (0--300\,GPa). 
  Theoretical predictions indicated in blue and experimental results in red.}\label{fig:Elemental_supra}
\end{figure*}

A new era of superconductivity was initiated by the discovery
of high-temperature conventional superconductivity in \sh~\cite{DrozdovEremets_Nature2015} and 
\lah~\cite{Nature_LaH_Eremets_2019,Hemley-LaH10_PRL_2019}. 
An unprecedented synergy between high-pressure experiments, theoretical methods and computational tools enabled these breakthroughs which are likely to bring many discoveries in the coming years.
This Review is meant to offer an up-to-date perspective on the new exciting field of 
high-temperature conventional superconductivity in hydrides at high pressure.

When, at the turn of the century, the Nobel laureate V.~L.~Ginzburg was asked
{\it What problems of physics and astrophysics seem now to be
especially important and interesting?}, room-temperature superconductivity 
and metallic hydrogen appeared as number two and three on his list~\cite{ginzburg1999problems,mukhin2003centenary}. 
Both problems had challenged and fascinated solid-state physicists for almost a century but seemed impossible to solve.  
Recently, the picture has changed and a path towards a common solution of these problems has been found, through  high-pressure hydrides.

Already in the mid-'30s, Wigner and  Huntington~\cite{wigner1935possibility} speculated 
that hydrogen could attain a  solid metallic phase under pressure; 
Neil Ashcroft and Vitaly Ginzburg~\cite{Ashcroft_PRL1968,Ginzburg_1969} at the end of the '60s discussed the possibility 
that such a phase could exhibit superconductivity at room temperature, but
only recently, thanks to key advances in theoretical and experimental techniques, 
it was possible to make the crucial step from 
general plausibility arguments to accurate predictions and observations. 
In 2017 a (still controversial) report of a solid metallic 
phase of hydrogen appeared in the literature~\cite{Dias_hydrogen_Science2017}. 
The observed metallization pressure is consistent with that predicted 
by the most accurate theoretical calculations, i.e. 400\,GPa~\cite{mcmahon2012properties}. 
Although it is still challenging to reach this pressure in experiments, 
these conditions are common in the core of celestial bodies, and now it is well established~\cite{Militzer_JupiterInterior_JGeoResPlanets2016} 
that a large portion of Jupiter's core is formed by hydrogen in a metallic liquid state.

At lower pressure, an alternative route to high-temperature hydrogenic superconductivity was suggested, through  chemical precompression of the hydrogen sub-lattice in 
hydride materials~\cite{Gilman_1971_PRL_LiHF,Ashcroft_PRL2004}. 
As a consequence, in 2018, a new record of superconducting transition temperature (\tc) 
of -23\,$^\circ$C~\cite{Nature_LaH_Eremets_2019,Hemley-LaH10_PRL_2019} was reported in a hydrogen-rich solid, \lah . 
Superconductivity at room temperature, conventionally set at 300\,K (27\,$^\circ$C), is only 50\,$^\circ$C away.

What makes solid hydrogen and high-pressure hydrides different from previously-known superconductors? 
In contrast to the previous record-holders for superconductivity, such as 
cuprates and Fe-based superconductors where the high-\tc\ is most likely due to a complex interplay between charge, spin and orbital fluctuations and lattice vibrations (phonons), superconductivity in hydrides 
is described by the conventional theory for superconductivity, which describes material such as 
aluminium, lead or niobium.
What is different are the superconducting parameters that determine the critical temperatures, 
that are exceptional compared to all previously studied compounds. 
This happens because extreme pressures  allow stabilizing chemical environments 
which would not exist spontaneously at ambient pressure.

Before delving longer into the problem of superconductivity, 
we briefly introduce a few key concepts and definitions related
to pressure, which will be one of the key actors of this Review. 
Pressure ($p$) is a thermodynamic quantity which plays an important role in modifying the properties of substances; 
for example, it can turn most semiconductors into metals. 
In this Review, we will distinguish two regimes: the first, {\bf high-pressure}, extends from 1 to 100\,gigapascal (GPa); 
the second, {\bf megabar pressure}, from 100\,GPa to the terapascal range (1,000\,GPa or 1\,TPa). 
For comparison, 1\,GPa is equivalent to 10,000 bar, or 9869 atmospheres. 
Thus, a pressure of 100\,GPa (1\,Mbar) equals roughly one million times the atmospheric pressure. 
Pressures in the megabar range are typically found in the interior of the planets: 
for example, the pressure in the core of the Earth ranges from 330 to 360\,GPa~\cite{earth_pressure-core_2002}, 
while that in Jupiter core is estimated to be above 1\,TPa.

In order to understand how a solid can change when subjected to megabar pressures, 
let us consider a substance compressed at 200 GPa (2\,Mbar). 
In most solids, this would correspond to a volume reduction of factor 1.5. 
In gases, the volume reduction can be more drastic~\cite{molodets2005scaling}; 
for instance, the volume of hydrogen is reduced by a factor 10 from 0 to 200 GPa~\cite{loubeyre1996x}. 
One can also estimate the free-energy change of the system induced by such compression. 
For this range of pressures, the $pV$ term in the Gibbs free-energy, $G=E+pV-TS$ 
is of the order of 10\,eV per two atoms, i.e. it can easily exceed the energy of any chemical bond at zero pressure~\cite{hemley2000effects,hermann2017chemical,yoo2013physical}. 
In other words, in a system subject to megabar pressures, chemical 
bonds are drastically altered, and new type of bonds may form. 
These arguments can provide a first hint on how high pressure may be used to form materials 
with physical and chemical properties different from ambient pressure and hence 
shed light on problems such as room-temperature superconductivity.  

The quest for room-temperature superconductivity is a rather old one. 
The first superconductor ever measured, found by H.~K. Onnes in 1911~\cite{onnes1913further}, 
was mercury (Hg), an elemental substance that possesses a very 
low \tc =4.2\,K. 
Such low temperatures require expensive helium cooling; hence, exploiting the characteristic 
properties of superconductors (vanishing resistivity, perfect diamagnetism) 
for large-scale applications would require materials with a substantially larger \tc . 
More than a century has passed since Onnes' discovery; over the years, the field has continuously
evolved, and materials with higher and higher \tc 's have been discovered.  
However, the rate of \tc\ increase has been extremely discontinuous: in fact, for almost 50 years (1911-1965), 
the maximum critical temperature was slowly increased with a rate of $\approx$ 0.3\,K/year, 
until a maximum \tc\ of 23.2\,K measured in an alloy of Nb and Ge (Nb$_3$Ge)~\cite{Nb3Ge_PRB1965}. 

Higher values of \tc\ seemed impossible to achieve, which led 
Cohen and Anderson to postulate in 1972 that a hard limit of 25\,K 
must exist for the \tc\ of conventional superconductors~\cite{Th:Cohen_anderson}. 
This idea, based on several misconceptions on material properties 
and stability limits, had a long-lasting detrimental impact in the field 
for the following decades. Over the last 20 years (2000-2020), the misconception has been proved false by three crucial discoveries: 
I) Superconductivity at 39\,K in MgB$_2$~\cite{SC:akimitsu_mgb2}, 
II) high-\tc\ superconductivity at 203\,K in \sh~\cite{DrozdovEremets_Nature2015} and 
III) near room-temperature superconductivity at 250\,K in \lah~\cite{Nature_LaH_Eremets_2019,Hemley-LaH10_PRL_2019}. 
High-\tc\ cuprates, discovered in 1986 (max \tc=133\,K~\cite{Schilling_cuprate_nature1993}), 
and iron-based superconductors, discovered in 2006 (Ref.~\cite{fesc:kamihara2006,fesc:kamihara_JACS_2008} max \tc=100\,K), 
technically did not contradict the Cohen-Anderson limit, since in these cases superconductivity is not of conventional electron-phonon origin. 

Empirical arguments such as the Cohen-Anderson limit (and the earlier Matthias' rules~\cite{matthiasrules}), 
became obsolete once accurate methods for calculating electron-phonon spectra of materials
from first-principles became available towards the end of the '90s~\cite{DFT:Savrasov_PRB_1996}. 
Indeed, the lack of a predictive theory for superconductivity was probably 
the main limiting factor to the discovery of new superconductors in the twentieth century. 
In 1966, Ted Geballe, after discovering a new compound with a
record \tc\ of 20\,K, declared that ``{\it there is no theory whatsoever for 
the high-transition temperature of a superconductor}"~\cite{geballe2006never}. 
This statement is founded on misconceptions and does not reflect the actual evolution of the field. 
A fully microscopic theory of superconductivity had in fact become available a few years earlier. 
The BCS theory of superconductivity (1957) states that electrons in conventional superconductors are pairwise coupled 
via excitations of bosonic character such as phonons (lattice vibrations)~\cite{BCS_1957}. 
The BCS theory assumes an instantaneous interaction between electrons and permits to describe quantitatively only a few, 
low-temperature superconductors. On the other hand, its strong-coupling diagrammatic 
extension developed in the '60s, i.e. Migdal-\'Eliashberg theory~\cite{Eliashberg}, 
was able to successfully interpret the experimental data for all superconductors known at the time, assuming a phonon-mediated interaction between electrons. 
However, at the time, Migdal-\'Eliashberg theory was mainly a semi-phenomenological theory, 
because its key ingredient, i.e. the electron-phonon spectral function, 
could not be estimated {\it a priory} for a given material, 
but only indirectly extracted from experiments. Thus, at that time, one could 
describe with a high accuracy an existing superconductor but not predict whether 
a given material would be a good (high \tc) or bad (low \tc) superconductor.

The situation has changed in this century, thanks to the development of quantitative methods for superconductivity~\cite{OGK_SCDFT_PRL1988,Lueders_SCDFT_PRB2005,Marques_SCDFT_PRB2005,Margine_anisoEliashberg_PRB2013,Arita_Nonempirical_AdvMat2017,sanna-flores_2018_Eliashberg}, and accurate and efficient computational tools to calculate phonon frequencies and electron-phonon 
coupling~\cite{Baroni_LinearResponse_PRL1987,Baroni_DFPT_RMP2001,DeGironoli_DFPTmetals_PRB1995,Savrasov_ep-PRL-1994,Savrasov2_PRB}. 
The accuracy reached by computational methods for conventional superconductors
is demonstrated by the periodic table of superconductivity for elemental solids, shown in the top panel of Fig.~\ref{fig:Elemental_supra} . 
Coloured in yellow are elements showing superconductivity at ambient pressure 
and coloured in blue are elements showing superconductivity under pressure.  
A large body of literature shows that, if computational-theoretical methods
are used to estimate the value of \tc\ for these 53 superconducting 
elements, the  deviations with respect to the experimental values are small, usually less than 20\% of \tc.

A strong impulse to the development of computational methods for
superconductivity  
came with the report of a \tc\ of 39\,K in \mg\ in 2001~\cite{SC:akimitsu_mgb2}.  
Although its critical temperature was not spectacular, compared to the record \tc\ 
of the cuprates, \mg\ possesses several features which had a significant impact on the future developments of the field:
Superconductivity is of the conventional electron-phonon type;
\tc\ is distinctively higher than the Cohen-Anderson limit;  
from a material perspective, \mg\ is 
chemically very simple and differs from the best conventional superconductors known until that time, which all contained heavy transition metals, such as vanadium or niobium.
Being formed by magnesium and boron, two of the lightest elements of the periodic table, \mg\ was a rare example of covalent metal~\cite{SC:akimitsu_mgb2} in which (relatively) high-\tc\ superconductivity could be obtained using high phonon frequencies and large electron-phonon matrix elements.

\begin{figure}[tbp!]
  \centering   
  \includegraphics[width=1.0\columnwidth]{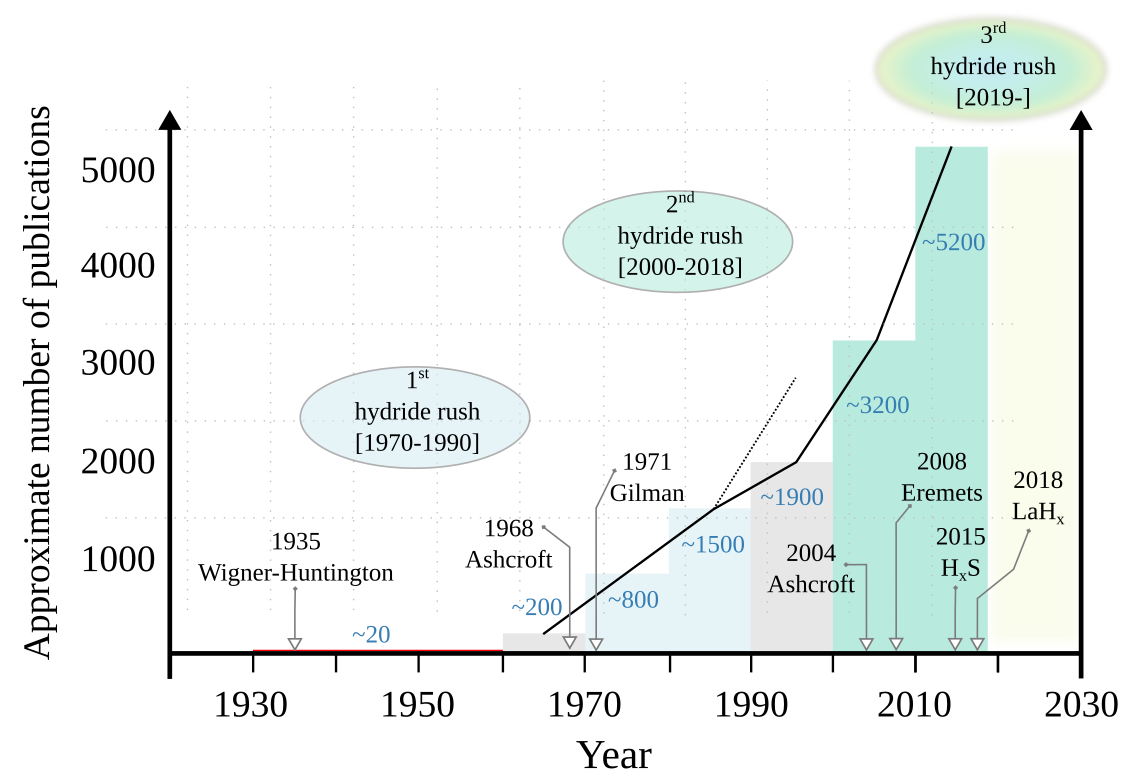}
  \caption{~The approximate number of publications in the field of hydrides per decade. 
  The first hydride rush took place right after Ashcroft's and Ginzburg's prediction of high-\tc\ 
  superconductivity in hydrogen at the end of the '60s~\cite{Ashcroft_PRL1968,Ginzburg_1969}.  
  The second hydride rush started at the dawn of the 2000s, 
  and the third is starting now (2019) after the discovery of \lah~\cite{Nature_LaH_Eremets_2019,Hemley-LaH10_PRL_2019}  
  and will continue through the following decade. 
  The expected number of publications in the following decade is above 10,000.}
  \label{fig:size_of_field}
\end{figure}

That high-\tc\ superconductivity could occur in compounds containing light elements was neither 
a new nor a particularly original idea. From a purely technical point of view, Migdal-\'Eliashberg
theory does not pose an upper bound of temperature for the emergence of superconductivity and 
even suggests compounds with light mass as the most obvious candidates for conventional superconductivity. 
Indeed, metallic hydrogen had been proposed to be a high-\tc\ superconductor as early as 1968. 
Metallic hydrides at ambient pressure, extensively studied during the '70s, 
on what we call the {\bf first hydride rush} in Fig.~\ref{fig:size_of_field}, 
achieved maximum \tc s around 16.6\,K, which fell just below the Cohen-Anderson limit. 

Experience had further shown that at ambient pressure, 
most of light-element compounds are insulating, while conventional superconductivity requires a good metallic ground state. Indeed, covalent metals like \mg\ are rare at ambient pressure; 
however, as we have seen, megabar pressures can actively modify the properties of materials and in particular, induce an insulator-to-metal transition in many substances.  
In 2004, Neil Ashcroft realized that the progress in high-pressure techniques, 
compared to the '70s, was such, that covalent hydrides could now be compressed to 
pressures so high to make them metallic, realizing an effective chemical precompression of the hydrogen sublattice~\cite{Ashcroft_PRL2004}. 
Although history proceeded otherwise, we believe that the origin of the idea of 
{\bf chemical precompression in hydrides}
can be traced back to 1971, 
when J.~Gilman studied the possibility of making a new form of hydrogen in 
a metallic state through the preparation of a covalent compound, LiH$_2$F under pressure~\cite{Gilman_1971_PRL_LiHF}. 

A {\bf second hydride rush} (see Fig.~\ref{fig:size_of_field}) started at the 
dawn of the 2000s with Ashcroft's paper and was characterized by an extremely rapid progress. 
The first metallic covalent hydride, SiH$_x$, was metallized in 2008~\cite{Eremets_silane_2008}; 
the first high-\tc\ high-pressure hydride, \sh, with a record \tc\ of 203\,K, 
was discovered in 2015~\cite{DrozdovEremets_Nature2015} 
and \lah, with a \tc\ of 250\,K, came three years later~\cite{Nature_LaH_Eremets_2019,Hemley-LaH10_PRL_2019}. 
The breakthrough discoveries of \sh\ and \lah\ followed a completely different paradigm compared to all previous
superconductor discoveries, i.e. they were the result of precise experimentation guided by accurate theoretical predictions. 
A full account of the circumstances is given in Sec.~\ref{Sec:Experiments}. 

How did this paradigm shift occur? 
We mentioned previously that the possibility of predicting
accurate critical temperatures (\tc ) from first principles played a vital role. 
Equally crucial was the development of computational tools to predict 
crystal structures and phase diagrams of materials under
given thermodynamical conditions, which was one of the most pressing open 
challenges for theoretical materials design~\cite{maddox1988crystals}.

To have and idea of the rapid progress in the field of superconductivity in high-pressure hydrides enabled by theoretical/computational methods for superconductivity and crystal structure prediction, the reader should take a look at the bottom panel of Fig.~\ref{fig:Elemental_supra}. This shows the periodic table of superconducting binary hydrides, constructed from all 
theoretical predictions available at the moment.
For each element, we show only compositions with the highest-\tc, 
 predicted in the pressure range of 0 to 300\,GPa (see Appendix for information on references). 
We count 61 superconducting binary hydrides; however, in order to assemble this table, 
many different pressures and compositions were analyzed, amounting to at least ten times as many distinct compounds. 
In red, we highlight those elements that form a hydride for which a superconducting transition under pressure has been experimentally measured, 
as of end of 2019: silicon, sulfur, selenium, phosphorus, lanthanum, yttrium and thorium. 
Notably, all measurements and predictions reported in the table were performed 
over the last ten years, and most of them in the last five.

Compared to the progress in the last century, the rate of material discovery in the last ten years is impressive. 
Given the complexity of high-pressure experiments compared to theoretical calculations, 
it is not surprising that predictions largely outnumber experimental realizations. 
For those hydrides where superconductivity has been measured, the agreement between theory and experiments 
is remarkably good (see Sec.~\ref{Sec:Trends}). 
This makes theoretical predictions extremely valuable, since they allow to focus experimental research on 
specific, pin-pointed chemical compositions and in some cases also suggest favourable 
thermodynamic conditions for the synthesis. Undoubtedly, we are currently witnessing 
the beginning of a new era of superconductivity, where hydrogen-rich materials are 
 the most promising candidates to deliver exciting discoveries.
Our periodic table of superconducting binary hydrides shows that predictions of  \tc\ even higher than the current record of 260\;K exist. 
Furthermore, the phase space left to explore is vast: ternary, 
quaternary and complex hydrides are still virtually unexplored. 
A {\bf third hydride rush} is likely to open at this point and will reserve many more surprises. 

Fig.~\ref{fig:size_of_field} summarizes the progress of the field of {\bf hydrides}, 
from the mid-'30s (Wigner-Huntington's paper) to 2019. 
The ordinate axis shows the approximate number of publications per decade (see Appendix for details). 
Although it is technically very challenging to filter the exact number of articles, 
these numbers serve as an estimate of the size and importance of the field. 
Between 1930 and 1960 the number of  articles related to superconducting or metallic hydrides 
is of the order of a few tens. By the '70s this number has increased to several hundreds publications 
and doubled during the '80s (first hydride rush). 
The research on hydride-superconductors was eclipsed by the discovery of 
unconventional high-\tc\ cuprates at the end of the '80s~\cite{Bednorz_Muller}. 
Consequently, during the '90s, the number of publications did not grow as expected. 
The field survived because hydrides appeared in other research contexts, 
predominantly as energy materials (hydrogen storage). The interest in 
hydrides was eventually revived at the beginning of the 2000's thanks to the maturity of high-pressure techniques 
and by the Gilman-Ashcroft's bold idea of chemical precompression~\cite{Gilman_1971_PRL_LiHF,Ashcroft_PRL2004}.  
More recently (2008-2019) the high-pressure hydride research was fueled 
by the experimental work of M. Eremets and collaborators (second hydride rush). 
The discovery of \lah\ will likely trigger the {\bf third hydride rush}.  
Based on the current publication rate, we foresee more than 10,000 publications over the next decades.

Although unveiling a new, potentially huge class of high-\tc\ superconductors is extremely exciting, 
the identification of new routes for material discovery opened by  accurate theoretical predictions has a much deeper significance. This work is consecrated to review the  theoretical tools underlying the latest hydride discoveries, 
as well as the experimental techniques and methodologies that permitted to extend the range of pressures to the megabar range. Several useful  reviews have appeared on the topic 
(see, for instance, 
Ref.~\cite{Review_Oganov-Pickard_2019,Duan_2019_review, 
wang2018hydrogen,Ma_NatRevMaterials_2017,wang2014perspective,Review_Zurek-2018,kresin_reviewH_RMP2018,Zurek_HiTcSC_Polyhydriedes_JCP2019,Pickard_Review_ARCMP2019}). 
However, we believe that some essential technical aspects have been, so far, overlooked.

The primary aim of this Review is to analyze the past, present and future of research in superconducting hydrides under pressure. 
The structure of the Review is as follows. 
Sec.~\ref{Sec:Experiments} reviews the essential developments in experiments, including high-pressure techniques. 
In Sec.~\ref{Sec:Theory}, we describe theoretical methods that made possible 
to understand and describe the mechanism of superconductivity at various levels and approximations. 
The computational tools that are currently used to predict crystal structures and phase diagrams of hydrides and, more in general, for materials are discussed in Sec.~\ref{Sec:Computational}. 
In Sec.~\ref{Sec:Trends}, we compiled a large number of studies, predictions and experimental results on these systems and analyzed representative examples of different classes of hydrides to understand their electronic structure and superconductivity features. 
Finally, in Sec.~\ref{Sec:Perspectives}, perspectives are presented, and different questions are addressed:
How vast is the chemical space of hydrides? What are the best strategies to optimize hydrides? 
How to decrease pressure and maximize \tc ?  
Are there other systems where, similarly to high-pressure hydrides, high-\tc\ conventional superconductivity could be realized? 
The last part is devoted to discussing the future developments in experimental techniques, 
theoretical models and computational tools which we believe will be necessary to tackle the challenges ahead. 
It is the hope of the authors that the Review will serve as a reference point for this exciting and rapidly-changing field.

\section{Experimental methods: High pressure Physics}\label{Sec:Experiments}

\begin{figure}[t]
  \centering   
  \includegraphics[width=1.0\columnwidth]{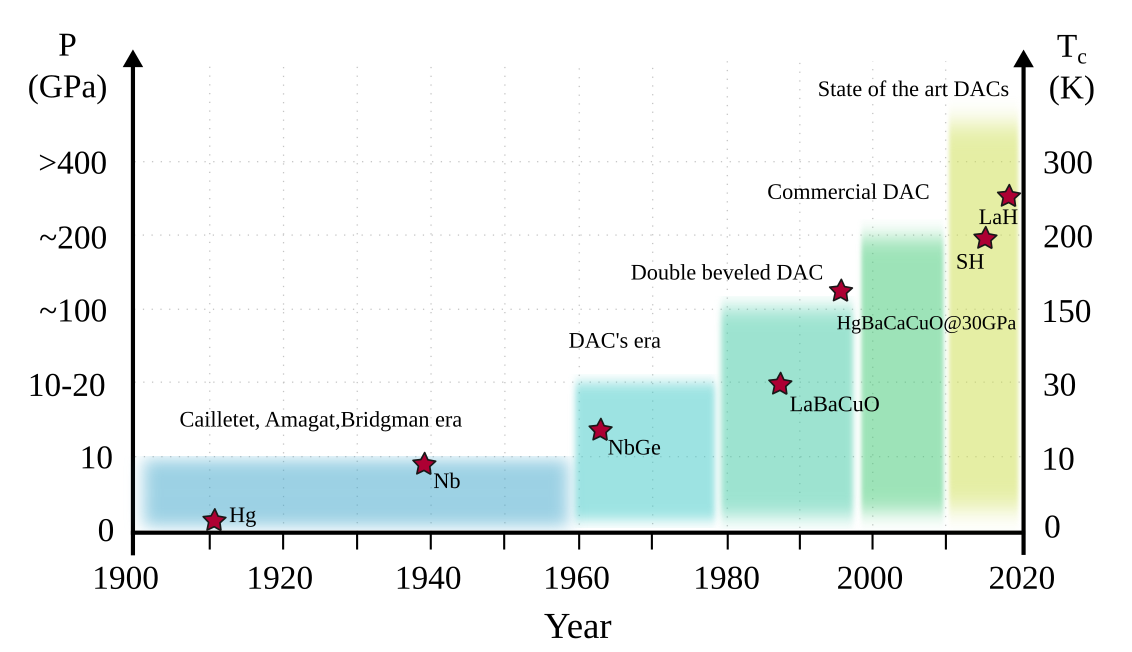}
  \caption{~Landmarks of experimentally accessible pressures from early stages 
  (Bridgman era) to the revolutionizing use of diamond as an anvil in the late '50s, 
  to the state-of-the-art DACs (left $y$-axis).
  At parallel, the increase in transition temperature of superconductors, 
  from simple metals to cuprates and high-pressure hydrides follows the same trend (stars symbol, right $y$-axis).}
  \label{fig:HP-story}
\end{figure}

High-pressure physics in its contemporary form started at the beginning of the 20${{\rm th}}$ century with the pioneering work of Cailletet, Amagat and Bridgman~\cite{bridgman1959way}. 
The field has evolved through the last century, racing to reach higher and higher pressures with a rate that, 
as shown in Fig.~\ref{fig:HP-story}, is surprisingly parallel to the increase 
of the maximum critical temperatures in superconductors. 

Bridgman's heavy, metal-made and gross anvils dominated high-pressure experiments for almost fifty years until 
a new era started in 1959 when diamond was used for the 
first time as an anvil by Wier, Van Valkenburg, Bunting and Lippincott~\cite{weir1959infrared}. 
This experiment introduced the diamond anvil cell (DAC) and marked 
the beginning of an era that would radically change the landscape of high pressure.
This is not only because higher pressures were allowed, although limited to about 20\,GPa in these early devices, 
but also because the innovative device permitted to have {\it in situ} characterization with spectroscopic 
techniques (being diamond transparent to a wide range of the electromagnetic spectrum).

Further advancement in DACs took place in 1978 when Mao-Bell~\cite{mao1978high} 
introduced beveled diamond anvils, cracking the limit of 100\,GPa static pressure. 
The full maturity of the DAC technology was reached in the mid-'90s with the rise of commercial DACs. 
These have introduced a broad array of characterization techniques and transformed the DAC 
from an exotic device, available only in a handful of centres worldwide, to a universal tool accessible to standard laboratories. 
By the year 2000s, pressures up to 200\,GPa could be reached routinely, 
and a number of applications had been published, improving our understanding of matter under extreme compression~\cite{nellis2017ultracondensed,dornheim2018uniform}. 

One of the first accomplishment of systematic investigations at high pressure has been 
to map the structural phase transformation and the emergence of superconductivity in elements under strong 
compression~\cite{SC:HAMLIN_highp_physicaC_2015,buzea2004assembling,Shimizu_EHP2005,shimizu2015superconductivity}. 
The first studies focused on simple diatomic molecules (O$_2$, N$_2$, H$_2$, among others) showing that, with applied pressure, rules of chemistry could change in unexpected ways. 
Oxygen is of particular interest because it shows magnetism at low temperatures and, 
under pressure exceeding 95\,GPa, solid molecular oxygen becomes metallic~\cite{akahama1995new}  
and superconducting with \tc\ of 0.6\,K~\cite{shimizu1998superconductivity}.

At ambient pressure, there are 29 elemental superconductors in the periodic table, none of which is an alkali metal~\cite{schilling2006superconductivity}. 
The first alkali metal discovered to become superconducting under
pressure was Cs~\cite{wittig1970pressure}, 
followed others~\cite{neaton1999pairing,shimizu2002superconductivity,deemyad2003superconducting,struzhkin2002superconductivity}.  
Years later, Lithium was confirmed to be a superconductor at ambient pressure 
but at mere 0.4\,mK~\cite{tuoriniemi2007superconductivity}. 
Alkali metals are exemplary nearly free-electron systems ~\cite{ma_2009_transparent-Na}.
One would expect that with pressure electronic bands should widen, and bandgaps close, 
leading in general to a ubiquitous metallic behaviour~\cite{lundegaard2009single,gregoryanz2005melting,richardson1997effective}. 
It was surprising, yet counter-intuitive to observe that under pressure, these metals exhibit 
marked deviations from the free-electron behaviour, transforming into semi-metallic or insulating states.  
Alkali metals represented a fertile ground of research for the scientist in the late ’90s to early 2000s to study the exotic chemistry under pressure, and superconductivity is still one of their most fascinating properties.

\begin{figure}[t]
  \centering     %======ok j
  \includegraphics[width=1.0\columnwidth]{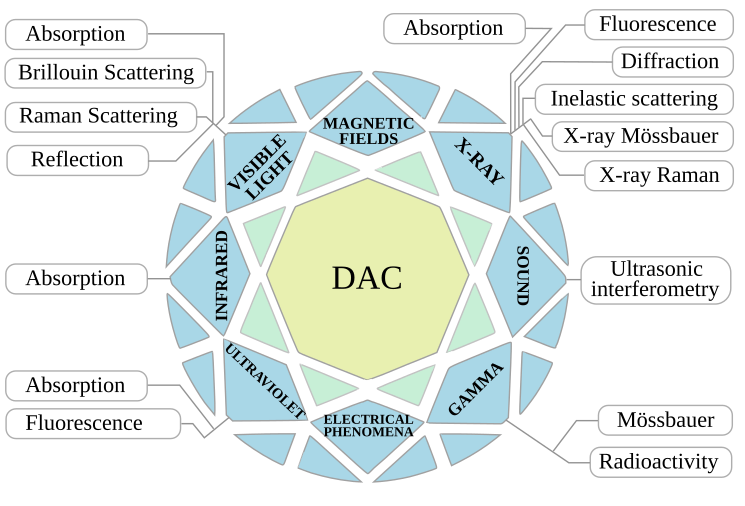}
  \caption{~Diamond anvil cell (DAC) is a universal tool vital to high-pressure physics 
  that can combine with a large number of spectroscopy techniques and {\it in situ} 
  characterization of electrical and magnetic properties.}
  \label{fig:DAC-techniques}
\end{figure}

Among Group-V elements, boron is particularly exciting. 
In 2001 with the report of \tc\ of 11.2\,K at 250\,GPa, set a record pressure for 
both electrical conductivity studies and investigations of superconductivity in a dense matter~\cite{eremets2001superconductivity}. 
Calcium is the element with the highest ever measured \tc\ that under pressure reaches 
29\,K~\cite{sakata2011superconducting,Katsuya_PRL-mafia_Ca_2013}. 
Interestingly, yttrium also shows a high \tc\ at substantially lower pressures~\cite{hamlin2006superconductivity}.  
Other elements which exhibit superconductivity upon compression 
are phosphorus~\cite{kawamura1985anomalous,karuzawa2002pressure,flores_interplay_2017},  
which at ambient conditions exhibits a complex phase diagram~\cite{Akahama_hexgonal_P_PRB1999,Phosphorus_IV_PRL2006,Incommesurate_P-IV-PRL2007,Marques_P-PRB-2008} 
and under pressure acquires a much simpler cubic 
structure~\cite{Toledano_a7structuresPRB-2008,Sugimoto_Psuperlattice-PRB-2012}. 
Other elements that undergo a long sequence of phase transformations are chalcogens~\cite{chalcogens_underpressure_2002} 
(sulfur~\cite{Akahama_S_PRB1993,Akahama_Se_PRB1993,struzhkin1997superconductivity} and selenium~\cite{Akahama_Se_met_PRB1997,kometani1997observation}), 
carbon~\cite{mao2003bonding,amsler-flores_Zcarbon_2012}, 
rubidium\cite{Schwarz_RbIV_PRL1999}, barium~\cite{Christensen_BaIV-P_PRB2004,Hamlin-Ba} 
and rhenium~\cite{Re_under_pressure_Vohra} to name a few~\cite{shimizu2015superconductivity}. 

Arguably, one of the most exciting surprises of the high-pressure chemistry/physics of 
materials~\cite{holzapfel1996physics,tonkov2004phase} has been the discovery that superconductivity becomes more common under pressure and occurs with substantially higher critical temperatures~\cite{Review_Oganov-Pickard_2019,Duan_2019_review,
wang2018hydrogen,Ma_NatRevMaterials_2017,wang2014perspective}. 
After this brief compendium of selected examples of high-pressure phase transformations 
and the emergence of superconductivity in the elements, 
we review, in the following, details on techniques and devices that made these discoveries possible.

\subsection{The diamond anvil cell}\label{sec:DAC}

The diamond anvil cell represents the icon~\cite{jayaraman1983diamond,bassett_50th-DAC_2009} of high-pressure research. 
In order to keep the pace with the progress of science in other fields, 
its design and capabilities are continuously improving, permitting to attain 
higher pressures and deeper levels of characterization~\cite{wang2014review,shen2016high}. 
This tool allows studying matter at extreme conditions of density, 
which has implications for planetary, biology, chemistry and materials science~\cite{mcmillan2005pressing,mao2018solids}.

Fig.~\ref{fig:DAC-techniques} summarizes the characterization techniques combined with the DAC. 
These cover a broad range of spectroscopy techniques~\cite{goncharov2012raman,goncharov2009laser} 
and probes~\cite{meier2018its,meier2017magnetic,eremets1996high}
including magnetic fields~\cite{mozaffari2019superconducting,kobayashi2007nonmagnetic,meier2018nmr}. 
Besides, modern DACs allow for synthesis and {\it in situ} characterization within temperature ranges 
from milli to thousands\,kelvin~\cite{boehler2000laser,gonchy_laser2010x2,mao2018solids}. 
The working principle of a DAC used in high-pressure experiments is simple: 
the sample is placed in-between two diamond surfaces and squeezed. 
However, in more elaborated experiments, a DAC can have a large number 
of interacting parts and minuscule components. In its essence, a diamond 
anvil cell consists of the body of the cell (base of the cell, piston, 
cap with screws for increasing pressure and springs to control separation),
anvil seats, diamond anvils, and a gasket. 
All these components are illustrated in Fig.~\ref{fig:DAC-parts}. 
In particular, the diamond geometry and the gasket are the crucial parts determining the
performance of the device. These are discussed in the following sections, 
together with other relevant aspects such as pressure measurements and loading.

\subsubsection{Diamond anvils}

\begin{figure*}[t!]
  \centering   %= j 
  \includegraphics[width=2.0\columnwidth]{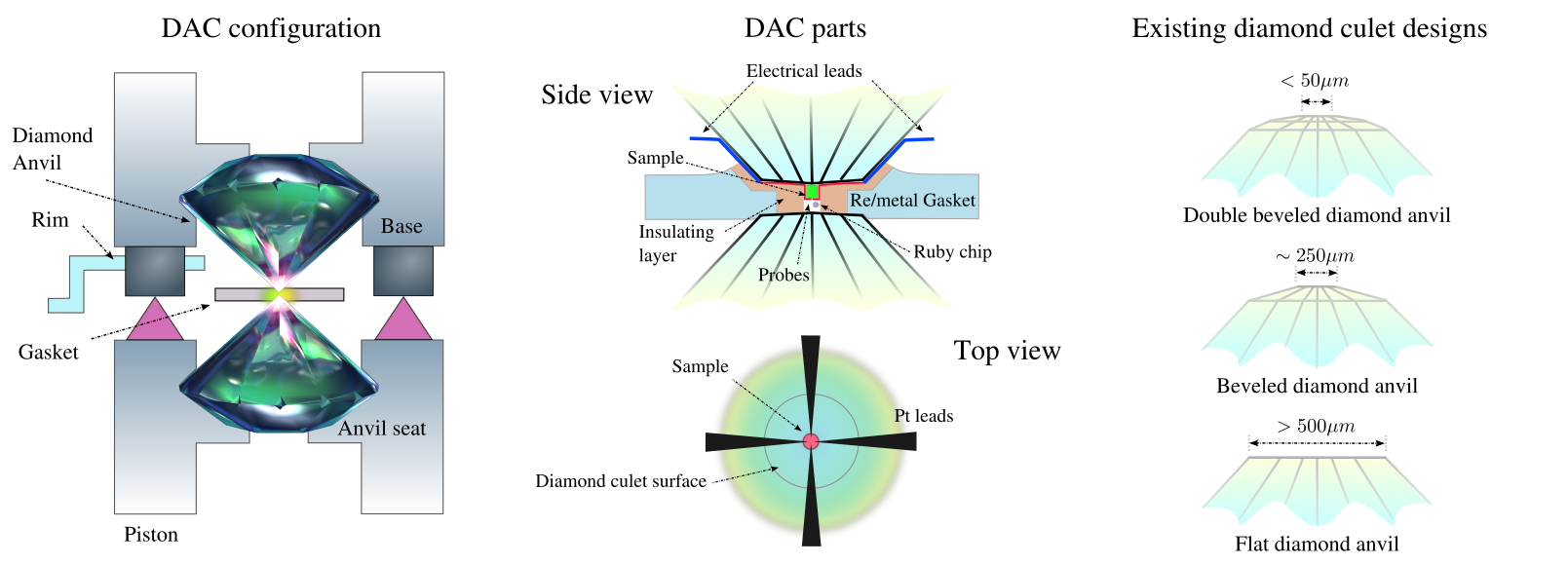}
  \caption{~Left panel: DAC schematic including anvil seats, gasket and pistons.  
          Center panel: detailed design of a DAC for transport measurements under pressure. 
           Right panel: different types of diamond culets used in high-pressure experiments.
           All drawings are not at scale.}
  \label{fig:DAC-parts}
\end{figure*}

As the name indicates, diamonds are the principal components of DACs. 
Diamond is known as the hardest material to occur naturally. 
This material is perfect for anvils because it also provides a window 
to observe the sample and study its properties with electromagnetic 
radiation from the far infrared (IR) to the ultraviolet (UV). 
In fact, diamond is transparent to nearly all visible electromagnetic 
radiation (with weak absorption in the IR range) and in the UV up to 220\,nm (corresponding to its 5.5\,eV bandgap). 
There is in principle no lower limit for the frequencies of the radiation. 
The features of the DAC strongly depend on the properties of the diamonds 
used as anvils. Diamonds can be categorized by 
a) their origin (natural or synthetic), 
b) the amount and type of impurities and 
c) the diamond cut or shape of the culet (see right panel in Fig.~\ref{fig:DAC-parts}).  
Natural diamonds are formed deep within the 
Earth below the core-mantle at depths of 140\,km, while synthetic diamonds are synthesized in laboratories at the pressure of 10\,GPa and temperatures 
over 2500\,$^{\circ}$C - High-Pressure-High-Temperature (HPHT). 
Additionally, diamonds can also be grown with a chemical vapour deposition (CVD) process from plasma at reduced pressures. 
Interestingly, diamonds from all three sources are used in high-pressure experiments. 
Synthetic diamonds produced by HPHT and CVD can be growth flawless. 
Such defect-free synthetic diamonds are typically useful for IR studies as they have deficient absorption, 
as well as for Raman studies, as they have low or negligible luminescence. 
Synthetic diamonds are currently the hardest known material.

Irrespective of the source, all diamonds are classified by the amount and character of impurities. 
Diamond can currently be doped only with small-size ions, such as H, Li, B, N, among others.
The main impurity found in natural diamonds is nitrogen, the so-called nitrogen-vacancy-center. 
Boron is relatively easily incorporated experimentally in diamond,
but boron-doped diamonds are rare in nature (known for their blue-pristine colour, 
which mainly finds use in the jewellery industry).  
Within the nitrogen-doped diamonds, there are two primary types of families, 
distinguished by the number of nitrogen impurities per million particles.
Type-I contain nitrogen inclusions of more than ten particles per million (ppm)
and type-II (high purity) contain less than 10 ppm. 
On the other hand, synthetic diamonds contain impurities of 
Ni, Co or Fe, elements used as catalysts. 
While for CVD diamonds, these often contain hydrogen as an impurity, the product from reactive plasma. 

Fig.~\ref{fig:DAC-parts} (center) shows in detail the essential parts that compose a DAC.  
Although its operating principle is simple and consists in bringing together two diamond surfaces, 
it requires highly precise machinery that works at micrometre accuracy. 
Piston-cylinder parts (left panel in Fig.~\ref{fig:DAC-parts}) are crucial 
to pressurize and control the applied pressure and, since diamond is a very 
brittle material, it easily damages during operation. In particular, 
any contact between the surfaces (diamond culet) should be avoided.  
Another crucial component are the anvil seats, which are made of a 
durable material because loads with forces larger than 1000\,kg  produce significant stress at the contact with the diamond anvils. 
Usually, these seats are made of tungsten carbide. 
Although for X-ray measurements, where a large aperture is crucial, 
seats made of boron nitride, are preferable because of its transparency to X-rays. 

The first models of DAC employed flat seats until Boehler~\cite{boehler2004new} 
introduced seats with conical support, which offer a significant advantage over their former counterparts. 
They have a larger aperture for optical and X-ray measurements, 
and diamonds are better protected once loaded 
(as they receive support from the sides). 
Over the years, mastering the cutting of diamonds in unusual shapes and developments in surface 
(culet) preparation have made it possible to expand the use of DAC to electrical 
and magnetic characterization at static pressures well above 200\,GPa~\cite{dubrovinskaia2016terapascal,vohra2015high,lobanov2015pressure,sakai2015high,mcmahon2018diamond}. 
The future of DACs foresees many more developments, mainly on two fronts: 
\begin{itemize}
\item[] I) Focus ion beam techniques will play a fundamental role in DAC, permitting to sculpt 
futuristic micro-anvils ($<$20\,$\mu$m) which will extend the range of pressure well above the 400\,GPa, 
the ceiling pressure for one-stage DAC. 
Recently, micro-milling gem-quality diamond tips shaped in toroidal form have been shown to 
extend the accessible pressure significantly up to $\approx$~550 to 600\,GPa on 
DACs~\cite{jenei2018single,dewaele2018toroidal,o2018contributed}. 
\item[] II) Thanks to the advancements in characterization techniques such as new X-ray sources in synchrotron and X-ray free-electron lasers, 
the availability of much brighter, sub$\mu$-beams will enable the study of compression volumes of only tenths of picolitres~\cite{mcmahon2018diamond}.
\end{itemize}

\subsubsection{Gasket preparation}

In the first DAC design, back in 1958, the sample was directly compressed between the two diamonds. 
It was in 1962 that Van Vakenbourg introduced the gasket in the high-pressure DAC technique.
A gasket is a piece of metal sheet with a drilled hole for sample loading which leads to several advantages. 
It allows for achieving better control of the sample. Protects the diamonds from direct contact that may cause them to fracture. 
It helps to develop a more homogeneous gradient of pressure across the sample by confining it in a relatively small area. 
It allows the possibility to add a chemically inert gas or a liquid as a pressure transmitting medium. 
Also, the gasket gives the possibility to load the DAC not only with solid samples 
but also with samples that are gaseous at room pressure, as in the case of hydrides. 

The thickness of the gasket plays a role in specific experiments, for instance, in the case of gaseous samples that are very compressible. 
A significant amount of experimentation has to be carried out before finding the ideal compromise between 
the size and the geometry of the anvil (especially concerning culet sizes) 
before an ideal gasket-thickness is found. 

For electrical measurements, 
it is necessary to introduce an insulating layer between the electrodes and the metallic gasket. 
The preparation of electrodes for electrical measurements is a crucial step in all high-pressure experiments. 
As mentioned above the preparation of the insulating gasket and the fabrication
of wires is not a standardized procedure and relies more on a trial-and-error approach. 
Different groups have mastered their approach to achieve the same experimental conditions. 

%%%%%%%

For what concerns the electrodes for electrical measurements, there are three different approaches. 
I) Manually crafted electrodes: for culets, less than 100\,$\mu$m wide, the best compromise 
is to prepare electrodes made of platinum strips or other metallic foil and carefully place them in the DAC. 
Mao-Bell~\cite{mao1981electrical} and Shimizu et al.~\cite{shimizu2002superconductivity} 
first introduced this relatively simple approach. 
A clear disadvantage of this method is represented by 
the constraints on the size of the culet and of the measured sample.
Although it takes many attempts before the tiny strip can be placed by hand inside a culet 
of around 20-30\,$\mu$m, this approach has been shown to pay-off, see Eremets et al.~\cite{eremets2001semiconducting}. 
II) Focused ion beam lithography: this is a more appropriate method for creating electrodes in tiny spaces. 
The technique consists of shaping metallic electrodes directly on the diamond surface with the aid of focused 
ion beam ultra-thin lithography. Recently, Rotondu et al.~\cite{rotundu2013high} show the capabilities of 
lithography and the possibility of reaching pressure up to 240\,GPa. 
The main advantage of this approach is the precision and control of the electrodes, 
while the focused ion beam does not have practical limitations on the size of the culet. 
Deposition of the electrodes with this technique can provide precision down to 20\,nm. 
It has also been shown that the created electrodes can sustain an indentation of the metal gasket. 
Unfortunately, this technique has a very high cost which makes it inaccessible to most laboratories. 
III) Sputtered electrodes: a third method is to sputter metals (Ti, Al, Ta, etc.)
to form ohmic contacts directly on the diamond culet. This technology has been proven 
to be robust and has worked since the pioneering works of 
Das~\cite{das1992review}, Evan~\cite{evans2009diamond}, and Werner~\cite{werner1995effect}. 
In practice, the electrodes are made with the aid of a mask with four slits which covers the diamond culet.
In some of the experiments in which one of us (M.E.) reported superconductivity in the H-S systems 
(see Sec.~\ref{sect:hs_superconductivity}) a mask of aluminium foil was used. 
The diamond surface was first ion-etched for 30-40 seconds and subsequently sputtered to form ohmic contacts.

\begin{figure}[t]
  \centering   % ok j
  \includegraphics[width=0.9\columnwidth]{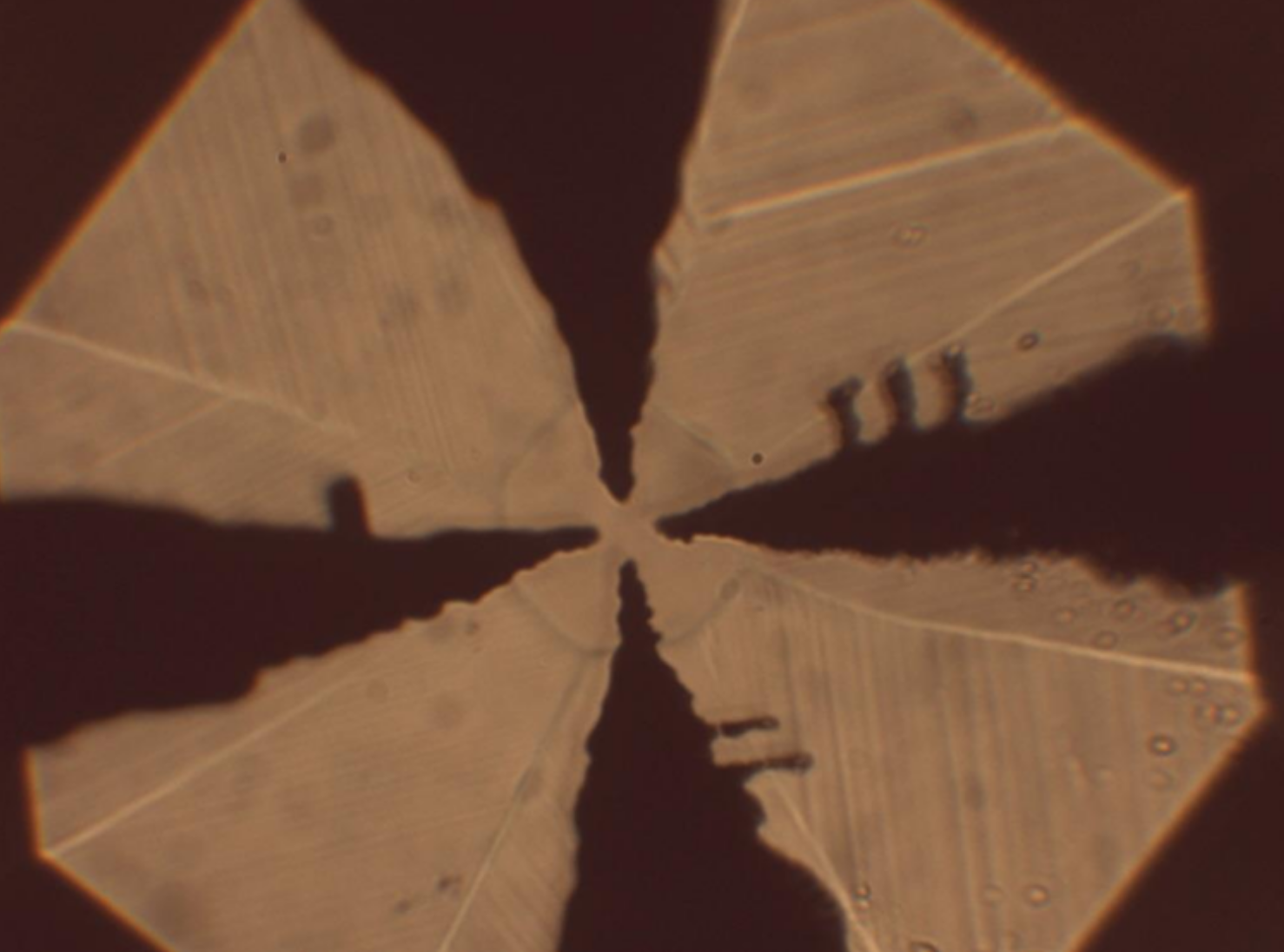}
  \caption{~Micrograph taken using optical microscopy showing in detail the sputtered electrodes 
  in a DAC prepared for high-pressure measurement in hydrogen-sulfide.}
  \label{fig:electrodes}
\end{figure}

After this, the metal contacts must be covered with gold or platinum to protect the electrodes from oxidation. 
In our (M.E. et al.) experiments --the electrodes in final assembling in a DAC can be seen in Fig.~\ref{fig:electrodes}-- 
the resistance in the electrodes is about 100\,$\Omega$, which is suitable for detecting any change upon temperature. 
The sputtered electrodes can be connected externally outside the DAC with copper wires, platinum foil or silver paint. 
It is essential to mention that the sputtered electrodes adhere conveniently to the surface of the diamond. 
However, they only last a maximum of 2 to 4 weeks and, after this period, resistivity increases dramatically. 

\subsubsection{Pressure measurements}

The most reliable procedure to measure pressures in the megabar range is to determine the volume of a known material and use its equation of state (EoS). The EoS of simple metals or other FCC compounds (NaCl, Au, Pt, Re) have been determined from shock-wave experiments and are a reliable pressure scale. The main disadvantage of this method is that it requires X-ray measurements, typically at synchrotron sources. 
In the low-pressure regime, a much more convenient technique for experimentation is to determine the pressure using the luminescence of ruby chips. Forman et al.~\cite{forman1972pressure} first introduced this technique using a small ruby. The shift of the sharp-line (R-line) in the luminescence of ruby under pressure was calibrated against the EoS and found to move uniformly up to 100\,GPa with high reproducibility. As ruby is chemically inert, it can be manufactured in micrometre-sized and produce a good signal. This technique has been a real breakthrough in the high-pressure community and made it possible to conduct high-pressure experiments with DAC in standard laboratories without accessing X-ray sources. 

Alternative methods are employed to estimate the pressure beyond the 100\,GPa limit. 
For a long time, it has been known that the Raman spectrum of the diamond anvil itself changes with pressure. 
Theoretically, the Raman peak evolution of carbon structures is well documented~\cite{flores2012raman}. 
Hanfland and Syassen~\cite{hanfland1985raman} reported in 1985 that the high-frequency diamond edge has a linear 
dependence on pressure, up to 300\,kbar (0.3\,GPa). Back in those days, they concluded that this shift could be 
hardly used for the determination of pressure mainly because the stresses at 
the tip of the loaded anvils are highly nonuniform and strongly depend on the geometry of the anvils, gasket among others. 
However, the accumulated experience has shown that the pressure dependence of the diamond edge 
is surprisingly universal and can be used for a reliable estimation of pressure. 
This pressure scale was introduced in 2003 by one of us (ME)~\cite{eremets2003megabar} 
and subsequently extended for pressures up to 410\,GPa by Akahama et al.~\cite{akahama2006pressure,akahama2010pressure}. 
Akahama also extended the study and analyzed different orientations in diamonds: 
the typical [0001] of the anvil axis but also [0111] and [0110]~\cite{akahama2007diamond}. 
This scale is necessary for the determination of pressure in the megabar range of pressures. 

\begin{figure}[t!]
\centering
\includegraphics[width=0.97\columnwidth]{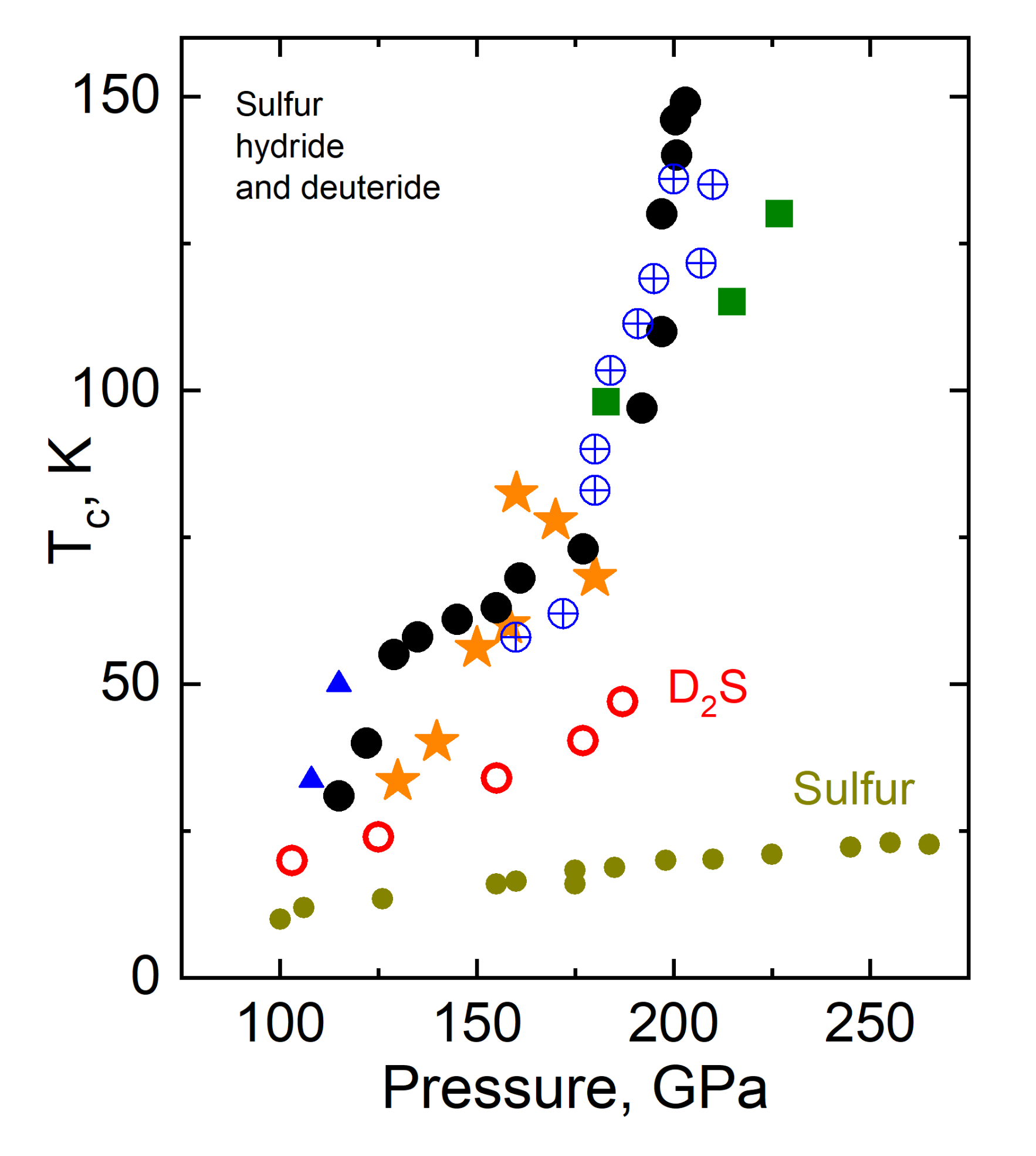}
\caption{~Pressure dependence of the critical 
temperature measured in different H$_2$S and D$_2$S loaded and pressurized at low temperatures (100-250\,K).
\tc\ for pure sulfur is plotted for comparison. 
Orange symbols mark calculations for H$_2$S (see Ref.~\cite{li2014metallization}).}
  \label{fig:new_Eremets}
\end{figure}

\begin{figure*}[ht!]
  \centering        % ok j
  \includegraphics[width=1.7\columnwidth]{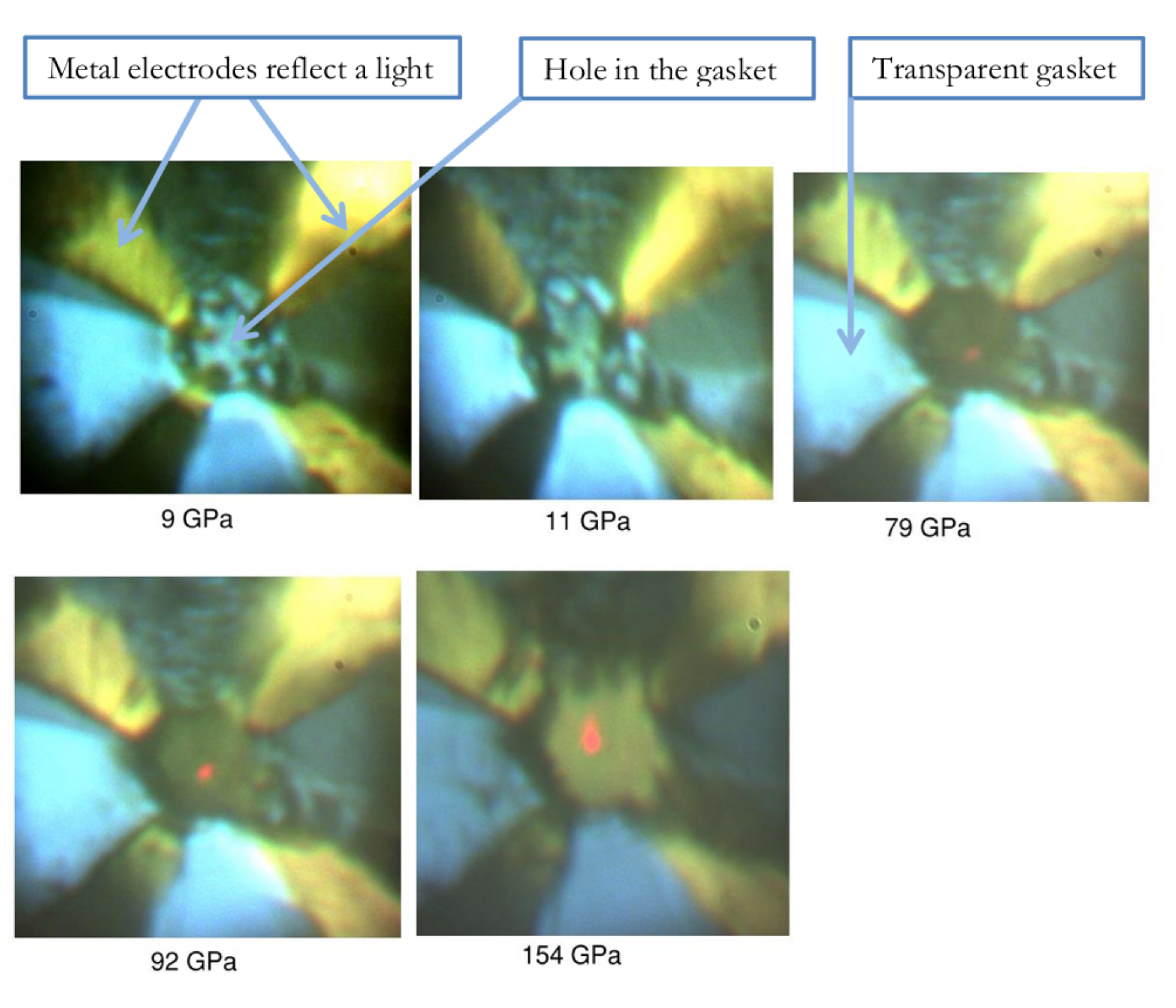}
  \caption{~Micrographs obtained using optical microscopy during the transformation of the sample upon compression. 
  Images are recorded with transmitting and reflective light. 
  Blue parts indicate the optically translucent gasket. 
  Yellow represents a reflection from the metal electrodes. 
  The sample is transparent at 9\,GPa. Darkening begins at about 11\,GPa, 
  and the sample becomes opaque at a pressure between 30--40 GPa. 
  Weak reflection (onset of metallicity) starts to appear at 79\,GPa. 
  Reflection of the sample becomes comparable with electrodes reflection at about 150\,GPa.}
  \label{fig:DAC_detail-transformation}
\end{figure*}

\subsubsection{Low-temperature loading}

A large number of interesting systems in the hydride family are 
gases at ambient temperature which complicates loading samples into the DAC. 
A particular loading procedure, described in the following, is required for H$_2$S and PH$_3$.
In order to prevent decomposition during loading, 
it is necessary to perform it at low temperatures. For this, the diamond anvil cell is 
to be placed into a cryostat and cooled down (with liquid He) to a temperature where 
most of the samples transform to their liquid phase. The left panel in Fig.~\ref{fig:DAC-parts} 
shows the inlet (rim) that introduces the gas into the chamber through a capillary.
At room temperature, the piston moves close to the base at about 100\,$\mu$m between the two anvils. 
Before loading, the air from this space should be pumped out through capillaries. 
In order to avoid any possible leaks at low temperature, 
the pistons move closer to a distance of about 20--50\,$\mu$m between the anvils. 
At temperatures lower than their boiling point, gaseous samples are put into the 
rim where they become liquid and completely fill the space  inside the rim, 
including the hole in the gasket. After, the sample is clamped between the anvils and the gasket.

\subsection{Superconductivity in sulfur-hydrogen compounds}\label{sect:hs_superconductivity}

In condensed matter physics, a substantial amount of research in the field of high pressure (post-Bridgman era) 
originated by the tantalizing idea of metallizing hydrogen, the Wigner-Huntington~\cite{wigner1935possibility} transition, dating back to the mid-’30s. 
The metallization of hydrogen is seen as the {\it holy grail} of high-pressure 
research and has been a compelling subject of interest for many scientists, ranging from experimental chemists and 
physicists~\cite{mao1994ultrahigh,LeToullec2002,Eremets_NatMat2011,Dias_hydrogen_Science2017,eremets2016low,mcmahon_high_2011,mcmahon2012properties}, 
to theoreticians~\cite{cazorla2017simulation,carbotte2018detecting,borinaga2018strong,monserrat2018structure}. 
However, despite the significant advancement in high-pressure techniques and tools, 
the metallization of hydrogen in its solid phase has proved to be 
challenging~\cite{Hemley_PRL2012,HRussell_hydrogenJACS2014,Salamat-Silvera_2016,eremets2017molecular,azadi2017role,magduau2017simple,liu2017comment,goncharov2017comment,zha2017melting,zaghoo2017conductivity,rillo2018coupled}. 
The landscape in the field broke in 2004 when Ashcroft suggested chemically precompressed hydrogen-rich materials~\cite{Ashcroft_PRL2004}
as an alternative route to decrease the tremendous pressure necessary to metalize hydrogen. 
Eremets et al.~\cite{Eremets_Science2008}
successfully proved the principle right in 2008 by metallizing silane (SiH$_4$). 
Although subsequent studies attributed these results to probable decomposition or the existence of other 
stoichiometries~\cite{Chen_disilane_PNAS2008,Wang_PNAS2009,Hanfland_PRL2011,PtH_PRL2011,Pt-H_2019experimental,flores_disilane_2012}, 
this work remains the first substantial evidence of superconductivity in chemically precompressed hydrides at high pressure.\\ 

%%%%%%%%%%%%%%%%%%%%%%%%%% serendipity starts here
{\bf Serendipity discovery of H$_x$S superconductors.} 
It is the wish of one of the authors (M.E.) to provide an accurate 
description of the context and circumstances that led to the measurements of 190--200\,K 
superconductivity in sulfur hydride. Very often in scientific conferences, seminars, 
talks and practically in 90\,\% of the publications that followed this discovery, 
the events are described with the wrong chronology.

This important discovery drove more by serendipity than is believed. 
Back in 2013--2014, a former PhD student, ``Sasha" Drozdov, and I started experiments on H$_2$S 
following ideas proposed by Li et al.~\cite{li2014metallization}, 
on the possibility of metallization and superconductivity in dense hydrogen sulfide. 
Adding as a side note, that at the time many shared a sceptical opinion about the possibility of 
achieving high-\tc\ in hydrides, and even doubted about the driving mechanism for superconductivity~\cite{hirsch2009bcs}. 
The calculations of Li et al.~\cite{li2014metallization} seemed simple to verify experimentally 
since highly pure liquid/gaseous H$_2$S is almost freely available. 
Within one month of experiments, we measured a \tc $\approx$ 60\,K at 170\,GPa, 
a significant result for those days. It was in apparent agreement with the theoretical prediction 
that we were following (see Fig.~\ref{fig:new_Eremets}). 

However, soon, we discovered that the \tc\ sharply increases 
with the further increase of pressure up to 150\,K (see Fig.~\ref{fig:new_Eremets}). 
This unexpected situation, which had not been predicted by theory, turned out to be more complicated to disentangle. 
We knew that it would be hard for this result to be accepted if we just plotted a resistivity drop at 150\,K. 
The community is plagued by many claims of high-\tc\ materials, most of them of doubtful nature. 
Therefore, we did not publish our first result but instead spent about eight months complicating our 
experiments to understand the reason for such scattered values of \tc . 
Our main goal was to find a reproducible thermodynamic path to obtain temperatures of 190\,K. 
We hypothesized that the abrupt emergence of \tc\ (refer to Fig.~\ref{fig:new_Eremets})
might be due to the increase in temperature (the pressure did not change much). 
Indeed, the samples pressurized $\approx$150\,GPa at low temperatures 
($<$100\,K) but warmed to room temperature showed a tremendous \tc\ of the order of 200\,K. 
These measurements, representing remarkable results blew away our minds.

The value of the critical temperature was clearly at odds with Yanming Ma's group original prediction~\cite{li2014metallization}. 
We came up with the guess that sulfur, known for its super-valency, 
could in principle have bound more hydrogen atoms than in H$_2$S, and presumably, 
our samples had transformed to another hydride (H$_4$S or H$_6$S). 

We were preparing the manuscript and solving the issues on the isotope effect when Duan's paper 
appeared online on November 10$^{\rm th}$, 
2014~\cite{Duan_SciRep2014} where theoretically another system (H$_2$S)$_2$H$_2$, 
different from pure H$_2$S, was considered. 
We uploaded our work to arXiv on December 1$^{\rm st}$, 2014~\cite{drozdov2014conventional}. 
It is important to remark that in the version submitted to arXiv, we (M.E. et al.) 
kept our original interpretation and that Duan's paper did not impact 
our work in any aspect, perhaps only by accelerating our submission. 
Note that the calculated \tc\ amazingly coincided with our result. 
This perfect agreement with the experiment was accidental because, as it will be revealed later, 
standard approximations in the calculations such as harmonicity 
turn out to be detrimental for the correct description of \tc . 
Immediately, computational and theoretical scientists associated our observation of superconductivity 
with the $Im\bar{3}m$ predicted structure.  
At that time, it was not apparent that the experimental and theoretical 
phases coincided since we did not have X-ray data of the high-\tc\ samples. 

The almost simultaneous announcement of experimental and theoretical 
works tremendously accelerated the understanding of the \tc\ in this compound. 
I believe that the unusual circumstances surrounding the rush raised by this material, 
which saw a theoretical and an experimental manuscript within a few weeks,
spread a view that Duan's prediction was first and our experiment followed. 
However, that was not the case.

%%%%%%%%%%%%%%%%%%%%%%%%%% serendipity ends here

In this section, we will briefly review some key aspects of the experimental 
work on sulfur hydride, focusing on the emergence of superconductivity at high pressure. 
The phase diagram of hydrogen sulfide was first studied by Shimizu et al.~\cite{shimizu1995pressure} back in 1995,
by Sakashita et al.~\cite{sakashita1997pressure} in 1997 and 
revisited by Fujihisa et al.~\cite{fujihisa2004molecular} in 2004. 
H$_2$S is a typical molecular compound with a rich phase diagram. 
Infrared studies under pressure show that hydrogen sulfide transforms into a metal at pressures of about 96 GPa~\cite{sakashita1997pressure}. 
The transformation is complicated by the partial dissociation of H$_2$S and the appearance of 
elemental sulfur at pressures larger than 27\,GPa at room temperature. 
In principle one could attribute the metallization of hydrogen sulfide 
to the disassociation into hydrogen and elemental sulfur, the latter known to become metallic above 95\,GPa~\cite{kometani1997observation}. 
It is worth mentioning that before 2014 there were no experiments or studies reported on hydrogen sulfide above 100\,GPa.  
Thanks to the advent of computational tools to explore energetically-stable crystal structures 
under pressure (see Sec.~\ref{Sec:Computational}), this material was revisited by Li et al.~\cite{li2014metallization}. 
Unexpectedly, the authors showed stable, metallic phases of H$_2$S, for pressures above 100\,GPa, 
with superconducting temperatures of the order of $\approx$80\,K at 160\,GPa. 
Most importantly, they showed that elemental decomposition into sulfur is highly unlikely, 
in apparent contradiction with previous experiments.  
To address this apparent disagreement between theory and experiment, back in 2014,  
we (M.E. et al.) performed a series of experiments in order to stabilize the predicted phases of H$_2$S under pressure. 

\begin{figure}[ht!]
  \centering  
  \includegraphics[width=1.0\columnwidth]{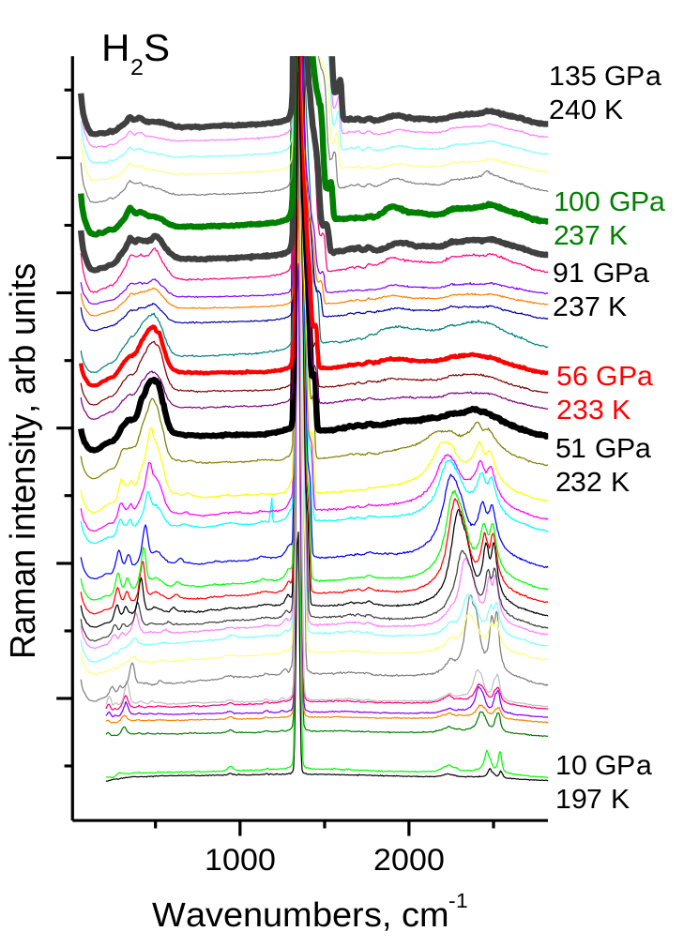}
  \caption{~Spectra of sulfur-hydride as a function of increasing pressure at 230 K. 
Spectra shifted to each other for ease of comparison. 
The phase transformation occurring at 51\,GPa is marked by the disappearance of
the characteristic vibron peaks in the 2100--2500 (wave-numbers) range. }
  \label{fig:Raman}
\end{figure}

\subsubsection{Sample preparation} 

In the H$_2$S experiments~\cite{DrozdovEremets_Nature2015}, the loading occurred in a cryostat at temperatures where hydrogen sulfide is liquid. The DAC was placed into a cryostat and cooled down to 191-213\,K (within the temperature range of liquid H$_2$S) 
and then the H$_2$S gas was inserted in the chamber through a capillary into the rim around the diamond anvil, where it is liquified. 
H$_2$S and D$_2$S gases of 99.5\% and 97\% purity, respectively, were used for all experiments.  
The filling of the chamber was monitored visually, and the sample was identified by measuring Raman spectra (see Fig.~\ref{fig:Raman}). 
After clamping the DAC, the system was heated to 220\,K to evaporate the rest of H$_2$S, and then the pressure was increased typically at this temperature. 
Pressure remained stable during the cooling within a 5\,GPa deviation. 
It was found~\cite{DrozdovEremets_Nature2015} that the low-temperature loading procedure seemed to be required to prepare samples with high \tc. For instance, if H$_2$S was loaded at room temperature, only sulfur was detected in Raman and X-ray scattering. 
Fig.~\ref{fig:DAC_detail-transformation} shows different images taken using optical microscopy during the transformation of the sample upon compression. These images are obtained with transmitting and reflecting light simultaneously. In the pictures, the blue parts indicate the optically-transparent gasket and yellow areas show reflection from the metal electrodes. Clearly, the sample is transparent at 9\,GPa, it starts darkening at about 11\,GPa and then becomes fully opaque at pressures between 30--40\,GPa. Hints of weak reflections, possibly pointing to the onset of metallicity, appeared at 79\,GPa. 
The reflection from the sample becomes comparable with that from the electrodes above 150\,GPa. 
  
\subsubsection{Raman measurements} 

The 632.8\,nm line of a He-Ne laser was used initially to probe the Raman spectra of hydrogen sulfide and to determine pressure. The Raman spectrometer is equipped with a nitrogen-cooled charge couple device and notch filters. The pressure was determined by a diamond edge scale as previously described~\cite{eremets2003megabar} at room and low temperatures. 
Ultra-low luminescence synthetic diamond anvils allowed us to record the Raman spectra at high pressures in the metallic state.
Fig.~\ref{fig:Raman} condenses a series of spectra taken at different conditions of pressure and temperature. 
For pressure above 50\,GPa, the characteristic molecular vibrations of hydrogen sulfide diminish considerably.  
Above this pressure, only a broadband at 1900\,cm$^{-1}$ remains. 
Vibron modes disappeared in D$_2$S at a higher pressure of about 100\,GPa. 
In general, the Raman spectra of hydrogen sulfide is complex and hard to analyze due to different factors: 
I) the decomposition in elemental phases occurs in some cases 
(vibration of sulfur appear at low frequencies (60-200\,cm$^{-1}$). 
One could also see that annealing at room temperature produced stronger sulfur peaks, while operating the sample close to 200\,K diminished the intensity of the peaks considerably. 
II) In contrast to the clear evidence of the presence of elemental sulfur, 
peaks belonging to elemental hydrogen (vibrons) were not observed despite the use of the ultra-low luminescence synthetic diamond anvils. 
Above 80\,GPa a new phase appeared, which persisted up to 150\,GPa in the metallic state. 
At 160\,GPa the Raman spectrum disappears, likely because of a transformation to the $\beta$-Po phase~\cite{DrozdovEremets_Nature2015}. In this figure (Fig.~\ref{fig:Raman}), the Raman spectrum of sulfur hydride is shown after the release of pressure from 208\,GPa at room temperature. 
The spectrum is much more intense than those of sulfur in the metallic
state at high pressures with non-confirmed evidence of phase transition at $\approx$180\,GPa.

\begin{figure}[t!]
  \centering    
  \includegraphics[width=1.0\columnwidth]{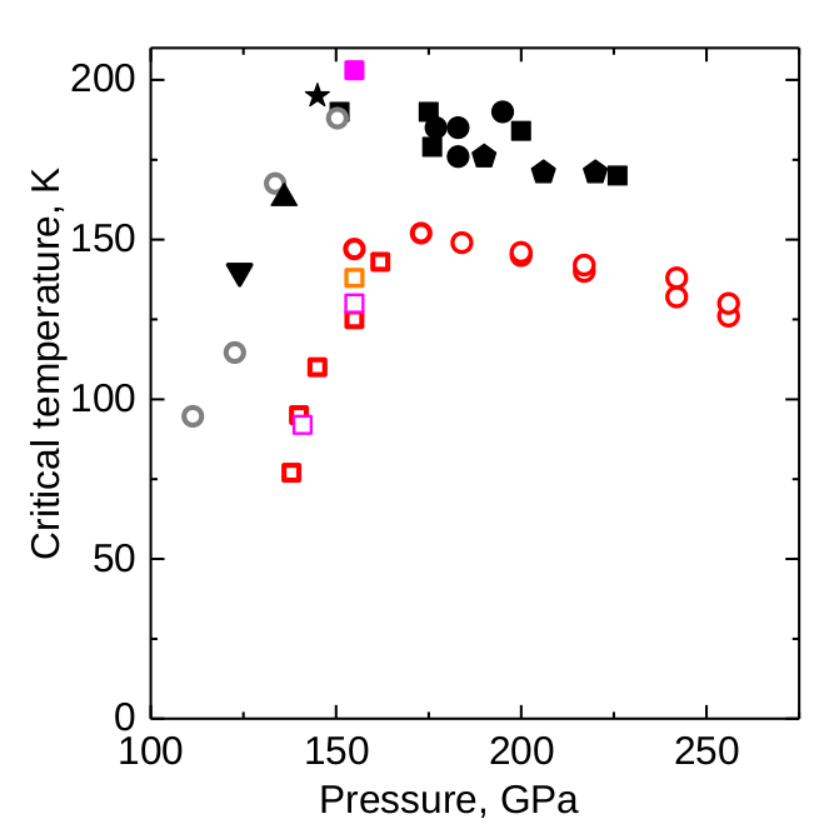}
  \caption{~Summary of critical temperatures of superconductivity upon pressure, 
  solid shapes (black and pink) shows for sulfur hydride (H$_x$S), and open shapes (except gray) represent sulfur deuteride (D$_x$S). Gray represent a different batch. Shown only data on annealed samples. 
  The highest measured \tc\ is 203\,K.}
  \label{fig:Tc_exp}
\end{figure}

\subsubsection{Electrical measurements} 

In the original experiment, the electrical measurement of resistivity was conducted using a four-probe Van der Pauw configuration with Tantalum electrodes sputtered on the diamond. A Keithley 6220 current source probed currents between 10 to 10,000\,$\mu$A and voltage was measured by a Keithley 2000 multimeter. Electrical measurements were also reproduced by the PPMS6000 Quantum Design physical property device. The insulating gasket was prepared from either Teflon, NaCl or CaSO$_4$ as these materials do not react with H$_2$S. To check a possible contribution of the diamond surface to the conductivity a configuration with a pair of electrodes on one diamond and another pair on the second diamond anvil was prepared, similarly to the technique reported in Ref.~\cite{Eremets_NatMat2011}. 
Temperature cycles were at a slow rate of $\approx$1\,K/min, which allowed the system to reach thermal equilibrium between measurements. 

All the samples started to conduct at pressures of $\approx$50\,GPa, in a semiconducting state indicated by the temperature dependence of resistance and by the presence of pronounced photoconductivity.
At 90--100\,GPa, the resistance drops further and the temperature dependence shows a metallic behaviour, 
confirmed by the absence of photoconductive response. 
It is noticeable that during the cooling of the sample for pressures above 100\,GPa the resistance dropped abruptly by three orders of magnitude indicating a transition to the superconducting state. 
A further increase in pressure resulted in consistently higher values of \tc.   

Fig.\ref{fig:Tc_exp} shows a summary of critical temperatures of superconductivity under pressure 
for sulfur hydride H$_x$S and sulfur deuteride (D$_x$S) samples. The highest measured \tc\ is 203\,K. 
A clear isotope effect, i.e. a shift of the \tc\ to lower temperatures in the case of sulfur 
deuteride points towards a phonon-assisted mechanism of superconductivity. 
The BCS theory gives a dependence of the critical temperature on the atomic mass $M$ as \tc\ $\propto$ $M^{\beta}$, 
for which experimentally we found $\beta=0.3$ at 175\,GPa. For further details see A. Drozdov's 
thesis work~\cite{Drozdov_thesis_2016} and Ref.\cite{DrozdovEremets_Nature2015}. 

\subsubsection{Measurements in magnetic field} 

In order to prove the actual occurrence of superconductivity, 
it is necessary but not sufficient to measure a drop in the electrical resistance. 
The isotope effect is also not conclusive because it may or not be present. 
However, to guarantee that superconductivity indeed occurs, the most crucial
signature which has to be present is the expulsion of the magnetic field (Meissner effect) below \tc. 

In the original work on H$_x$S, magnetic susceptibility 
measurements were performed using the MPMS Quantum Design setup. 
In order to carry out this type of complicated measurements, a specialized
miniature non-magnetic cell made of Cu:Ti alloy, working up to 200\,GPa, had to be designed 
(see Drozdov et al.~\cite{DrozdovEremets_Nature2015,Drozdov_thesis_2016}). 
The first series of experiments consisted on performing resistivity measurements upon a variable magnetic field, 
which confirmed the shift of \tc\ as a function of the magnetic field (roughly, a shift of 60\,K in 0--7\,Tesla). 
Furthermore, magnetic susceptibility was recorded as a function of temperature for sulfur hydride at a pressure of 155\,GPa in a the zero-field-cooled (ZFC) and 20\,Oe field cooled (FC) modes.

\begin{figure}[t!]
  \centering   % ok j 
  \includegraphics[width=1.0\columnwidth]{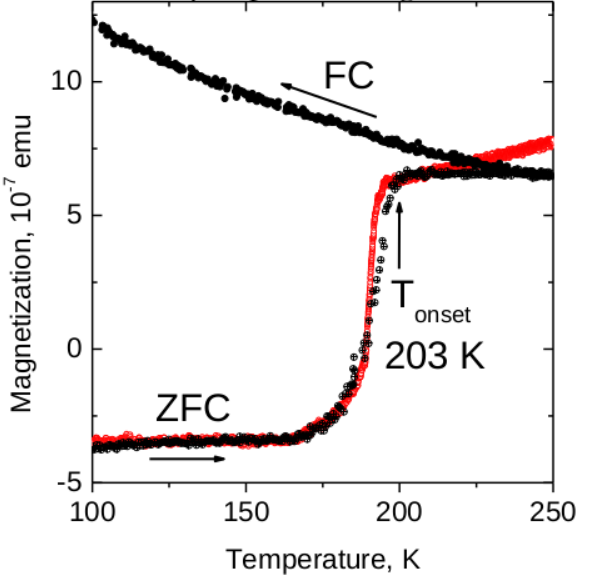}
  \caption{~Temperature dependence of the magnetization of sulfur hydride at 155\,GPa in zero-field cooled (ZFC) and 20\,Oe field cooled (FC) modes (black circles). The onset temperature is \tc = 203\,K. For comparison, the superconducting step obtained for sulfur hydride from electrical measurements at 145\,GPa is shown by red circles.}
  \label{fig:Tc_Chi}
\end{figure}

Fig.~\ref{fig:Tc_Chi} shows the temperature dependence of the magnetization of sulfur hydride at 155\,GPa in the zero-field-cooled (ZFC) and 20~Oe field cooled (FC) modes. The ZFC curve demonstrates a rather sharp transition from the diamagnetic to the paramagnetic state, which classified as a superconducting one. The onset temperature is \tc\ = 203\,K. For comparison, the superconducting step obtained for sulfur hydride from electrical measurements at 145\,GPa is depicted in red colour. Furthermore, the superconducting transition with \tc\ = 203\,K was confirmed by magnetization $M(H)$ measurements at different temperatures. The clear diamagnetism in the Meissner phase is responsible for the initial decrease of the magnetization until the first critical field H$_{c1}$ of 300\,Oe is attained. A subsequent increase of the magnetization as the external field increases (due to penetration of magnetic vortices) confirmed H$_3$S to be a Type-II superconductor. 
As the sign of the field change reverses, the magnetic flux in the Shubnikov state remains trapped, and the returning branch of the magnetic cycle runs above the direct one leading to the irreversibility of the magnetization curve. Hysteric behaviour of the magnetization becomes more clearly visible as temperature decreases.

Moreover, in-depth characterization of magnetization curves evidenced hysteresis, 
confirming a type-II superconductor (not shown). 
Magnetic hysteresis measurements were performed using subtraction of the background, 
determined in the normal state at $T=250$\,K. 
The magnetization curves, however, are distorted by possible paramagnetic impurities, 
often present in many other superconductors. 
We (M.E. et al.) reported the first critical field H$_{c1} \approx$ 300\,Oe, 
estimated roughly from where the magnetization deviates from the linear behaviour. 
At higher fields, magnetization increases due to the penetration of magnetic vortices. 
We have summarized the key points on the first experimental evidence of superconductivity 
in sulfur hydride and evidenced the record-breaking temperature of
203\,K, using both electrical resistance and magnetic measurements. 
However, as discussed in the following sections (Sec.\ref{Sec:Trends}) hydrogen sulfide possesses 
a rather complicated phase diagram on which different compositions and structure phases emerge. 
Decomposition from the original H$_2$S phase, loaded in our experiments, to H$_3$S is likely to occur upon pressure.
Experimental measurements of X-ray data by Einaga et al.~\cite{Einaga_H3S-crystal_NatPhys-2016} show that the superconducting phase is in good agreement with the theoretically predicted hexagonal and body-centred cubic ($bcc$) structures and coexists with elemental sulfur. 
Nevertheless, a phase mixture is also a possible scenario, since the precipitation of elemental sulfur upon decomposition could be expected~\cite{fujihisa2004molecular} and 
we (M.E. et al.) measured the \tc\ of sulfur at significantly lower temperatures ($<$\,20\,K). 
Another expected product of the decomposition of H$_2$S is hydrogen. 
However, the body of experimental evidence does not support this scenario, since the steady characteristic vibrational 
stretching mode of the H$_2$ molecule is absent in Raman measurements. A complicated path of transformation is thus likely to happen to account for hydrogen-rich compositions. 
The key features of this superconductor and other hydrogen-rich compositions will be addressed in detail with the aid of theoretical and computation methods in Sec.~\ref{Sec:Trends} and Sec.~\ref{Sec:Perspectives}.

\begin{figure}[t!]
  \centering    
  \includegraphics[width=1.0\columnwidth]{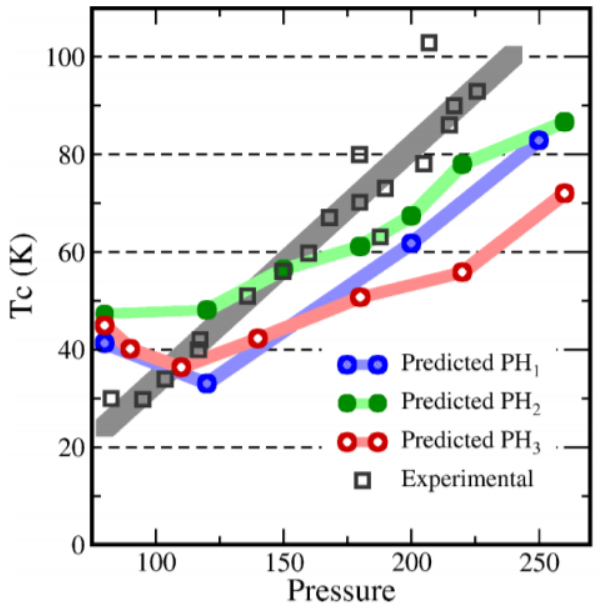}
  \caption{~SCDFT calculated critical temperatures \tc for P-H composition as 
  function of pressure as presented in Ref.~\cite{flores-sanna_PH3_2016}. 
  Experimental values of \tc extracted from Drozdov et al.~\cite{Drozdov_ph3_arxiv2015} are shown in black squares.}
  \label{fig:PH}
\end{figure}

\subsection{Superconductivity in P-H and La-H Hydrides} 

Sulfur-hydride is not an isolated example of conventional high-\tc\ superconductivity at high pressures~\cite{Ivan_Science-H2S-2016,Einaga_H3S-crystal_NatPhys-2016,Errea_anha-SH3_PRL2015,Heil-Boeri_PRB2015,flores-sanna_HSe_2016,flores_accelerated_2017,akashi_mangeli-phases}. 
Less than a year after the discovery of superconductivity in 
sulfur hydride~\cite{DrozdovEremets_Nature2015}, Drozdov and coworkers have reported high-\tc superconductivity 
in a second hydrogen-rich compound at extreme pressures: resistivity measurements on phosphine (PH$_3$) 
show that the samples, which are semiconducting at ambient pressure, metallize above 40\,GPa and become superconducting at around 80\,GPa, exhibiting a maximum \tc\ of 100\,K at about 200\,GPa~\cite{Drozdov_ph3_arxiv2015}. 
In the original publication, the composition responsible for the superconducting phase and 
its crystal structure, nor the mechanism responsible for \tc\ were addressed.

Analogies with the superconducting sulfur-hydride, obtained from a H$_2$S precursor, suggest that the superconductivity in the P-H system is conventional, but that the composition of the superconducting phase might be different from the original
PH$_3$ stoichiometry. In order to shed light on this matter, some of us used {\it ab initio} techniques to map out the high-pressure phase diagram of the P-H binary system by exploring the compositional 
and configurational space of PH$_n$ with structure prediction methods, and estimated the superconducting properties of the most promising phases (summarized in Fig.~\ref{fig:PH}).~\cite{flores-sanna_PH3_2016}). 
Interestingly, our results, as well as similar theoretical studies, reported that all the high-pressure binary phases of P and H as metastable to elemental decomposition in the pressure range 100--300\,GPa. 
However, the critical temperatures of the three phases closest to the convex hull (PH, PH$_2$, and PH$_3$) reproduce to a good approximation the experimental values. Possible ways to reconcile our results with experiments are addressed in Sec.~\ref{Sec:Trends}.

The finding of superconductivity in the H-S system at 200\,K under pressure and, later, in P-H, 
stimulated a large part of the community to explore other systems.
As we will discuss later in Sec.~\ref{Sec:Trends} and Sec.~\ref{Sec:Perspectives}, 
an entirely novel family emerges, {\bf superhydrides}, 
for hydrogenic compositions roughly higher than 6, i.e. XH$_{(\geq 6)}$. 
These structures leave the ``low" number of bonds 
(as in H$_3$S and PH$_x$) and form a highly symmetric motif identified as cages. 
A substantial amount of computational studies have been published and in some of them 
spectacular values close or above room temperature have been 
predicted~\cite{Ca_PNAS_2012superconductive,Clathrate_REHX_PRL_2017,PNAS_LaHx_2017_Hemley}. 
Many other systems have been theoretically proposed to form superhydrides, 
reviving also the interest in structural transformations of individual elements and their tendency to form metallic structures~\cite{struzhkin2002superconductivity,Li_PNAS2010,Shimizu_EHP2005,flores_interplay_2017}.

\begin{figure}[ht]
  \centering    % 
  \includegraphics[width=0.9\columnwidth]{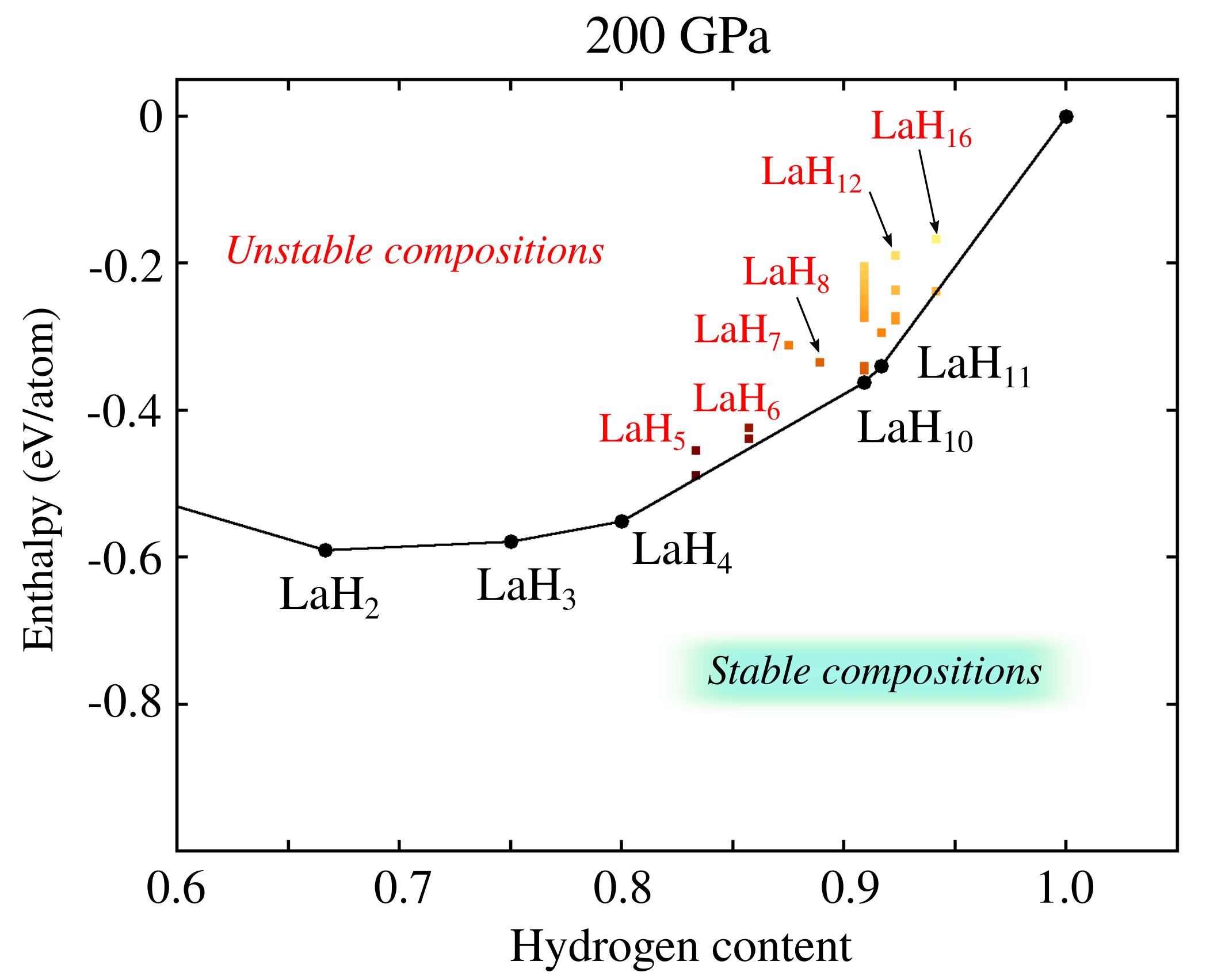}
  \caption{~Computed phase diagram of La-H compositions under pressure at 200\,GPa (see Ref.~\cite{work_on_LaHx_2019}).
  For low pressures, the LaH$_{10}$ composition is not stable (not shown) and only develops as a stable point in the classical convex hull of enthalpy formation above 175\,GPa. 
  Interestingly many other phases are stable points and could be accessible by controlling thermodynamic conditions.}
  \label{fig:LaH}
\end{figure}

In 2018, the work on H-rich compositions in 
the La-H system set a new superconducting record, representing the first measurement of \tc\ near room temperature. 
{\it Ab initio} techniques have been vital to predict the initial phase  diagram~\cite{Clathrate_REHX_PRL_2017,PNAS_LaHx_2017_Hemley}. 
Fig.~\ref{fig:LaH} shows the calculated convex hull of formation enthalpy without zero-point energy at 200\,GPa. In this figure, one can see that there are many stable compositions in the convex hull. 
An exciting composition, predicted to be responsible for the nearly room-temperature superconductivity, 
is LaH$_{10}$~\cite{PNAS_LaHx_2017_Hemley,Clathrate_REHX_PRL_2017}, 
which becomes enthalpically stable for pressure above 175\,GPa and 
remains on the convex hull well above 300\,GPa, 
according to Errea--Flores-Livas et al.~\cite{work_on_LaHx_2019}. 

This first lanthanum superhydride was synthesized under pressure above 160\,GPa and upon heating to 1000\,K. 
The X-ray data indicated that the stoichiometry corresponds to LaH$_{10+x}$ (-1$<x<$2) 
which is close to the predicted high-\tc\ LaH$_{10}$~\cite{PNAS_LaHx_2017_Hemley,Clathrate_REHX_PRL_2017}. 
In 2018, Somayazulu et al.~\cite{Hemley-LaH10_PRL_2019} measured the temperature dependence of the resistivity 
of La heated with NH$_3$BH$_3$ as hydrogen source under similar pressure and observed a drop in the
resistance at $\approx$260\,K upon cooling and 248\,K upon warming the sample, which was assigned
to the superconducting transition of LaH$_{10+x}$ ($-1 < x < 2$). 
In their original claim, they also observed a series of resistance anomalies at
temperatures as high as 280\,K. However, neither a zero-resistance state nor additional confirmations
(such as the Meissner effect or isotope effects or the effect of an external magnetic field on the transition
temperature) were provided. Simultaneously, an independent research measured a superconducting transition \tc\ $\approx$\,215\,K in LaH$_x$~\cite{drozdov2018_215} and 
subsequently a \tc\ of 250\,K~\cite{drozdov2018_250,Nature_LaH_Eremets_2019}. 

Apart from setting a new \tc\ record, high-pressure experiments in the La-H system played a 
significant role also in establishing methods to control the high-pressure synthesis of materials.
This is a crucial point since many --if not all-- of the structures observed in high-pressure 
experiments are metastable phases~\cite{flores_thesis_2012} (or near thermodynamic equilibrium), and it 
represents one of the most significant challenges when using high pressures 
in practical applications~\cite{grochala2007chemical}. Different strategies to help
the stabilization of HP-materials have been proposed~\cite{hot_Ma_view_point_PRL2019}. 
A major role is played by the precise control of thermodynamic conditions~\cite{nguyen2006high,degtyareva_formation_2009,Li_PNAS2010,flores_interplay_2017,Strobel_PRL2011,levitas2018high}  and the case of La-H represents a step towards mastering it. 
For each of the routes followed by Geballe el al.~\cite{geballe2018synthesis}, 
Somayazulu et al.~\cite{Hemley-LaH10_PRL_2019} and Drozdov et al.~\cite{drozdov2018_215,drozdov2018_250}, 
the factor that played a crucial role in ensuring a successful synthesis was temperature. 
The stable compositions measured by experiments were accessed using pulsed laser heating, 
starting from different initial constituents. In Geballe and Somayazulu work's ammonia 
borane was used as the hydrogen source and elemental lanthanum as a starting
material, while Drozdov et al. used (LaH$_2$ + H$_2$) as source materials. 
In both experiments, and independently of the P-T path followed, the elevated 
temperature was necessary to access compositions having the highest \tc.

So far experimentally, eight remarkable results on superconducting 
hydrides at very high pressure have been confirmed to occur at the time of writing this Review (ordered by date):  
\begin{center}
\setlength{\tabcolsep}{3pt}
\begin{tabular}{r l l l l}
I\,)&    SiH$_x$,   & \tc\ $\approx$ 17\,K & (2008)&\cite{Eremets_Science2008}. \\
II\,)&   BaReH$_9$, & \tc\ $\approx$  7\,K & (2014)&\cite{Maramatsu-Hemley_2015}. \\
III\,)&  H$_3$S,    & \tc\ $\approx$200\,K & (2015)&\cite{DrozdovEremets_Nature2015,mozaffari2019superconducting}.  \\
IV\,)&   PH$_3$,    & \tc\ $\approx$100\,K & (2015)&\cite{Drozdov_ph3_arxiv2015}. \\
V\,)&    LaH$_{10}$,& \tc\ $\approx$250\,K & (2018)&\cite{drozdov2018_250,Hemley-LaH10_PRL_2019}.\\
VI\,)&   ThH$_{10}$,& \tc\ $\approx$160\,K & (2019)&\cite{Semenok_arxiv_ThH10_2019}.\\ 
VII\,)&  YH$_6$,    & \tc\ $\approx$220\,K & (2019)&\cite{Troyan_arxiv_YH6_2019}.\\
VIII\,)& YH$_9$,    & \tc\ $\approx$240\,K & (2019)&\cite{Kong_arxiv_YH9_2019}.\\
\end{tabular}\setlength{\tabcolsep}{6pt} 
\end{center}
For compressed hydrides not showing superconductivity, see Sec.~\ref{Sec:Trends_Nosupra}. 
We conclude this section of the Review, with the firm belief that in the 
coming years, we will witness measurements of even higher \tc, 
at room temperature (300\,K) or possibly above. 
In fact, theory (Migdal-\'Eliashberg)~\cite{Eliashberg,Margine_anisoEliashberg_PRB2013,sanna-flores_2018_Eliashberg}) 
predicts no upper limit for the critical temperature~\cite{Carbotte_RMP1990}. 

%%%%%%%%%%%%%%%%%%%%%%%%%%%%%%%%%%%%%%%%%%%%%%%%%%%%%%%%%%%%%%%%%%%%%%%%%%%%> muy bonito!  

\section{Theoretical Methods for superconductivity}\label{Sec:Theory}
\subsection{Phonons and electron phonon coupling}\label{sec:DFPT}

    \subsubsection{Phonons}\label{subsec:phonons-LR} 

The underlying assumption which allows the calculation of vibrational 
properties in solids is the {\it adiabatic} approximation 
of Born-Oppenheimer (BO)~\cite{born1927quantentheorie}. 
By decoupling the vibrational from the electronic degrees of freedom, 
the lattice-dynamical properties of a system are determined by 
the eigenvalues $\varepsilon$ and eigenfunctions $\Phi$ of the Schr\"odinger equation:
\begin{equation}
\left[-\sum_i \frac{\hbar^2}{2M_i}\nabla^2_i + E({\rm\bf \underline R}) \right]\Phi({\rm\bf \underline R}) =\varepsilon\Phi({\rm\bf \underline R})
\end{equation}
where ${\rm\bf \underline R}\equiv$\{${\rm\bf R_1},..,{\rm\bf R_n}$\} 
are the nuclear coordinates, $M_i$ are nuclear masses, and 
$E({\rm\bf \underline R}$) is the energy of the whole system at fixed ion positions, 
which defines the Born-Oppenheimer energy surface~\cite{born1954dynamical}. 
In particular, $E({\rm\bf \underline R}$) is the ground state eigenvalue 
of the electronic Hamiltonian:
\begin{equation}\label{eq:BO-KS-normal}
H_{{\rm BO}}({\rm\bf \underline R})=\frac{-\hbar^2}{2m}\sum_i\nabla^2_i + \frac{e^2}{2}\sum_{i\neq j}\frac{1}{\left| {\rm\bf r}_i - {\rm\bf  r}_j\right|} + V_{\bf \underline R}({\rm\bf \underline r}) + E_N({\rm\bf \underline R})
\end{equation}
with  
\begin{equation}V_{\bf \underline R}({\bf \underline r})= - \sum_{i,h}\frac{Z_he^2}{\left| {\rm\bf r}_i - {\rm\bf  R}_h\right|}\end{equation}
\begin{equation}E_N({\bf \underline R})  = \frac{e^2}{2}\sum_{i\neq j}\frac{Z_iZ_j}{|{\rm\bf R}_i- {\rm\bf R}_j|}\end{equation}
where  ${\rm\bf \underline r}\equiv$\{${\rm\bf r}_1,..,{ {\rm\bf r}_n}$\} 
are the electronic coordinates, {$Z_h$} are the atomic numbers of the ion, 
$E_N$ the bare nuclear electrostatic energy and $V_{\bf \underline R}$ the electron-nuclear interaction. 
The BO approximation neglects possible electronic transitions induced by the ionic motion, and more importantly, the nuclear interaction is assumed to be instantaneous.  
The equilibrium geometry of the system is given by the condition 
that the forces acting on individual nuclei vanish 
(the first derivative of $E({\rm\bf \underline R})$ is zero).  
The diagonalization of the Hessian matrix of $E({\rm\bf \underline R})$
gives the vibrational frequencies in the harmonic approximation, according to the relation  \begin{equation}\label{eq:phonons1}
{\rm det}\left|\frac{1}{\sqrt{M_iM_j}}\frac{\partial^2E({\rm\bf \underline R})}{\partial{{\rm\bf R}}_i\partial{{\rm\bf R}}_j} -\omega^2 \right|=0.
\end{equation}  
That in terms of periodic displacements from equilibrium can
be rewritten~\cite{Baroni_DFPT_RMP2001} as: 
\begin{equation}
{\rm det}\left|\frac{1}{\sqrt{M_iM_j}}\tilde C_{ij}^{\alpha\beta}({\rm\bf q}) -\omega^2({\rm\bf q}) \right|=0,
\end{equation}
where 
\begin{equation}\label{eq:dynamical_matrix_q}
\tilde C_{ij}^{\alpha\beta}({\rm\bf q})=\frac{\partial^2E}{\partial u_i^{*\alpha}({\rm\bf q})\partial u_j^{\beta}({\rm\bf q})},
\end{equation}
is the interatomic force constant (IFC)~\cite{Baroni_DFPT_RMP2001} 
in momentum space, and where $u_i^{\alpha}({\rm\bf q})$ 
is the normalized displacement of the atom $i$ and wavevector ${\bf q}$. 
$\alpha$ and $\beta$ are the Cartesian components of the displacement.
Phonon eigenvalues and eigenvectors are obtained by diagonalizing 
$\tilde C_{ij}^{\alpha\beta}({\rm\bf q})/{\sqrt{M_iM_j}}$, 
also called the dynamical matrix. 

\subsubsection{Density Functional Theory}\index{DFT}\label{sec:DFT}
 
Due to its excellent cost/accuracy ratio, 
density functional theory (DFT) is the method of choice 
for the calculations of electronic and lattice properties of materials~\cite{DreizlerGross_DFT1990}. 
In particular, it is the universally used approach to compute, from first principles, 
the dynamical matrix in Eq.~\ref{eq:dynamical_matrix_q}. 
DFT is most conveniently employed in practice using the Kohn and Sham (KS) 
construct~\cite{KohnSham_PR1965}\index{Kohn-Sham system!DFT}, i.e. introducing
an auxiliary system of non-interacting electrons which reproduces exactly the
ground-state density of the fully interacting system. 
The Kohn-Sham system is obtained from the solution of the Schr\"odinger equation: 
\begin{equation}\label{eq:KSscf}
H_{s}\psi_n({\rm\bf r})\equiv-\left(\frac{\hbar^2}{2m}\nabla^2 + v_s({\rm\bf r})\right)\psi_n({\rm\bf r})=\epsilon_n\psi_n({\rm\bf r})
\end{equation}
where the effective potential $v_s$ is to be computed self-consistently and reads:
\begin{equation}\label{eq:Vscf}
v_s({\rm\bf r})=v_{ext}({\rm\bf r}) + e^2\int\frac{\rho({\rm\bf r}')}{\left|{\rm\bf r}-{\rm\bf r}'\right|} + v_{xc}({\rm\bf r})
\end{equation}
where $v_{ext}({\rm\bf r})$ is the potential due to the bare nuclei, 
the second term on the right-hand side is the Hartree potential, and $v_{xc}$ is the functional derivative of $E_{xc}[\rho]$ with respect to
the electronic density. $E_{xc}[\rho]$ is a universal (not material-dependent) functional of the density;
it is  a central object in DFT, for which many good approximations  exist~\cite{PBE_PRL1996,PerdewZunger_LDA_PRB1981,SunRuzsinskyPerdew_SCAN_PRL2015}. 
The electronic density is obtained by occupying the orbitals that are solutions of Eq.~\ref{eq:KSscf}, $\psi_n({\rm\bf r})$ with the lowest 
eigenvalues ($\epsilon_n$) according to the Pauli principle:
\begin{equation}\label{eq:KSdensity}
\rho({\rm\bf r})=2\sum_{n=1}^{N/2}\left|\psi_n({\rm\bf r})\right|^2,
\end{equation}

For a metallic system the occupation of the states can be smeared around the Fermi level, with a distribution function $S_{\sigma}\left(\epsilon\right)$: \begin{equation}
\rho({\rm\bf r},\epsilon)=\sum_n S_{\sigma}\left(\epsilon\right)\left|\psi_n({\rm\bf r})\right|^2.
\end{equation}
Among other functions, $S$ can be chosen as the Fermi-Dirac distribution with a fictitious electronic temperature that helps to stabilize numerical calculations.  
Using DFT in this scheme, the total energy of a given system is expressed a the sum of terms: 
\begin{eqnarray}\label{eq:DFT_energy}
E[\rho]&=&2\sum_n S_{\sigma}\left(\epsilon\right)\epsilon_n-\frac{e^2}{2}\int\frac{\rho({\rm\bf r})\rho({\rm\bf r}')}{\left|{\rm\bf r}-{\rm\bf r}'\right|}d{\rm\bf r}d{\rm\bf r}' \nonumber\\ &+& E_{xc}[\rho] -\int \rho({\rm\bf r})v_{xc}({\rm\bf r})d{\rm\bf r}. 
\end{eqnarray}
This equation allows us to directly compute the dynamical matrix and the phonon spectrum by performing numerical derivatives through small finite displacements of the atoms. 
Alternatively, assuming small amplitudes, phonons can be studied as infinitesimal perturbations, computing Eq.~\ref{eq:dynamical_matrix_q} in the linear response regime, using density functional perturbation theory (DFPT) applied to the phononic perturbation~\cite{Zein-1984,Baroni_DFPT_RMP2001,Baroni_LinearResponse_PRL1987,DeGironoli_DFPTmetals_PRB1995}. A great advantage of this method consists in its capability to treat perturbation of whatever wavelength without the need to simulate the lattice distortion on a unit cell commensurate with the $\bf{q}$-vector of the phonon.

\subsubsection{Electron-phonon interaction}\label{sec:elphME}

So far, we have treated phonons only within the adiabatic approximation introduced 
with the Born-Oppenheimer approach, neglecting any electron-phonon (\ep) scattering process and assuming that electrons respond instantaneously to the ionic motion. Due to the small phonon energies (10\,--\,100 meV) 
as compared to typical electronic energies (1\,--10\,eV), in metals \ep\ processes involve only a fraction of electronic states, within a small (of the order of the Debye energy) window around the Fermi energy $E_F$.  
Since the phonon frequencies depend on the  whole valence charge density, this argument easily explains why highly accurate phonon frequencies are obtained from the unperturbed electronic structure~\cite{DeGironoli_DFPTmetals_PRB1995}.  
On the other hand, many important properties of metals, such as the normal state resistivity or superconductivity, are strongly affected by \ep\ interactions. Both these properties arise from elemental processes in which an electron with momentum ${\rm\bf k}$ emits (absorbs) a phonon of wavevector ${\rm\bf q}$ and goes in a state of wavevector ${\rm\bf k}+{\rm\bf q}$. 
To develop a theory for superconductivity, we need to estimate a coupling factor for this process. 

The idea for determining this coupling factor is to consider the phonon, dressed with the electronic charge that adiabatically follows, 
as a perturbing potential that introduces a transition probability between the KS eigenstates. 
This potential takes the form:
\begin{equation}
\Delta V = \sum_i \eta_i \frac{\partial v_{s}({\bf r})}{\partial \eta_i}
\end{equation}
where $\eta_i$ is the position operator for the $i$-th atom. 
Transforming to phonon coordinates:
\begin{equation}\label{eq:ph-amplitude}
\eta_i=\sum_{\bf q}\sqrt{\frac{\hbar}{2M_i\omega_{{\bf q}\nu}}} e^{i{\bf q}\cdot{\bf R^i_0}}\left(b^{\dagger}_{{\bf q}\nu} - b_{{\bf q}\nu}\right) 
\end{equation}
where ${\bf R}^i_0$ are the unperturbed lattice positions 
and $b^{\dagger}_{{\bf q}\nu},b_{{\bf q}\nu}$ are the phonon raising and lowering operators. 
If we calculate the expectation value of the operator above for a process in which 
a single phonon is created or absorbed, we get a coupling matrix element:  
\begin{equation}\label{eq:elphME}
g_{m{\bf k}+{\bf q},n{\bf k}}^{\nu}=\sqrt{\frac{\hbar}{2\omega_{{\bf q}\nu}}}\left<\psi_{m{\bf k}+{\bf q}}|\Delta v_{s}^{{\bf q}\nu} e^{i{\bf q}\cdot{\bf r}}|\psi_{n{\bf k}} \right>,
\end{equation}
where $\Delta v_{s}^{{\bf q}\nu}e^{i{\bf q}\cdot{\bf r}}$ is the finite variation in the self consistent potential corresponding to a phonon displacement of wavevector ${\bf q}$ and mode index $\nu$.  
 The phase factor $e^{i{\bf q}\cdot{\bf r}}$ cancels out with 
 a corresponding one stemming from the Bloch wavefunction. 
The $\Delta v_{s}^{{\bf q}\nu}$ is lattice periodic and can be evaluated within the periodic unit cell. 
The \ep\ interaction for the Kohn-Sham system is described by the Hamiltonian:
\begin{eqnarray}
~\hspace{-0.8cm}\tilde H_{e-ph}\!\!&\!\!=\!\!&\!\!\sum_{mn\sigma}\sum_{{\vect k}{\vect q}\nu}g_{m{\vect k}+{\vect q},n{\vect k}}^{\nu}\sum_{\sigma}\psi^{\dagger}_{\sigma m{\vect k}+{\vect q}}\psi_{\sigma n{\vect k}}\left(b_{{\vect q}\nu}+b^{\dagger}_{-{\vect q}\nu}\right)  \\~\hspace{-0.8cm}\!\!&\!\!=\!\!&\!\!\sum_{{\vect q}\nu}\sqrt{\frac{\hbar}{2\omega_{{\vect q}\nu}}}\int d^3r\Delta V_{scf}^{{\vect q}\nu}\left({\vect r}\right)\psi^{\dagger}_{\sigma}\left({\vect r}\right)\psi_{\sigma}\left({\vect r}\right)\left(b_{{\vect q}\nu}+b^{\dagger}_{-{\vect q}\nu}\right)\nonumber
\end{eqnarray}
where $\psi^{\dagger}_{\sigma n{\vect k}}$ and $\psi_{\sigma n{\vect k}}$ are creation and destruction operators for Kohn-Sham states. This form of the \ep\ interaction is by construction an approximation of the true \ep\ interaction. 
It is valid while the spectrum of the Kohn-Sham system remains 
a good approximation to the real quasi-particle spectrum of the material. 
Whenever this is not the case, one has to rely on perturbation approaches~\cite{MariniPonceGonze_MBPTandElectronPhonon_PRB2015}.

\subsubsection{Anharmonic effects}

The harmonic theory of lattice vibrations discussed so far is based on the second-order  
expansion of the BO energy surface~\cite{born1954dynamical} 
around the ionic equilibrium positions~\cite{Born1955}. 
The harmonic approximation assumes implicitly that the amplitude of atomic displacements from equilibrium are relatively small, 
and describes lattice vibrations as excitations of 
non-interacting quasiparticles (phonons), with an infinite lifetime and a 
temperature-independent energy spectrum. 
The harmonic theory has proven to be exceptionally good 
in describing the vibrational properties of most solids. 
However, important cases are known where this approximation fails. 
Anharmonic effects, caused by higher-order terms in the expansion of the energy surface, 
introduce interactions between phonons, and thus finite scattering rates and finite lifetimes~\cite{Ashcroft_book_yes,ziman2001electrons}. 
Systems containing light mass atoms, like hydrides under pressure, exhibit intrinsically large vibrational displacements and hence show-case a variety of effects due to strong anharmonicity, including
phonon softening, finite linewidths, anharmonic stabilization,
among others. 

The most intuitive approach to include anharmonic corrections, which
was also the first to be introduced historically, is to consider higher-order terms in the potential expansion as a small perturbation of the harmonic potential~\cite{maradudin1962scattering,koehler1966theory}. 
Perturbative approaches, however, are restricted to conditions in which the displacements of the atoms are small enough that higher-order terms of the potential are considerably smaller than the harmonic one. Unfortunately, there are situations in which perturbative approaches are not suitable, for example in the presence of lattice instabilities, which may be caused by the proximity to a structural transition.

\begin{figure}
  \centering  
  \includegraphics[width=1.0\columnwidth]{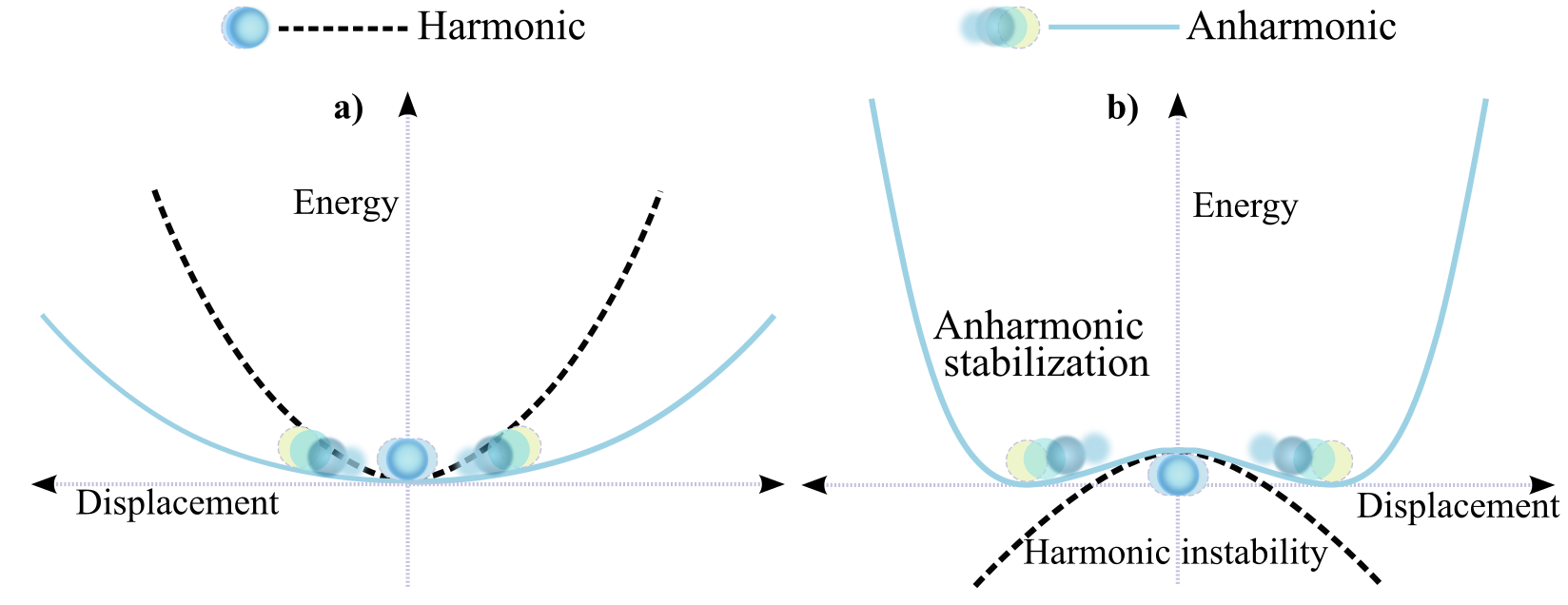}
  \caption{~Possible scenarios of a) weak and b) strong anharmonicity. 
  Harmonic vibration in the left panel leads to substantial softening and wrong determination of phonon frequencies. In the right panel, the harmonic potential is unstable (imaginary frequencies) and stabilises only via anharmonicity.}
  \label{fig:ana_V}
\end{figure}

An example of two different regimes of anharmonicity is depicted in Fig.~\ref{fig:ana_V}. In the first case (left plot), the anharmonic effect occurs within the minimum of the harmonic solution. A situation that is conceptually much simpler, as anharmonic corrections only amount to a renormalization of the phonon frequency. This scenario is likely to be accurately captured by a perturbative expansion. 
In the second case (right plot), the energy versus displacement profile has a shallow double minimum (non-perturbative regime). In this scenario, the harmonic approximation brakes down and the lattice would drift into a distorted configuration corresponding to one of the two minima. 
The anharmonic correction stabilizes the harmonically unstable phonons 
when the thermal amplitude or zero-point amplitude is sufficiently large. 
In high-pressure hydrides, the second scenario is often present. 

Several approaches have been developed to deal with the  problem of strong anharmonicity and are classified into two broad categories: those that rely on molecular dynamics (MD), and those that rely on the self-consistent field theory. 
Anharmonic effects at a non-perturbative level are treated within {\it ab initio} molecular dynamics (AIMD)
methods~\cite{wang1990tight,hellman2011lattice,hellman2013_PRB-1,hellman2013_PRB-2}. 
These approaches are computationally expensive 
since they require long simulation times to converge renormalized phonon energies. 
Another intrinsic limitation is that they are based on Newtonian dynamics, 
limiting their application to temperatures above the Debye temperature. 
Path-integral molecular dynamics~\cite{ceperley1995path} overcome this situation by incorporating 
the quantum character of atomic vibrations, however at cost of increasing the computational load.  
To overcome the aforementioned limitations in AIMD methods, 
several DFT-based methods have recently been developed for the non-perturbative treatment of
anharmonic effects in solids. These methods rely on the vibrational self-consistent-field 
theory~\cite{bowman1978self,monserrat2013anharmonic} and 
the self-consistent phonon (SCP) theory
~\cite{hooton1955lviii,hooton1958_book,Werthamer_SCHA_PRB1970,koehler1966theory,CochranCowley_phononsInPerfectCrystals_Book1967}, and can incorporate the effect of lattice anharmonicity at the mean-field level. 
In these methods, anharmonic phonon frequencies are calculated through 
the construction of effective harmonic force constants obtained self-consistently 
by repeatedly calculating atomic forces in supercells with suitably chosen atomic configurations. 
In both variants, perturbative expansions and self-consistent field approaches, 
anharmonic corrections can be computed by a diagrammatic expansion~\cite{Cardona_RamanAnharmonic_PRB1984,souvatzis2008entropy,SCHA-method_PRB_Errea_2014,Terumasa_PRB_2015}. For example, in the self-consistent harmonic approximation the leading contributions can be sketched diagrammatically as:  
  \begin{center}
  \includegraphics[width=1.0\columnwidth]{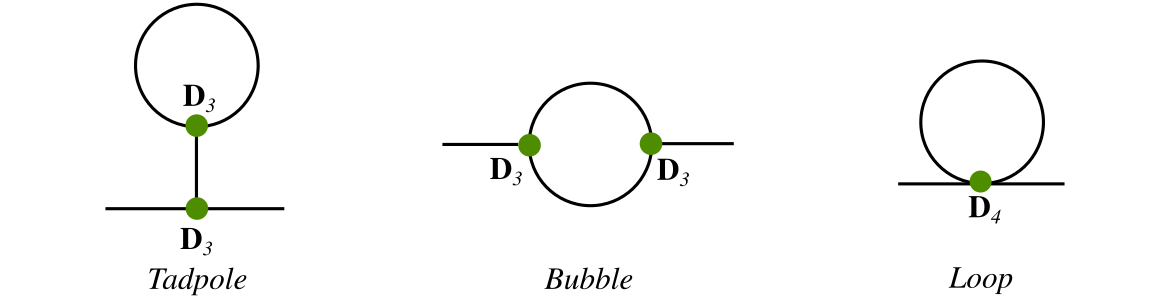}
  \end{center}
Lines represent the (dressed) phonon propagator and points are different types of interaction vertices. 
This perturbative approach has the undeniable beauty that anharmonic effects are seen and connected to a clear physical picture of interacting phonons and phonon decay. 
The first term (tadpole, three-phonon) is a Hartree-type diagram in which phonons are coupled with the average phononic distortion. 
The term becomes identically zero for optical modes whenever Wyckoff positions are locked by symmetry~\cite{MaradudinFein_ScatternigAnharmonicCrystal_PR1962,LazzeriCalandraMauri_AnharmonicRamanMgB2_PRB2003}, while for the coupling with acoustic modes gives the thermal expansion. 
The last two terms (bubble and loop) describe respectively, 
three-phonon and four-phonon correlated interaction. All terms contribute to the real part of 
the phonon self-energy and hence give rise to a shift of the phonon frequency, 
but only the bubble diagram, which is non-hermitian, gives a finite contribution to the phonon linewidth
~\cite{LazzeriCalandraMauri_AnharmonicRamanMgB2_PRB2003,CochranCowley_phononsInPerfectCrystals_Book1967,Cardona_RamanAnharmonic_PRB1984}.

An excellent approach to deal with all regimes of anharmonicity, developed by Hooton in 1955~\cite{hooton1955lviii}, 
is the self-consistent harmonic approximation (SCHA). 
The theory starts from the true anharmonic free energy: 
\begin{equation}
\mathcal{F}_H\left[\rho\right]={\rm Tr}\left[\rho H\right] +\frac{1}{\beta}{\rm Tr}\left[\rho{\rm ln}\rho\right]
\end{equation}
with $\beta$ the inverse temperature, $H$ the ionic Hamiltonian and $\rho$ the density matrix.  
The fundamental idea behind the SCHA is to use a variational principle, the Gibbs-Bogoliubov (GB) principle, 
in order to approximate the free energy of the true ionic Hamiltonian with the free energy calculated 
with a trial harmonic density matrix for the same system. 
According to the Gibbs principle $\mathcal{F}_H\left[\rho\right]$ is minimized by the equilibrium density matrix $\rho_H$ such that:
\begin{equation}
\mathcal{F}_H\left[\rho\right]\geq \mathcal{F}_H\left[\rho_H\right] \label{eq:trial_F}
\end{equation}
The above minimization is performed restricting the search for $\rho$ to a set of density matrices generated by the support harmonic Hamiltonian ($H_{Harm}$). That is particularly convenient because this auxiliary system consists of non-interacting phonons:
\begin{equation}\label{eq:Hscha}
H_{Harm}=E_0\left[{\bf x}^0\right] +\frac{1}{2}\sum_s^{3N} {\bf p}^2_s +\frac{1}{2}\sum_s^{3N} \omega^2_s {\bf q}^2_s
\end{equation}
where ${\bf x}^0$,  ${\bf p}_s$,  $\omega_s$ and  ${\bf q}_s$ are the vibration centers, momenta, frequencies and coordinates of the modes ($s$) to be variationally optimized.
The auxiliary system reduces to the original harmonic one in the absence of anharmonicity, but in general, it may have different frequencies, eigenmodes of vibration and centres of vibration from the harmonic one. However, the computational cost of the SCHA is relatively high because at each self-consistent iteration expensive high-order derivatives of the energy surface are required in the minimization over $\rho$.   
This problem has been recently tackled by 
Errea and coworkers who introduced a stochastic approach to sampling the energy surface~\cite{Errea_PdH_PRL2013,SCHA-method_PRB_Errea_2014,Bianco_PRB2017,Monacelli_PRB2018}. 

In the stochastic self-consistent harmonic approximation (SSCHA), 
the free energy is calculated from a set of (supercell) configurations. 
These are chosen stochastically from the probability distribution defined by the trial Hamiltonian (Eq.~\ref{eq:Hscha}). 
The minimization problem with respect to the coefficients of the harmonic trial potential is solved with an efficient conjugate gradient algorithm. In the minimization, new stochastic configurations may be added to improve the sampling. 
However, the use of an ingenious re-weighting procedure allows 
re-using all previously generated supercell calculations, tremendously saving on computational cost. 
The SSCHA is non-perturbative as it samples the system even far 
from the harmonic minimum (right plot in Fig.~\ref{fig:ana_V}) 
and, owing to its excellent system-size scaling properties, 
it is applied in conjunction with {\it ab initio} calculations to study complex systems as hydrides. 

Another efficient implementation, similar to SSCHA, based on SCP theory~\cite{Werthamer_SCHA_PRB1970} to treat anharmonicity from first principles is the (deterministic) algorithm ({\sc alamode} code) developed by Tadano and Tsuneyuki~\cite{Terumasa_PRB_2015}. 
Both methods minimize the trial free energy (Eq.~\ref{eq:trial_F}). 
Both are highly successful approaches, however, differ in technical details. For instance, SSCHA samples anharmonicity by systematically displacing atoms in a supercell, while the latter one uses inputs of anharmonic constants (usually up to fourth-order) 
computed efficiently~\cite{Terumasa_PRL_2018}  adopting a modern compressed sensing approach~\cite{CSLD_2014_ozolins}. 
This second algorithm is very efficient at calculating anharmonic phonon at various temperatures. 
While SSCHA, computationally slightly demanding in comparison, is highly accurate since it does not truncate higher orders anharmonic terms. 
It is noticeable that both approaches lead to essentially the same results 
(see Ref.~\cite{work_on_LaHx_2019}). 
Some practical applications of this self-consistent field method will be discussed in Sec~\ref{Sec:Trends}, as they are fundamental to properly characterize the superconducting state of hydrides under pressure.

\subsubsection{Dielectric Screening in Metals}\label{sec:screening}

Condensation of Cooper pairs via an attractive lattice interaction is universally accepted to be the microscopic mechanism of conventional superconductors. Electron-phonon interaction, despite being weak compared to Coulombic electron-electron repulsion, becomes dominant at very low energy (within the Debye frequency). Nevertheless, Coulomb forces are in action and are important in the formation of the superconducting state as they partially balance the phononic mediation reducing the overall coupling and with it, the critical temperature.

Unlike the electron-phonon coupling, this (screened) Coulomb interaction turns out to have an overall effect that is usually weakly material dependent. Trusting a consolidated evidence, % over the years, 
one often bypasses the calculation of Coulomb interactions. 
As discussed in Sec.~\ref{sec:McMillan}, 
its role is encompassed by the use of an 
adjustable/semi-empirical $\mu^*_c$ parameter. 
Nevertheless, Coulomb interactions can be incorporated from first principles. 
As we will discuss later in Sec.~\ref{sec:SCDFTSetUp}, 
these enter the superconducting problem in the form of a screened effective interaction within 
a GW~\cite{Hedin_GW_PR1965,AryasetiawanGunnarsson_GWMethod_RepProgPhys1998} electronic self-energy form 
in Nambu space~\cite{VonsovskySuperconductivityTransitionMetals}. 

Screening can be reliably computed in the random phase approximation (RPA) 
or within the framework of time-dependent DFT (TD-DFT)~\cite{Gross_TDDFT_PRL1984,DreizlerGross_DFT1990}. 
The resulting interaction has a simple and intuitive form: \begin{equation}
w\left({\bf r},{\bf r}_1,\omega\right)=
\int d^3 r_2 \frac{\epsilon^{-1}\left({\bf r}_2,{\bf r}_1,\omega\right)}{\left|{\bf r}-{\bf r}_2\right|}. \label{eq:screened Coulomb}
\end{equation}Where the direct Coulomb repulsion is screened by the inverse dielectric function $\epsilon^{-1}$. 
The interaction usually enters the superconducting problem in the form of its scattering matrix elements between Kohn-Sham electrons: 
\begin{equation}
W_{n{\bf k},n'{\bf k}'}\left(\omega\right)\!\equiv\!\!\int\!\! d^3r d^3r' \psi^*_{n{\bf k}}({\bf r})\psi_{n'{\bf k}'}({\bf r})w\left({\bf r},{\bf r}'\!,\omega\right)\psi^*_{n'{\bf k}'}({\bf r}')\psi_{n{\bf k}}({\bf r}').\label{eq:CoulombME}
\end{equation}
The average of $W_{n{\bf k},n'{\bf k}'}\left(0\right)$ at the Fermi level 
($\epsilon_{\bf k}=\epsilon_{{\bf k}'}=E_F$) multiplied by the density of states 
is known as $\mu_c$, the Coulomb potential parameter. 
As will be discussed in Sec.\ref{sec:CoulombMorelAnderson}, $\mu^*_c$ is derived from $\mu_c$ by integrating out high energy electronic states~\cite{MorelAnderson_1962}. 

The dynamical $\omega$ dependence becomes relevant when significant retardation effects in the screening process occur. 
Due to the fast nature of Coulomb forces, the dynamical part of the interaction is often negligible, justifying the common use of a static approximation. Nevertheless, under certain conditions some materials may feature strong plasmonic effects 
(see for instance the work of Akashi et al.~\cite{Akashi_PlasmonSCDFT_PRL2013} in Li under pressure).
Further theoretical details on plasmon effects are addressed in Sec.~\ref{sec:SCDFTSetUp}. 
It is important to observe at this point that, since the interaction is computed in a GW-type of approximation, vertex corrections are neglected. This implies that some Coulomb effects such as paramagnetic spin fluctuations are not included. 
However these type of interactions are not relevant to the physics of hydrides superconductors 
as they usually occur in the proximity of magnetic phase transitions
~\cite{Stewart_UncnventionalSuperconductivity_Review2017,Scalapino_UconventionalPairing_RevModPhys2012,Essenberger_SpinFluctuationsTheory_PRB2014}.   

\subsection{Formalism for conventional superconductivity} 

\subsubsection{BCS theory}\label{sec:BCS}

The theory of Bardeen, Cooper and Schrieffer 
(BCS)~\cite{BCS_1957} has been the first theory of superconductivity able to explain the microscopic nature of the superconducting state.
BCS is perhaps one of the most successful theories in condensed matter physics. 
Although its predictive power is limited, compared to modern {\it ab initio} methods, it is still today an essential theoretical and computational framework for the characterization and the understanding of superconducting materials. 

BCS theory assumes the following electronic Hamiltonian: 
\begin{eqnarray}
H&=& H_0 + H_{\rm int} \label{HBCS1}\\
H_0&=&\sum_{{\bf k}\sigma} \epsilon_{\bf k} c^\dagger_{{\bf k}\sigma}c_{{\bf k}\sigma} \label{HBCS2}\\
H_{\rm int}&=&-\sum_{{\bf k}{\bf k'}} V_{{\bf k}{\bf k'}} (c^\dagger_{{-\bf k'}\downarrow}c^{\dagger}_{{\bf k'}\uparrow})
(c_{{\bf k}\uparrow}c_{{-\bf k}\downarrow}) \label{HBCS3}
\end{eqnarray}
where $c_{{\bf k}\sigma}$ and $c_{{\bf k}\sigma}^\dagger$ are annihilation and creation operators for the 
Bloch-wave with a wavevector ${\bf k}$ and spin $\sigma$, respectively. $V_{{\bf k}{\bf k'}}$ are 
matrix elements of the leading superconducting interaction. Under the assumption of a momentum independent 
phononic coupling, it can be written as:
\begin{equation}
V_{{\bf k},{\bf k'}}=-V\theta(\hbar\omega_D-|\epsilon_{\bf k}|)\theta(\hbar\omega_D-|\epsilon_{\bf k'}|)
\label{interaction}
\end{equation}
where $\omega_D$ is the Debye frequency. BCS solved this model by an ansatz for the many-body wavefunction 
and found an exact solution. The same can be achieved following Gor'kov~\cite{Gorkov_EnergySpectrumSC_JETP1958,AllenMitrovic1983,Abrikosov_QuantumFieldTheory_Book1975} 
and solving for the Green's function:
\begin{equation}
G({\bf k},\tau)=-\langle T_\tau c_{{\bf k}\sigma}(\tau)c_{{\bf k}\sigma}^\dagger(0) \rangle
\end{equation}
where the imaginary-time dependence of the operators are defined as $X(\tau)=\exp(H\tau)X\exp(-H\tau)$ and $T_\tau$ denotes time ordered product, and the expectation value of physical quantity $A$ is calculated as:
\begin{equation}
\langle A \rangle =\frac{{\rm Tr} \left(\exp(-\beta H)A\right)}{{\rm Tr} \left(\exp(-\beta H)\right)}.
\end{equation}
$G$  satisfies the equation of motion 
\begin{align}\label{eqmotion}
\left(-\frac{d}{d\tau}-\epsilon_{\bf k}\right)&G({\bf k},\tau)=\\&\delta(\tau) -\sum_{\bf k'}V_{{\bf k}{\bf k'}}\langle T_\tau c_{-{\bf k}\downarrow}^\dagger(\tau)c_{{\bf k'}\uparrow}(\tau)
c_{-{\bf k'}\downarrow}(\tau) c_{{\bf k}\uparrow}^\dagger(0) 
\rangle \nonumber
\end{align}
That can be solved by introducing the mean-field approximation:
\begin{equation}
\langle
T_\tau c_{-{\bf k}\downarrow}^\dagger(\tau)c_{{\bf k'}\uparrow}(\tau)
c_{-{\bf k'}\downarrow}(\tau)
c_{{\bf k}\uparrow}^\dagger(0)
\rangle
\rightarrow F({\bf k},0)F^*({\bf k},\tau)
\end{equation}
with the definition of the anomalous Green's function: 
\begin{eqnarray*}
F({\bf k},\tau)&=&-\langle T_\tau
c_{{\bf k'}\uparrow}(\tau)
c_{-{\bf k'}\downarrow}(0)
\rangle 
\end{eqnarray*}
This leads to:
\begin{eqnarray*}
G({\bf k},i\omega_n)&=& \frac{-i\omega_n-\epsilon_{\bf k}}{\omega_n^2+E_{\bf k}^2} \\
{F}({\bf k},i\omega_n)&=&\frac{\Delta^*_{\bf k}}{\omega_n^2+E_{\bf k}^2}
\end{eqnarray*}
where $\omega_n=(2n+1)\pi k_B T$ are the Matsubara frequencies, $E_{\bf k}^2\equiv \epsilon_{\bf k}^2+|\Delta_{\bf k}|^2$, and the function $\Delta_{\bf k}$ is defined as:
\begin{equation}
\Delta_{\bf k}\equiv \sum_{{\bf k'}}
V_{{\bf k},{\bf k'}}F({\bf k'},0)
\label{eq4}
\end{equation} 
This quantity carries the meaning of a (superconducting) gap and satisfies the self-consistent BCS equation:
\begin{equation}
\Delta_{\bf k}=
\frac{1}{\beta}
\sum_{\bf k'}
\sum_{n=-\infty}^{\infty}
\frac{V_{{\bf k}{\bf k'}}\Delta_{\bf k'}}{\omega_n^2+E_{\bf k}^2}
=\sum_{\bf k'}\frac{V_{{\bf k}{\bf k'}}\Delta_{\bf k'}}{2E_{\bf k'}}
\tanh
\left(
\frac{E_{\bf k'}}{2k_B T}
\right)
\label{bcsgap}
\end{equation}
If one assumes the gap function to be isotropic:
$\Delta_{\bf k}=\Delta\neq 0$, 
then the gap expression (Eq.~\ref{bcsgap}) allows to extract the following limits:
\begin{enumerate}
\item For $T\to0$ and $VN(E_F)\ll 1$
\begin{equation}
\Delta \sim 2\hbar \omega_D \exp \left( \frac{-1}{VN(E_F)}\right)
\label{gapweak}
\end{equation}
where $N(E_F)$ is the density of states at the Fermi level.
\item For $T\sim T_c$, then $\Delta\rightarrow 0$, one obtains:
\begin{equation}
T_c=1.13 \omega_D \exp \left(\frac{-1}{VN(E_F)}\right)
\label{tcweak}
\end{equation}
\end{enumerate}
By comparing (\ref{gapweak}) and (\ref{tcweak}), we find the universal BCS ratio:
\begin{equation}
\frac{2\Delta}{T_c}\sim \frac{4}{1.13}\sim 3.54
\end{equation}.

The main drawback of BCS theory is that the pairing interaction between electrons assumed in Eq.~\ref{HBCS3} is an instantaneous effective field, while, as we will see from the {\it ab initio} density-functional framework, the true nature of the phononic pairing of Sec.~\ref{sec:elphME} and Coulomb interactions of Sec.~\ref{sec:screening} is dynamical. 
The intrinsic timescale should be taken into account to achieve a quantitative accuracy using many-body perturbation theory including explicitly the lattice degrees of freedom.

\subsubsection{\'Eliashberg Theory}\label{sec:eliashberg}

\'Eliashberg theory is a many-body perturbation approach to superconductivity. 
In a modern perspective, it can be seen as a GW$_0$ approximation applied in the presence of a superconducting proximity effect field and including in the screened interaction (W$_0$) both Coulomb and phonon propagators. 
The derivation follows similar steps as in Sec.~\ref{sec:BCS}, where instead of a model BCS pairing the interactions are computed from first principles. 

The starting Hamiltonian $H$, is one for interacting electrons and ions, 
in which the lattice dynamics are decoupled, and the electron-phonon coupling 
is described within the Kohn-Sham theory as defined in Sec.~\ref{sec:elphME}. 
The decoupling procedure will be discussed in more detail in the framework of 
density functional theory for superconductors in Sec.~\ref{sec:SCDFTSetUp}. 
Here to set up the \'Eliashberg perturbative approach, 
$H$ is split in a zero-approximation $H_0$ plus an interaction part $H_I$. 
A convenient choice for the zeroth-order Hamiltonian includes the 
coupling with the external field, $H_{ext}$, and 
the Kohn-Sham Hamiltonian entering in Eq.~\ref{eq:KSscf}: 
 \begin{equation}
 H_s=\sum_{\sigma}\int d{\vect r}\psi^{\dagger}_{\sigma}\left({\vect r}\right)\left[-\frac{\nabla^2}{2}+v_s\left({\vect r}\right)-\mu\right]\psi_{\sigma}\left({\vect r}\right),
 \end{equation} 
 while the rest is absorbed into $H_I$ (see also Sec:~\ref{sec:SCDFTSetUp}). 
 Thus, it conveniently writes as, \begin{eqnarray}
 H_0&=& H_s + H_{ext} \\
 H_I&=& H_{ee} + \tilde H_{e-ph} - H_{DC}
 \end{eqnarray}
 where the last term (as in conventional GW theory) 
 removes extra $xc$ effects already included in $H_s$ therefore avoiding any \textit{double counting}:
 \begin{equation}
 H_{DC}=\sum_{\sigma}\int d{\vect r}\psi^{\dagger}_{\sigma}\left({\vect r}\right)v_s\left({\vect r}\right)\psi_{\sigma}\left({\vect r}\right).
 \end{equation}

 Unfortunately conventional many-body perturbation theory~\cite{FetterWalecka_QuantumTheoryOfManyPatricleSystems_Book1971} 
 can not be directly applied to $H_0+H_I$ because the superconducting condensation $H_{ext}$ should contain a proximity
field~\cite{DeGennes_SupercondMetalsAlloys_Book1966,VonsovskySuperconductivityTransitionMetals} 
(see also sec:~\ref{sec:OGK}):
\begin{equation}
H_{\Delta_{ext}}=\int d^3rd^3r'\Delta^*_{ext}\left({\vect r},{\vect r}'\right)\psi_{\uparrow}\left({\vect r}\right)\psi_{\downarrow}\left({\vect r}'\right)+ h.c.
\end{equation}
that introduces extra processes forbidden in a particle conserving theory. 

In the Nambu-Gor'kov formalism \index{Nambu-Gor'kov formalism}  one defines two new electronic field operators:
 \begin{eqnarray}
 \bar \psi\left({\vect r}\right)&=&\left(\begin{array}{c}\psi_{\uparrow}\left({\vect r}\right) \\ \psi^{\dagger}_{\downarrow}\left({\vect r}\right)\end{array}\!\right)\\
 \bar\psi^\dagger\left({\vect r}\right)&=&\left(\begin{array}{cc} \psi^{\dagger}_{\uparrow}\left({\vect r}\right) & \psi_{\downarrow}\left({\vect r}\right) \end{array} \right).
 \end{eqnarray}
 that still obey Fermionic commutation rules.
 With these two-component fields $\bar\psi$ one can rewrite $H_0$ and $H_I$ as:
 \begin{eqnarray}
 ~\hspace{-1cm}H_0\hspace{-0.15cm}&\hspace{-0.15cm}=\hspace{-0.15cm}&\hspace{-0.3cm}\int\hspace{-0.15cm} d{\vect r} \bar \psi^{\dagger}\left({\vect r}\right) \bar H_0\left({\vect r},{\vect r}'\right) \bar \psi\left({\vect r}'\right)\label{eq:H0}\\
 ~\hspace{-1cm}H_I\hspace{-0.15cm}&\hspace{-0.15cm}=\hspace{-0.15cm}&\hspace{-0.3cm}\int\hspace{-0.15cm} d{\vect r} \bar \psi^{\dagger}\left({\vect r}\right) \left[ \sum_{\nu{\vect q}}\sqrt{\frac{\hbar}{2\omega_{{\vect q}\nu}}}\int\hspace{-0.15cm} d{\vect r}\Delta V_{scf}^{{\vect q}\nu}\left({\vect r}\right)\bar\sigma_3 b_{\nu{\vect q}} - \bar\sigma_0 v_s\left({\vect r}\right)\right] \bar \psi\left({\vect r}\right) \nonumber\\
    \hspace{-0.4cm}&\hspace{-0.4cm}+\hspace{-0.15cm}&\hspace{-0.15cm} \frac{1}{2}\hspace{-0.1cm} \int\hspace{-0.15cm} d{\vect r} d{\vect r}'  \left[\bar \psi^{\dagger}\left({\vect r}\right)\bar\sigma_3\bar \psi\left({\vect r}\right) \right] \frac{1}{\left|{\vect r}-{\vect r}'\right|}\left[\bar \psi^{\dagger}\left({\vect r}'\right)\bar\sigma_3\bar \psi\left({\vect r}'\right) \right],\label{eq:HI}
 \end{eqnarray} 
 where $\bar\sigma_3$ is the Pauli matrix $\left(\!\begin{array}{cc} 1 & 0 \\ 0 & \!\!-1\end{array}\!\! \right)$ 
 and $\bar H_0$ is defined as:
 
 \begin{equation}
 \bar H_0\left({\vect r},{\vect r}'\right)\!=\!\left(\!\!\!\begin{array}{cc} \left[-\frac{\nabla^2}{2}+v_s\left({\vect r}\right)-\mu\right]\delta\left({\vect r}-{\vect r}'\right) \hspace{-1cm}&\hspace{-1.2cm} \Delta_{ext}\left({\vect r},{\vect r}'\right) \\ \Delta^*_{ext}\left({\vect r},{\vect r}'\right) \hspace{-1cm}&\hspace{-1.2cm} -\left[-\frac{\nabla^2}{2}+v_s\left({\vect r}\right)-\mu\right]\delta\left({\vect r}-{\vect r}'\right) \end{array}\!\!\!\right)
 \end{equation}
 
The Hamiltonian in this new form does not feature source terms for the field $\bar \psi$ any longer. 
Therefore, the perturbation expansion for $H_I$ will have the same contributions (diagrams) 
as in conventional perturbation theory. The difference is that Green's functions and self-energy will 
have a $2\times2$ matrix structure and vertices will carry the extra $\bar \sigma_3$ term~\cite{VonsovskySuperconductivityTransitionMetals}.
$\bar G$ is obtained by solving of the Dyson equation, that is:
 \begin{equation}\label{eq:dyson0}
 \bar G\left({\vect r},{\vect r}',\omega_i\right)=\bar G_0\left({\vect r},{\vect r}',\omega_i\right) + \bar G_0\left({\vect r},{\vect r}',\omega_i\right)\bar\Sigma\left({\vect r},{\vect r}',\omega_i\right)\bar G\left({\vect r},{\vect r}',\omega_i\right)
 \end{equation} 
where $\bar G_0\left({\vect r},{\vect r}',\omega_i\right)$ is the Green's function corresponding to the non interacting Hamiltonian $H_0$. 
Like in GW theory~\cite{Hedin_GW_PR1965,AryasetiawanGunnarsson_GWMethod_RepProgPhys1998} it is possible to increase the order of the approximation by dressing propagators and Green's functions (the phonon propagator is already dressed 
as it is computed externally) therefore defining the following approximation:
 \begin{equation}\label{eq:sigma}
 \hspace{0.7cm}\begin{minipage}{0.08\textwidth}$\bar\Sigma\hspace{0.1cm}=$\\\vspace{0.4cm}\end{minipage}\underbrace{\includegraphics[width=0.37\columnwidth]{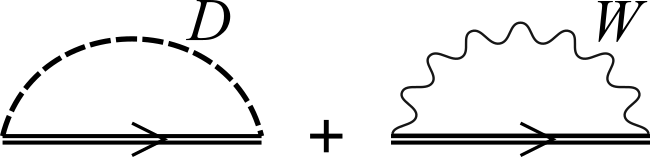}}_{{\bar\Sigma_{xc}}}\begin{minipage}{0.14\textwidth}-$\hspace{0.4cm}\bar\Sigma_{DC}$\\\vspace{0.4cm}\end{minipage}\hspace{-0.7cm}
 \end{equation}
where $\bar\Sigma_{DC}$ is simply $\bar\tau_3 v_{xc}$. 
This approximation neglects vertex corrections; the assumption is justified, for the phonon interaction, by the same argument given by Migdal for the normal state~\cite{Migdal1958,Eliashberg,AllenMitrovic1983}. 
The absence of vertex corrections in the Coulomb vertex instead 
neglects the most important magnetic contributions to the superconducting pairing~\cite{VonsovskySuperconductivityTransitionMetals,Essenberger_SpinFluctuationsTheory_PRB2014}.  

The computational cost for the solution of Eq.~\ref{eq:dyson0}, even within the approximation above mentioned, is still relatively expensive. 
By looking at the Coulomb diagram, it corresponds to the self-consistent GW approach in a 
$2\times2$ Nambu-Gor'kov space. It is better instead to rely on the same approximations discussed in Sec.~\ref{sec:band_decoupling}, assuming that electronic states are already well described by 
the Kohn-Sham Hamiltonian and neglect inter-band hybridization. 
The self-energy is the same as Eq.~\ref{eq:sigma}, but the diagonal part of the second 
diagram is removed together with $\bar\Sigma_{DC}$ (that was inserted in the first place to avoid the double-counting of $xc$ terms). 
Besides, the Coulomb interaction in Eq.~\ref{eq:CoulombME} can be taken to be static.

In the basis of the Kohn-Sham states ($\psi_{n{\vect k}}$) the self-energy then takes the form:
\begin{eqnarray}\label{eq:sigmaGW_explicit}
 ~\hspace{-0.8cm}&\bar\Sigma_{n{\vect k}}\left(\omega_i\right)&=-\frac{1}{\beta}\sum_j\sum_{m{\vect q}'} \Bigl\{ 
 \bar\sigma_3 \bar G_{n{\vect k}}\left(\omega_i\right)\bar\sigma_3   \sum_{\nu}g_{ { n{\vect k},m{\vect k}+{\vect q}}}^{\nu}D^{ij}_{{\vect q}\nu} \nonumber \\ 
 ~\hspace{-0.8cm}& & +\bar G_{n{\vect k}}\left(\omega_i\right)\circ \bar\sigma_1 W_{n{\vect k},m{\vect k}+{\vect q}}\Bigr\}
  \end{eqnarray}
where $\omega_i$ are the Matsubara frequencies, $\bar\sigma_1$ is the Pauli matrix $\left(\!\begin{array}{cc} 0 & 1 \\ 1 & 0\end{array}\!\! \right)$, $D^{ij}_{{\vect q}\nu}=-2\omega_{\nu{\vect q}}/\left[\left(\omega_i-\omega_j\right)^2+\omega^2_{\nu{\vect q}}\right]$ is the phonon propagator, $\circ$ is the element-wise product and $W$ the screened Coulomb interaction. $\bar G_{n{\vect k}}\left(\omega_i\right)$ is the Nambu-Gor'kov Green's function 
that in momentum space is:
 \begin{equation}
 \bar G_{n{\vect k}}\left(\omega_i\right)=\int_0^{\beta}d\tau e^{-\ii\omega_i\left(\tau-\tau'\right)}\int d{\vect r}{\vect r}'\varphi^*_{n{\vect k}}\left({\vect r}\right)\bar G\left(\tau{\vect r},\tau'{\vect r}'\right)\varphi_{n{\vect k}}\left({\vect r}'\right)
 \end{equation}
 and is the solution of the Dyson equation\index{Dyson equation}:
 \begin{equation}\label{eq:dyson}
 \bar G_{n{\vect k}}\left(\omega_i\right)=\bar G_{0\,n{\vect k}}\left(\omega_i\right) + \bar G_{0\,n{\vect k}}\left(\omega_i\right)\bar\Sigma_{n{\vect k}}\left(\omega_i\right)\bar G_{n{\vect k}}\left(\omega_i\right)
 \end{equation}
where $\bar G_{0\,n{\vect k},}\left(\omega_i\right)$ is the Green's function 
corresponding to the non-interacting Hamiltonian $H_0$. 
The above equations are the central result of the \'Eliashberg theory of superconductivity. 
Their solution is achieved by first expanding this matrix equation 
into Pauli matrices and separating it into components. 
The decomposition leads to an intuitive form for 
$G_{n{\vect k}}\left(\omega_i\right)$ which reads as: 
  \begin{equation}\label{eq:G_Eli}
   \frac{\left(\!\begin{array}{cc}\ii\omega_iZ_{n{\vect k}}\left(\omega_i\right)+\left[\xi_{n{\vect k}}+\chichi_{n{\vect k}}\left(\omega_i\right)\right]\hspace{-1.2cm}&\hspace{-1.2cm}
  \phi_{n{\vect k}}\left(\omega_i\right) \\ 
  \phi_{n{\vect k}}\left(\omega_i\right) \hspace{-1.2cm}&\hspace{-1.2cm}  
  \ii\omega_iZ_{n{\vect k}}\left(\omega_i\right)-\left[\xi_{n{\vect k}}+\chichi_{n{\vect k}}\left(\omega_i\right)\right] \end{array}\!\right)}
  {\left[\ii\omega_i Z_{n{\vect k}}\left(\omega_i\right)\right]^2 -\left[\xi_{n{\vect k}}+\chichi_{n{\vect k}}\left(\omega_i\right)\right]^2 -\phi^2_{n{\vect k}}\left(\omega_i\right) }
  \end{equation}
where $\xi_{n{\vect k}}$ are the eigenvalues of $H_s$ and $\chichi$ (not to be confused with the superconducting order parameter) 
shifts the non-interacting energies, $Z$ behaves as a mass term and $\Delta=\phi/Z$ 
is the function giving the superconducting gap (this interpretation is evident by
analytically continuing $\bar G$ to the real frequency axis ($\ii\omega_i\to \omega$). 
$\Delta$, $\chichi$ and $Z$ are now scalar functions and can usually be assumed to be real-valued~\footnote{One can show that for the Hamiltonian Eq.~\ref{eq:H}, the $\Delta$, 
$Z$ and $\chichi$ functions satisfy a set of equations with real coefficients. 
Note that complex solutions can be found as in the famous three crystal experiment~\cite{Tsuei_3crystalexperimentYBCO_PRL1994}. 
However, this situation is rather unusual and is generally neglected.}. 
The set of equations satisfied by these functions are the \'Eliashberg   equations~\cite{Eliashberg,Ummarino_Eliashberg2013}. 
Due to the presence of nested Matsubara and momentum integrations, their computational cost is non-negligible. Thus, these are usually solved by imposing further approximations. 
In particular on the Coulomb interaction that are treated in Sec.~\ref{sec:CoulombMorelAnderson}. Neglecting for the moment the Coulomb term in the expression of the self-energy, the \'Eliashberg equations read: 
\begin{eqnarray}
~\hspace{-0.5cm}Z_{{\bf k}}(i\omega_i)\hspace{-0.15cm}&\hspace{-0.2cm}=\hspace{-0.2cm}&\hspace{-0.15cm}1\hspace{-0.1cm}-\hspace{-0.1cm}
\frac{1}{\beta}\sum_{m{\vect q}\nu,j}\left[
|g_{n{\vect k},m{\vect k}+{\vect q}}|^2D^{ij}_{\nu{\vect q}}
\right] 
\frac{i\omega_{j}Z_{m{\vect k}+{\vect q}}(i\omega_j)}{i\omega_i\Theta_{m{\vect k}+{\vect q}}({\bf },i\omega_j)} 
\label{eliashbergZ}
\\
~\hspace{-0.5cm}\chichi_{n{\vect k}}(i\omega_i)\hspace{-0.15cm}&\hspace{-0.2cm}=\hspace{-0.2cm}&\hspace{-0.15cm}
\frac{1}{\beta}\sum_{m{\bf q}\nu,j}\left[
|g_{n{\vect k},m{\vect k}+{\vect q}}|^2D^{ij}_{\nu{\vect q}}
%+W({\bf k}-{\bf k'})
\right] %\nonumber\\
%&&\times
\frac{\chichi_{m{\vect k}+{\vect q}}(i\omega_j)+\xi_{m{\vect k}+{\vect q}}}{\Theta_{m{\vect k}+{\vect q},}(i\omega_j)}  
\label{eliashbergchi}\nonumber\\
~\hspace{-0.5cm}\Delta_{n{\vect k}}(i\omega_i)\hspace{-0.15cm}&\hspace{-0.2cm}=\hspace{-0.2cm}&
\hspace{-0.15cm}\frac{1}{\beta}\sum_{m{\vect q}\nu,j}\left[
|g_{n{\vect k},m{\vect k}+{\vect q}}|^2D^{ij}_{{\vect q}\nu}
%+W({\bf k}-{\bf k'})
+ W_{n{\vect k},m{\vect k}+{\vect q}}
\right] %\\
%&&\times
\frac{\Delta_{m{\vect k}+{\vect q}}(i\omega_j)}{\Theta_{m{\vect k}+{\vect q}}(i\omega_j)},  \nonumber
\label{eliashbergdelta}
\end{eqnarray}
where $\Theta_{m{\vect k}+{\vect q}}$ is the denominator of Eq.~\ref{eq:G_Eli}.

The position of the Fermi level is determined by calculating the number of particles per unit cell:
\begin{eqnarray*}
~\hspace{1.5cm}1-\frac{2}{\beta}\sum_{{\vect k},i\omega_j}
\frac{\chichi_{n{\vect k},}(i\omega_j)+\xi_{n{\vect k}}}{\Theta_{n{\vect k}}(i\omega_j)}.  
\end{eqnarray*}

As mentioned above, due to the nested Matsubara and Brillouin zone (BZ) summation,  
these equations are cumbersome and numerically expensive to solve. 
A recent attempt by Sano et al.~\cite{PRB_Sano_Van-Hove_H3S_2016} 
for a full solution on a realistic system (H$_3$S) evidenced its complexity. 
In particular, the BZ summation is troublesome because it requires at the same time 
to perform a meticulous integration in ${\vect k}$ around the Fermi level, 
and the high energy integration to converge the Coulomb contribution 
(that extends to energies up to several Rydberg). 
Two approximations that are conventionally used to overcome this difficulty are: 
\begin{enumerate} 
\item to replace the Coulomb interaction by an effective one that acts only near the Fermi level. 
\item to assume that the only relevant ${\vect k}$ dependence occurs through $\xi_{n{\vect k}}$.
\end{enumerate}
The first approximation has been extensively used by Giustino and coworkers to develop 
an anisotropic \'Eliashberg approach~\cite{Margine_anisoEliashberg_PRB2013}. 
The second approximation is used in Ref.~\cite{sanna-flores_2018_Eliashberg}. 
In the following, we focus only on a combined approximation in which both are performed at the same time.  

\paragraph{Isotropic Approximation}
\'Eliashberg equations are simplified enormously by expressing any  
${\vect k}$-dependence of the coupling kernels by the energies $\xi_{n{\vect k}}$. 
At the same time one assumes that all interactions, including the Coulomb interaction, can be 
replaced by their Fermi surface average. This leads to the simpler form of Eqs.~\ref{eliashbergZ}:
\begin{eqnarray}\label{eq:IsoEli}
~\hspace{-0.1cm}Z(i\omega_i)\hspace{-0.2cm}&\hspace{-0.2cm}=\hspace{-0.2cm}&\hspace{-0.2cm}1+\frac{\pi}{\omega_n \beta}
\sum_j\frac{\omega_j}{\sqrt{\omega_j^2+\Delta^2(i\omega_j)}}\lambda(i\omega_j-i\omega_i)\nonumber\\
~\hspace{-0.5cm}\Delta(i\omega_i)Z(i\omega_i)\hspace{-0.2cm}&\hspace{-0.2cm}=\hspace{-0.2cm}&\hspace{-0.2cm}\frac{\pi}{\beta}
\sum_{j}\frac{\Delta(i\omega_j)}{\sqrt{\omega_j^2+\Delta^2(i\omega_j)}}\left[\lambda(i\omega_j-i\omega_i)-\mu_c\right]
\end{eqnarray}
where 
\begin{eqnarray*}
\lambda(i\omega_i-i\omega_j)&=&\int  \frac{2\omega\alpha^2F(\omega)}{(\omega_i-\omega_j)^2+\omega^2}d\omega,
\end{eqnarray*}
with 
\begin{equation}
 \alpha^2F(\omega)= \frac{1}{N(E_F)} \sum \limits_{{\vect k} {\vect q},\nu} |g_{n{\vect k},m{\vect k}+{\vect q},\nu}|^2 \delta(\xi_{n{\vect k}}) \delta(\xi_{m{\vect  k}+{\vect q}}) \delta(\omega-\omega_{{\vect q}\nu})~. \label{eq:a2F}
\end{equation}
% where $N(E_F)$ is the density of states (DOS) at the Fermi level. 
In this approximation, the Fermi shift function $\chichi$ becomes zero. 
In Eq.~\ref{eq:IsoEli} the entire role of the Coulomb interaction 
is eventually assumed by the single number $\mu_c$ defined as the average 
of $W_{n{\vect k},m{\vect k}+{\vect q}}$ at the Fermi level, times $N(E_F)$ itself. 
This approximation will be discussed in the next section.
   
By assuming that $\Delta$ is small enough (near \tc), one can derive a linearized form of the \'Eliashberg equations:
\begin{eqnarray}
Z(i\omega_i)\!\!&\!=\!&\!\!1+\frac{\pi}{\omega_i \beta}
\sum_{\omega_j}\lambda(i\omega_j-i\omega_i) \frac{\omega_j}{|\omega_j|} \label{simpleeliash1}\\
\Delta(i\omega_i)Z(i\omega_i)\!\!&\!=\!&\!\!\frac{\pi}{\beta}
\sum_{\omega_j}\frac{\Delta_g(i\omega_j)}{|\omega_j|}\left[\lambda(i\omega_j-i\omega_i) -\mu_c\right]
\label{eq:Eliashberg_linearized}
\end{eqnarray}
particularly useful to compute \tc.

\subsubsection{Morel-Anderson theory}\label{sec:CoulombMorelAnderson}

Coulomb interaction, as discussed in Sec.~\ref{sec:screening},
play a major role in the mechanism of phonon-driven superconductors~\cite{ScalapinoSchriefferWilkins_StrongCouplingSC_PR1966,MorelAnderson_1962,AllenMitrovic1983}. 
In Eqs.~\ref{eq:IsoEli} and Eq.~\ref{eq:Eliashberg_linearized} the role of 
Coulomb interaction has been collapsed to a single quantity $\mu_c$. 

This can be justified by assuming $W_{n{\vect k},m{\vect k}+{\vect q}}$ to be almost constant in a wide window around the Fermi level ($-\epsilon_F,\epsilon_F$) and zero outside of it~\cite{ScalapinoSchriefferWilkins_StrongCouplingSC_PR1966}.
On the other hand $\lambda$ is finite (attractive) only when $i\omega_j-i\omega_i$ is small, practically below a much shorter cutoff frequency $\omega_c$. %the first cutoff is in energy (\xi) while the other is in frequency ... but better not to dig too much 
For high frequencies we can also assume that $\Delta(i\omega_i)$ is constant ($=\Delta^{\infty}$) 
and $\lambda=0$. Then the linearized gap equation for low frequencies is simplified as, 
\begin{eqnarray}
~\vspace{-0.4cm}\Delta(i\omega_i)&=&
\frac{\pi}{\beta Z(i\omega_i)}
\sum_{\left|\omega_j\right|<\omega_c}\frac{\Delta(i\omega_j)}{|\omega_j|}
\left(\lambda(i\omega_j-i\omega_i)-\mu_c\right) \nonumber\\
~\vspace{-0.4cm}&-&\mu_c\Delta^{\infty}\frac{\pi}{\beta Z(i\omega_i)}
\sum_{\omega_c < 
\left|\omega_j\right| 
< \epsilon_F}\frac{1}
{\left| \omega_j \right|}
\label{simpleeliash5}
\end{eqnarray}
Since we can assume that $Z(i\omega_i)=1$ (that is, the mass renormalization effect due to the electron-phonon 
coupling is absent for large $\omega_i$),
\begin{eqnarray*}
\Delta_g^\infty&=&
-\mu_c\frac{\pi}{\beta}
\sum_{\left|\omega_j\right|<\omega_c}\frac{\Delta(i\omega_j)}{|\omega_j|}
-\mu_c \Delta^{\infty}\frac{\pi}{\beta}
\sum_{\omega_c < 
\left|\omega_j\right| 
< \epsilon_F}\frac{1}
{\left| \omega_j \right|}
\end{eqnarray*}
In the limit of low temperature, replacing the summation with integration, this gives
\begin{eqnarray}
\Delta^\infty=
-\left(\mu_c\frac{\pi}{\beta}
\sum_{\left|\omega_j\right|<\omega_c}\frac{\Delta_g(i\omega_j)}{|\omega_j|}
\right)/\left(
1+\mu_c \log\left(\epsilon_F/\omega_c\right)
\right)
\label{simpleeliash6}
\end{eqnarray}
If we plug Eq.~\ref{simpleeliash6} into Eq.~\ref{simpleeliash5}, we obtain
\begin{eqnarray}
\Delta(i\omega_i)=
\frac{\pi}{\beta Z(i\omega_i)}
\sum_{\left|\omega_j\right|<\omega_c}\frac{\Delta(i\omega_j)}{|\omega_j|}
\left(\lambda(i\omega_j-i\omega_i)-\mu_c^*\right)
\label{simpleeliashlast}
\end{eqnarray}
where
\begin{eqnarray}\label{eq:MorelAnderson2}
~\hspace{1.5cm}\mu_c^*=\frac{\mu_c}{1+\mu_c \log\left(\epsilon_F/\omega_c\right)}
\end{eqnarray}
which, for all practical reasons, we call the pseudo-Coulomb potential~\cite{MorelAnderson_1962,ScalapinoSchriefferWilkins_StrongCouplingSC_PR1966}. Moreover, if we assume that $\lambda(i\omega_j-i\omega_i)$ is constant
\begin{equation}\label{eq:lambda_a2F}
   \lambda\equiv\lambda(0) =2\int  \frac{\alpha^2F(\omega)}{\omega^2} d\omega
\end{equation}
for $|\omega_j|<\omega_D$ and $Z=1+\lambda$, Eq.~\ref{simpleeliashlast} becomes
\begin{eqnarray}
~\hspace{0.5cm}(1+\lambda)\Delta(i\omega_i)=(\lambda-\mu_c^*)\sum_j^{|\omega_i|<\omega_D}
\frac{\Delta(i\omega_j)}{|2j+1|}.
\end{eqnarray}
When $\omega_i$ dependence of $\Delta$ can be neglected, Eq.~\ref{simpleeliashlast} is further simplified as
\begin{eqnarray}
~\hspace{1.0cm}\frac{1+\lambda}{\lambda-\mu_c^*}=\sum_{i=0}^{\omega_D/2\pi T_c -1/2}\frac{1}{i+1/2}
\end{eqnarray}
which once solved gives an analytically estimation for  $T_c$  :
\begin{eqnarray}
~\hspace{1.0cm}T_c=1.13\omega_D \exp\left(-\frac{1+\lambda}{\lambda-\mu_c^*}\right) \label{eq:Tc_BCS_strong_coupling}
\end{eqnarray}
By comparing this expression with that obtained with the BCS weak coupling theory,
\begin{eqnarray}
~\hspace{1.3cm}T_c=1.13\omega_D \exp\left(-\frac{1}{\lambda}\right),
\end{eqnarray}
we see that both the mass enhancement effect ($Z=1+\lambda$) and the Coulomb repulsion ($\lambda-\mu_c^*$), reduce superconductivity.

\subsubsection{Empirical models for Tc}\label{sec:McMillan}

Equation~(\ref{eq:Tc_BCS_strong_coupling}), as it stands is not reliably accurate for predictions, but it can be greatly improved by introducing extra parameters to be fitted to the solution of the \'Eliashberg equations (Eqs.~\ref{eq:Eliashberg_linearized}).
%========================== 
This procedure was introduced by McMillan in 1968~\cite{McMillanTC_PR_1968} which 
used, as a reference, the experimental phonon spectral function of Nb for 
different scaling and $\mu_c^*/\lambda$ ratios. 
He obtained the so-called McMillan equation for $T_c$, which depends on a small number  of simple parameters:
\[
T_c=\frac{\omega_D}{1.45}\exp\left[-\frac{1.04(1+\lambda)}{\lambda-\mu_c^*(1+0.62\lambda)}
\right]
\]

\begin{figure}
\centering
\includegraphics[width=1.0\columnwidth]{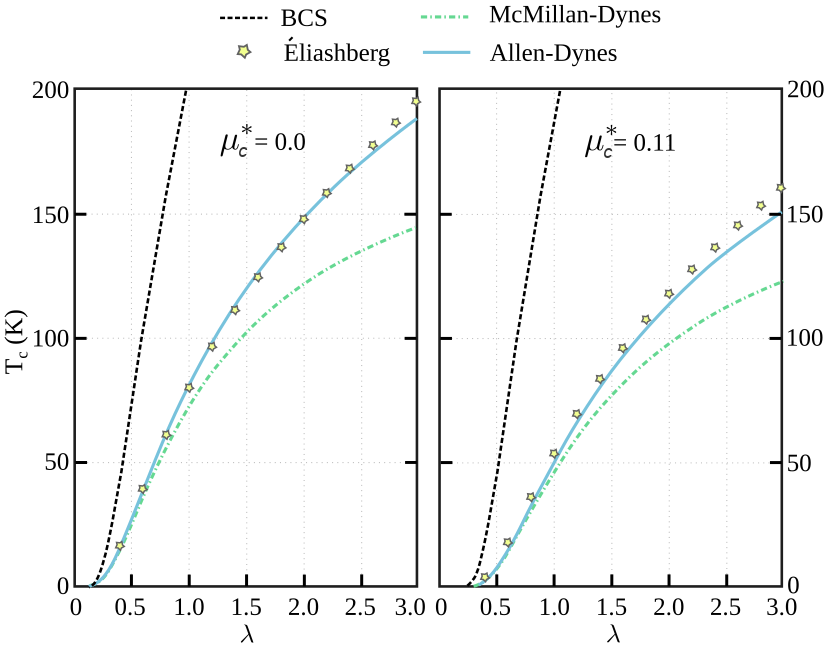}
\caption{~Superconducting critical temperatures calculated with 
the BCS equation, 
the solution of the \'Eliashberg equations, 
the McMillan-Dynes formula~\cite{dynes1972mcmillan} 
and the Allen-Dynes parameterization~\cite{AllenDynes_PRB1975}. 
The phononic coupling used in this model is an Einstein mode 
coupling at 60\,meV of energy ($\omega_{\text log}$=\,60\,meV\, $\sim$596\,K)}\label{fig:Tc_models}
\end{figure}

In 1972, Dynes~\cite{dynes1972mcmillan} modified the prefactor to $\omega$/1.20 instead of $\omega_D$/1.45. 
Subsequently, in 1975, Allen and Dynes~\cite{AllenDynes_PRB1975} performed a thorough 
analysis on the dependence of \tc\ on material properties ($\lambda$, $\mu_c^*$, phonon spectrum) 
and this equation was re-parameterized in a slightly different form:
\begin{eqnarray}
~\hspace{1.cm}T_c=\frac{\omega_{\text log}}{1.2}\exp\left[-\frac{1.04(1+\lambda)}{\lambda-\mu_c^*(1+0.62\lambda)}
\right]
\label{eq:McMillanAllenDynes}
\end{eqnarray}
where, in place of the Debye energy, a logarithmic average is introduced:
\begin{equation}\label{eq:omlog}
~\hspace{1.cm}\omega_{\text log}=\exp
\left[
\frac{2}{\lambda}\int \log\left(\omega\right) \frac{\alpha^2 F(\omega)}{\omega}d\omega 
\right].
\end{equation}
This term corrects for the low energy phonons that are relevant for the superconducting pairing. 
The McMillan equation in Allen-Dynes form~(\ref{eq:McMillanAllenDynes}) is the most widely used approach 
for the calculation of the superconducting critical temperature from first principles. 
One of the aspects that makes it so accurate, despite its formal simplicity, 
is that in this parameterization $\mu_c^*$ results to be 
largely material independent with a typical value of 0.11. 
Note that this universality of $\mu^*_c$ is only valid for the McMillan formula. 
In the original \'Eliashberg equations the value of $\mu_c^*$ is linked to the 
cutoff frequency $\omega_c$ and it should be computed from equation~\ref{eq:MorelAnderson2}: 
Its value may differ significantly from 0.11. 

The McMillan approach starts to deviate from the \'Eliashberg formulation at very 
strong coupling ($\lambda > 1.5$, see Fig.~\ref{fig:Tc_models}), above which it tends to saturate, underestimating 
the true critical temperature. 
In this regime a more accurate approach was introduced by Allen and 
Dynes~\cite{AllenDynes_PRB1975} involving the additional parameter,  
\begin{eqnarray}
~\hspace{1.5cm}\omega_2=\left\{\frac{2}{\lambda}\int \omega\ \alpha^2F(\omega)\ d\omega\right\}^{\frac{1}{2}}\ ,
\end{eqnarray}
and the prefactor
\begin{eqnarray}
f&=&\left\{1+\left[\frac{\lambda}{2.46\ (1+3.8\mu_c^*)}\right]^{\frac{3}{2}}\right\}^{\frac{1}{3}} \\
&\times&\left\{1+\frac{(\omega_2/\omega_{\text log}-1)\lambda^2}{\lambda^2+3.31(1+6.3\mu_c^*)(\omega_2/\omega_{\text log})}\right\} \nonumber,
\end{eqnarray}
that multiply the right hand side of Eq.~(\ref{eq:McMillanAllenDynes}), 
\begin{equation}
T_c=f\frac{\omega_{\text log}}{1.2}\exp\left[-\frac{1.04(1+\lambda)}{\lambda-\mu_c^*(1+0.62\lambda)}
\right].
\label{eq:McMillanAllenDynes2}
\end{equation}
This extended form of the McMillan equation (Allen-Dynes) is very accurate (see Fig.~\ref{fig:Tc_models}) and for conventional isotropic superconductors 
predicts critical temperatures that are usually identical to those obtained by the solution of the \`Eliashberg equations.

\subsection{Density functional theory for superconductors}\label{sec:SCDFTSetUp}

Density functional theory for superconductors (SCDFT) is an extension of DFT to account for the very peculiar symmetry breaking that occurs in a superconductor~\cite{BCS_1957,DeGennes_SupercondMetalsAlloys_Book1966}. 
Proposed by Oliveira, Gross and Kohn in 
1988~\cite{OGK_SCDFT_PRL1988}, it was later revised and  extended~\cite{PhDKurth,Lueders_SCDFT_PRB2005,Marques_SCDFT_PRB2005} in 2005 to merge with the multi-component DFT of Kreibich and Gross~\cite{KreibichGross_MulticomponentDFT_PRL2001} to include nuclear motion. 
This version of SCDFT has been extremely successful in predicting superconductivity in a wide variety 
of materials~\cite{flores-Sanna_honeycombs_2015,sanna2018superconductivity} and proved especially 
useful for the investigation of superconductivity in high pressure environment~\cite{flores-sanna_HSe_2016}. 

\subsubsection{SCDFT Hamiltonian and OGK theorem}\label{sec:OGK}

The starting point of SCDFT is the non relativistic Hamiltonian for interacting electrons \textit{and} nuclei subjected to external fields:
\begin{equation}\label{eq:H}
H=H_e+H_{e-n}+H_n+H_{ext},
\end{equation}
where $e$ stands for electrons, $n$ for nuclei and $ext$ for external fields.  The electronic Hamiltonian reads as:
\begin{eqnarray}
~\hspace{-0.6cm}H_e\!\!&\!\!=\!\!&\!\!\sum_{\sigma}\int d^3r\psi^{\dagger}_{\sigma}\left({\vect r}\right)\left[-\frac{1}{2}\nabla^2-\mu\right]\psi_{\sigma}\left({\vect r}\right) \\
   ~\hspace{-0.6cm}\!\!&\!\!+\!\!&\!\!\frac{1}{2}\sum_{\sigma\sigma'}\int d^3r d^3r'\psi^{\dagger}_{\sigma}\left({\vect r}\right)\psi^{\dagger}_{\sigma'}\left({\vect r}'\right)\frac{1}{\left|{\vect r}-{\vect r}'\right|}\psi_{\sigma'}\left({\vect r}'\right)\psi_{\sigma}\left({\vect r}\right)    \nonumber
\end{eqnarray}
with $\psi$ the electronic field operators and $\mu$ the chemical potential. Nuclei need to be considered explicitly (not just as source of an external potential like in conventional DFT~\cite{HohenbergKohn_DFT_PR1964}) because in electron-phonon driven superconductors 
the ion dynamics provides an essential part of the superconducting coupling:
\begin{eqnarray}
~\hspace{-0.6cm}H_n\!\!&\!\!=\!\!&\!\! -\int d^3R\Phi^{\dagger}\left({\vect R}\right)\frac{\nabla^2}{2M}\Phi\left({\vect R}\right) \\
\!\!&\!\!+\!\!&\!\! \frac{1}{2}\int d^3R d^3R' \Phi^{\dagger}\left({\vect R}\right)\Phi^{\dagger}\left({\vect R}'\right) \frac{Z^2}{\left|{\vect R}-{\vect R}'\right|}\Phi\left({\vect R}'\right)\Phi\left({\vect R}\right)  \nonumber \\
~\hspace{-0.6cm}H_{e-n}\!\!&\!\!=\!\!&\!\! - \sum_{\sigma} \int d^3R d^3r\psi^{\dagger}_{\sigma}\left({\vect r}\right) \Phi^{\dagger}\left({\vect R}\right) \frac{Z}{\left|{\vect R}-{\vect r}\right|}\Phi\left({\vect R}\right)\psi_{\sigma}\left({\vect r}\right) 
\end{eqnarray}
where $\Phi$ are ionic field operator, $M$ the mass of the nuclei and $Z$ the atomic number (for simplicity in notation, we only consider monoatomic systems).

The Hamiltonian needs to include an external symmetry breaking field~\cite{DeGennes_SupercondMetalsAlloys_Book1966} 
that for singlet superconductivity can be chosen as: 
\begin{equation}\label{eq:Delta_ext}
H_{\Delta_{ext}}=\int d^3rd^3r'\Delta^*_{ext}\left({\vect r},{\vect r}'\right)\psi_{\uparrow}\left({\vect r}\right)\psi_{\downarrow}\left({\vect r}'\right)+ h.c.
\end{equation}
In addition, one should also add an external field coupling with the electronic density:
\begin{equation}\label{eq:v_ext}
H_{v_{ext}}=\int d^3r ~ v_{ext}\left({\vect r}\right) \sum_\sigma \psi^{\dagger}_{\sigma}\left({\vect r}\right)\psi_{\sigma} \left({\vect r}\right)
\end{equation}
and an (different) external field that couples with the nuclei:
\begin{equation}\label{eq:Gamma_ext}
H_{W_{ext}}=\int \left[\prod_j d^3R_j \Phi^\dagger\left({\vect R}_j\right) \Phi\left({\vect R}_j\right)\right] W_{ext}\left(\underline{\vect R}\right) .
\end{equation}

In its modern form~\cite{Lueders_SCDFT_PRB2005,Marques_SCDFT_PRB2005}, SCDFT is based on the three densities: 
\begin{eqnarray}\label{eq:densities}
\rho\left({\vect r}\right)&=&\Tr\left[\varrho_0\sum_{\sigma} \psi^{\dagger}_{\sigma}\left({\vect r}\right)\psi_{\sigma}\left({\vect r}\right)\right]\\ %\equiv\sum_{\sigma}\left<\psi^{\dagger}_{\sigma}\left({\vect r}\right)\psi_{\sigma}\left({\vect r}\right)\right>
\chi\left({\vect r},{\vect r}'\right)&=&\Tr\left[\varrho_0\,\psi_{\uparrow}\left({\vect r}\right)\psi_{\downarrow}\left({\vect r}'\right)\right]\\%\equiv\left<\psi_{\uparrow}\left({\vect r}\right)\psi_{\downarrow}\left({\vect r}'\right)\right>\\
\Gamma\left(\left\{{\vect R}_i\right\}\right)&=&\Tr\left[\varrho_0\prod_j \Phi^\dagger\left({\vect R}_j\right) \Phi\left({\vect R}_j\right) \right]
\end{eqnarray}
where $\varrho_0$ is the grand canonical density matrix. 

A recent paper by Schmidt et al.~\cite{schmidt2019representability} 
has evidenced the problem of non-interacting $(v,\Delta)$-representability 
of the superconducting densities in SCDFT. 
Indeed, their work proves that, strictly at zero temperature, such a non-interacting system does not exist. As an alternative, the same group has proposed a reduced density matrix functional theory for superconductors~\cite{RDM_SCDFT_schmidt2019} and proved the existence of a Kohn-Sham system at finite temperature with a corresponding Bogoliubov-de Gennes-like single particle equation.
On the other hand, at any finite temperature i.e. for any practical purpose, no represantability problem exists for SCDFT and it can be used reliably. 

\begin{figure}
  \centering  
  \includegraphics[width=1.05\columnwidth]{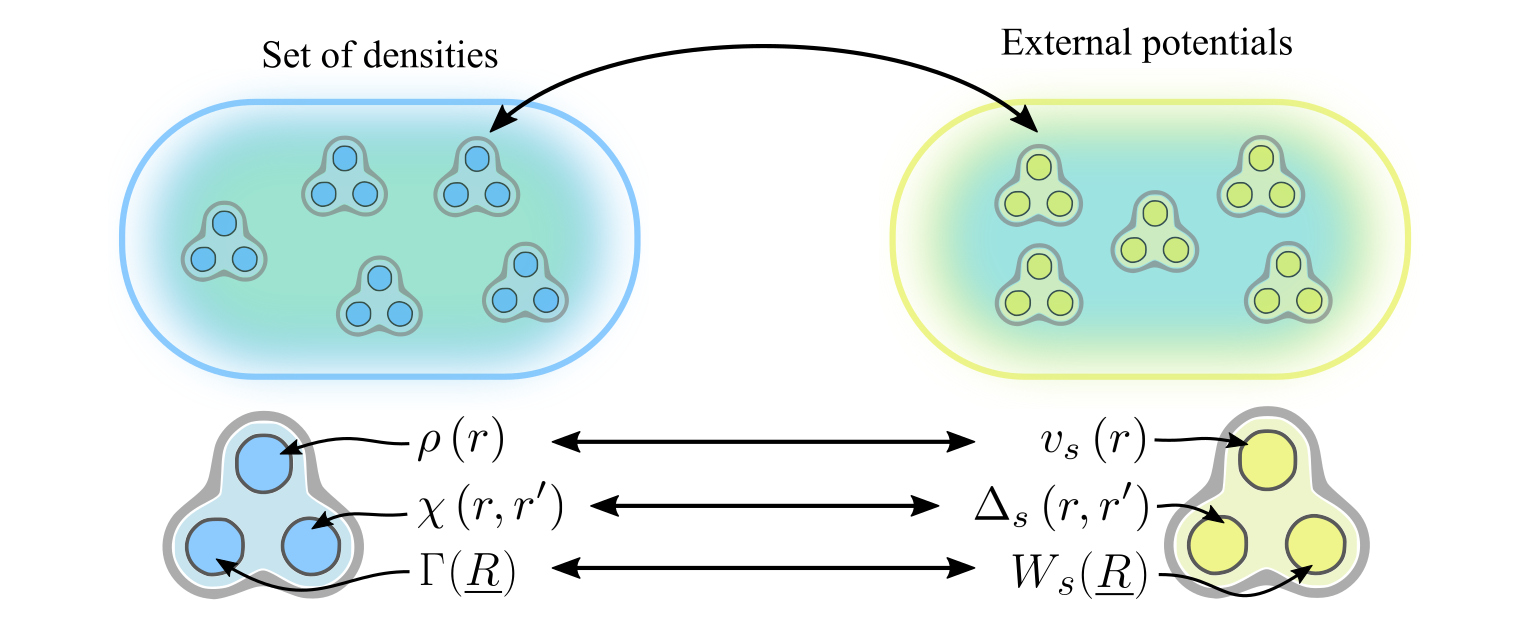}
  \caption{~Illustration to show the Oliveira-Gross-Kohn (OGK) theorem that 
  guarantees a one-to-one mapping between the set of densities 
  ($\rho,\chi,\Gamma$) onto external potentials (${v,\Delta,W}$) in SCDFT.}
  \label{fig:HK}
\end{figure}

The SCDFT generalization of the Hohenberg-Kohn theorem~\cite{HohenbergKohn_DFT_PR1964} (at finite temperature~\cite{Mermin_ThermalElectronGas_PR65}) states:
\begin{enumerate}
\item There is a one-to-one mapping between the set of densities $\rho\left({\vect r}\right)$, $\chi\left({\vect r},{\vect r}'\right)$, $\Gamma\left(\underline {\vect R}\right)$ onto the set of external potentials $v_{ext}\left({\vect r}\right)$, $\Delta_{ext}\left({\vect r},{\vect r}'\right)$, $W_{ext}\left(\underline {\vect R}\right)$
\item There is a variational principle so that it exists a functional $\Omega$ that:
\begin{eqnarray}
\Omega\left[\rho_0,\chi_0,\Gamma_0\right]&=&\Omega_0 \\
\Omega\left[\rho,\chi,\Gamma\right]&>&\Omega_0  \hspace{0.5cm}\textrm{for}\hspace{0.5cm} \rho,\chi,\Gamma\ne\rho_0,\chi_0,\Gamma_0\nonumber
\end{eqnarray}
where $\rho_0,\chi_0,\Gamma_0$ are the ground state densities and $\Omega_0$ the grand canonical potential.
\end{enumerate}

The fact that all observable are functionals of the densities and that $H$ is the sum of internal interactions (Eq.~\ref{eq:H}) and couplings with external fields (Eq.\ref{eq:Delta_ext} + Eq.~\ref{eq:v_ext} + Eq.~\ref{eq:Gamma_ext}) allows $\Omega\left[\rho,\chi,\Gamma\right]$ to be written as: 
\begin{eqnarray}\label{eq:Omega}
 \Omega\left[\rho,\chi,\Gamma\right]&=&F\left[\rho,\chi,\Gamma\right] +\int d^3rv_{ext}\left({\vect r}\right) \rho \left({\vect r}\right)\\ &+&\int \Gamma\left(\underline {\vect R}\right) W_{ext}\left(\underline {\vect R}\right) \prod_j d^3R_j\nonumber \\ &+&\int d^3rd^3r'\Delta^*_{ext}\left({\vect r},{\vect r}'\right)\chi\left({\vect r},{\vect r}'\right)+c.c. \nonumber
\end{eqnarray}
that defines the universal functional $F\left[\rho,\chi,\Gamma\right]$. 

\subsubsection{The Kohn-Sham system}\label{sec:KS}\index{SCDFT!Kohn-Sham system}

As for conventional DFT, it is useful to introduce the Kohn-Sham system~\cite{KohnSham_PR1965}, a non-interacting system whose free energy is minimized by the the same densities of the physical (interacting) one, under the three external potentials: 
\begin{eqnarray}\label{eq:s_potentials}
v_s\left({\vect r}\right)&=&v_{ext}\left({\vect r}\right)+v_{H}\left({\vect r}\right)+v_{xc}\left({\vect r}\right) \\
\Delta_s\left({\vect r},{\vect r}'\right)&=&\Delta_{ext}\left({\vect r},{\vect r}'\right)%+\Delta_{H}\left({\vect r},{\vect r}'\right)
+\Delta_{xc}\left({\vect r},{\vect r}'\right) \nonumber\\
W_s\left(\underline {\vect R}\right)&=&W_{ext}\left(\underline {\vect R}\right)+W_{H}\left(\underline {\vect R}\right)+W_{xc}\left(\underline {\vect R}\right).\nonumber
\end{eqnarray}
The subscript $H$ stands for Hartree terms and $xc$ are the exchange-correlation potentials obtained by a functional derivative of the xc functional of the theory $F_{xc}\left[\rho,\chi,\Gamma\right]$, that can be obtained from perturbation theory~\cite{Lueders_SCDFT_PRB2005,Marques_SCDFT_PRB2005}.

The electronic part of the Kohn-Sham equations are then derived by diagonalizing the Kohn-Sham Hamiltonian with a Bogoliubov-Valatin transformation~\cite{Bogoljubov_NewMethodSC_FdP1958,DeGennes_SupercondMetalsAlloys_Book1966}: 
\begin{equation}
\psi_{\sigma}\left({\vect r}\right)=\sum_i\left[u_i\left({\vect r}\right)\gamma_{i\sigma}-{\rm sgn}\left(\sigma\right)v_i\left({\vect r}\right)\gamma^{\dagger}_{i\sigma} \right]
\end{equation}
leading to the diagonalization conditions:
\begin{eqnarray}
~\hspace{-0.9cm}\left[-\frac{\nabla^2}{2}+v_s\left({\vect r}\right)-\mu\right]u_i\left({\vect r}\right)+\int \Delta_s\left({\vect r},{\vect r}'\right)v_i\left({\vect r}'\right) d^3r'\!\!\!&\!\!\!=\!\!\!&\!\!\! E_i u_i\left({\vect r}\right) \nonumber \\ 
~\hspace{-0.9cm}-\left[-\frac{\nabla^2}{2}+v_s\left({\vect r}\right)-\mu\right]v_i\left({\vect r}\right)+\int \Delta^*_s\left({\vect r},{\vect r}'\right)u_i\left({\vect r}'\right) d^3r' \!\!\!&\!\!\!=\!\!\!&\!\!\! E_i v_i\left({\vect r}\right)\nonumber \\ \label{eq:BdG1}
\end{eqnarray}
that are the electronic Kohn-Sham equation for SCDFT. Their mathematical form is well known in superconductivity literature as Bogoliubov-deGennes (BdG) equations~\cite{DeGennes_SupercondMetalsAlloys_Book1966} which are mostly used, within the BCS model, to describe superconducting structures in real space. In SCDFT, these equations become exact for the calculation of  the total energy and the normal [$\rho\left({\vect r}\right)$] and anomalous [$\chi\left({\vect r},{\vect r}'\right)$] densities:
\begin{eqnarray}
~\hspace{-0.9cm}\rho\left({\vect r}\right)\!\!&\!\!=\!\!&\!\!2\sum_i \left[\left|u_i\left({\vect r}\right)\right|^2 f\left(E_i\right)+\left|v_i\left({\vect r}\right)\right|^2 f\left(-E_i\right)\right] \label{eq:rho}\\
~\hspace{-0.9cm}\chi\left({\vect r},{\vect r}'\right)\!\!&\!\!=\!\!&\!\!\sum_i u_i\left({\vect r}\right)v^*_i\left({\vect r}'\right)f\left(-E_i\right) -  v^*_i\left({\vect r}\right) u_i\left({\vect r}'\right)f\left(E_i\right).  \label{eq:chi}
\end{eqnarray}

In the absence of superconductivity, both $\chi$ and $\Delta$ are zero and the Kohn-Sham equations~\ref{eq:BdG1} become the usual Kohn-Sham equation of conventional DFT:
\begin{equation}\label{eq:DFT_KS}
\left[-\frac{\nabla^2}{2}+v_s\left({\vect r}\right)-\mu\right]\varphi_{n{\vect k}}\left({\vect r}\right)=\xi_{n{\vect k}} \varphi_{n{\vect k}}\left({\vect r}\right).
\end{equation}
This form is slightly more general because it would still include the full effect of temperature and ionic motion since it is still coupled with the ion dynamics via the potentials in Eq.~\ref{eq:s_potentials}.

\subsubsection*{Transformation to momentum space}
Eq.~\ref{eq:DFT_KS} can be solved in the superconducting state (i.e. keeping the non-zero $\chi$ in the functional $v_s[\rho,\chi,\Gamma]$) and the corresponding eigenfunctions $\varphi_{n{\vect k}}\left({\vect r}\right)$ can be used as a basis set to express the BdG equations in ${\vect k}$ space. Introducing the expansion:
\begin{eqnarray}
u_i\left({\vect r}\right)&=& \sum_{n{\vect k}} u_{i,n{\vect k}}\,\varphi_{n{\vect k}}\left({\vect r}\right) \label{eq:ur_uk}\\
v_i\left({\vect r}\right)&=& \sum_{n{\vect k}} v_{i,n{\vect k}}\,\varphi_{n{\vect k}}\left({\vect r}\right) \label{eq:vr_vk}\\
\Delta_s\left({\vect r},{\vect r}'\right)&=& \sum_{nn'{\vect {kk}}'} \Delta_{s,{nn'{\vect {kk}}'}}\,\varphi_{n{\vect k}}\left({\vect r}\right)\varphi_{n'{\vect k}'}\left({\vect r}'\right) \label{eq:Drr_Dkk}
\end{eqnarray}
that when inserted into Eq.~\ref{eq:BdG1} and using the orthogonality of the basis set gives:
\begin{eqnarray}\label{eq:kBdG}
\xi_{n{\vect k}}\,u_{i,n{\vect k}}+\sum_{n'{\vect k}'} \Delta_{s,{nn'{\vect {kk}}'}} \,v_{i,n'{\vect k}'}&=&E_i\,u_{i,n{\vect k}}\\
-\xi_{n{\vect k}}\,v_{i,n{\vect k}}+\sum_{n'{\vect k}'} \Delta^*_{s,{nn'{\vect {kk}}'}} \,u_{i,n'{\vect k}'}&=&E_i\,v_{i,n{\vect k}} \nonumber
\end{eqnarray}
which is a form of the BdG equations particularly useful for introducing approximations. 
At this stage, the problem to solve is still very complicated and can not be tackled without introducing approximations.
The fundamental approximation is to decouple as much as possible the many degrees of freedom (and densities) of the problem:
\begin{enumerate}
\item Decouple electrons from ions separating static and dynamic part of the interaction, including the latter in a perturbative fashion. 
\item Decouple the high energy chemical scale (responsible for bonding) from low energy pairing interactions (responsible for superconductivity). 
\end{enumerate}

\subsubsection{Electron-phonon interaction}\label{sec:phonons}\index{electron-phonon interaction}

The problem of correlated electron-nuclear dynamics is enormously complex, 
however for systems close to equilibrium, theoretical methods can 
describe accurately and efficiently the nuclear dynamics and \ep\ coupling
~\cite{Baroni_DFPT_RMP2001,Giustino_elecronphonon_RMP2017,MariniPonceGonze_MBPTandElectronPhonon_PRB2015}.  
A key approximation is to ignore the effect of superconductivity 
on the lattice dynamics and on the \ep\ interaction. 
As the superconducting transition is usually of second order, this approximation is exact near the critical temperature, where the superconducting density becomes 
infinitesimally small. This allow us 
to use the lattice dynamics of the normal state as in Sec.~\ref{sec:DFPT}. 

The step one needs to perform, out of computational convenience, is to approximate the dynamic part of $H_{e-n}$ with $\tilde H_{e-ph}$ defined in Sec.~\ref{sec:elphME}. 
This step can be certainly justified empirically by its success in many practical  applications~\cite{Savrasov_ep-PRL-1994,Baroni_DFPT_RMP2001,DeGironoli_DFPTmetals_PRB1995} 
but is theoretically less rigorous. 
The most compelling justification is 
that if the Kohn-Sham band structure is close to the interacting one, 
so will likely be its response to a lattice motion. 
Clearly if Kohn-Sham bands are far from the interacting ones 
(like in strongly correlated systems) then use of Kohn-Sham 
\ep\ coupling is expected to be a poor approximation.

\subsubsection*{Band decoupling approximation}\label{sec:band_decoupling}

The electronic BdG Kohn-Sham equations~\ref{eq:kBdG} can be further simplified 
by assuming that the superconducting condensation is a small perturbation on the 
non-superconducting system. As already pointed out in the previous subsection, 
since the superconducting transition is of second order, the assumption becomes 
exact close to \tc . Therefore, it does not affect the estimation of \tc\ itself. 

This assumption implies that the superconducting transition will not induce a structural one;  therefore $\Delta_s\left({{\vect r},{\vect r}}\right)$ should keep the original lattice periodicity and the quantum number ${\vect k}$ in Eq.~\ref{eq:DFT_KS} must be maintained~\cite{PhDKurth,PhDLueders}. 
In other words, the summations in equation Eq.~\ref{eq:ur_uk},\ref{eq:vr_vk} 
should only run over the band index $n$ and not over ${\vect k}$.

The summation over $n$ means that the superconducting transition 
can still induce {\it hybridization} between different bands corresponding to the same ${\vect k}$-point. However, unless bands are degenerate (or close to degeneracy with respect to the energy scale set by $\Delta_s$ that is of the order 10\,meV), 
this hybridization must be extremely small.  
Therefore, apart from anomalous cases, 
one can introduce a second and stronger approximation 
by ignoring this superconductivity-induced band hybridization effect. 
Eqs.~\ref{eq:ur_uk},\ref{eq:vr_vk} reduce to:
\begin{eqnarray}\label{eq:decoupling}
u_i\left({\vect r}\right)\equiv u_{n{\vect k}}\left({\vect r}\right)&=& u_{n{\vect k}}\varphi_{n{\vect k}}\left({\vect r}\right) \\
u_i\left({\vect r}\right)\equiv v_{n{\vect k}}\left({\vect r}\right)&=& v_{n{\vect k}}\varphi_{n{\vect k}}\left({\vect r}\right) \nonumber,
\end{eqnarray}
that implies $\Delta_{s,{nn'{\vect {kk}'}}}\to\delta_{n{\vect k},n'{\vect k}'}\Delta_{s,n{\vect k}}$.

 Inserting Eq.~\ref{eq:decoupling} into Eq.~\ref{eq:kBdG} one can formally solve these equations obtaining:
 \begin{eqnarray}
 u_{n{\vect k}}&=&\frac{1}{\sqrt{2}} {\rm sgn}\left(E_{n{\vect k}}\right) e^{\phi_{n{\vect k}}}\sqrt{1+\frac{\xi_{n{\vect k}}}{\left|E_{n{\vect k}}\right|}} \\
 v_{n{\vect k}}&=&\frac{1}{\sqrt{2}}\sqrt{1-\frac{\xi_{n{\vect k}}}{\left|E_{n{\vect k}}\right|}}
 \end{eqnarray}
 with $e^{\phi_{n{\vect k}}}=\Delta_s\left(n{\vect k}\right)/\left|\Delta_s\left(n{\vect k}\right)\right|$ and $E_{n{\vect k}}=\pm\sqrt{\xi^2_{n{\vect k}}+\left|\Delta_s\left(n{\vect k}\right)\right|^2}$.
 While the densities in Eqs.~\ref{eq:rho},\ref{eq:chi} take the simple form:
 \begin{eqnarray}
 ~\hspace{-0.8cm}\rho\left({\vect r}\right)\!\!&\!\!=\!\!&\!\! \sum_{n{\vect k}} \left[1-\frac{\xi^2_{n{\vect k}}}{\left|E_{n{\vect k}}\right|}{\rm tanh}\left(\frac{\beta \left|E_{n{\vect k}}\right|}{2}\right) \right]\left|\varphi_{n{\vect k}}\left({\vect r}\right)\right|^2\\
~\hspace{-0.8cm}\chi\left({\vect r},{\vect r}'\right)\!\!&\!\!=\!\!&\!\!\frac{1}{2}\sum_{n{\vect k}} \frac{\Delta_s\left(n{\vect k}\right)}{\left|E_{n{\vect k}}\right|}{\rm tanh}\left(\frac{\beta \left|E_{n{\vect k}}\right|}{2}\right)\varphi_{n{\vect k}}\left({\vect r}\right)\varphi^*_{n{\vect k}}\left({\vect r}'\right)\label{eq:chi_Deltaxc}
  \end{eqnarray}
 
The entire superconducting problem is now reduced to the 
construction of the matrix elements of 
the Kohn-Sham potential $\Delta_{s}\left(n{\vect k}\right)$ 
that are obtained by the solution of the equation:
\begin{equation}
 \Delta_s=\frac{\delta F_{xc}\left[\rho,\chi{\left[\Delta_{s},\rho,\Gamma\right]},\Gamma\right]}{\delta\chi}. \label{eq:SCDFTgap1}
\end{equation}
Several approximations for $F_{xc}$ have been proposed and tested~\cite{Marques_SCDFT_PRB2005,PhDMarques,Sanna_Migdal,Akashi_PlasmonSCDFT_PRL2013,Akashi_SCDFTplasmons_JPSJ2014}, and all lead to a BCS-like form of the equation above:
\begin{eqnarray}
\Delta_{s}\left(n{\vect k}\right)&=&\mathcal{Z}\left(n{\vect k}\right)\Delta_{s}\left(n{\vect k}\right)\label{eq:SCDFTgap2}\\ &+& \frac{1}{2}\sum_{n'{\vect k}'}\mathcal{K}\left(n{\vect k},n'{\vect k}'\right)\frac{\tanh\left(\frac{\beta}{2}E_{n'{\vect k}'}\right)}{E_{n'{\vect k}'}}\Delta_{s}\left(n'{\vect k}'\right), \nonumber
\end{eqnarray}
where the two kernels $\mathcal{K}$ and $\mathcal{Z}$ depend on 
the chosen $F_{xc}$ functional and contain all the key information about electronic states, electron-electron and electron-phonon coupling. 
The main strength of the theory is its low computational cost, 
as compared to Green's function methods, mainly because the gap equation~\ref{eq:SCDFTgap2}, while being fully dynamical and including strong coupling effects, 
does not involve cumbersome Matsubara integration 
(all the complexity is absorbed in the process of functional construction). 
This implies that the equation can be easily solved 
in its full ${\vect k}$ resolution and a full energy window. 
Therefore, all anisotropic effects~\cite{Floris_Pb_PRB2007,Cudazzo_PRL2008} as well as high-energy Coulomb interactions 
can be included fully {\it ab initio}~\cite{Massidda_SUST_CoulombSCDFT_2009}, 
adding also the paring induced by exchange of spin-fluctuations~\cite{Essenberger_SpinFluctuationsTheory_PRB2014,Essenberger_FeSe_PRB2016}.

\subsubsection{Plasmon-Assisted Superconductivity}\label{plasmonic:effects}

 \begin{figure}[t!]
  \centering  
  \includegraphics[width=\columnwidth]{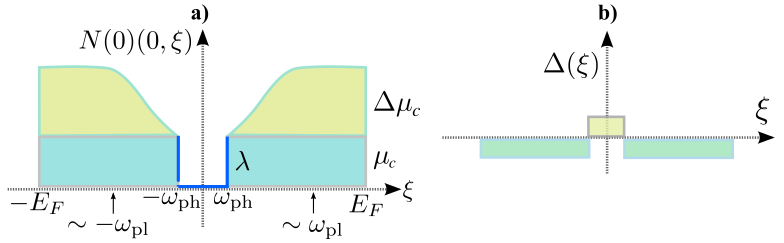}
  \caption{~Schematic illustration on the energy dependence of 
  a) the kernel $\mathcal{K}$ and 
  b) the gap function $\Delta$ in the SCDFT equation.}
  \label{fig:plasmon}
\end{figure}

The consequences of electron-electron interaction in superconductors beyond the approximations discussed in Sec.~\ref{sec:screening} have been extensively studied~\cite{Scalapino_UconventionalPairing_RevModPhys2012},  
for instance, to include magnetic effects in cuprates and Fe-superconductors. 
A Coulomb effect that is often overlooked is 
the plasmon mechanism, that exploits the dynamical structure of the screened Coulomb interaction represented by the frequency-dependent 
dielectric function $\epsilon(\omega)$~\cite{Takada_plasmonicSC_JPSJ1978,radhakrishnan1965superconductivity,Rietschel1983,frohlich1968superconductivity}. 

In the theoretical treatment presented so far, 
we have neglected the frequency dependence 
of the screened Coulomb interaction $W$ (Eq.~\ref{eq:screened Coulomb}). 
However, it has been argued that the dynamical 
structure in $W$ can influence or even enhance superconductivity. 
Takada was the first to show that superconductivity 
could emerge in the electron-gas model even 
in the absence of phonon-mediated attractive interaction~\cite{Takada_plasmonicSC_JPSJ1978}. 
A natural theoretical background to treat on an equal footing the phonon-mediated interaction 
and dynamical screened Coulomb interaction is the {\it ab initio} 
density functional theory for superconductors as presented 
by Akashi et al.~\cite{Akashi_PlasmonSCDFT_PRL2013,Akashi_SCDFTplasmons_JPSJ2014} in full detail.

To give a clear physical picture of the plasmon mechanism, we will consider here a simplified model. 
Let us divide the kernel $\mathcal{K}$ of the SCDFT gap 
Eq.~\ref{eq:SCDFTgap2} into the contributions of the \ep\ coupling $\mathcal{K}^{\rm e-ph}$ and the Coulomb interaction $\mathcal{K}^{\rm e-e}$:
\begin{equation*}
     \mathcal{K}(n{\vect k},n'{\vect k'})=\mathcal{K}^{\rm e-ph}(n{\vect k},n'{\vect k'})+\mathcal{K}^{\rm e-e}(n{\vect k},n'{\vect k'}).
 \end{equation*}
For $\mathcal{K}^{\rm e-ph}$, it is finite and attractive within the
a phonon energy $\omega_{\rm ph}$, but it becomes negligibly small for high energy states. 
Thus it can be approximated as:
\begin{eqnarray*}
N_{F}\mathcal{K}^{\rm ph}(n{\vect k},n'{\vect k'})=
\begin{cases}
-\lambda &(\xi_{n{\vect k}},\xi_{n'{\vect k'}}<\omega_{\rm ph})\\
0 & ({\rm otherwise})
\end{cases}
\end{eqnarray*}
On the other hand $\mathcal{K}^{\rm e-e}$, has large energy range. Adopting the simplified  Morel-Anderson theory of Sec.~\ref{sec:CoulombMorelAnderson}
\[
N_{F}\mathcal{K}^{\rm e-e}(n{\vect k},n'{\vect k'})=\mu_c.
\]
This situation is schematically shown in Fig.~\ref{fig:plasmon}. 
As was discussed in Sec.~\ref{sec:CoulombMorelAnderson}, 
the sign of $\Delta$ is different for a low $\xi$ and a high $\xi$, 
so that the scattering due to $\mu_c$ between high energy 
states and low energy states becomes effectively attractive.
%%%%%%%%%%%%%%%%%%%%%%%%%%
Let us now consider the frequency dependence of $W$ or $\xi$
dependence of $\mathcal{K}^{\rm e-e}$. 
Since the screening effect on $W$ is less effective, 
$\mathcal{K}$ is large for high energy states 
($\Delta \mu_c$ in Fig.~\ref{fig:plasmon}).
Here, $\Delta\mu_c$ gives an additional attractive interaction between 
low energy states and high energy states in the SCDFT gap equation 
(Eq.~\ref{eq:SCDFTgap2}) so that \tc\ is enhanced. 
This plasmon mechanism for superconductivity becomes important when the ratio between $\omega_{\rm ph}$ and the energy scale of the frequency dependence of 
$W$ is not negligibly small. 
In H$_3$S, for instance it has been shown that the plasmon mechanism has a substantial effect. However, since $\lambda$ is extremely large, the plasmon effect does not dominate over 
the phonon mechanism~\cite{Akashi_PRB2015,PRB_Sano_Van-Hove_H3S_2016}.

 %%%%%%%% pum pum bam bam my baby one mortime... Nov. 2019
 
 %% january 2020. ohh yea babay one day I will know.

\section{Computational methods for structural prediction}\label{Sec:Computational}

Since the pioneering ideas of Sch{\"o}n and Jansen~\cite{Schon-Jensen_1996_angewandte} on global structure optimization, 
the solid-state community has gathered considerable 
expertise on how to identify the stable phases of materials from numerical simulations. 
Over the last 25 years (1995--2020), the tremendous development in the field 
has introduced computational tools that efficiently explore the potential energy 
(or enthalpy) landscape of complex systems --often yielding accurate predictions 
of the experimental results. 

These tools had significant repercussions in many fields, but in particular on high-pressure research, on which several successful predictions occurred, including high-temperature superconductivity. 
In this section, we will concisely review the fundamental concepts of the most widely used methods for crystal structure prediction (CSP) that have been used at the forefront of high pressure research in hydrides. 

\subsection{{\it Ab initio} crystal structure prediction}

The determination of the correct crystal structure of a material is fundamental to determine its physical properties. In particular, one typically seeks the thermodynamically most stable structure at a given temperature and pressure, although low-energy metastable structures are also of interest. It is often necessary to search over structures with many different stoichiometries to determine the most stable ones. In practice, only a finite number of stoichiometries can be searched, and only a finite number of structures with a particular stoichiometry can be calculated, whereas both the number of stoichiometries and number of structures are in principle infinite. 
Stable and metastable structures of a given system correspond to global and local minima of its potential energy surface (PES). Locating the global minimum of the PES of a complex system is a central problem in physics, chemistry and biology.

Finding the most stable structure for a given composition is a challenging task that quickly becomes computationally prohibitive as the number of atoms (or degrees of freedom) under consideration increases. In the case of high-pressure hydrides, like in other materials, the number of local minima increases exponentially with the number of atoms in the system~\cite{stillinger1999exponential} and the unconstrained structure prediction based solely on the chemical composition is considered an NP-hard problem (non-deterministic polynomial-time hard) in computer science. For a periodic system of $N$ atoms, the number of degrees of freedom is $3N$ for the atomic coordinates, plus nine for the translational vectors that define the unit cell. 
Considering that six degrees of freedom account for the translational invariance of the crystal and the rotational invariance of the cell, the remaining degrees of freedom define a $3N + 3$ dimensional optimization problem. 

\subsubsection{Potential energy surface}

The Born-Oppenheimer energy surface (or potential energy surface - POS) 
is defined as the ground state eigenvalue $E({\rm\bf \underline R})$ of Eq.~\ref{eq:BO-KS-normal}.  
% I changed this sentence because Eq. BO-KS-normal defines H not E. E is i turn defined by the eigenvalues of H. that is conceptually different. But I cold not change the concept without rewriting the sentence. 
% in short the potential energy surface (PES), is the function that associates 
% the total energy of a set of atoms to their positions: 
% note that $E({\rm\bf \underline R})$ is formally the ground-state eigenvalue 
% of this Hamiltonian. 
% There is a second tricky point the energy surfaces and POS are ALL the eigenvalues of H we only care about the ground state ... but in principle we are giving a very unusual definition of POS
In a non-periodic system, the coordinates ${\rm\bf \underline R}$ are a finite set, whereas 
in periodic crystals the PES depends on three angles $(\alpha, \beta,\gamma)$ 
and three lattice vectors  $(a,b,c)$ that describe the unit cell, 
plus the internal coordinates of the atoms in the cell. 
The free energy, function of the internal degrees of freedom of the system, is modelled with computational methods. 
Its study is fundamental for the characterization of a material at given external conditions.

\begin{figure}[ht!]
  \centering  
  \includegraphics[width=1.0\columnwidth]{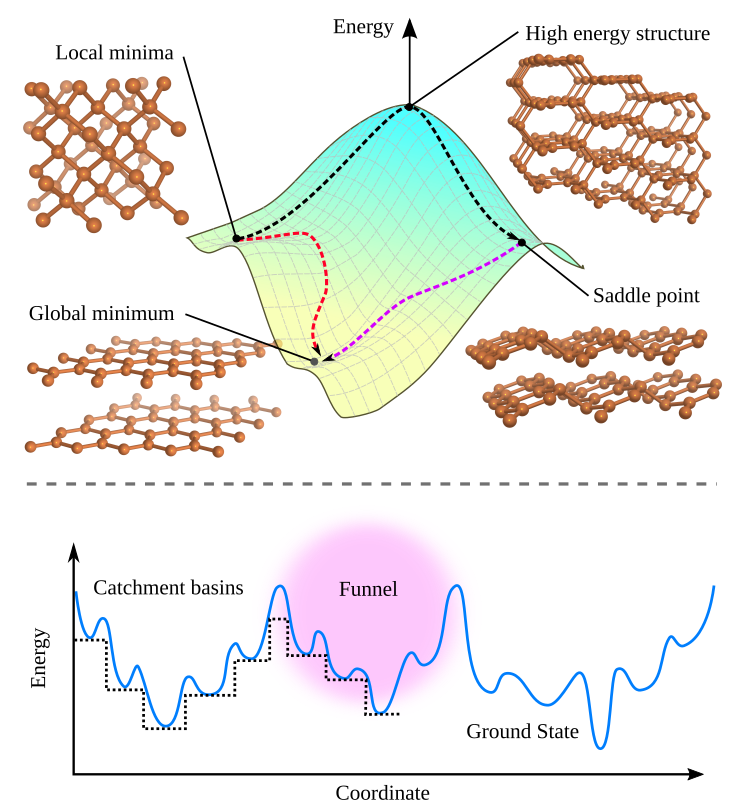}
  \caption{~Top: Conceptual illustration of a potential energy surface and hypothetical structures ranked in energy. 
  Bottom: a one-dimensional energy plot as a function of the reaction coordinate.}
  \label{fig:PES}
\end{figure}

Fig.~\ref{fig:PES} shows a schematic view of a PES with hypothetically energy ranked crystalline structures. We recall that each point on the PES represents a single atomic configuration ${\rm\bf \underline R}$. The stationary points on the PES correspond to a set of internal atomic positions and cell configurations for which forces acting on atoms vanish. These points are of utmost importance to predict the metastable structures and the thermodynamics of the system. These stationary points can be minima, where the system is locally stable, maxima (not energetically viable) or saddle points (transition points). 
Thus, the PES divides into regions each of them uniquely related to a local minimum~\cite{Wales_PES_2003_book}. These regions are called \textit{basins of attraction} or catchment basins, defined as the set of configurations from which the steepest descent relaxation converges to the basin minimum. The bottom panel of Fig.~\ref{fig:PES} shows a schematic one-dimensional energy profile on which a set of neighbouring basins form a super-basin structure, called funnel. Funnels are collections of basins from which the global minimum of the set is reached without crossing very high barriers. This means that the barrier to be crossed should not be larger than the average difference in energy between neighbouring basins within the funnel.

Energy landscapes are described by analyzing the energetic distribution of local minima and transition states connecting them. 
In systems characterized by PESs with staircase-like shapes, where only one funnel exists, and very few local minima have competitively similar energies separated by low barriers, the ground state can easily be found with any CSP scheme. 
On the other hand, systems characterized by energy landscapes that contain multiple funnels are a great challenge for global optimization algorithms. 

\subsubsection{Methods to explore structures}

In structural prediction methods, the goal is primarily to find the correct energetic ordering of dynamically stable local minima on the PES (a dynamically stable energy minimum is one whose Hessian is positive definite). This is equivalent to minimizing the Gibbs free energy $G=U-TS+pV$, whose components are the internal energy, temperature, entropy, pressure and volume of the system --$U,T,S,p$ and $V$ respectively. The stable configuration with the lowest Gibbs free energy corresponds to the ground-state structure and represents the most likely structure that a system will acquire at equilibrium conditions.

The main task of a global optimization method is thus to find the lowest-energy structure among all local minima. As mentioned above, the number of such local minima on a high-dimensional PES increases exponentially with the system size~\cite{stillinger1999exponential}.
It is impossible to identify the global minimum even for moderately-sized systems with simple enumerative search methods. 

Limited computational time restricts all global optimization algorithms to perform a search on a (small) fraction of the total number of local minima. These minima are usually chosen by making some assumptions on the characteristics of the PES. The lowest-energy structure found after a thorough search over visiting a vast number of configurations is identified as the putative ground state. 
Global optimization methods divide into two main groups: Those based on thermodynamic (Boltzmann) approaches and the metaheuristics ones, which include evolutionary and stochastic global optimizers. Simulated annealing, metadynamics and basin hopping are based on thermodynamic principles; genetic algorithms and particle swarm optimization, for example, belong to the class of evolutionary
methods; random search, database searching and data mining schemes are stochastic approaches which rely primarily on chemical intuition and structure-chemistry correlation models. We will briefly describe the main characteristics of those methods and how these have been used for searching for new high-pressure hydrides. 

\subsubsection*{Random sampling} 

A methodology based on the premise that a large part of the PES of a reasonably large assembly of atoms corresponds to very high-energy structures, in which some atoms are much closer or far apart than their equilibrium bond length. Indeed, for most cases, the empirical evidence suggests that large portions of the PES correspond to fragmented structures, while local minima correspond to niches (localized areas)  around basins of attraction~\cite{pickard_PRB-2019_hyperspatial}.  
All these undesired portions are avoided by imposing {\it a priory} constraints on the configuration space. 

Random structure searching requires very few external parameters and is relatively easy to implement. The biases are primarily controllable, logical and based on physical principles. It is also possible to condition or steer the searches by adding information from experiments, hypothetical candidates suggested by chemical or structural considerations on the system in question, as well as information generated by previous searches. 

Technically, a random search starts by generating a set of random atomic arrangements based on educated guesses of bond lengths. A random set of lattice parameters and angles between lattice vectors of the unit cell is generated, and the cell volume is renormalized to yield a random value within 50\% of a chosen mean volume. Together with this initial random guess, an important step in these algorithms is the ``shake", in which one performs a random displacement of the atoms and, if appropriate, a random adjustment of the unit cell. 
These atomic displacements (shakes) of a fraction of a bond length have a non-negligible chance of pushing the system into a nearby basin of attraction. This methodology has proven to be robust and highly reliable for finding (in an unbiased way) the global minima of systems with unit cells containing up to $\approx$12 atoms. Just before the price for the combinatorial complexity becomes too high to be acceptable. Imposing suitable constraints, however, searches have been successfully conducted on larger systems. 

When the method was first proposed, the searches were conducted mostly at DFT level, but nowadays force-fields and other methods are coupled with random searches to improve the final efficiency. The most successful implementation of the random search methodology is the
Ab initio Random Structure Searching ({\sc AIRSS}) by Pickard-Needs~\cite{original-idea_AIRSS_2006,AIRSS_2011_Pickard-Needs}. 
In the context of high pressure, it has been largely used to propose and uncover new stable phases of 
hydrogen~\cite{pickard_NatPhys_2007_structure-H}, aluminum~\cite{pickard_NatMat_2010_Al}, 
ammonia~\cite{pickard_NatMat_2008_amonia} and many other materials~\cite{Needs_APL_2016_perspective-review}. 
The corresponding code is released under the GPL2 licence 
and is tightly integrated in the {\sc CASTEP} first-principles total energy code~\cite{CASTEP_2002}. 
However, it is relatively straightforward to modify the scripts provided 
to adapt the method to other DFT-codes to obtain the core functionality.

\subsubsection*{Particle swarm optimization}

Particle swarm optimization (PSO) is a methodology inspired by the choreography of a bird flock. It is a distributed-behaviour algorithm that performs multidimensional searches~\cite{eberhart_PSO_1995}. PSO belongs to the class of metaheuristic methods, as it makes few or no assumptions about the solutions. The algorithm works by moving particles (structures) in the search space based on efficient algorithms that use the position and velocity of the particles. Hence, all the individuals in the swarm can quickly converge to the global position and a near-optimal geographical position following the behaviour of the flock and their flying histories. 

In practice, in PSO the behaviour of each individual, a structure, is affected by either the best local or the best global individual which helps it to fly through a hyperspace (i.e. PES). 
A structure can be optimized by a feedback mechanism and 
learn from its past experiences to adjust its flying speed and direction (search areas). 
This methodology exploits symmetry constraints during structure generation, which tend to reduce the search space and enhance structural diversity. In particular, during the searches, 
a certain percentage of new structures is introduced at each generation to enhance structural diversity. 
PSO also uses structural characterization techniques to eliminate similar structures from the swarm. 
The algorithm then employs local structural optimization, which reduces the noise of the energy surfaces and ensures the generation of physically justified structures.

Particle Swarm Optimization exists in two variants, 
based on global and local PSO algorithms, which have both been successfully used in many applications~\cite{wang_CALYPSS-2012_CPC}. The global PSO has the advantage of fast convergence, while the local PSO is good at avoiding premature convergence and thus enhances the capability of dealing with more complex systems. 
{\sc CALYPSO}, which stands from Crystal structure AnaLYsis by Particle Swarm Optimization, is the main implementation of this methodology. It is an efficient structure prediction method which is available as a free package that can be used to predict/determine the crystal structure and design multi-functional materials~\cite{wang_CALYPSO-2010_PRB}. This package is popular and has been interfaced to several local structural optimization codes (VASP, QE, GULP, SIESTA, CP2K CASTEP) varying from highly accurate DFT methods to fast semiempirical approaches that can deal with large systems. Notably this PSO-based algorithm~\cite{lv2012particle,zhang2013first,lu2014self,gao_XRD-searches-assisted_2017} 
combined with a fingerprint and matrix bond analysis has been successfully used in solving numerous structural 
problems~\cite{shi2016investigation,Ma_CALYPSO_database_2017,Machine-larning_pot_CALYPSO_2018,PNAS_Oxygen-chain_pressure_Ma,amonia_PNAS_2017_Hermann}, including the prediction of new high-pressure superconducting 
hydrides~\cite{clathrate_CaH6_PNAS_Ma,PNAS_LaHx_2017_Hemley,superconductivity_ScHx_2018,nonmetallic_FeH6_JPCC_2018}. 

\subsubsection*{Evolutionary algorithms}
    
Evolutionary methods are a modern class of algorithms that rely on operations or 
principles inherited from genetic algorithms. 
In evolutionary algorithms, a population of candidate structures is evolved over successive iterations of random variation and selection operations (as in genetic algorithms). Random variation provides a mechanism for discovering new solutions. Selection determines which individuals are retained for further evolution. Solids represented trough six lattice parameters (three unit-cell vectors and three angles), encoded as the lengths of the three lattice vectors and fractions of 2$\pi$, respectively. Each atom, represented by three coordinates, is expressed as a fraction of the corresponding lattice vector. 
In modern evolutionary algorithms, a given set of values defines one structure, and a locally optimized structure is called an individual. A set of individuals is called a population or, depending on the context, a generation. The comparison between different individuals is based on the corresponding {\it fitness} function (in this case, the fitness is the free energy or enthalpy). The simulation technically starts with individuals (structures) generated by some educated guess or in a completely random fashion. These structures then undergo evolution through a series of different genetic operations. New candidate individuals are obtained applying one or more variation operators to selected individuals. For every operation one or two individuals, depending on the type of operation, are chosen stochastically from the population. 
The probability of a given individual being chosen for a given operation is a function of the individual's fitness rank. To a predefined number of worst-performing individuals, a probability of zero is assigned. Once it has been selected for an operation, an individual is not removed from the pool, and thus it can be selected multiple times. The generated candidates  are then scaled to a given unit cell volume, and those who do not fulfil some hard constraints are discarded. 
The remaining candidates are locally optimized (at DFT level), and at this moment new individuals are created. Each operation is repeated until the user-requested number of new individuals for this operation is produced. The total number of individuals generated by different operations equals the population size. After the calculation of the fitness value for each new individual, a new population is obtained by taking the best individuals from the combination of offspring and a user-defined number of best individuals from the parental population. 

{\sc USPEX}, which stands for Universal Structure Predictor: Evolutionary Xtallography, 
is perhaps the most popular and by far the most used method among all those presented in this section. 
This methodology, originally described in Ref.~\cite{Glass_2006_uspex_original,oganov_boook-2011} by Glass et al., 
introduces a unique number of features or variation operators: heredity, mutation and 
permutation~\cite{oganov2006crystal,oganov2008evolutionary,kruglov2017refined}. In heredity, two individuals are selected and used to produce one new candidate. This is achieved simply by taking a fraction of each individual and combining these fractions to create new individuals. The fraction of each individual should be chosen to contain as much information as possible. The main information in a crystal structure is the relative position of the neighbouring atoms. 
Thus, in order to conserve information, the fraction of an individual is selected, taking a spatially-coherent slab. The two slabs, one of each individual, are then fitted together, and the resulting structure is then rearranged by adjusting the number of atoms of each type to the requirements dictated by the chemical formula.
In mutation, an individual is selected and used to produce a single new candidate: technically, this is done transforming the lattice vectors to new vectors applying a strain matrix. 
A permutation is used to produce a single new candidate; in this operation, two atoms of different types are interchanged. 
This operation, which is of course only possible for systems with different types of atoms, facilitates finding the correct atomic ordering.
Another appealing feature is volume scaling; according to this operation, 
every produced candidate is scaled to a particular unit cell volume, before testing it against hard constraints and to local optimization. 
Thus, enhancing the performance for systems where the initial value of the volume is not sufficiently accurate. 
For each new generation, the new volume is obtained as a weighted average between the old and the average volume of the
best individuals of the previous generation. A fundamental step, as in other methodologies, is to perform an efficient local optimization for each individual; this increases the computational cost paid for each individual, but reduces the search space to local optima, enhances comparability between different structures and provides locally optimal structures for further usage~\cite{ma2008high,zaleski2011high,lyakhov2013new}. 
The {\sc USPEX} software is free of charge and is provided as a robust and straightforward black-box tool to
find the stable crystal structure of systems with up to several dozens of atoms/cell. 
Due to local optimization and the exploration of promising regions, 
many highly competitive metastable structures are found during the search. 
This method is interfaced with several {\it ab initio} codes, such as VASP, SIESTA, GULP, Quantum Espresso, CP2K, CASTEP, 
LAMMPS, etc. and has been used in the context of materials discovery~\cite{oganov_boook-2018} 
and high-pressure materials for 
superconductivity~\cite{martinez2009novel,ma_2009_transparent-Na,duan2015pressure,Twisted_paper_LaH10_2018_Oganov}.

Another package which implements evolutionary algorithms for crystal structure prediction is {\sc XtalOpt }, 
An Open-Source Evolutionary Algorithm for Crystal Structure Prediction, 
by Lonie et al.~\cite{lonie_2012_Xtalcomp,lonie_2011_xtalopt,avery2017xtalopt}. 
{\sc XtalOpt} features a periodic displacement (ripple) operator which is ideally suited for extended systems. 
It features hybrid operators, which combine two pure operators, thus reducing the number of duplicate structures in the search. It allows for better exploration of the potential energy surface of a given system. 
This software has been applied to explore hydrides under pressure~\cite{Zurek_FeH5_JPCC_2018,Zurek_CaH2_mix_JPCC_2018} and is available under the GNU Public License. One main advantage for beginners is that it possesses an intuitive graphical interface and interfaces to force-fields and various DFT codes. 

Although it is rarely used, a genetic algorithm that can, in principle, explore novel structures for a given stoichiometry is implemented in the geometry optimization routines of {\sc abinit}~\cite{abinit_2009}.

\subsubsection*{Minima Hopping method}

The minima hopping algorithm is primarily conceived to quickly climb out of wrong funnels in a potential energy surface with a multi-funnel structure. It is achieved by abandoning the standard Markov-based Monte Carlo methods and introducing a feedback mechanism which, based on the whole simulation history, enforces the exploration of new regions of the configuration space.

\begin{figure}[t!]
  \centering  
  \includegraphics[width=1.0\columnwidth]{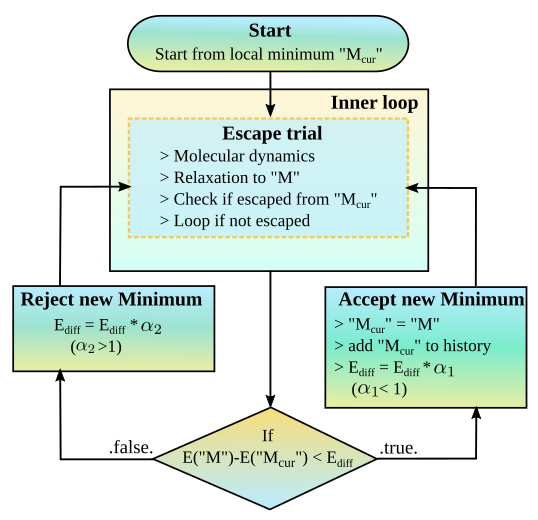}
  \caption{~Flowchart of the minima hopping algorithm.}
  \label{fig:mhm}
\end{figure}

The minima hopping method (MHM)~\cite{Goedecker_mhm_2004,Goedecker_mhm_2005} consists of an inner part, which performs jumps to the local minimum of another basin, and an outer part, which will accept or reject this new local minimum. Fig.~\ref{fig:mhm} shows the flowchart of this method.  
The acceptance/rejection decision is performed based on an energy threshold, i.e. a step will be accepted if the energy of the new local minimum $E_{new}$ is less than a given $E_{diff}$ higher than the current energy $E_{cur}$. The energy difference threshold is continuously adjusted during the simulation so that half of the moves are accepted, and half are rejected. 
This condition, implemented in the outer part, introduces a preference for steps that go down in energy. 
However, if the inner part proposes only steps that go up in energy, 
in the end, also these steps will be accepted because after many rejections $E_{diff}$ will become sufficiently large.
The inner part consists of a more complex escape mechanism that aims at moving from the current local minimum, 
followed by a geometry relaxation which will bring the system into another neighbouring local minimum or will make it escape to another basin. 
The geometry relaxation is performed by a combination of standard steepest-descent and conjugate-gradient methods or more involved methods~\cite{FIRE_algorithm_2006}.  
Initially, a velocity with random direction and a Gaussian-distributed magnitude is chosen for each atom. Amsler and Goedecker implemented a generalized version of the algorithm of minima hopping to periodic systems using variable cell shape molecular dynamics for the escape step~\cite{Amsler_mhm_2010}.  The escape mechanism, in this case, is represented by a short molecular dynamics simulation run that starts from the current minimum. In order to optimize the escape steps, the initial atomic and cell velocities are aligned to low-curvature directions of the current local minimum. Thus, the system has enough energy to overcome energy barriers which are lower than the kinetic
energy measured relative to the current minimum. If the kinetic energy is small, the system will arguably fall back into the current minimum; otherwise, if it is sufficiently large, the system will most likely be ejected from the current basin and end up in a different minimum. Softening works by aligning velocities towards soft directions, therefore biasing the MD simulation to efficient escape trials~\cite{Amsler_Thesis_2012}. 

Another essential feature that MHM exploits, is the Bell-Evans-Polanyi principle~\cite{jensen2017introduction},
which states that ``highly exothermic chemical reactions have low activation energy". Interpreted as ``it is more likely to find a low-energy local minimum if one goes from the current basin into a new basin crossing a low barrier than if one has to overcome a high barrier"~\cite{Amsler_MHM_handobook}. 
Since the early stages of its development, the MHM has proven not only to provide reliable solutions for the global minimum but also for nearby local minima, which can be of interest for many applications. 
It has been tested on a large set of materials classes~\cite{Amsler_FeBi_HP_ChemScien_2017,Clarke_ChemMat_CuBi_2017}, including superconducting hydrides at high pressure~\cite{flores_disilane_2012,Amsler_Ternary_HSeS_PRB2019}.  
Although this package is not as popular as previously described algorithms, it is compelling and reliable for a large variety of situation and systems, including ternaries~\cite{Amsler_ternary_mixed_ML_2019}, Heusler-systems~\cite{Heusler_HT_ChmMat_Wolverton_2018}, etc.  
The code has an interface with several local structural optimization codes such as ABINIT, VASP, QE, GULP, SIESTA, LAMMPS, among others.  

Another method, similar at least in the name, is basin hopping~\cite{Wales_PES_2003_book}, which exploits powerful Monte Carlo methods for the determination of the global minimum. This methodology advantageously transforms the potential energy surface: 
the value of the potential energy within one basin is replaced by the value of the basin minimum.
The basin hopping method virtually eliminates the barriers between the basins of different minima, making it easy to connect them or to jump from one to the other. It is important to note that the basin hopping method does not eliminate the barriers between super-basins or funnels. On the other hand, the transformed piece-wise constant potential energy surface of the basin hopping method still exhibits barriers that have to be overcome by Monte Carlo steps. This methodology has not been used in the context of periodic solids, or to explore potential superconducting structures of hydrides at high pressure. 

\subsubsection*{Simulated annealing}

Stochastic simulated annealing-type algorithms to optimize global and local problems were introduced more than 35 years ago (1985). 
Being based on Boltzmann statistics, from a purely historical perspective, it is understandable that they were the first methods developed to tackle optimization problems. 
Nevertheless, it was not until the 90's that these techniques started 
to be used to help structure determination and to predict solid phases of elements. 
As a side note, the early development of these methods happened almost perfectly in parallel with the availability of commercial diamond anvil cells (see Sec.~\ref{Sec:Experiments}, which gave access to high pressure to many laboratories, opening the era of characterization of solid-phase transformations under pressure. 
However, at that time, addressing theoretically solid-phase transformations was considered a major, if not unsolvable, problem. 
There were a few brave attempts in this direction, see for example Refs.~\cite{PRB_Cohen_oldschool_Si3N4,pannetier_simulated_ann-1990_nature,PRL_SiO-Tsuneyuki_1988}. We refer in this context to the famous 1988 editorial note in Nature by John Maddox: ``one of the continuing scandals in the physical sciences is that it remains in general impossible to predict the structure of even the simplest crystalline solids from a knowledge of their chemical composition"~\cite{maddox1988crystals}. Back to our original storyline, simulated annealing algorithms were first introduced by 
Kirkpatrick et al.~\cite{kirkpatrick1983optimization} and {\v{C}}ern{\`y} et al.~\cite{vcerny1985thermodynamical}. 
The great advantage of these methods lies in their relative ease of implementation and the very general applicability, 
regardless of the specific optimization problem~\cite{van1987simulated}. 
In general, a simulated annealing algorithm employs random walker(s) to explore energy landscapes based on a few fundamental features that make it stochastic: First, a set of walkers that may or not interact with each other and learn from each other's steps. Second, a configuration space with an energy cost function which may remain unchanged or evolve as the algorithm proceeds. Third, a movement class, which for each state returns a neighbouring structure that may or not be accessed with a certain probability (also this probability can evolve along with the algorithm or remain unchanged). 
Fourth, an acceptance criterion according to which the walker makes moves to the selected neighbour. 

In principle, the algorithm starts from the current configuration $i$ (random structure or well-educated guess). 
A neighbouring configuration $i+1$ is chosen randomly according to a set of rules (defined in the "move class"). 
If the energy $E_{i+1}$ is below or equal to $E_i$, the move is always accepted, 
that is, $i +1$ becomes the new current configuration. 
Otherwise, the move is accepted with a probability $e^{- (E_{i+1} + E_i)/C}$, 
where $C$ is a control parameter of the random walk. 
Thus during a sequence of such Monte Carlo steps, the system can climb over
barriers of the energy surface. It can be shown that in the long-time limit ($t \rightarrow \infty$) 
for an ergodic system the probability $p(i)$ of visiting the configuration $i$ 
is given by the Boltzmann distribution for the system at a temperature of $T=C/k_B$.

The general prototype of this type of algorithms is the so-called Monte Carlo Metropolis algorithm,
which describes a single walker at a constant temperature $T$. 
These algorithms explore ergodic regions, especially if $T$ is varied, or if the landscape evolves during the simulation. 
The generalization of the Metropolis algorithm consists in varying the temperature during the run; this is what is called simulated annealing. Simulated annealing works by slowly lowering the temperature,
which implies that the walker will move from average states (ergodic regions) to states with lower and lower energy. 
The premise of this algorithm is that if ones proceed slowly enough, and
waits for a sufficiently long time, at the end of the simulation
the walker will have found the global minimum of the energy landscape. 
Of all methodologies introduced so far, this is the only one for which one can is guaranteed to find the optimal solution, i.e. the global minimum (structure), at the end of the run.

Although relatively easy to implement, the method is computationally very demanding, especially for large systems. 
Applications to predicting solid compounds using simulated annealing are lead by Sch{\"o}n, Doll, and Jansen~\cite{doll2007global,schon2010predicting}, with important successful cases like the prediction of the structure of lead sulfide at standard and elevated pressures~\cite{zagorac2011ab}.

\subsubsection*{Data mining and high-dimensional correlations}

A frequently used approach to reveal hidden correlations is the use of data-mining methods on large databases of computational~\cite{curtarolo2013high} or experimental~\cite{hattrick_APL_2016_datamining} material data. 
The approach is justified since compounds which are chemically similar also tend to exhibit similar crystal structures. 
Only by examining the behaviour of neighbouring elements in the periodic table, one may derive hints on stable compositions and crystal structures for hypothetical systems. Properties like electronegativity, atomic radii, number of valence electrons, valence electron energy or electron configuration could be exploited to discern trends in the search space. 

In this context, starting from the experimental data contained in the inorganic crystal structure database, 
Glawe et al.~\cite{glawe_2016_pettifor_data-minig} used a statistical analysis to determine the likelihood that a chemical element $A$ may be replaced by another $B$ in a given structure. 
This information can be stored matrix form, associating to every pair $(A, B)$ the corresponding likelihood of substitution. By ordering the rows and columns of this matrix to reduce its bandwidth, a one-dimension scale can be used to sort analogous chemical elements, i.e. the Pettifor scale. Alternatively, Goldschmidt's rules of substitution describe, for example, simple recipes to exchange atomic species in ionic crystals. 

{\bf The Materials Project} initiated by Ceder et al.~\cite{PYMATGEN_2013-paper} has been a pioneering attempts in this direction. Ceder's group developed several approaches based on data mining, combining high-throughput with {\it ab initio} calculations to predict ground-state structures of novel compounds~\cite{NatMat_Ceder_datamining_2006}. This Data Mining Structure Predictor (DMSP) scheme relies on the idea that in alloys, a strong structure-structure correlation exists among different compositions due to underlying physical properties. Therefore, it is possible to obtain a probability distribution for the crystal structures of an alloy at unknown composition based on the already known crystal structures at other compositions. The data-mining techniques are limited only to structures already contained in the underlying database, limiting their use in crystal structure prediction.
For example, applicability to high-pressure or high-temperature phases is restricted due to limited available data from theory and experiments. We expect that, thanks to new and rapid developments in the field, it will be soon possible to study high-pressure materials databases, specifically in the case of hydrides.

An interesting methodology that aims at decreasing the high computational 
need for the creation of large-scale databases, for instance in materials under pressure, 
was proposed by Amsler et al.~\cite{Amsler_PRX_2018}. 
Although simple in spirit, this method estimates the enthalpy of stable compounds at 
ambient conditions with a linear approximation to the enthalpy (in the high-pressure limit) 
of compounds contained in databases of calculated materials properties. 
This work has successfully demonstrated that it is possible to explain the occurrence in nature 
of phases which are metastable at ambient conditions and to estimate the pressures at 
which such phases become thermodynamically accessible. 

\subsection{Convex hull and phase diagrams}

\begin{figure*}[t!]
  \centering  
  \includegraphics[width=2.0\columnwidth]{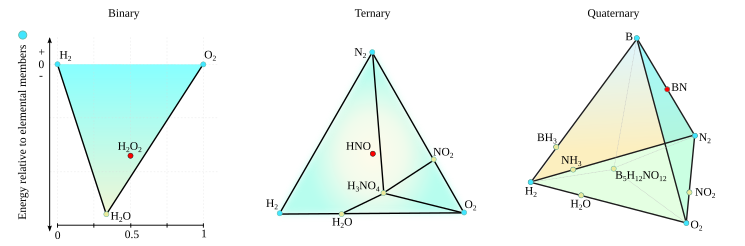}
  \caption{~Examples of binary, ternary and quaternary construction of convex hull type diagram,  for H-O, H-O-N, and H-O-N-B, respectively. Blue nodes on the phase diagrams represent elemental phases; yellow nodes are compositions that are calculated to be stable and red phases that are not stable. The chemical space becomes quickly challenging to explore, adding a one-dimension (another element) the complexity in compositions is evident: B$_5$H$_{12}$NO$_{12}$.}
  \label{fig:Hull-description}
\end{figure*}

As previously mentioned, the prediction of computer-designed materials is a formidable task, since it requires a thorough structure exploration with the computation of enthalpy and entropy functions, and also barriers for phase transformations between all competing phases under given conditions. The convex hull is used as a guide to estimate the formation energy (or enthalpy) and to determine the thermodynamic stability for a given composition to its forming elements.

Fig.~\ref{fig:Hull-description} shows examples of binary, ternary and quaternary convex hulls. To determine the ground state of a system at a given composition, one needs to identify the ordered compounds that have an energy lower than any other structure or any linear combination of structures corresponding to that composition. 
Mathematically, the convex hull of a set of $x$ points, in the Euclidean plane or a Euclidean space, 
is the smallest convex set that contains $x$. 
In our example, the set of points representing ground-state structures forms a convex hull, 
while other structures (points) having an energy above the set of tie lines connecting 
the energy of the ground state ones are not stable. 
The convex hull thus reflects the relative Gibbs free energy of the compounds at zero temperature. 

The convex hull construction permits to visualize very valuable information, i.e. it can be used to derive a "phase diagram" of the compositions that in principle have the lowest energy and hence are accessible to experiments. Thus, a convex hull of the enthalpy in material science represents a rudimentary compositional phase diagram, in which the boundaries and eutectic points are not displayed. Fig.~\ref{fig:Hull-description} shows real systems represented as convex hull phase diagram; an example of binary compositional phase diagrams is shown for H$_2$O, in which the $y$-axis represents the formation energy per atom and the $x$-axis the composition. The stable composition obtained by combining H$_2$ and O$_2$, represented by a green dot, is one of the most stable molecules in the universe, H$_2$O. The black line shows the convex hull construction, which connects stable phases (H$_2$ to H$_2$O to O$_2$). This line is also called the line of stability. The H$_2$O$_2$ composition, which has a negative formation enthalpy to elemental constituents but lies above the line of stability, is not stable. Thus, the convex hull suggests that, likely, the only composition that will spontaneously form is H$_2$O. 

In the ternary case, one dimension is added; the energy axis is removed so that the entire compositional space can be represented. 
The ternary system (H, N, O) is now represented by an equilateral triangle, with three elements at its vertices and the nodes represent compositions for which the decomposition energies are zero (see the middle panel in Fig.\ref{fig:Hull-description}). 
In the quaternary case, the chemical space becomes exceptionally challenging to explore and even to represent; we now use a pyramid to represent the B, N, O, H system. 
The complexity in compositions is evident: we have indicated on the plot an interesting stable point: B$_5$H$_{12}$NO$_{12}$. 
The black lines in the ternary phase diagrams are projections of the convex hull construction into compositional space. 
The lines form Gibbs triangles, which can be used to find stable phases at arbitrary compositions. 
At any point in the phase diagram, other than the stable nodes, the equilibrium phases are given by vertices of the triangle bounding that composition. 
For example, the equilibrium phases for a composition with H:N:O ratio of 3:1:4, 
is predicted to be H$_2$O, O$_2$, NO$_2$ and N$_2$.
We will revisit the implications that this has for hydrides and superconductivity in a subsequent section. 

In order to calculate the convex hull of formation energy of a given system to its forming elements, the following expression is used for binaries, $\Delta^{sys}$=$ (E^{sys} - (E^{a} +E^{b}))/N$, 
and for ternaries $\Delta^{sys}$=$(E^{sys} - (E^{a} + E^{b} + E^{c}))/N$, 
where $ E^{sys}$ represent the energy for a given structure, $N$ the number of atoms and $E^{a}, E^{b},... E^{n}$  the energies of elemental decompositions in their lowest-energy structure. For multi-components systems, as evidenced in Fig.~\ref{fig:Hull-description}, when the number of components is larger than four, it is no longer convenient to try and visualize the system graphically. Although one may, in principle, try to reduce the number of dimensions employing mathematical reduction of dimensions or exploring fixed compositions by mapping via function composition. In practice, this has never been attempted in the case of hydrides.

Furthermore, it is, of course, reasonable to ask what the accuracy of a prediction based on convex hull formation energies is. 
For points (structures) which correspond to clear minima of the formation energy, 
like  H$_2$O, which has formation energy of a few eV/atom, it is arguable that this composition is the most stable. However, when one deals with challenging compositions and complex structures under high pressure, where configurations are energetically very close, it is much more problematic to discern whether or not a structure will form based on its formation enthalpy. What is the limit where we could consider stability or metastability for secondary phase structures or compositions? 

Sun et al.~\cite{Ceder_thermoscale_2016_SciAdv} were the first to address this question and suggested an energy differences of the order of 100\,meV/atom among polymorphs of the same stoichiometry.  
This criterion is based on differences in properties such as entropy, volume, or surface energy and convoluted information from pressure, temperature, or the surface area between two systems. An energy difference of 100\,meV/atom is sufficient to overcome a $\Delta $S of 10\,J/mol\,K, a $\Delta$V of 2\,\AA$^3$/mol, or a $\Delta \gamma$ of 2\,J/m$^2$. More recently, Aykol et al.~\cite{aykol_2018thermodynamic} also evidenced the lack of a rigorous metric to identify which compounds may or may not be synthesized, and suggested a thermodynamic upper limit on the energy scale, above which the laboratory synthesis of a polymorph is highly unlikely. They defined this limit on the basis of the amorphous state and validated its applicability by effectively classifying more than 700 polymorphs in 41 common inorganic material systems of the {\sc Materials Project} for synthesizability. 

These estimates are relatively safe for drawing conclusions at zero pressure, and for cases analogous to those contained in the original dataset, however, it could be highly risky to extrapolate them to the high-pressure regime. In fact, both works reported that these values are strongly chemistry-dependent. In particular, the data sets of materials studied in the two works above do not contain hydrides under pressure. In hydrides under pressure, the existence of polymorphs stabilized via quantum effects (vibrational entropy) and the anharmonic effects of the hydrogen-related modes make predictions further challenging. 

\subsection{Artificial intelligence and machine learning}

Material scientists seem to be gradually embracing the inclusion of advanced computational methods such as artificial intelligence and machine learning in their research. The number of reported applications in material science grew at an extraordinary rate 
(see for instance Ref.~\cite{mueller2016machine,ward2017atomistic,Togo_book_2018}). In hydrides and materials science under pressure, this approach is still mostly under development. However, it is expected that, in the future, these methodologies will lead to a tremendous progress. 

In structure prediction these methods can be used at two fronts: 
A) crystal structure prediction via deep learning~\cite{deep_learning_2015}. Neural-network models can be trained to effectively distinguish chemical elements based on the topology of their crystallographic environment/properties~\cite{xie2018crystal}. This approach has been used to effectively guide the synthetic efforts in the discovery of new materials, especially in the case of systems composed by three or more chemical elements~\cite{deep-learning_ryan2018crystal}. B) Machine learning can also be employed at the level of structural exploration to speed-up the searches by learning inter-atomic model potentials~\cite{behler2016perspective,podryabinkin2019accelerating}. 

In principle, the construction of a generalized convex hull is a type of problem perfectly accessible for machine learning~\cite{faber2015crystal} and neural networks~\cite{ye2018deep}. Forces are relatively straightforward for machine learning, and in the chemistry community, there is ample evidence that this is possible for molecular systems. Once accurate forces are learned; also vibrational properties (phonons) are accessible. 
The transfer of these technologies to the solid-state community is underway and, likely, progress in the field of periodic solids at high pressures will soon occur. 

Other extraordinary examples in the solid-state community are those led by Curtarolo and his team that exploits repository of {\it ab initio} calculations combined with the Quantitative Materials Structure-Property Relationship models to predict properties such as metal/insulator classification, bandgap energy, bulk/shear moduli, Debye temperature and heat capacities~\cite{curtarolo_universal}. 

Finally, in the field of superconductivity, there have been recent efforts to tackle this puzzling problem irrespective of the underlying mechanism of superconductivity~\cite{Olle_klintenberg2013possible,owolabi2015estimation,norman2016materials}, 
for instance to machine learning models of superconducting critical temperatures~\cite{ML_superconducting_curatolos_2018}. 
In the field of hydrides under pressure a recent work correlated structural fingerprint, primary characteristics in the energy landscape of H$_3$S and H$_3$Se with their superconducting properties. This type of correlations between crystal structure and electronic structure serve as fundamental steps to further employ machine learning and neural network material design~\cite{mehta2019high,flores_accelerated_2017}. Generally, we see the field of hydrides under pressure as green and vast terrain ready to flourish, on which modern computational techniques such as deep learning and statistical models can play a significant role (see Sec.~\ref{Sec:Perspectives}, perspectives). 
\subsection{First-principles calculations}

Quantum mechanical modelling using density functional theory~\cite{KohnSham_PR1965} allows investigating the electronic structure of many-body systems, in particular of atoms, molecules, and their condensed phases with relatively affordable computational cost and high accuracy. 
Nowadays, DFT is perhaps the primary computational tool in terms of sheer number of users worldwide and represents one of the driving forces behind many applications in physics, chemistry and materials science. More than 20,000 publications using results based on DFT are published every year. 

DFT is increasingly used in an automated fashion to construct large databases or repositories of {\it ab initio} calculations or combined with multiscale techniques, limiting the direct human intervention. We estimate that more than 100,000 calculations are executed every day worldwide, producing more than 36 million DFT total energies a year. This has been possible thanks to the existence licensed and free-to-use codes, which are well documented, easy to use and efficient.

A substantial number of DFT computer codes are available, and many of them differ considerably in the implementation details; this means that each of them tend to attain a different ``precision".
However, reproducibility of DFT results generated with different codes and approximations was recently addressed based on different benchmarks by an extensive worldwide collaboration. Studies of this type are critical to reassure the scientific community on the reliability of DFT results~\cite{DFT_Delta-test_Science_2016}. The study convincingly showed that despite the large number of codes, basis sets and different implementations solving the same equations, all codes lead to the same results, of course within a threshold error. Thus the community has a reliable and extensive array of available first-principles DFT codes (see for instance Ref.~\cite{Reference_wiki_dft-codes}). %Irrespective of the code, 
All of them can provide the DFT quantities used in crystal structure prediction that are, primarily, the total energy of the system and the forces acting on the ions, together with stresses on the unit cell in the case of periodic crystals. 

\paragraph{Other methods}

Structure prediction for condensed matter systems containing a large number of atoms (more than hundreds) is at present out-of-reach at the DFT level, even with the most efficient basis sets. If a many-atoms system is studied, it is preferable to decrease the computational cost for estimating energies and forces by one order of magnitude or more. There are packages such as GULP~\cite{gale2003general} and LAMMPS~\cite{GULP,LAMMPS} (commonly used in the solid-state community) that implement accurate force-fields (FF), which permit to reach these goals. The term "force-fields" refers to the functional form of interatomic potentials used to calculate the potential energy of a system of atoms. The parameters of the energy functions may be derived from experiments in physics or chemistry, {\it ab initio} quantum mechanical calculations, or both. 
The obvious downside in the use of force-fields is the availability of transferable interatomic potentials for all atomic species under consideration. The density functional based tight-binding (DFTB) method represents an alternative method which combines the accuracy of DFT with the scalability of force-fields. It is based on a second-order expansion of the Kohn-Sham total energy in DFT to charge density fluctuations. The zeroth-order approach is equivalent to a standard non-self-consistent (TB) scheme, while at second-order a transparent, parameter-free, and readily calculable expression for generalized Hamiltonian matrix elements can be derived~\cite{TBDFT}. 
However, also, in this case, one needs parameters, the so-called Slater-Koster files (the matrix elements of the Hamiltonian operator and the overlap between basis functions centred on two atoms). Codes available with DFTB are {\sc CP2K}, {\sc DFTB+}, {\sc AMBER }, {\sc Gaussian}, among others. 

\begin{figure*}[t!]
\centering  
 \includegraphics[width=2.0\columnwidth]{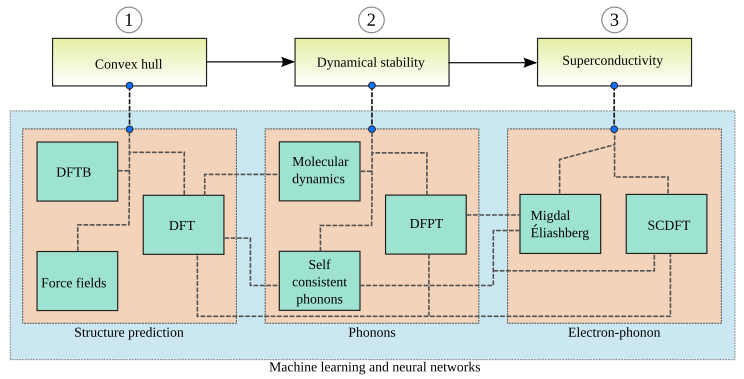}
  \caption{~Computational steps used to predict superconductors at high pressure. 
{\bf The first step} is to explore the configurational space of selected compositions and build the convex hull. 
Many structure prediction algorithms are available and can be used at various levels of accuracy from classical to quantum mechanics. 
{\bf The second step} is verification and consists of characterizing the vibrational properties of the thermodynamically-stable compositions. Available methodologies include among other MD, DFPT, and self-consistent phonons. 
{\bf The third step} is accessing the superconducting properties for candidates. The great effort by the community has made it possible to predict room temperature superconductors without prior experimental knowledge. 
The extensive palette of available methods reflects the maturity of the field and further development of these methodologies will guarantee many more exciting discoveries. 
Presently the bottleneck is the computational cost involved in state of the art simulations. 
Strategies to reduce it are being investigated; the most promising appears to be artificial intelligence approaches.}\label{fig:comp_steps}
\end{figure*}

\paragraph{Basis sets}

In Hartree-Fock or DFT methods the electronic wavefunction is expanded on a basis set to transform the partial differential equations (see Sec.~\ref{Sec:Theory}) of the model into algebraic equations suitable for efficient implementation on a computer. The basis set used in a DFT code will largely determine its use and applicability to a particular class of problems and hence target a specific scientific community. Historically, Slater-type orbitals (STO) were the first type of basis sets available. However, STO proved to be extremely inefficient because required cumbersome numerical calculation of integrals. STO was followed by Numerical Atomic Orbitals (NAO) and basis sets based on the Linear Combination of Atomic Orbitals LCAO ansatz which provided a much better trade-off and ease of implementation. Gaussian-type orbitals (GTO) are by far the most often used in terms of numbers of codes and users, especially in the quantum-chemistry community. This type of basis set allows efficient implementations by exploiting the property that the product of two GTOs can be written as the linear combination of GTOs. Thus integrals with Gaussian basis functions can be written in closed form, and this leads to a considerable saving of computational time. In addition to localized basis sets, plane-wave basis sets can also be used in quantum-chemical simulations, and find vast applicability in periodic solids. Plane waves are in principle, a complete basis set. However, in practical implementations, the Hilbert space is truncated and only a finite number of plane-waves, typically defined by a kinetic energy cut-off,  are included in the basis. By increasing the number of plane waves the solution is guaranteed to variationally converge to the exact one. In order to obtain convergence with a reasonable number of basis functions, in plane-wave codes, the full electron-ion Coulomb potential is usually replaced by a pseudopotential, which cuts all small-scale oscillations of the nuclear wave function close to the nuclei (see next section). Analogously to the plane wave basis sets, where the basis functions are eigenfunctions of the momentum operator, there are basis sets based on eigenfunctions of the position operator, that is, points on a uniform mesh in real-space. Actual implementations of this idea may use finite differences, finite elements or Lagrange sine-functions or wavelets~\cite{wavelets_2008daubechies}. 

Other mixed-basis sets, such (linear) augmented plane waves, have proven to be able to achieve rigorous accuracy in solids and are used as standard benchmark providing reference values~\cite{Andersen1975_Linearmethods,Gulans_muHa-prec_solids_2018,Elephant_paper_2017}. All-electron, full-potential (opposed to pseudopotential) approaches are in principle the most suitable to be used in high-pressure calculations, where the definition of a cut-off radius for the pseudopotential construction may be problematic. However, due to the expertise necessary to conduct these type of calculations, together with their elevated computational cost, they are inconvenient to be used for crystal structure prediction.

\paragraph{Pseudopotentials and benchmarks}

As alluded before, in the last years, the scientific community has made concentrated efforts to improve the efficiency and reliability of DFT codes. Plane-wave codes employ a pseudopotential approximation in which an effective potential replaces the full Coulomb potential between nuclei and electrons. This pseudo-wavefunction reproduces the correct behaviour of the valance electrons wavefunctions, in the region where the chemical bonds are formed, but cuts out all short-range oscillations  close to the nuclei, i.e. below a suitably-chosen cutoff radius. Core electrons are not included explicitly in the calculation, and this permits to reduce the number of plane-wave basis functions used to represent the density and the potential in the Kohn-Sham equations. Thus the pseudopotential approximation introduces an additional source of inaccuracy compared to all-electron approaches. However, for most solids, most of the pseudopotentials employed nowadays have achieved an accuracy practically comparable to that of reference all-electron calculations.
This was possible thanks to years of pseudopotential developments, which have substantially increased the accuracy compared to the early norm-conserving pseudopotentials. A substantial step forward in the wide-spread use and improvement of pseudopotentials has been the development of website platforms in charge of producing and distributing standardized pseudopotentials, intensively tested and open to public scrutiny. The first in this effort was the {\bf PseudoDojo.org} website, which gives access to the latest released version of pseudopotentials~\cite{pseudodojo_2018}. Another standard protocol proposed later on to verify publicly available pseudopotential libraries, based on several independent criteria including verification against the all-electron equations of state, as well as plane-wave convergence tests for several properties, including phonon frequencies, band structure, cohesive energy and pressure resulted in the Standard Solid State Pseudopotential (SSSP) efficiency and precision libraries~\cite{SSPS_marvel_2018}. 

\paragraph{$k$-point sampling and other sensitive parameters of DFT calculations}
A method for Brillouin zone sampling, which allows for fast, yet accurate, estimates of the total energy is particularly important when dealing with crystal structural explorations. It has been demonstrated that non-converged DFT calculations can lead to substantial biases in structure searches since the quality of the $k$-sampling determines the final accuracy of forces.  The community primarily uses Monkhorst-Pack (MP)~\cite{Monkhorst_pack} meshes of $k$-points. A criterion that can be used, when comparing structures with different geometries, is to employ for each of them a MP mesh where the  smallest separation among the $k$-points is below a fixed threshold.
A rule-of-thumb, based on practical experience, is to use a grid spacing of $2\pi \times$ 0.07 \AA$^{-1}$ for structure searches and preliminary results, and $2\pi \times$ 0.04 \AA$^{-1}$ to refine structures and obtain good energies. These rules apply to systems with a gap in the electronic spectrum (insulators and semiconductors), while metals require denser grids, $2\pi \times$ 0.05 \AA$^{-1}$ for searching and preliminary results, and $2\pi \times$ 0.02 \AA$^{-1}$ to refine the structures and obtain reasonable estimate of total energy. For the cut-off energy in the plane wave expansions, preliminary tests have to be carried out at the beginning of the investigation; however, if one works with pseudopotential libraries where benchmark tests have been carried out before, it is advisable to start from these recommended values. 

\paragraph{The nitty-gritty of predicting superconductors at HP}
Finally, we would like to summarize in Fig.~\ref{fig:comp_steps} the computational steps which are regularly followed to uncover new hydride superconductors. One starts by picking a composition or set of compositions. This choice is usually based either on chemical or physical intuition. In the case of hydrides, stable H-rich molecules that exist at ambient conditions are, for example, a good starting point. 

The first step is to explore the configurational space of selected compositions and construct the convex hull. 
As described in previous sections, one can choose from a vast array of available methods. 
{\it Despite different claims, in reality, all the methods are equally good, and there is not a method/software 
that is superior to others}. Differences reside more in the tools that each package offers to process the data. 
Mainstream packages are robust, and users can certainly use them as black-boxes. On the other hand, less developed software packages typically provide only basic functionalities, perhaps less efficient implementations, that require the users to write their scripts for post-processing. Virtually all existing software packages for structure prediction are interfaced to DFT, DFT tight-binding or force-fields codes. In synthesis, the main aim in this step is to build a phase diagram (convex hull) to identify low-enthalpy compositions, once the PES has been sampled thoroughly with the method of choice.

The second step is a verification one, which consists in characterizing the vibrational properties of the thermodynamically-stable compositions. This step in most cases is performed using linear-response theory at DFT level, 
which assumes harmonic vibrations of the atoms around their equilibrium positions. 
Conversely, molecular dynamics or other methods to estimate anharmonic interatomic force constants are available. 
If phonon frequencies are real in the entire Brillouin zone, i.e. the predicted compound is dynamically stable, 
one can estimate the vibrational free energy and the zero-point energy contribution. 
Especially in hydrides, these terms have proven to have a significant impact on the energy ranking of configurations.

The third step consists in characterizing the superconducting properties of the best candidates. 
For this, one needs to estimate the electron-phonon coupling spectral function ($\alpha^2 F(\omega)$), 
the electron-electron Coulomb repulsion, and solve the superconductivity equations. 
For a first, approximate estimate of the transition temperature, a safe choice is to use empirical formulas 
for \tc\ such as the McMillan-based expression, Eq.~\ref{eq:McMillan}. 
However, as described in Sec.~\ref{Sec:Theory}, this approach becomes less accurate in the limit of sizeable electron-phonon coupling or in the case of systems with strongly anisotropic electronic properties. In this case, it is more appropriate to use other methods based on a perturbative Green's function approach such as the fully {\it ab initio} Migdal-\'Eliashberg approach 
(see Sec.~\ref{sec:eliashberg}, or advanced methods such as SCDFT (see Sec.~\ref{sec:SCDFTSetUp}). 

In order to conduct DFPT calculations and electron-phonon properties of superconductors, 
the two most common choice are the popular, free-to-use plane-wave basis set codes: 
{\sc abinit}~\cite{abinit_2016} and {\sc Quantum ESPRESSO}~\cite{Quantumespresso_2017}. 
Both can calculate also calculate higher-order derivatives and anharmonic (phonon-phonon scattering). 
A licensed code that is very popular and features excellent scaling together with a robust structural optimization routine is {\sc vasp}~\cite{VASP_Kresse} which also provides its well-tested pseudopotential libraries. 
Recently the use of graphical processing units (GPU) in conjunction with DFT calculations, has been proven feasible, resulting in the speed-up of computational time for systems with up to a thousand atoms~\cite{GPU_accelerated-paper_2018}. 
However, this requires state-of-the-art supercomputers (CPU+GPU) that are still not available in most supercomputer centres. 
Moreover, force-field codes such as {\sc GULP} or {\it ab initio} molecular dynamics methods are popular and accurate while being computationally efficient. These are often used to perform preliminary scans of the PES. 

In addition to the steps mentioned above, in Fig.~\ref{fig:comp_steps}, one can distinguish an additional box that encloses them. 
This large box symbolizes methodologies such as machine learning (ML) and neural networks, 
both belonging to artificial intelligence. 
The recent progress in the field of machine learning for molecular 
and materials science was reviewed by Butler et al.~\cite{butler2018machine}.

\section{Trends in hydrides under pressure}\label{Sec:Trends}

Historically, the first hint on the possible metallization of hydrogen dates back to 1926~\cite{Philip_Ball_Note_Nature-Materials}, when J. D. Bernal suggested that any element should become a metal at high enough pressures. In particular, for a diatomic molecule such as hydrogen, the intermolecular distance decreases with pressure, until a single atom cannot be assigned clearly to one molecule or another,
and the material becomes metallic, with all atoms approximately equidistant and all electrons uniformly distributed. Years later, in 1935, Eugene Wigner and Hillard Bell Huntington published the first paper in which they predicted that for pressure above 25\,GPa hydrogen would become an alkali metal-like solid~\cite{wigner1935possibility}. 
Decades later, in 1968, Ashcroft and independently Ginzburg in 1969~\cite{Ginzburg_1969}, evidenced a fundamental implication of hydrogen metallization, by predicting the possibility of high-\tc\ 
superconductivity~\cite{Ashcroft_PRL1968,Ginzburg_1969}, based on BCS theory~\cite{BCS_1957}.  

Ashcroft's prediction can be readily understood using the McMillan-Allen-Dynes expression 
for \tc\  (Eq.~\ref{eq:McMillanAllenDynes}),  
\begin{eqnarray}
~\hspace{1.cm}T_c=\frac{\omega_{\rm log}}{1.2}\exp\left[-\frac{1.04(1+\lambda)}{\lambda-\mu_c^*(1+0.62\lambda)}
\right]\label{eq:McMillan}
\end{eqnarray}
together with the simplified Hopfield's expression~\cite{AllenDynes_PRB1975,Th:Hopfield_PR_1969} 
for the \ep\ coupling parameter: 
\begin{equation}
\label{eq:lambdaHopfield}
\lambda=\frac{N_{F}\,I^2}{M \omega^2},
\end{equation}
obtained from Eq.~\ref{eq:a2F}, Eq.~\ref{eq:lambda_a2F} and Eq.~\ref{eq:elphME} for an Einstein mode of effective mass $M$ 
and frequency $\omega$. $N(E_{F})$ is the density of states at the Fermi level and $I=g\sqrt{M\omega}$, 
with $g$ the average \ep\ matrix element in Eq.~\ref{eq:elphME}, the deformation potential.

Certainly, a metallic phase of hydrogen would optimize several 
factors appearing in McMillan's expression: 
\begin{itemize}
\item The average phonon frequency, which appears as a prefactor in Eq.~(\ref{eq:McMillan}), 
would be high due to the light hydrogen mass.  
\item The matrix element $I$ would also be large, since hydrogen 
does not have core electrons that screen the Coulomb electron-ion potential.  
\item Assuming that hydrogen at HP crystallizes in a densely-packed structure, 
as initially postulated by Wigner and Huntington, then electronic density and
the density of states at the Fermi level, $N(E_{F})$, would be large. 
The latter would imply a large \ep\ coupling ($\lambda$), 
which appears in the exponent of Eq.~\ref{eq:McMillan}. 
\end{itemize}

Although the suggested pressure to metallize hydrogen at that time was merely 25\,GPa (1935), it was far beyond reach for the technology of the epoch (see Fig.~\ref{fig:HP-story}). As more accurate calculations and experiments were performed,  this estimate was quickly outdated and replaced by higher values, i.e. the pressure threshold was shifted first to 1 megabar and subsequently to 3 megabar (300\,GPa). In fact, as discussed in the experimental section, the search for metallic hydrogen was one of the main impulses to develop high-pressure techniques. 
The pressure to transform H to a solid metallic substance is nowadays unambiguously set to be above 450\,GPa. 

In the '70s, the idea that hydrogen could achieve high superconducting critical temperatures, as explained in the Introduction, was simply too attractive and this stimulated the study of superconductivity in metal hydrides throughout the whole decade. One can clearly identify that the field of superconducting hydrides was born at the beginning of the '70s after Ashcroft's bold prediction of high-\tc\ in elemental hydrogen. 
The rationale of the epoch was to focus on hydrogenated intermetallic compounds~\cite{mueller_metal-H_ed1968} hoping to find evidence of the exciting idea of hydrogen metallization. As a side note, we recommend the reader to read the Review of Maksimov and Pankratov of 1975~\cite{Maksimov_1975}, 
which is an excellent piece of work that provides a deeper detail on the development of metallic hydrides during the '70s. The first metal-hydride ever synthesized and confirmed to be superconductor was Th$_4$H$_{15}$ in 1970, which at atmospheric pressures had a non-negligible \tc\ of 7.6\,K~\cite{Satterthwaite_ThH_1970}. This compound is also special historically because according to our bibliographic investigation, it was also the first metal-hydride to be a subject of high-pressure studies, back in 1974. An increase in \tc\ to 8.5\,K was meticulously measured upon compression~\cite{Th4H15_pressure_supra1974}. 

The discovery of high superconducting transition temperatures in palladium~\cite{PdH1,PdH2} 
and other noble metal-hydrogen systems further fueled the interest in this problem~\cite{oesterreicher1976studies} 
(see Fig.~\ref{fig:HP-story}). We recall that the max \tc\ at that time was $\approx$19\,K for NbGe. 
As a matter of fact, the highest \tc\ ever measured and confirmed 
for a hydride material at zero pressure is 16.6\,K, measured in 1974 for PdCuH$_x$~\cite{PdCuH_Stritzker1974_16K}. 
Other hydrides subject of studies were HfV$_x$H$_x$ (\tc\ 4.8\,K)~\cite{HfVH_PRL_1976}, 
niobium-hydrogen (\tc\ 9.4\,K)~\cite{NbHx_Welter1977} 
and hydrides of vanadium, zirconium, titanium and lanthanum, for which superconductivity was not reported, 
at least in the temperature range explored~\cite{ponyatovskii1985new,tonkov1998compounds}.  

Important developments and predictions took place during the first hydride rush~\cite{Maksimov_1975}. We believe that the origin of {\bf chemical precompression} dates back to 1971 and should be attributed to J.~J. Gilman, who studied the possibility of making a new form of hydrogen in a metallic state through the preparation of a covalent compound, LiH$_2$F, under pressure~\cite{Gilman_1971_PRL_LiHF}. In his very instructive work, Gilman described how the reduction of the distance between molecules under pressure could be reached in lithium dihydrogen fluoride. 

The second hydride rush was triggered in 2004 by Ashcroft~\cite{Ashcroft_PRL2004}, who
proposed to look for new superconductors in molecular hydrides such as SiH$_4$ and SnH$_4$. Although the idea was not new, Ashcroft's paper provided novel perspectives and revived the field. 
Important concepts to guide the search for superconducting materials were re-introduced. First, in order to obtain an effective metallization of the hydrogen sublattice, it would be more convenient to start from an existing hydrogen-rich molecule, since in this case hydrogen does not have to be incorporated into a host metal lattice, as it was done in the '70s. Second, a chemical precompression (as Gilman suggested in 1971) mechanism can be exploited. In molecular hydrides, the characteristic distances between hydrogen atoms are reduced, as compared to the large values found in metal hydrides. Thus, covalent systems are of prime interest (as LiH$_2$F).  Third, the chemical precompression further reduces the pressure necessary for metallization and, more importantly, this pressure readily accessible to experiments. %!!

Eremets gave the successful proof-of-concept of these ideas in 2008. His team reported the metallization of silane (SiH$_4$) at 50\,GPa and the subsequent appearance of a superconducting phase with a maximum \tc\ of 17\,K at 120 GPa~\cite{Eremets_Science2008}. 
Despite this initial success, the field was not yet ready to wake up. 
It was only in 2015, when Eremets' team reported convincing evidences 
of high-temperature (\tc $\simeq$203\,K) superconductivity at HP (P$\gtrapprox$ 170\,GPa) 
in H$_x$S~\cite{DrozdovEremets_Nature2015}, that the field truly erupted~\cite{mazin2015superconductivity}.  
The field has evolved at such an incredible pace that in less than three years a 250\,K superconducting transition 
in LaH$_x$ was reported by two independent teams~\cite{drozdov2018_250,Hemley-LaH10_PRL_2019} 
and novel claims in systems such as Y-H and Th-H have also appeared. 
Computational and theoretical predictions played a central role behind this great success in discovering high-\tc\ materials. 
As shown in Fig.~\ref{fig:size_of_field}, the number of works published in the field increased dramatically during the last years, especially computational studies. It is clear that this field is flourishing and entering a new era, the third hydride rush. We expect that by the end of the next decade the number of publications will be well above 10,000. 

\subsection{Periodic table of binary hydrides}
    
\begin{figure*}[ht!]
\begin{center}
\includegraphics[width=1.55\columnwidth]{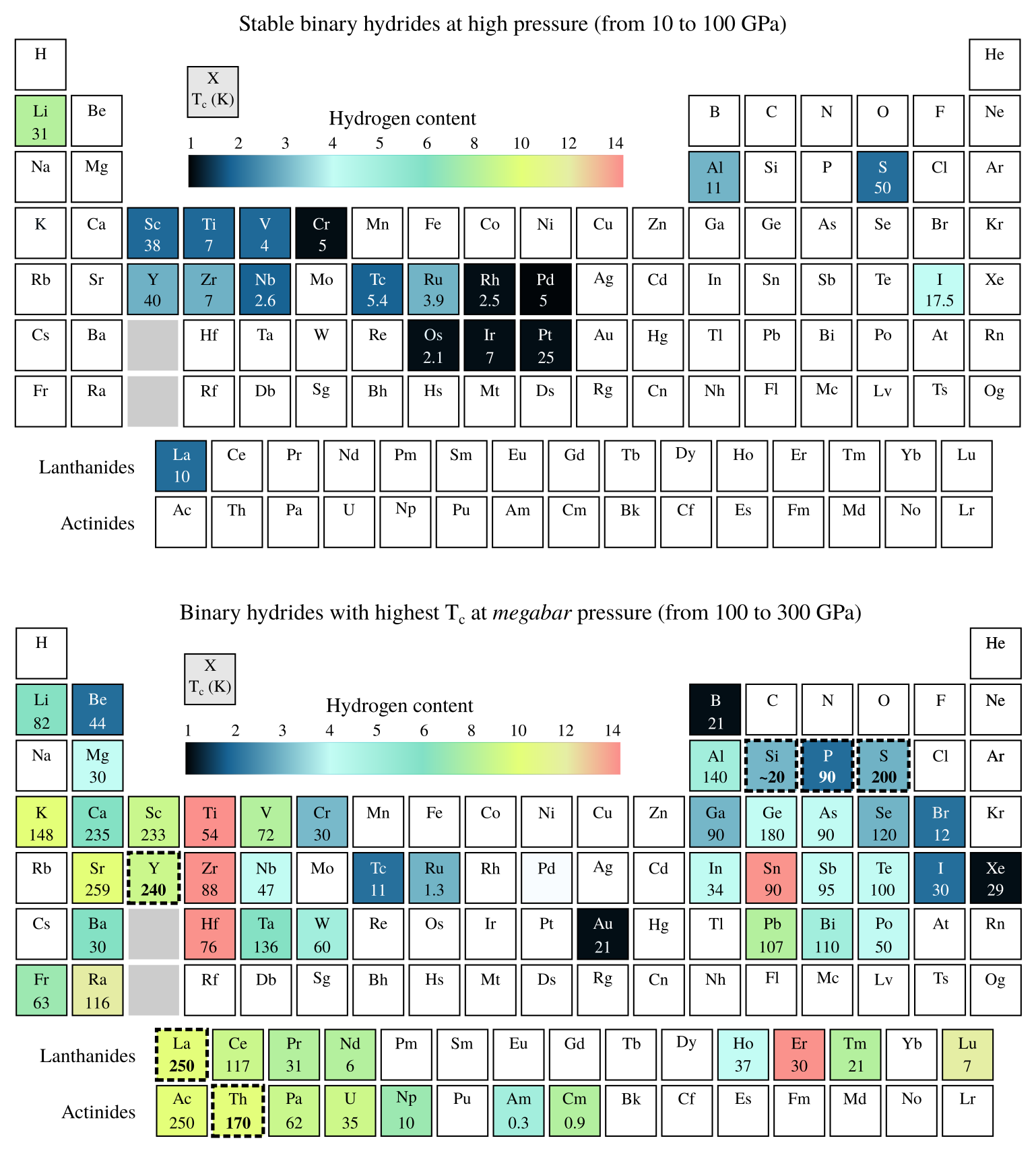}
\end{center}
\caption{~Computational predictions reported in the literature on binary hydrides. 
Top periodic chart prioritize stability: \tc\ is reported for the most stable configuration 
(lowest energy structure and composition in the convex hull) in the 10-100\,GPa. 
The periodic bottom chart highlights highest \tc, independently of weather composition and structure are stable in the convex hull. 
In both panels, the number below the symbol element displays \tc\ and the colour scale correlates to hydrogen content. 
A dash-dotted in some elements indicates that the compound has been subjected to experimental work.}\label{fig:Periodic-table_hydrides}
\end{figure*}

As mentioned before, the number of studies involving hydrides on the field of superconductivity has considerably increased. In little less than five years (2014-2019), the community has conducted computational explorations on practically all binary hydrides. A convenient form to summarize these results is to show the corresponding periodic table of superconducting hydrides. The following sections describe superconducting hydrides, 
as well as binary hydrides that do not show superconductivity. 

   \subsubsection{Superconducting hydrides} 

In order to disentangle the relatively large and uncorrelated information that the community has published on hydride superconductors under pressure, first, we would like to distinguish two regimes of pressure.  
From 10 to 100\,GPa (relatively low pressure) and above 100\,GPa up to 300\,GPa (megabar pressures). The collected data is shown in Fig.~\ref{fig:Periodic-table_hydrides} in the form of two periodic tables, reporting for each element-hydride the predicted \tc\ and chemical composition. The numbers below the element symbols indicate the value of \tc, while the colour of the box refers to the chemical composition. The darkest colors (black and blue) correspond to a low H content (1, 2 and 3). 
The top periodic-table in Fig.~\ref{fig:Periodic-table_hydrides} reports \tc\ for the thermodynamically stable composition (lowest energy structure and composition in the convex hull) in the 10-100\,GPa range. 
The bottom periodic-table in Fig.~\ref{fig:Periodic-table_hydrides} 
shows the highest \tc\ in the 100-300\,GPa range, independently of whether the composition and structure are stable in the convex hull. See Appendix for references used to construct these tables (Sec.~\ref{Sec:Appendix}).

Let us first focus on the low-pressure regime. The highest \tc\ reported in this range of pressure is 50\,K for H$_2$S. We observe that at low pressures, the hydrogen to guest-atom ratio is dictated by the valence of the guest atom and superconductivity is found in three categories: 
Hydrides that show critical temperatures of the order of 40\,K or higher are, besides H$_2$S, the ones formed by Sc, Y and Li. These compositions are thermodynamically stable and in principle should also be accessible to experiments. The hydrides formed by I, La and Pt show superconductivity below 20\,K. Interestingly, there are not many stable compositions in the low-pressure range as one could expect, and the total number of stable hydrides is merely 19. The absence of superconducting hydrides formed with group 1 (alkali), 2 (alkaline earth metals), post-transition metals groups 11, 12, 13 and group 14 elements is noticeable. 

On the other hand, most of the transition metals tend to form hydrides 
that are stable in the low-pressure range. 
At low pressures, the stability of hydrides is dominated by compositions with {\bf low H-content}. One can notice that darker colours dominate the periodic table in the top panel.

Let us now examine the HP regime. The bottom periodic-table in Fig.~\ref{fig:Periodic-table_hydrides} shows the highest \tc\ in the 100-300\,GPa range. Compared to low pressures, the most noticeable features are: a) there are many more hydrides reported with considerable higher \tc\ and b) the table is cheerfully-coloured, indicating that at high-pressure hydrides with {\bf high H-content} are favoured. The highest \tc\ predicted in this range of pressure is for YH$_{10}$ with 326\,K (not shown). However, recent experiments have set the experimental \tc $\approx$240\,K (see Appendix for references). 

Note that, due to our choice to restrict the table to chemically-accessible pressures, some high-\tc\ predictions such as Mg-hydrides are not included.  It is fascinating that there are also many more hydrides predicted with astonishingly  
high-\tc, including the whole La-series and group 2. On the opposite, transition metals seem doomed in the HP range and neither \tc\ nor the H-content increase compared to low pressures. This evidences that the strategy followed during the '70s to maximize \tc\ was bound to fail. 

A word of caution should also be added at this point: these tables contain predictions extracted from a vast collection of published articles from many groups. Not all the studies reported in these articles were conducted with the same level of accuracy, convergence etc. By no means, the numbers reported here are guaranteed to be conclusive for a given compound (see for instance Ref.~\cite{zunger2019beware}). Therefore, we cannot guarantee specific predictions and the use of these tables should be restricted to discuss trends among the elements. Circumstances or cases in which theory/computation fails are well documented, and the reader can find them discussed elsewhere. Specifically, rare-earth-metals and actinides are theoretically challenging to describe, and one can quickly obtain wrong results applying standard density-functional theory. 

Moreover, we can observe that the megabar range of pressures is substantially richer, superconductivity appears in most regions of the periodic table, including noble gases (Xe) and elements belonging to groups 14, 15 and 16. There is a broad spread of critical temperatures from low to high, as well as in chemical compositions, from low to high H content. The high pressure allows overcoming enthalpy barriers to form bonds and stabilize several {\bf superhydrides}, i.e. systems with a high-H content ($>$~6). 

\subsubsection{Non-superconducting hydrides}\label{Sec:Trends_Nosupra}

At high pressure, superconductors are eclipsing other materials that either do not metallize or do not show superconductivity, and yet form stable compositions. 
There is a clear trend in this sense, as elements of the second period such as C, N, O, F and 3$d$ transition metals such as Mn, Fe, Co, Ni and Zn do not form metallic hydrides under pressure. 
Notably, H$_2$O, which is perhaps the most abundant and best-studied hydride, 
remains insulating up to the terapascal range of pressures~\cite{H2O_terapascal_PRL,PhysRevB.84.220104,PhysRevB.87.024112,PhysRevLett.110.245701,RevModPhys.84.885}. 
Noble metals (Cu, Ag, and Au) show an outstanding resilience to hydride formation at ambient pressure. 
Experimentally there are no reports of stable compounds with a hydrogen molar ratio $> 1$ 
at room temperature~\cite{donnerer2013high,Cu-H_binns2019structural}. Through high pressure and {\it in situ} 
laser heating, H can eventually be incorporated in ratios higher than 1, but these compounds usually remain either magnetic~\cite{antonov2002phase} or/and insulating at HP~\cite{donnerer2013high}.

Studies of hydrides at HP dates back to 2003, 
when Eremets et al. studied B$_{10}$H$_{14}$~\cite{eremets2003exploring} and its transformation under pressure. 
Other systems studied for their potential transformation to a metallic substance are:  
He-H~\cite{He-H_Gonchy_shit}, 
Li-H~\cite{H:Pepin_PNAS_Li_2015},
Na-H~\cite{H:Struzhkin_Na_Natcomm_2016}, 
H-O~\cite{ponyatovsky1992pressure,goncharov1996compression}, 
AlH$_3$~\cite{AlH_experiment_PRL_2008}, 
Fe-H~\cite{PhysRevLett.113.265504,H:Pepin_FEH_2017,H:Heil_FEH_2018,FeH5_meier2019pressure,PhysRevB.97.214510}, 
Co-H~\cite{Co-H_wang2018high}, 
Ni-H~\cite{NiH-hydrides_under_pressure_exp,Ni-H_PRM_2018}, 
Si-H~\cite{PRL_SiH4,PhysRevB.83.144102,Eremets_silane_2008}.
Cu-H~\cite{Cu-H_binns2019structural}, 
SeH$_3$~\cite{Gonchy_PRB_SeH3-2018}, 
Nb-H~\cite{Nb-H_experiments_2017}, 
Rh-H~\cite{Rh_H_exp_li2011rhodium}, 
I-H~\cite{I-H_experiments_2018_gregoryanz}, 
Ta-H~\cite{Ta-H_PRB_exp}, 
Ce-H~\cite{Ce-H_salke2018synthesis}, 
Eu-H~\cite{Eu-H_Matsuoka_2011}, 
La-Ni-H~\cite{La-Ni_hydrides_1976} 
and other rare earth trihydrides under high pressure~\cite{REM-H_MENG201729344}. 
These systems were experimentally studied under pressure, and phase transformation occurred, 
but no clear evidence of either metallicity or superconductivity was given, or transport measurements were not carried out. 
We should not rule out the possibility that under other thermodynamic conditions, some of these systems may show superconductivity. 

%-----------------------------------------------------------------------------
%----------------------------------------------------------------------------- que bello.... me gusta 
%-----------------------------------------------------------------------------

\subsection{Overview of selected hydrides}

Within the broad classes of low and high-pressure hydride superconductors, several classification schemes have been proposed in the literature, among others, we would like to  mention studies based on structural motifs~\cite{Papaconstantopoulos_2019}, inter-atomic distances, guest to host atom ratio~\cite{sennikov1994weak,Pickett_2019_PRB}, 
valence band analysis~\cite{PRB_2015_Abe-Ashcroft}, 
electron localization function, among others~\cite{gupta1981trends,burger1984electron,Review_Zurek-2018,semenok2018distribution,wang2018hydrogen,Review_Oganov-Pickard_2019,Duan_2019_review,kim_general_2010}.  The main scope of these analyses has been to identify general trends to devise material optimization strategies. It is, of course, desirable to reach a general classification of hydrides. However, these types of global perspectives may oversimplify reality and eventually turn out to be misleading.  

One may realize that even two important theoretical arguments, such as the Ashcroft-Ginzburg (metallic hydrogen) and the Gilman-Ashcroft (chemical precompression) one offer an oversimplified picture of hydrides. The precompression paradigm assumes that the guest atom would only be exerting "pressure" on the hydrogen sublattice but not alter bonding or chemical properties. In this perfect scenario, superconductivity would be dominated by H states and occur at very high-\tc\ (as it was predicted for solid metallic hydrogen). In practice, hydrides are incomplete realizations of this concept: 
guest atoms affect the {\it chemistry} of hydrogen and provide additional electronic and vibrational states strongly entangled with those of hydrogen, ultimately playing a non-trivial role in superconductivity. 

\begin{figure*}[t]
\centering
\includegraphics[width=1.7\columnwidth]{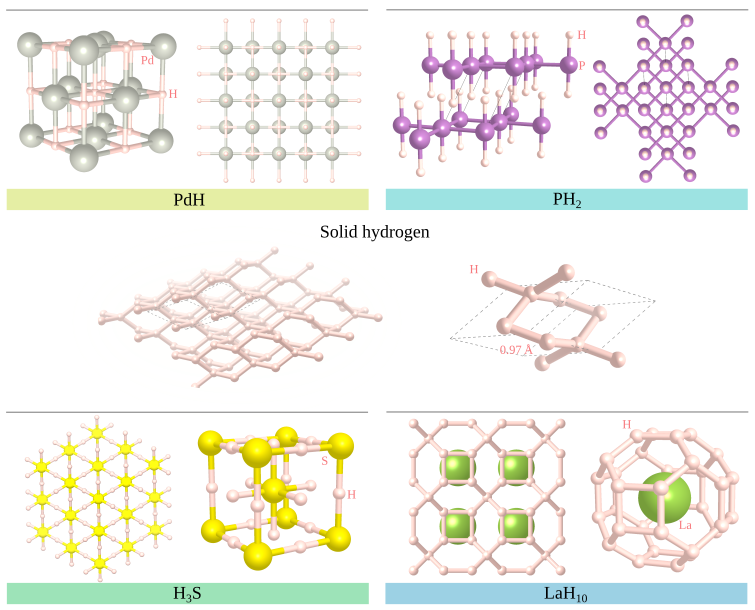}
\caption{~Crystal structures of representative binary-hydrides and elemental hydrogen. PdH: Metal hydride at ambient pressure (0\,GPa).   
PH$_2$: a molecular  hydride at intermediate pressure (120\,GPa). 
H$_3$S: a covalently-bonded hydride at high pressure (200\,GPa). 
LaH$_{10}$: a superhydride at very high pressures (300\,GPa) 
and H$_2$: solid (atomic) metallic hydrogen at megabar pressure (500\,GPa).}\label{fig:Crys_struc}
\end{figure*}

In this Review, we do not attempt to derive a general trend for hydrides or propose any classification scheme. Instead, we identified four binary hydrides that cover a spectrum of pressure and hydrogen contents 
and exemplify the increasing role that H plays in superconductivity with pressure. In the following, these are treated as representatives of different hydride classes. In addition to these systems, we consider metallic hydrogen, in the so called atomic (non-molecular) phase. 
Crystal structures of the representative systems are shown in Fig.~\ref{fig:Crys_struc}, and discussed in order of {\it increasing} hydrogen content: 
\begin{itemize} 
  \item[I)] PdH, representing metal hydrides at ambient pressure (0\,GPa).
 \item[II)] PH$_2$, an example of metallic molecular crystals of covalent hydrides at intermediate pressures (120\,GPa).
\item[III)] H$_3$S, represents covalently-bonded hydrides at high pressure (200\,GPa). 
 \item[IV)] LaH$_{10}$, representing superhydrides at very high pressures (300\,GPa).  
  \item[V)] Solid metallic hydrogen, a non-molecular structure stable above 500\,GPa.
\end{itemize}
For the representative compound in each class, we will address the following questions: How does this material form? What are its main structural features? How good is it as a superconductor? What is the connection between its \tc\ and specific aspects of its electronic structure, such as bonding, electronic and phononic band dispersion? What is the role of hydrogen? 

To address these questions, we will analyze for each material the basis of its electronic structure, bonding properties, phonons and electron-phonon spectral functions, extracted from the best existing literature. 

Superconductivity is discussed with the help of the McMillan-Allen-Dynes Eq.~\ref{eq:McMillan} using \mus=0.1.  
Such an approach avoids the complexity of a full-scale theory (Sec.~\ref{Sec:Theory}), 
while not quantitatively excellent, it is sufficiently reliable to discuss trends and capture physical effects. 
The \ep\ coupling $\lambda$ is decomposed with the help of the 
Hopfield Eq.~\ref{eq:lambdaHopfield}, with  $\omega$=\omlog .

For our analysis, it is crucial to have an understanding of the chemical bonding. We studied it based on the Crystal Orbital Hamilton Population (COHP) introduced by Dronskowki and Bloechl in 1993~\cite{DFT:COHP_Dronskowski_JPC1993}.\footnote{The COHP method permits to decompose the complicated energy-dependent electronic structure of a material into a sum of pairwise interactions between atoms, in a local basis framework, like tight-binding. The interaction between two atom-centred orbitals (say $|\phi_{\mu}>$ and $|\phi_{\nu}>$) is described by the Hamiltonian matrix element $H_{\mu, \nu}=<\phi_\mu|\hat{H}|\phi_\nu>$, which, when multiplied by the density-of-states matrix, measures the bonding, anti-bonding or non-bonding character of the corresponding electronic states if it lowers, rises or does not change the band-structure energy, respectively.  This simple real-space interpretation of otherwise abstract reciprocal space information contained in the band structure, can result in a useful descriptor to understand the physical and chemical mechanisms through which (high-\tc) superconductivity is realized in different hydrides.}
This type of analysis is extremely useful because it permits to decompose the complicated energy-dependent 
electronic structure of a material into a sum of pairwise interactions between atoms, in a tight-binding like local framework, and hence identify which atom pairs participate in a given bond and whether their contribution is of bonding, antibonding or non-bonding nature. 
In addition, the COHP encodes information on the
band dispersion, not included in other types of real-space bond analysis, and hence can be easily connected to the two electronic factors in the Hopfield formula (Eq.~\ref{eq:lambdaHopfield}). 

To understand the role of phonons in the coupling, we will rely on the \'Eliashberg spectral function (Eq.~\ref{eq:a2F}) and its two moments \omlog\ and $\lambda$,  Eqs.~\ref{eq:omlog} and Eq.~\ref{eq:lambda_a2F}, respectively. 
The \'Eliashberg function is like a phonon density of states weighted by electron-phonon matrix elements, 
therefore high peaks on the spectrum are associated with stronger coupling.

\subsubsection{Palladium hydride}\label{PdH} 

\paragraph{Background}
After Ashcroft's suggestion of possible high-temperature superconductivity in metallic hydrogen~\cite{Ashcroft_PRL1968}, the hope to induce or increase the superconducting critical temperature of metals through hydrogen alloying lead to the discovery of superconductivity in Th~\cite{Satterthwaite_ThH_1970}
and Pd~\cite{PdH1, PdH2} metal hydrides. Hydrogen in these systems is at a relatively high concentration so that these hydrides can be thought of as alloys in which hydrogen (guest) and metal (host) atoms are both embedded in a common metallic environment. 

Elemental palladium is a transition metal which is not superconducting at ambient pressure. Interestingly, its moderate \ep\ interaction ($\lambda=0.36$) and characteristic frequencies (\omlog=$180 K$), would lead to a small, but non-zero superconducting \tc . However, superconductivity is suppressed by paramagnetic spin fluctuations, with a coupling constant $\lambda_{SF}$ comparable to the \ep\ one~\cite{Savrasov_ep-PRL-1994,Th:Berk_Schrieffer_PRL_1966}. 

Hydrogen incorporation results in \tc\ as high as 9\,K for Pd-H\cite{PdH2,PdH1}, indicating a low or intermediate \ep\ coupling. Indeed, the phase diagram of PdH$_x$ is quite complicated: the crystal is stable for $x > 0.03$ and superconductivity appears at $x\simeq$0.7 with \tc $\simeq$ 5\,K and further increases up to 9\,K with a seemingly linear dependence on the H content. The proper determination of the structure of the crystalline phase of PdH$_x$ at various concentrations is a crucial question. In Pd-$fcc$ lattice, there are two inequivalent interstitial sites, the octahedral site ($O$) and the tetrahedral site ($T$). 
While at $x$=1 both neutron experiments and first-principles calculations indicate that the $O$ site is stable~\cite{PdH-neutron,Alavi,houari_PdH-stability_JAP_2014}, a substantial $T$ occupation is likely to occur at lower H concentrations.

The existence of two different  tetrahedral or octahedral
absorption sites for hydrogen can be related to early~\cite{PdH-300K} 
and more recent~\cite{PdH-HTc-arxiv} reports of superconducting phases 
in PdH with very high-\tc\ (300 and 62\,K, respectively). 
Given metastable superconducting phases recently observed for other hydrides, the small energy difference between $O$ and $T$ site occupations can be accessed via thermodynamic conditions. Perhaps, slow or rapid quenching from the high-T phase may stabilize different a phase with substantial tetrahedral occupations, and improved superconducting properties. However, recent first-principles calculations~\cite{PdH-tetra} indicate that a full occupation of tetrahedral sites is detrimental for superconductivity since it decreases the electron-phonon coupling with H-derived modes, which is at odds with previous studies. 

A quite peculiar aspect of PdH is the existence of an inverse isotope effect. The isotope substitution of H with deuterium results in an increase of \tc\ up to 11\,K in PdD\cite{PdH2}. The inverse isotope effect relates to a strong anharmonicity of the phonon modes, which is also supported by resistivity, photoemission, inelastic neutron scattering and tunneling experiments~\cite{PdH-resistivity,PdH-photoemission,PdH-phonons1,PdH-phonons2,burger_1981_PdH_tunneling}.

Only recently the complicated physics of PdH was put on firm grounds by first-principles calculations of phonon frequencies and electron-phonon coupling which include the anharmonic phonon contributions non-perturbatively. 
These calculations permitted to reconcile severe discrepancies between theory and experiments in the structural stability and the superconducting properties~\cite{Errea_PdH_PRL2013}. 
In particular, the inclusion of anharmonic corrections on the phonon spectrum strongly renormalizes the frequencies of the optical branches, yielding a much better qualitative agreement with experiments and stabilizes the lowest acoustical branch, 
which is predicted to be unstable at the harmonic level. 
More importantly, anharmonicity correct the severe overestimation of the superconducting \tc\ found by harmonic calculations; 
\tc\ reduces from 47\,K (harmonic) to 5.0\,K for PdH and 6.5\,K for PdD (anharmonic). 
Calculations including phonon anharmonicity 
correctly reproduce the anomalous inverse isotope effect observed by experiments, 
although the calculated value of the isotope coefficient, $\alpha$=0.38, is about 50\% lower than the experimental one~\cite{Errea_PdH_PRL2013}. 
The remaining quantitative differences may be related to Coulomb effects, which in Ref~\cite{Errea_PdH_PRL2013} 
are treated in the $\mu_c^*$ approximation, or to anharmonic effects in the deformation potential, which were also disregarded. 

\begin{figure}
\centering
\includegraphics[width=0.9\columnwidth]{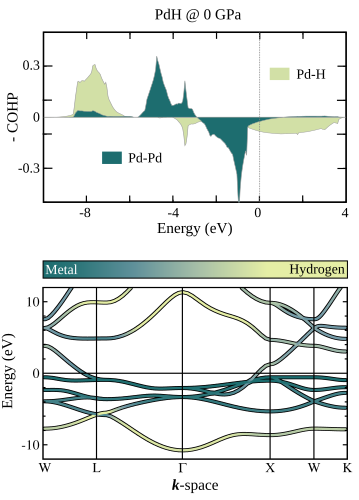}
\caption{~Electronic structure of PdH at 0\,GPa. 
Top: shows the COHP bonding analysis. Bottom: the KS-DFT electronic band structure, on which H projection of the electronic states are shown by a colour-code scale: from green (low) to yellow (high).}\label{fig:COHP_PdH}
\end{figure}

\paragraph{Electronic structure and superconductivity}

For the following discussion, we consider PdH with hydrogen in the $O$-sites (octahedral) which, according to first-principles calculations and experiments, should be the ground-state structure for $x \sim 1$.
The crystal structure is shown in the top left corner of Fig.~\ref{fig:Crys_struc}: Pd and hydrogen atoms create two interpenetrating $fcc$ lattices, with a lattice constant of 4.04\,\AA, which is around 5\% larger than in elemental Pd.  

In Fig.~\ref{fig:COHP_PdH}, we report the atom-projected DFT band structure and COHP. As shown, palladium-derived bands dominate the region near the Fermi level. These are $d$-states octahedrally split into $e_g$ and $t_{2g}$ manifolds; the latter can form $\sigma$-type bonds with hydrogen in the rocksalt structure. Indeed, we observe a precise splitting of Pd-H hybridized states into a bonding ($\simeq$ -10 eV at the $\Gamma$-point) and an antibonding ($\simeq$ 10\,eV at the $\Gamma$-point) manifold. Compared to pure Pd, in which states sit at the Fermi level, in Pd-H, due to the presence of an additional H 
electron, the Pd-$d$ states are full, and hence the Fermi surface consists of a single Pd-H band, which also has additional Pd-$sp$ character. Note that the filling of the $d$-bands compared to pure Pd has the additional effect of reducing the spin fluctuations, which in Pd suppress superconductivity~\cite{Th:Berk_Schrieffer_PRL_1966}. In the top panel of Fig.\ref{fig:COHP_PdH}, we present the COHP as a function of energy for two different types of interactions (Pd-Pd and Pd-H). In this plot, and in all COHP figures that follow, we plot the negative of the COHP, so that positive and negative peaks indicate bonding and antibonding interactions, respectively. 

The bonding Pd-Pd interaction is evident from the (positive) peak at  $\simeq$ -4.5\,eV, while the antibonding (negative) peak at $\simeq$ -0.4\,eV is filled due to electrons donated by hydrogen. 
The first dispersing band (from -10 to -6\,eV) has Pd-H bonding character, and the corresponding antibonding region is located from -2 to 3\,eV. Bands crossing at the Fermi energy are responsible for electron-phonon interaction and superconductivity and are antibonding states derived both from Pd-H  (the most significant contribution), and Pd-Pd interactions. The hydrogenic fraction at the Fermi level is estimated to be of the order of 5\%, a rather small but non-negligible value. 

The presence of hydrogenic states crossing the Fermi level implies that the corresponding phonon modes should provide, 
to some extent, a sizable deformation potential. 
This seems to be the case (see Tab.~\ref{table:eph_table}): the Hopfield ratio $I^2/M=0.08$ 
that accounts for the part of the coupling that is independent of the phonon energy and of the electronic DOS, 
is three times larger than that of niobium (\tc =9.3\,K). 
That hydrogenic mode are responsible for a significant part of the coupling can also be seen from the \'Eliashberg function (Fig.~\ref{fig:a2ftotal}) which is weighted towards high energies with \omlog =53\,meV. 
However the total electron-phonon coupling has a very small value $\lambda=0.4$ and the computed critical 
temperature is very low (2.7\,K~\footnote{Our estimation is lower than that in Ref.~\cite{Errea_PdH_PRL2013} 
because we have opted for a conventional value \mus =0.1. This choice was made to allow a comparison between all selected hydrides on the same ground.}). This highlights a small drawback of hydrogenic superconductivity: high-frequency phonon modes provide potentially a large energy scale for the phonons that provide scattering (prefactor of Eq.~\ref{eq:McMillan}). However, since the same phonon energies also affect the coupling strength  ($\lambda$), according to Eq.~\ref{eq:lambdaHopfield}, unless the deformation potential $I$ is strong enough, superconductivity is killed by the sublinear dependence of \tc\ at weak coupling (Fig.~\ref{fig:Tc_models}). One could also make a step further and use Eq.~\ref{eq:lambdaHopfield} to compute the optimal phonon energy that would maximize \tc. The optimal \tc\ for PdH would be as high as 40\,K, if \omlog\ could be reduced to 20\,meV. The  critical temperature of 2.7\,K computed in Ref.~\cite{Errea_PdH_PRL2013} including anharmonic corrections is however quite off with respect to the experimental value of 9\,K. This inaccuracy can be ascribed in part to the \mus=0.1 assumption and to the sensitivity of \tc\ to the phonon energies, which are strongly affected by anharmonicity. It is, in fact, possible that anharmonic corrections to the phonon frequencies have been overestimated resulting in modes that are too stiff, or that corrections to the deformation potential, not included in the calculation of the \'Eliashberg function, are relevant. As mentioned in the previous section, this suggests that more accurate calculations of anharmonic \ep\ coupling and Coulombic screening would be necessary to obtain quantitative agreement with experiments in PdH.

To summarize, the presence of hydrogen in PdH is crucial for superconductivity for several reasons: i) Hydrogen efficiently incorporates in the $fcc$ Pd matrix at large stoichiometries ($x\approx$1). ii) The hybridization between H and Pd states is sufficiently large to promote an anti-bonding Pd-H band at the Fermi energy. iii) This band couples with H-derived phonon modes showing significant anharmonic effects. iv) The additional electrons donated by hydrogen fill the Pd-$d$ bands suppressing spin fluctuations that would otherwise prohibit superconductivity. However, the coupling provided by hydrogen is relatively weak. $\lambda$ is comparable to the one in elemental Pd due to the considerable (anharmonic) phonon energies, effectively limiting the coupling strength.  

\subsubsection{Phosphorus hydride}\label{PH2}

\paragraph{Background}

Less than one year after the \sh\ result, the Eremets' group reported the discovery of high-temperature superconductivity in phosphines under pressure~\cite{Drozdov_ph3_arxiv2015}. 
PH$_3$ (phosphine) molecules have already the same hydrogen to host-atom ratio as \sh . Experiments reported that a mixture of this gas and hydrogen, loaded in a diamond anvil cell, becomes metallic at 40\,GPa and eventually superconducting at 80\,GPa, with a maximum \tc\ of 103\,K at 207\,GPa (see Sec.~\ref{Sec:Experiments}). 
At the time of the experiments, there were no indications on the possible crystalline phases of phosphine under pressure and, without available diffraction experiments, only first-principles methods could give hints about what phases could become stable at high pressure.  
Independently, two groups reported a complete 
high-pressure phase diagram of PH$_x$ with  $x=1,2,3,4,5,6$ compositions~\cite{shamp_decomposition_2016,flores-sanna_PH3_2016}. Quite surprisingly, both studies found a thermodynamical instability of all the compositions upon dissociation towards pure phosphorus and pure H. The only metastable phases sufficiently close to the convex hull in the range 100-200\,GPa were those with PH, PH$_2$, and PH$_3$ compositions. The ground-state structures at a pressure of 120\,GPa have a $I4/mmm$ space group for both PH and PH$_2$ and $C2/m$ for PH$_3$. At this pressure, 
the P-H bond length is 1.4\,\AA\ (of the same order of the S-H bond length in the superconducting H$_3$S), 
while the P-P bonds are $\simeq$ 2.1\,\AA\ long in all three structures. 
Besides, and at variance with other hydrides, it was found that P atoms tend to form 1D polymeric structures or 2D layers, connected by weak interactions among them (see the structural sketch in the top right panel of Fig.~\ref{fig:Crys_struc}). Assuming that metastable phases may actually be stabilized in experiments, the authors of Refs.~\cite{shamp_decomposition_2016, flores-sanna_PH3_2016} found that PH, PH$_2$, PH$_3$ are dynamically stable, and their calculated critical temperatures 
and evolution with pressure are in such a good agreement with the experimental results (see Fig.~\ref{fig:PH})  that the hypothesis of metastability is probably valid~\cite{flores-sanna_PH3_2016}.   

\begin{figure}[t]
\centering
\includegraphics[width=0.9\columnwidth]{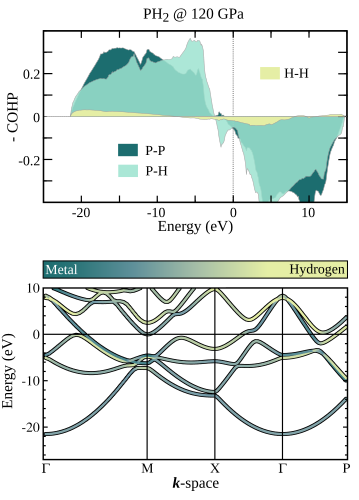}
\caption{~Electronic structure of PH$_2$ at 120\,GPa. 
Top: shows the COHP bonding analysis. Bottom: shows the KS-DFT electronic 
band structure on which H-projected electronic states are shown by a colour-code scale: 
from green (low) to yellow (high).}\label{fig:COHP_PH2}
\end{figure}

\paragraph{Electronic structure and superconductivity}

The structural motif of PH$_2$ at 120\,GPa (shown in Fig.\ref{fig:Crys_struc}) is characterized by a two-dimensional square lattice of linear PH$_2$ molecules linked together by P-P bonds.
The two-dimensional layers are stacked in an AB fashion along the direction perpendicular to the layers. This layered crystalline phase is unique in the realm of high-pressure hydrides, and this is probably at the origin of its metastability. 

The projected band structure of PH$_2$ at 120\,GPa, reported in Fig.~\ref{fig:COHP_PH2}, shows that occupied bands and, in particular, the states at the Fermi level, have a strong P-H hybridization. 
The integrated H projection at the Fermi level accounts for about 50\% of the DOS (see Tab.~\ref{table:eph_table}).

The bonding properties of the 2D network can be analyzed in terms of the COHP, also shown in Fig.~\ref{fig:COHP_PH2}. The P-H interaction shows well-separated bonding and antibonding regions (roughly below and above the  Fermi level, respectively). However, antibonding states are partially occupied with a Fermi surface formed by 
antibonding states which extend 2\,eV below the Fermi energy.
These states originate from molecular orbitals of the PH$_2$ units extending within the layer, linked together by P-P bonds, which show clear occupied bonding states. A key aspect is that the system is made of layered units that are weakly bound, as seen from the low values of H-H COHP. The difference with the bonding character in H$_3$S (see next section) is evident: in the latter, the covalent, metallic bond is stronger and extends to the entire solid. 

The vibrational spectrum is characterized by a large energy scale dictated by the hydrogenic modes, a scale that is about three times larger than that of PdH. In addition to this, the hydrogen contribution to the Fermi DOS is estimated to be much larger ($\approx$~55\%). The electron-phonon coupling is distributed homogeneously over all the modes~\cite{flores-sanna_PH3_2016}. 

A predicted \tc\ of 53\,K indicates that the Gilman-Ashcroft mechanism is at play, but not acting at its full potential. The total electron-phonon coupling ($\lambda=1$) is in the mid-coupling regime and the \omlog\ of 59\,meV is quite low, not much higher than that of PdH (53\,meV). The rationale for this is that, while hydrogenic phonons and hydrogen-related electronic states at the Fermi level are both available, the latter is not efficiently scattered by the former. This aspect can be traced back to the layered nature of the compound, which implies that the distortion in the electronic potential caused by the H modes takes place outside the conductive network of the P-layers. 

\subsubsection{Sulfur hydride}\label{H3S} 

\paragraph{Background}

At ambient pressure, the sulfur hydride is a molecular solid formed by H$_2$S molecules, with a relatively complex phase diagram~\cite{li2014metallization}. When placed in a hydrogen-rich atmosphere, H$_2$S easily absorbs hydrogen. In 2015, experiments showed that at pressure higher than 90\,GPa the H$_2$S+H$_2$ mixture becomes metallic and at about 100\,GPa superconductivity appears with a critical temperature reaching $\approx$\,8\,K at $p\approx$\,150GPa~\cite{DrozdovEremets_Nature2015} (see Sec.~\ref{Sec:Experiments} for details). 
Upon further increasing pressure, while measuring the resistance up to room temperature, (to obtain reliable results), experiments showed a rapid increase of the critical temperature up to $\approx$ 200\,K at $p\gtrapprox$\,170\,GPa, 
indicating a new superconducting phase. The measured shift of the critical temperature with applied magnetic field from 0 to 7 Tesla endorsed the superconducting nature of the phase transition and the measure of an isotope coefficient of $\alpha=0.3$ upon H/D substitution, its conventional (electron-phonon) origin. 

The initial interest in the sulfur-hydrogen system was motivated by a key theoretical prediction by Li and coworkers~\cite{li2014metallization}, pointing to metallization and superconductivity of H$_2$S. While this work exposed the field, it was later discovered that in experiments a second phase with
H$_3$S stoichiometry formed, with an unexpectedly high \tc.
Duan and coworkers~\cite{Duan_SciRep2014} gave a correct and complete prediction of stability and superconductivity in H$_3$S under pressure (independently and short before the publication of the experimental results~\cite{drozdov2014conventional}).  
For contextual history, readers are referred to the section {\bf Serendipity of the discovery of superconducting H$_x$S} in the Sec.~\ref{Sec:Experiments}. 

In their work, which can certainly be considered a milestone for {\it ab initio} superconductivity,
Duan et al.~\cite{Duan_SciRep2014} performed a study of the phase stability of the (H$_2$S)$_2$H$_2$ system under pressure. Their study was motivated by a previous prediction and observation of hydrogen-rich SiH$_4$(H$_2$); 
SH$_2$, analogously to silane, had been shown to easily incorporate hydrogen under pressure. According to DFT calculations, H$_2$ molecules start to incorporate into the H$_2$S molecular crystal already at around 10\,GPa, forming structures in which, at the lowest pressures where they are stable, the two molecular species are still clearly recognizable. 
However, upon increasing pressure, H$_2$S and H$_2$ molecules are pushed closer until, at around 90\,GPa, intramolecular bonds break and new bonds start to form between sulfur and hydrogen, giving rise to distinct H$_3$S stoichiometric units.

The first predicted H$_3$S phase, which is stable from 111\,GPa up to 180\,GPa, has a $R3m$ symmetry. In this phase, sulfur atoms occupy the lattice points of a (rhombohedral distorted) conventionally 
$bcc$ cell and each of them is surrounded by three hydrogens, forming isolated H$_3$S units~\cite{Duan_SciRep2014}.
The formation of a H$_3$S phase is supported experimentally 
because the predicted superconducting critical temperature for this phase is $\approx$160\,K at 130\,GPa and rapidly increases with pressure, exactly in line with the experimental evidence.
Starting at 180\,GPa another structural transition is predicted, towards a $Im\bar{3}m$ ($bcc$) phase, characterized by the symmetrization of the bonds between sulfur and H atoms, thus not forming isolated H$_3$S units anymore. In this structure, shown in Fig.\ref{fig:Crys_struc}, the H atoms are symmetrically placed between two S atoms resulting in the formation of hydrogen-shared SH$_6$ units. Interestingly, the calculated~\cite{duan2015pressure,flores-sanna_HSe_2016,Nature_Errea_2016} 
\tc\ for the $Im\bar{3}m$ structure at 200\,GPa is $\approx$ 200\,K, 
making it a strong candidate to be the high-\tc\ phase observed in the experiments.

The final confirmation of the crystal structure of the superconducting phase was given by synchrotron X-ray diffraction measurements~\cite{Einaga_H3S-crystal_NatPhys-2016}, which
showed the H$_3$S units sit on a $bcc$ lattice, as theoretically predicted. Besides, infrared optical spectroscopy measurements observed optically active phonons at 160\,meV with a dramatically enhanced oscillator strength~\cite{H:Capitani_SH3_nphys_2017} 
as well as evidence of superconducting gap at 76\,meV, interpreted with the \'Eliashberg theory of electron-phonon driven superconductivity.
As in PdH, however, the proper theoretical interpretation of all the experimental results requires the inclusion of anharmonic effects. Indeed, proper treatment of these effects~\cite{Nature_Errea_2016} 
is fundamental to predict the correct critical  pressure for the $R3m\rightarrow Im\bar{3}m$ phase transition,
the anharmonic phonon frequencies of the hydrogen-dominated phonon modes and the shift of \tc\
upon isotope H/D substitution~\cite{H3S-anharm-Bianco}.

In addition to the high-\tc\ phase, experiments identify one
or more low-pressure phases. For these, many structural models have been proposed theoretically, based on thermodynamic stability considerations and estimates of  \tc\ \cite{Duan_SciRep2014,Yanming_JCP2014,flores-sanna_HSe_2016}.  
In particular, a modulated structure model with long periods of alternating slab-like H$_2$S and H$_3$S regions forming a sequence of {\it Magn\'eli} phases were proposed~\cite{akashi_mangeli-phases}. 
Even though the pressure trend of the comparatively low-\tc\ phase may be reproduced assuming these structures, the calculated XRD patterns of these {\it  Magn\'eli} phases do not match the experimental 
data~\cite{Tse_magneli-non,HxS_majumdar2019mechanism}. 

\begin{figure}[t]
\centering
\includegraphics[width=0.9\columnwidth]{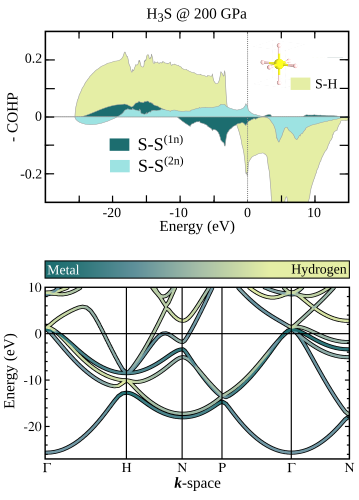}
\caption{~Electronic structure of H$_3$S at 200\,GPa. 
Top: shows the COHP bonding analysis. 
Bottom: the KS-DFT electronic band structure on which H-projected electronic states are shown by a colour-code scale: 
from green (low) to yellow (high).}\label{fig:COHP_H3S}
\end{figure}

\paragraph{Electronic structure and superconductivity}

Rather than as a dense hydrogen sublattice, high-\tc\ $bcc$\ \sh\ should be more accurately described as a dense solid of covalently-bonded hydrogen and sulfur. In this structure, shown in the left bottom panel in Fig.~\ref{fig:Crys_struc}, sulfur sits on a $bcc$ lattice with lattice constant $a=2.96$\,\AA\ (at 220\,GPa), while hydrogen fills the positions in the middle of the edges and faces. 
Thus, each hydrogen has two sulfur and four hydrogens nearest neighbours, at an equal distance of 1.48\,\AA.

Electronic band structure and COHP analysis plots for H$_3$S are shown in Fig.~\ref{fig:COHP_H3S}. The relevant bandwidth extends 25\,eV below the Fermi energy, with the lower band originating from S-3$s$ orbitals with a (small) H-1$s$ contribution. 
The hybridization between H and S states is strongly enhanced in a broader region around the Fermi energy (from -5 to +5\,eV), where the S-3$p$ and H-1$s$ interaction dominates. 
The origin of one of the van Hove singularities~\cite{van_ovfs-singuluarity} can be traced back to the maximum of the band dispersion along the H-N direction of the Brillouin zone, which is below the Fermi energy~\cite{PRB_Sano_Van-Hove_H3S_2016,SH3-vHs-Pickett,H3S-vHs-Marsiglio,H3S-vHs-Bianconi}. 
The presence of van Hove singularities close to the Fermi level has stimulated discussions of new and alternative superconducting mechanisms or refinements of the conventional electron-phonon theory of superconductivity~\cite{ortenzi_TB_2015,PRB_Sano_Van-Hove_H3S_2016,SH3-vHs-Pickett,H3S-vHs-Marsiglio,H3S-vHs-Bianconi}.  Hydrogen contribution is predominant on the higher energy part of the electronic spectrum.

The COHP, shown in the top panel in Fig.\ref{fig:COHP_H3S}, permits to interpret the same electronic structure in terms of real-space interactions. 
A bonding S-H interaction dominates the majority of the valence region. 
The corresponding energy states are fully occupied, indicating strong stability of the S-H network. The Fermi level falls into the region of antibonding (negative COHP) S-H orbitals with the van Hove singularity visible close to the Fermi energy.  
Sulfur-Sulfur interactions give a lower but non-negligible contribution considering that they extend 
up to the next nearest neighbour, while direct H-H interactions do not contribute to the bonding.
In PdH, the stable Pd-Pd bonding/antibonding levels were fully occupied due to the additional electrons donated by H, so that the states at the Fermi level were eventually of Pd-H antibonding character.
Also in H$_3$S, the dominant contribution to the states at the Fermi level is of antibonding H-host character, but the S-S interaction is significantly reduced. The much larger electron-phonon coupling and 
phonon frequencies in H$_3$S result from the fact that the metal-H distances are sensibly shorter (d$_{S-H}$=1.4\,\AA, d$_{Pd-H}$= 2.02\,\AA).

The vibrational properties of H$_3$S are characterized by low-energy ($\lessapprox$\,75\,meV) S-derived modes,  these are separated from H-derived modes which extend up to $\approx$\,250\,meV, once anharmonic corrections are included~\cite{Nature_Errea_2016}. 
The latter show a clear separation into bending modes, ranging from $\approx$ 100 to $\approx$\,200\,meV, and high-energy stretching modes from $\approx$\,200 to $\approx$\,250\,meV, which are the modes most strongly coupled to electrons at the Fermi energy~\cite{Nature_Errea_2016}. Calculation including anharmonic corrections predicts an isotope effect $\alpha$=0.35, sensibly reduced to the BCS value and in good agreement with experiments~\cite{Nature_Errea_2016}. The total electron-phonon coupling constant (see Fig.~\ref{fig:a2ftotal}) is $\lambda$=1.84, which results in a critical temperature of 194\,K (using \mus=0.1) at 200\,GPa 
which agrees with the experimentally reported critical temperature of 190\,K~\cite{DrozdovEremets_Nature2015}. 

The large \tc\ is due to the key role played by hydrogen vibrational modes and electronic states. Hydrogen contributes to about 45\% of the states at the Fermi level, slightly less than it does in PH$_{\rm 2}$, but still quite significantly. The Fermi level is pinned at a peak in the density of states, although the Fermi DOS is similar to that of other hydrides and not significantly higher. 
Hydrogen-related vibrational modes spread to high energies ($\approx$ 250\,meV) and yet, despite these large scales, these high-energy modes show up in the \'Eliashberg function as a broad  high peak (Fig.~\ref{fig:a2ftotal}) overcoming the negative effect of the $\omega^{-1}$ factor in the electron-phonon coupling (Eq.~\ref{eq:elphME}). 
One can disentangle the effect of phonon frequencies in the electron-phonon coupling considering the $I^2/M$ factor in Hopfield formula Eq.~\ref{eq:lambdaHopfield}: for \sh\ one gets $I^2/M=1.46$ a tremendous value, one order of magnitude larger than in PH$_2$ and about 40\% of that computed for pure hydrogen (See Tab.~\ref{table:eph_table}). 
In short, \sh\ is exceptional because, while hydrogen only accounts for 50\% of the Fermi DOS, 
these hydrogens are fully covalently bound together with sulfur, forming a full 3D network of conductive bonds.  Besides, one should also consider that sulphur itself provides extra coupling, as S modes significantly enhance $\lambda$. Notably, elemental sulphur is a superconductor in the high-pressure regime (see Fig.~\ref{fig:Elemental_supra}).

\subsubsection{Lanthanum hydride}\label{LaH10}

\paragraph{Background}

\lah\ currently (end of 2019) holds the record for the highest \tc\ ever measured among all known superconductors ($\approx$250\,K). 
It is a superhydride, which forms upon laser heating of lanthanum in a hydrogen-rich atmosphere, 
and belongs to a large class of clathrates-like structures~\cite{PNAS_LaHx_2017_Hemley,Clathrate_REHX_PRL_2017}.  

Everything started, two years ago when in 2017, the Argonne group reported the synthesis of lanthanum superhydrides under pressure, 
through laser heating of lanthanum in a hydrogen-rich atmosphere~\cite{geballe2018synthesis}. The hydrogen to lanthanum ratio was estimated to be $x \in [9,12]$, based on the unit cell volume. 
Due to the small X-ray cross-section of hydrogen, it is not possible to resolve the H sublattice directly, but only the La one, which was indexed as a $Fm\bar{3}m$, $fcc$ lattice. 
The high-\tc , $fcc$ phase of LaH$_{10}$, for which Li et al. predicted a \tc\ of 280\,K at 210\,GPa, 
was thus suggested as the most likely candidate for the observed high-\tc\ phase~\cite{PNAS_LaHx_2017_Hemley}. 
The central structural motif of the structure is a cage of hydrogens enclosing central La atoms, linked together by direct H-H bonds, see Fig.~\ref{fig:Crys_struc}.

A year later, in 2018, two groups reported superconductivity measurements on lanthanum hydrides under pressure. 
The first group reported a maximum \tc\ of 215\,K, around 200\,GPa~\cite{drozdov2018_215}. 
The second set of measurements reported \tc's as high as 260\,K, in samples heated up to 1000\,K 
at 180\,GPa~\cite{Hemley-LaH10_PRL_2019}.  
Finally, the same authors as in Ref.~\cite{drozdov2018_250} confirmed a few months later a \tc\ as high as 250\,K 
in LaH$_x$ samples, but also reported the existence of several phases with lower and distinct \tc's, depending on the synthesis conditions~\cite{drozdov2018_250}. 

The same study contains other measurements, which are crucial to establish superconductivity and its conventional nature in \lah: isotope effect (hydrogen vs deuterium substitution) and measurements of \tc\ under magnetic field. Due to the problematic synthesis conditions, the linear dimensions of the current samples of \lah\ (10---20 $\mu m$) are much smaller than those of \sh\ (up to 100 $\mu m$ ). This does not allow a proper measurement of the Meissner effect. 
However, the authors show that indeed magnetic field suppresses superconductivity.

The dependence of the critical temperatures on pressure reported by the two studies are also conflicting: while the Argonne measurements report an abrupt increase in \tc\ up to 290\,K~\cite{Hemley-LaH10_PRL_2019}, Ref.~\cite{drozdov2018_250} reports a dome-like shape of the \tc\ vs $p$ curves. All the observations above suggest that, depending on the synthesis conditions, different phases with different stoichiometries and crystal structures may be stabilized by laser heating. 

\begin{figure}[t]
\centering
\includegraphics[width=0.9\columnwidth]{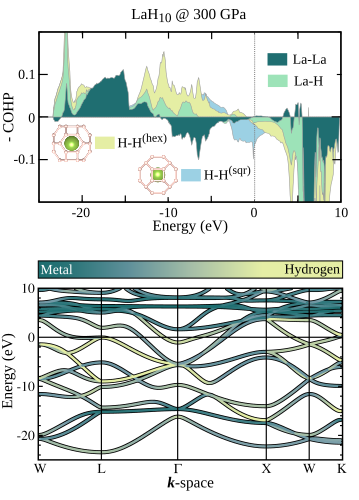}
\caption{~Electronic structure of LaH$_{10}$ at 300\,GPa. 
Top: shows the COHP bonding analysis. Bottom: the KS-DFT electronic band structure on which H-projected electronic of states 
are shown by a colour-code scale: from green (low) to yellow (high).}\label{fig:COHP_LaH10}
\end{figure}

\paragraph{Electronic structure and superconductivity}

The $fcc$ phase of LaH$_{10}$, shown in the right panel of the lower row of Fig.~\ref{fig:Crys_struc}, 
can be considered one of the closest realization of the Ashcroft-Gillan chemically-precompressed hydrogen sublattice in hydrides. 
The central structural motif is a highly symmetric cage formed by 32 hydrogens, with 12 hexagonal and 6 square faces, 
which surround a central La atom. These two interconnected cages form a dense hydrogen sublattice, 
with bond lengths of $d_{HH}$= 1.07\,\AA\ for square and $d_{HH}$=1.15\,\AA\ for hexagonal edge.

The electronic band structure of \lah\ in the high-\tc\ ($fcc$ phase) at 300\,GPa is shown in the lower panel of  Fig.~\ref{fig:COHP_LaH10}. 
As a confirmation of the general trend, the states at the Fermi level derive from both the H sublattice and the La sublattice, even if in this case there is a more precise separation between La- and H- derived Fermi surfaces.

The Fermi surface comprises six distinct pockets, centred around the BZ centre and boundaries, 
formed by H $1s$ and La states bands, which hybridize along some directions of the BZ.  
The hybridization is mainly due to the presence of low-lying $f$-states of La, centred around 5\,eV above the Fermi level. The surprising presence of a non-negligible contribution of $f$ electrons to the Fermi level was already pointed out by Liu et al.~\cite{PNAS_LaHx_2017_Hemley}, who observed that in \lah\ 
not only $s$ and $p$ La electrons contribute to the DOS at \ef, but also the $f$ electrons.  
We remark that these states, being strongly localized, are usually not well described by DFT-like methods, 
and hence this seems to cast doubts about the reliability of the DFT results on this system. 
On the other hand, the fact that the high-\tc\ properties computed in this system are very similar 
to other hydrides which do not contain $f$ electrons, indirectly confirms the reliability of existing predictions.
As evident from the band structure, at the $\Gamma$-point an electron-like H-derived band crosses E$_F$ and two other bands, of mixed La and H character, form electron pockets. 
States with dominant La character form hole pockets at the zone boundary. At the L-point of the BZ, we observe a Van Hove singularity very close to the Fermi level which, 
similarly to what we discussed for H$_3$S, may be partially responsible for enhancing the superconducting properties~\cite{Liu_MicroscopicMechanismLaH10_PRB2019}. 

In this case, the COHP decomposition (see Fig.~\ref{fig:COHP_LaH10}) turns out to be fundamental to analyze the bonding properties of the system, which are quite intricate.  First, we consider the H-H interactions. In the system, there are two nonequivalent H-H bonds, 
the nearest - neighbour  H-H (1.06\,\AA) connecting H atoms forming the hexagons, and the next-nearest neighbours H-H bonds (1.15\,\AA) among H atoms forming the squares (see Figure \ref{fig:Crys_struc}). The former interaction is covalent with occupied bonding states well separated from the unoccupied antibonding states (brown curve). 
The next-nearest neighbours (1.15\,\AA, square bonds) 
H-H interaction is different, showing that the antibonding states are populated (this feature is similar to what observed in other hydrides). The bonding states between La and H (averaged between nearest and next-nearest neighbours) are located at very low energies (they extend only up to -10\,eV), while at the Fermi level the (hybridized) states are mainly non-bonding (COHP is $\simeq$ 0.0). The states derived from the direct La - La interaction have a non-bonding character at \ef, with populated bonding and antibonding states.

\paragraph{Phonons and electron-phonon coupling}

The phonon dispersion of \lah\ extend up to $\approx$3000\,cm$^{-1}$. 
Due to the very different phonon mass of La and hydrogen, La and H modes split into low-energy modes with distinct La character, and high-energy modes with H character. The \ep\ coupling is quite uniformly spread over all phonon modes, resulting in $\lambda=1.8$ and \omlog=1490\,K (values at 300\,GPa from Ref.~\cite{PNAS_LaHx_2017_Hemley}), which yields a \tc\ of 215\,K using the modified McMillan-Allen-Dynes formula or higher values solving Migdal-\'Eliashberg equations, assuming a standard value of \mus\ (0.1-0.13).  In principle, anisotropic effects and anomalous Coulomb screening could strongly modify the calculated \tc\ values. A first-principles study by Heil et al.~\cite{heil_YH_2019superconductivity} on the closely-related YH$_{10}$ system seems to rule out both possibilities. A more recent study, Ref.~\cite{work_on_LaHx_2019}, confirm that these effects for \lah\ can be neglected. 

An interesting issue is the dynamical stability of the high-\tc , $fcc$ structure. 
According to harmonic DFPT calculations, the sodalite-like structure, which is the most plausible candidate for superconductivity is only stable above 230\,GPa, while experiments measure a finite \tc\ already at 130\,GPa. 
The two most likely explanations are that: 
either anharmonic effects stabilize the structure to lower pressures (by an almost 100\,GPa~!); or that this structure represents a metastable phase synthesized by peculiar thermodynamical conditions~\cite{liu2018dynamics}. 

Electron-phonon coupling calculations~\cite{PNAS_LaHx_2017_Hemley} are available for $p$=300\,GPa. \'Eliashberg spectral function and coupling parameters, reported in Fig.~\ref{fig:a2ftotal}, picture a system that is quite similar to H$_3$S. Like in H$_3$S, $\lambda$ is in the strong-coupling regime. Furthermore, hydrogen significantly contributes to high energy phonons. The scenario is illustrated by the Hopfield decomposition of $\lambda$ and the use of the McMillan-Allen-Dyes equation for \tc. 
Tab.~\ref{table:eph_table} shows that \lah\ is akin to \sh: comparable Fermi DOS, and hydrogenic fraction, as well as an almost identical deformation potential. The only difference between the two is that the coupling is slightly shifted towards lower energies (see Fig.~\ref{fig:a2ftotal}), so that the coupling constant $\lambda$ is larger, leading to an increase in \tc\ to 234\,K. From this argument, one is tempted to conclude that \lah\ is a better superconductor than \sh\ because of a slightly better tuning of the phonon energy to coupling ratio. The calculation of the optimal \omlog\ using the Hopfield formula in the McMillan equation would indicate for both \sh\ and \lah\ value of about 100\,meV. 

However, one has to be careful because actual calculations refer to a pressure (of 300\,GPa) that is way higher than the pressure at which the system is measured $\simeq$160\,GPa. At this pressure, \lah\ results to be neither thermodynamically nor dynamically stable (harmonic phonons). It is safe to say that, unlike for \sh\ the physics of \lah\ is not fully unveiled. The disagreement with experiments could be possibly ascribed to a different structural phase, to the theoretical model or too significant anharmonic effects (as compared to \sh).  \\ 

Recently, Errea and Flores-Livas et al.~\cite{work_on_LaHx_2019}, addressed these issues showing that quantum atomic fluctuations stabilize, in the experimentally relevant pressure range, a high-symmetry $Fm\bar{3}m$ crystal consistent with experiments. 
A colossal electron-phonon coupling with $\lambda$=3.5 was attributed to the LaH$_{10}$ phase 
(the first time that such high value was reported). 
The paper also pointed out that {\it ab initio} classical calculations predict a set of distort structures yielding a complex energy landscape, while quantum fluctuations ``wash out" all the landscape complexity. 
This result questions many ({\it if not all}) of the crystal structure predictions made for hydrides within a classical approach. 

\subsubsection{Elemental hydrogen at HP}\label{atomh}

\paragraph{Background}

The metallization of hydrogen has been seen as a compelling subject of study by many scientists, ranging from experimental 
chemists and physicists~\cite{LeToullec2002,Eremets_NatMat2011,eremets2016low,mcmahon_high_2011,mcmahon2012properties,dalladay2016evidence,Dias_hydrogen_Science2017}, to theoreticians
~\cite{carbotte2018detecting,borinaga2018strong,cazorla2017simulation}. However, despite the great advancement in high pressure-techniques and tools, the metallization of hydrogen in its solid phase has proven to be very challenging, and still debated~\cite{Hemley_PRL2012,HRussell_hydrogenJACS2014,Salamat-Silvera_2016,eremets2017molecular,azadi2017role,magduau2017simple,liu2017comment,goncharov2017comment,zha2017melting,zaghoo2017conductivity,rillo2018coupled}. The $p-T$ phase diagram of hydrogen comprises many structures, including solid and molecular crystals and is so complex that could be subject of a Review by itself~\cite{dalladay2016evidence}.  

Here, we study the $\beta$-Sn structure, which, according to first-principles calculations, should be the first atomic phase of metallic hydrogen, stable between 500\, GPa and 1\,TPa~\cite{pickard_NatPhys_2007_structure-H,mcmahon2012properties}. In the atomic phases of hydrogen, the  0.7\,\AA\ bond, typical in all molecular phases occurring at lower pressures, is broken. However, both the $\beta$-Sn and the next stable structure, the $R\bar{3}m$ one are relatively open, and H forms bonds of different types and lengths. Only at very high pressures ($\ge 3.0$\,TPa), hydrogen is predicted to form genuinely close-packed ($bcc$ or $hcp$) phases. 

\begin{figure}
\centering
\includegraphics[width=0.85\columnwidth]{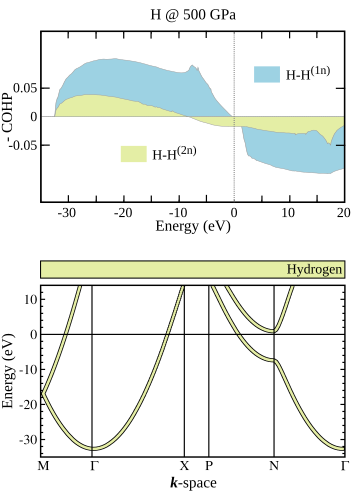}
\caption{~Electronic structure of H at 500\,GPa. 
Top: shows the COHP bonding analysis. 
Bottom: the KS-DFT electronic band structure.}\label{fig:COHP_H}
\end{figure}

\paragraph{Electronic structure and superconductivity}

The $I4_1/amd$ structure, commonly known as $\beta-$Sn or Cs-$IV$ structure, 
comprises two nonequivalent H atoms in a $bct$ unit cell, 
each of which forms four short (0.98\,\AA) and four long (1.21\,\AA) bonds.

The electronic structure at 500\,GPa, is shown in Fig.~\ref{fig:COHP_H}. 
Hydrogen in the $\beta$-Sn structure has an enormous bandwidth ($\sim 40$\,eV), reflecting a considerable electronic kinetic energy. The band dispersion is very close to the free-electron one, except for $\approx 5$\,eV gaps opening close to the zone boundaries due to the strong unscreened electron-ion potential, responsible for the \ep\ interaction. 
The corresponding Fermi surface is an almost perfect Fermi sphere, 
refolded into the Brillouin zone, with strong H-s electronic 
character, which exhibits tubular deformations around the zone boundary.

The COHP function is shown in Fig.\ref{fig:COHP_H}, 
resolved into the 1$^{st}$ and 2$^{nd}$ nearest neighbours H-H contributions. 
The plot clearly shows that the bonding is characterized by filled bonding states for nearest neighbours H-H interactions, clearly separated at E$_F$ from the antibonding states. Antibonding states from next-nearest interactions are on the contrary occupied starting from -10\,eV;
the  3$^{rd}$ n.n. contribution is non-bonding (not shown).

\begin{figure}[t]
\begin{center}
\includegraphics[width=\columnwidth]{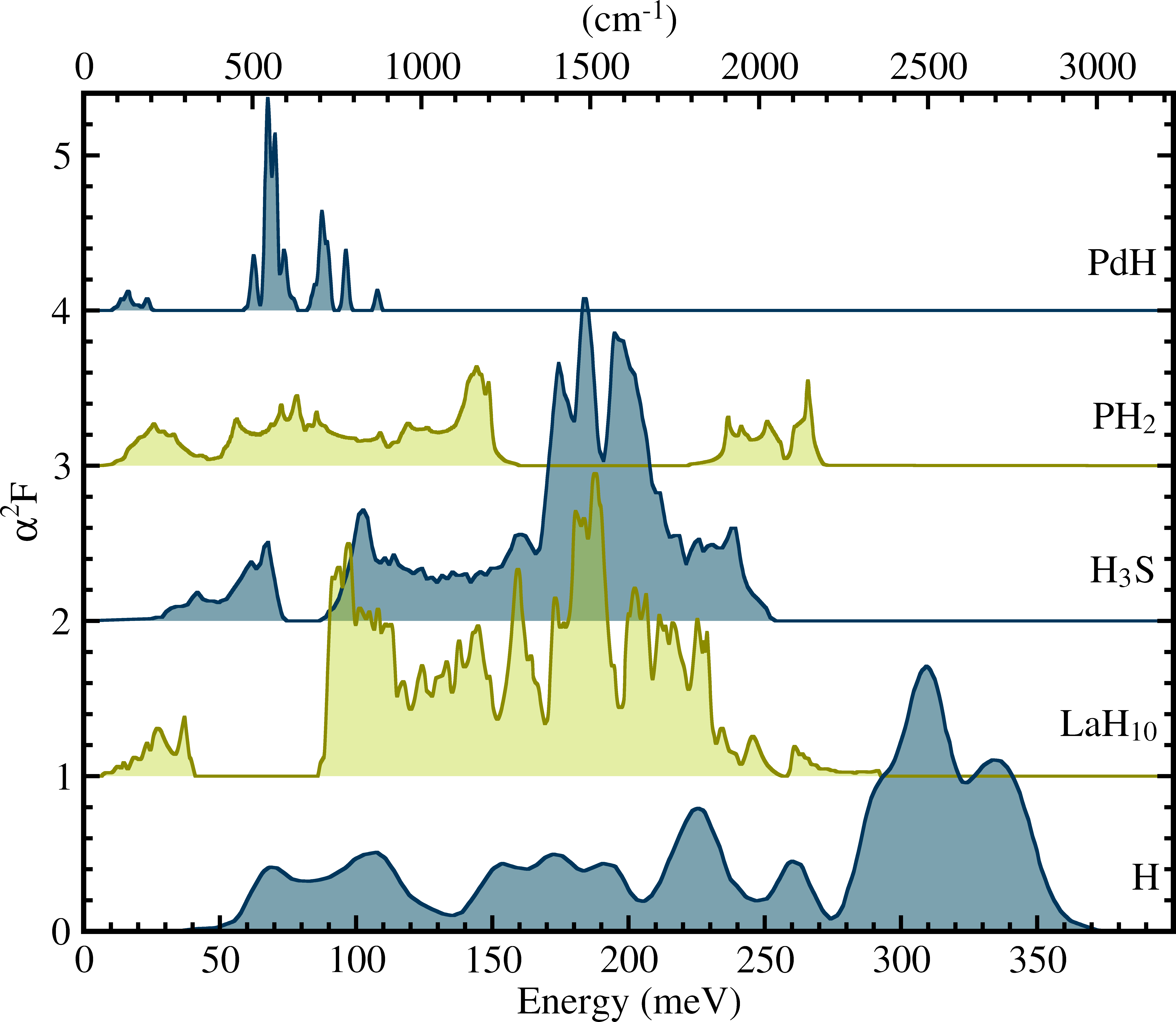}
\caption{~\'Eliashberg $\alpha^2F$ function computed for selected hydrides: 
         PdH (0\,GPa)~\cite{Errea_PdH_PRL2013}, 
    PH$_2$ (120\,GPa)~\cite{flores-sanna_PH3_2016}, 
    H$_3$S (200\,GPa)~\cite{Nature_Errea_2016}, 
LaH$_{10}$ (300\,GPa)~\cite{Liu_MicroscopicMechanismLaH10_PRB2019} 
 and H$_2$ (500\,GPa)~\cite{H:Borinaga_PRB_2016}. 
A vertical shift is applied for plotting the functions.  
The value of the integrated coupling constant $\lambda$ and other key parameters are computed and collected in Tab.~\ref{table:eph_table}.}
\label{fig:a2ftotal}
\end{center}
\end{figure}

At 500\,GPa, the phonon dispersion (from Ref.~\cite{H:Borinaga_PRB_2016}) extends up to 370\,meV, 
a range comparable to the frequencies calculated for clathrate hydrides at similar pressures. 
The \'Eliashberg function reported in Fig.~\ref{fig:a2ftotal}, 
shows that the coupling is spread quite homogeneously over the entire frequency range. 
The logarithmic-averaged phonon frequency (\omlog=166\,meV) is larger than in other hydrides, 
partially reflecting stronger bonds and the massive  applied pressure. 
Despite this, the \ep\ coupling constant $\lambda$ is the same as in H$_3$S, leading to a larger \tc=282\,K. 
The reason why $\lambda$ can be high notwithstanding the sizeable average phonon energy lies in the larger $I^2/M$ ratio. 
An analysis of the optimal \omlog, as done for PdH and other hydrides, 
indicates a maximum achievable \tc\ of 300\,K if \omlog\ cold be softened down to 140\,meV (see Tab.~\ref{table:eph_table}).

Even in this case, the \tc\ of metallic-solid hydrogen would remain in the same ball-park as those predicted for the best clathrate hydrides. 
In order to increase \tc\ further, the only possible way would be to increase pressure. McMahon and Ceperley extended the study of superconductivity of 
hydrogen in its $\beta$-Sn structure up to 1.5\,TPa (1,500\,GPa)~\cite{mcmahon_high_2011}. 
At those pressures, \tc\ is predicted to increase up to 500\,K at $\approx~$700\, GPa and then experience a sharp drop. \tc 's up to 750\,K (at 2\,TPa) have been predicted for the next stable H polymorph, the $R\bar{3}m$ phase, which is not discussed here.

\subsection{Summary}

We have shown on the previous sections that {\it ab initio} calculations 
closely reproduce the superconducting properties of the five representative hydrides that we have chosen as examples. These compounds are in many ways all realizations of the Aschcroft-Gilman's arguments, but with very different outcomes. In fact, the actual superconducting properties vary from being marginal (as in PdH) 
to astonishing high, as in \sh\ or \lah, depending on specific aspects of electronic structure and coupling -- see Table~\ref{table:eph_table}.

PdH, our first example, is a prototypical metallic hydride, where H is incorporated in the original $fcc$-Pd matrix at a small stoichiometry ($x\approx$1).  This amount of hydrogen is sufficient to
promote an antibonding Pd-H band at the Fermi energy, yielding a small and non-negligible fraction of H states at the Fermi level. 
These states are coupled with H-derived phonon modes showing significant anharmonic effects.  However, the hydrogen contribution 
is not large enough to lead to a substantial \ep\ coupling.
The Hopfield factor $I^2/M$, which gives an estimate of the \ep\ coupling fraction 
independent of the electronic DOS and phonon spectrum,
is merely 0.08, as opposed to $\approx 1.5$ in \sh\ and \lah.
The resulting \ep\ coupling constant is relatively weak ($\lambda=0.4$),
comparable to that of elemental Pd ($\lambda=0.36$). However,
compared to elemental Pd, the presence of hydrogen increases
the phonon energy scale and yields a finite \tc.
Furthermore, due to a substantial anharmonic hardening, 
the interplay between phonon frequencies and electron-phonon matrix elements are not optimal. Therefore \tc\ is substantially reduced with respect to the optimal value \tc=40\,K.

\begin{table}[t]
\caption{~Collected electron-phonon parameters computed for the set of discussed materials. We tabulated: Pressure ($p$), $N_{F}$ that is the density of states at the Fermi level per spin (sp.) and N$^{\rm H}_{F}$ the fraction of it that projects on the H site (percentual). $\omega_{log}$ and $\lambda$ are electron phonon coupling parameters defined in Eq.~\ref{eq:omlog} and Eq.~\ref{eq:lambda_a2F}. $I^2/M$ is computed from Eq.~\ref{eq:lambdaHopfield} for $\omega=$\omlog. \tc$^{McM}$ is the McMillan critical temperature (Eq.~\ref{eq:McMillanAllenDynes}) at \mus=0.1; \tc$^{AD}$ is the more accurate Allen-Dynes critical temperature (Eq.~\ref{eq:McMillanAllenDynes2}), also computed for \mus=0.1; Max\tc$^{AD}$ is the maximal \tc that can be achieved by tuning the value of \omlog (at \mus=0.1).}
\begin{center}
\begin{tabular}{l| b | y | b | y | b}
&PdH   &PH$_2$   &\sh\   & \lah\   &  H   \\\hline
$p$ (GPa)                & 0    & 120   & 200   & 300   & 500  \\  
$N_{F}$(eVA$^3$sp.)$^{-1}$& 0.014& 0.021 & 0.019 & 0.016 & 0.011\\
N$^{\rm H}_{F}$(\%)  & 5    & 55    & 45    & 50    & 100  \\
$\omega_{log}$(meV)    & 53   & 59    & 129   & 104   & 166  \\
$\lambda$              & 0.4  & 1.0   & 1.6   & 2.2   & 1.63 \\
$I^2/M$ ($eV^3A^3$)    & 0.08 & 0.16  & 1.46  & 1.48  & 4.30 \\
\tc$^{McM}$ (K)        & 2.6  & 48    & 182   & 183   & 237  \\
\tc$^{AD}$ (K)         & 2.7  & 53    & 214   & 234   & 282  \\
Max\tc$^{AD}$  (K)    & 40.3  & 78    & 225   & 235   & 296  \\
\hline
\end{tabular}
\end{center} \label{table:eph_table}
\end{table}

Hydrogen incorporation in a metal framework is very far from the idea of chemical precompression, which implies a rearrangement of bonds into a dense hydrogen sublattice, only possible by overcoming energy barriers in the megabar range.
PH$_2$ represents an intermediate step towards this idea.  With respect to phosphine, which at ambient pressure forms an open crystal of PH$_3$ pyramidal molecules, a significant rearrangement of bonds occurs. 
At high pressure, the new structural motif is characterized by a two-dimensional square lattice of linear PH$_2$ molecules linked together by P-P bonds, but not H-H bonds. Due to the relatively high concentration of hydrogen, H contributes around half of the electronic states at the Fermi level. 
However, the open 2D structure is very inefficient in terms of \ep\ coupling, and the Hopfield factor, although twice as large as in PdH, is almost one order of magnitude smaller than in the two other high-pressure hydrides, \sh\ and \lah. 

In the high-\tc\ phases of \sh\ and \lah, the effect of megabar pressure on the crystal structure is dramatic. In both structures, hydrogen is embedded in a closely-packed metallic network held together by strong covalent bonds. The superconducting properties are comparable to those of solid hydrogen. The COHP analysis shows that the states at \ef\ have an antibonding character. In \sh, where sulfur is incorporated in the hydrogen lattice, the dominant interaction is of S-H type, while in \lah\ the geometry favours direct H-H interactions. One may argue that \lah, where only hydrogen is involved in covalent bonding, is a better realization of the Ashcroft-Gilman concept than \sh, and hence should exhibit a higher \tc.
However, the presence of another atom involved in the bonding
seems to be marginal in this sense. In both compounds, hydrogen contributes roughly 50\% of the total DOS at E$_F$ and dominates the phonon spectrum. This yields very similar values for the Hopfield factor and \omlog. A $\sim$10\% difference in the superconducting properties comes from a 
better optimization of the phonon spectrum. Differences in \tc\ of this order can be easily be compensated by other effects, such as doping or pressure.

It is interesting to compare \ep\ coupling values of \sh\ and \lah\ 
with those of atomic hydrogen at 500\,GPa --see Tab.~\ref{table:eph_table}. Indeed, in the latter, the Hopfield factor is about a factor of three larger, but since the shape of the phonon spectrum is not optimized and the electronic DOS is low, the superconducting properties do not improve appreciably. 
Significantly higher \tc\ is found in hydrogen only at much higher pressures, where the H lattice is denser and fully symmetric.

Our analysis confirms that many of the empirical parameters that are used to identify promising hydrides, such as H-H distances and hydrogenic fraction, do not merely correlate with \tc. For instance, once hydrogen is embedded in a covalent, metallic structure, \tc\ can only be predicted by a detailed analysis that takes into account all full aspects of electronic structure. 

This means that it is probably very hard, if not impossible, to formulate general rules. On the other hand, this implies that hydride superconductivity is highly tunable: by acting on the bonding and the density of charge carriers by doping, for example, it could be easy to optimize the superconducting properties within a given material class. 

Rather than a set of precise rules, our analysis suggests a
modern set of {\bf necessary} conditions to maximize \tc\ in hydrides and similar materials. 
\begin{itemize}[leftmargin=*]
\item Electronic states originating from hydrogen or light-elements should cross the Fermi level.
\item Hydrogen or light elements should form covalent bonds among them or with the host, 
preferably forming an extended (non-molecular) bonding structure. 
\item Phonon energies should be optimally tuned to the strength of this coupling: 
A stronger coupling requires higher energy phonon modes.  
\end{itemize}

Arguably, the above requirements are likely to be fulfilled at extreme pressures,
where thermodynamic conditions permit to stabilize metallic phases. 
It is also highly possible that extending the search space to {\it poly}-hydrides or by 
devising suitable thermodynamical conditions one could overcome the issue 
of stabilizing these metallic high-\tc\ phases at substantially lower pressures.

\section{Discussion and Perspectives}\label{Sec:Perspectives}

Over the last five years (2014--2019), we witnessed two practical realizations of conventional high-\tc, in \sh\ (2015) and \lah\ (2019), both compositions stabilized at HP and with critical temperatures above 200\,K. 
This hints to the imminent possibility of reaching room temperature (300\,K) superconductivity in the coming years.  
However,  practical applications of this type of materials are clearly hindered by the high pressures needed to stabilize their crystal structure and stoichiometry. The primary and most pressing question thus becomes:
How can pressure be lowered, retaining high-temperature superconductivity? and ultimately, how can we find a superconductor at {\bf ambient conditions} of pressure and temperature? 

In this section, we will discuss possible strategies to realize these goals, based on the present understanding of superconductivity and currently available materials and synthesis techniques. 
In the first two parts (Sec.~\ref{Sec:Perspectives_chemSpace}~and~\ref{sec:StategiesHydrides}) we will maintain the focus on hydrides, which at the moment represent an excellent hunting ground for room temperature superconductivity, and most likely will continue to yield exciting surprises in the coming years. In the third part (Sec.~\ref{Sec:Perspectives_others}) we will go beyond hydrides and briefly discuss a more general perspective involving other possible superconducting materials of conventional type.

\begin{figure}[t]
\centering
\includegraphics[width=1.0\columnwidth]{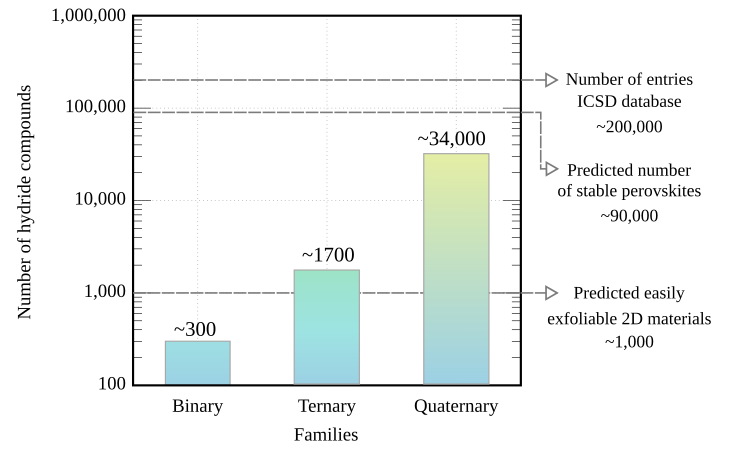}
\caption{~Estimated size of the hydrides chemical space. For comparison, the number of entries in the Inorganic Crystal Structure Database (ICSD) is over 200,000 structures, and the estimated chemical space for drug discovery spans up to 10 million (structures and compositions).} 
\label{fig:chem-space}
\end{figure}

\subsection{Chemical space of hydrides}\label{Sec:Perspectives_chemSpace}

The first question to ask is, how big is the chemical abundance of hydrides? This may give an order of magnitude on the number of potential candidates for superconductivity. 
Let us start with binary hydrides ($X$H), which have been studied extensively. A reasonable estimate is to consider that, for each element $X$, in a pressure range between 50 and 300\,GPa at least five compositions would lie in the convex hull of stability. 
Assuming that 60 elements could form a stable hydride and show superconductivity (a generous estimate), this gives a safe upper limit of {\bf 300 binary hydrides} that exist, are thermodynamically stable and superconduct.  

Providing an equally meaningful estimate for ternaries is considerably complicated because, at the moment, the knowledge of their phase stability and energy landscape is very limited. 
As the number of elements increases, the number of possible configurations rapidly grows and, at the same time, 
the number of channels for structural decomposition also increases. In addition, it is plausible that the role of 
metastability will be more crucial than in binary hydrides, i.e. metastable phases will prevail in some regions 
of the phase diagram, as decomposition paths become more complex. 
A very crude estimate can be obtained (at a single pressure) by a mere combinatorial argument. 
Similarly to binaries, we assume that ternaries can be formed with 60 elements of the periodic table. 
At a single pressure, not considering alloying, the number of possible, stable {\bf ternary hydrides compounds is 1,770}. 
However, there is no sufficient data to estimate how many structures will be on the hull of stability in the 50 to 300\,GPa range. 
It is even harder to provide a reasonable estimate for quaternary hydrides, but the same analysis yields 34,220 combinatorial compounds 
at a single pressure, neglecting alloying, metastability and configurational disorder. 
The number of conceivable structures in complex hydrides is large and grows rapidly with the number of elements involved. 
Arguably, any experimental attempt for synthesis without prior knowledge from computation or theory is bound to fail. 

We report our estimated numbers in Fig.~\ref{fig:chem-space} where, in order to give a perspective on the size of this chemical landscape, we compare it to other interesting classes of compounds. For instance, {\bf perovskites} that 
are one of the most versatile and chemically diverse families~\cite{FloresLivas_MagneticNitridePerovskites_2019}. 
There are currently approximately 2,000 known compositions, but the number of perovskites predicted to be stable is close to 90,000~\cite{filip2018geometric,xu2018rationalizing} at ambient pressure. Nevertheless, the number of accessible structures could potentially be much larger, especially if we include hybrid metal-organic perovskites under pressure~\cite{PhysRevMaterials.2.085201}. {\bf Two-dimensional (2D) materials}, have emerged as promising candidates for next-generation electronics applications. 
Although only a few dozens 2D materials have been successfully synthesized~\cite{gong2017discovery} or exfoliated~\cite{novoselov2004electric}, a recent study identified $\approx$1,000 easily exfoliable compounds~\cite{mounet2018two}. 
{\bf Inorganic 3D lattices} contained in the ICSD are above 200,000.  
The AFLOW  database (http://aflowlib.org/) lists $\approx$1,600 binary systems, $\approx$30,000 ternaries and 150,000 quaternaries. Finally, the estimated chemical space for {\bf drug discovery} comprises several million structures and compositions~\cite{reymond2012exploring}, and this shows that, put in perspective, 
the challenge we face for material discovery in hydrides is relatively accessible.  

\subsection{Optimizing Tc and pressure in hydrides}\label{sec:StategiesHydrides} 

Having observed that the hypothetical chemical space in hydrides is vast, it is clear that computational explorations are and will be essential to help steer research in the field towards promising directions. 
We now describe practical strategies targeted at increasing the \tc\  and reducing the pressure to stabilize hydrides.

$\bullet$ {\bf Selected ternary phases:} 
Due to the vast chemical space of ternary hydrides, a systematic search can hardly be afforded, and thus only a few studies, in literature, explore selected phases and compositions. For example, one can focus on ternary compositions stable at ambient pressure or attempt substitutions in binary hydrides, based on concepts such as chemical substitutability~\cite{Goldschmidt_Krystallochemie1926,glawe_2016_pettifor_data-minig}, or electronic structure arguments.
A good starting point is to consider ternary doping of existing 
high-\tc\ binary hydrides, as done by several authors who investigated different strategies of doping for \sh. 
Heil and Boeri~\cite{H:Heil_PRB_2015}, studied isovalent substitution of sulfur in \sh, with chalcogens (O, S, Se, Te) using the virtual crystal approximation to identify trends that would affect the \tc . 
They evidenced that isovalent substitution of sulfur with oxygen or selenium, for example, modifies the superconducting properties by acting on the bonding rather than on the Fermi level position, which hints towards a general trend in hydrides under pressure. 
In particular, the authors predicted that a moderate oxygen substitution in \sh\ might enhance the \tc, while substitution with heavier elements should be detrimental to superconductivity. 
In 2018, Liu et al.~\cite{Binbin_PRB_2018_Ternary-wrong} performed crystal structure prediction calculations at a fixed composition of S$_{0.5}$Se$_{0.5}$H$_{3}$ and found that the lowest-enthalpy structures indeed correspond to different decorations of the cubic SH$_3$ lattice. Based on their electron-phonon calculations, 
the superconducting temperature decreases when S is replaced by Se, which the authors attribute to a decreasing strength of the covalent H-S or H-Se bonds. More recent calculations of thermodynamical stability of H$_3$S$_{1-x}$Se$_x$ compound by Amsler~\cite{Amsler_Ternary_HSeS_PRB2019} revealed that, when stable, the mixed S/Se compositions does not show improved superconducting properties. A similar strategy applied to superhydrides, this time inspired by the discovery of \lah\, brought to the prediction of superconductivity at 258\,K in CaYH$_{12}$ under pressure~\cite{CaYH_12_ternary2019,CaYH_12_clathrate2019}. 

The resurgence of hydrides as energy materials could  have profound implications also for the field of superconductivity. 
The field of energy materials is in fact much more mature than that of high-pressure superconductivity, and the phase diagram and thermodynamics of many hydrides at ambient pressure are known in great detail. This knowledge can be exploited to design new superconductors. Large families such as borohydrides, amides, alanates, gallates, transition metal-hydrides and possibly active materials for lithium‑ion storage are potential candidates for high-pressure studies. The evolution of ternary hydrides as energy materials in a broader context has been recently  reviewed in an excellent work by Mohtadi-Orimo~\cite{mohtadi_Henergy-material-2017_review}. 
Two examples of high-pressure studies in ternary hydrogen-storage materials were recently reported by Kokail et al.~\cite{H:Kokail_LIBH_PRM_2017}, on lithium borohydrides, and by Hermann et al.~\cite{H-O-N_Hermann_2018,Amonia_Hermann_2017_icy}, on H-O-N. 

$\bullet$ {\bf Doping under pressure:} 
The study of ternary phases discussed above still focuses on  thermodynamical stability. However, this is likely a sufficient but not necessary condition for a material to be physically synthesizable. One could abandon the pursuit of thermodynamic stability and directly search for long-lived metastable structures,  especially if a synthesis path can be conceived. This is for example the case of doping into a stable crystal, realized inserting (substitutionally or interstitially) extra atoms, at low concentrations.

An obvious starting point is to perform charge doping of existing metallic phases at high pressure through partial replacement of guest atoms. This strategy can be used, for example, to optimally position the Fermi level in a region of high density of states, modify the shape of Fermi surfaces, or induce structural or electronic phase transitions which increase the electron-phonon coupling. However, high-pressure doping presents inherent experimental difficulties, and thus this possibility has been explored so far only by computational studies. For example, it has been shown that hole doping of \sh\ with 6.25\% of phosphorus at 200\,GPa increases \tc\ from 189 to 212\,K~\cite{Shimizu-Pdoping-SH3} due to the positive effect of the van Hove singularity. On the other hand, electron doping achieved by 6.25\%\ chlorine doping reduces \tc\ to 160\,K but has a positive impact on the pressure, lowering the stabilization threshold for the  $Im\bar{3}m$ phase. In a similar spirit, Ge et al.~\cite{Yungui_PRB_2016_ternaryCE} proposed doping \sh\ with elements from neighbouring groups in the periodic table and concluded that a \tc\ as high as 280\,K could be reached at 250\,GPa in S$_{0.925}$P$_{0.075}$H$_3$. However, the stability of other phases on the ternary convex hull could hinder doping of binary phases --a still unexplored possibility.

A viable strategy to use doping in the search for new superconducting hydrides is to induce metallicity in insulating crystals. It is well known that by introducing enough electron- or hole-donating impurities a semiconducting system
can be made metallic and even superconducting~\cite{ekimov_superconductivity_2004,Boeri_BdopedDiamond_PRL2004,bustarret_superconductivity_2006}. As a matter of fact, the class of molecular crystals containing light elements like carbon and hydrogen has a huge potential. Following this idea, Flores-Livas et al.~\cite{flores-sanna_ice_2017,Flores_polyethylene} studied H$_2$O and polyethylene, identifying promising substitutions/dopants and showed that for realistic levels of doping at high pressure, these covalent phases might exhibit superconductivity with a critical temperature of about 60\,K at moderate pressures.  Given the vast number of hydrides that are covalently-bonded but remain insulating up to rather high pressures, this venue represents a very promising route to hunt for high-\tc\ hydrides (see for instance Ref.~\cite{hot_Ma_view_point_PRL2019} for an overview). 

Despite a few promising theoretical hints, the study of doping at high pressure is at an early stage both experimentally and computationally. Experimentally one faces the problem that methods for doping under pressure have not been developed~\cite{willardson_HP-semiconductors-book_1998}. 
Computationally, one faces the problem of accessing the doping space of materials, which requires to apply systematic and well-converged protocols to scan for suitable dopants~\cite{grauzinyte2017computational}, and establish the properties and stability of dopants. 
At zero pressure and for semiconducting systems accurate  estimates~\cite{Freysold_PointDefectsInSolids_RMP2014,Bruneval_FormationVolumeChargedDefects_PRB2012,PCCP_defects_2019}  have been obtained, but this type of computations are intrinsically complicated and have not been applied to high-pressure hydrides so far.

$\bullet$ {\bf Mechanical precompression: nano scaffolds:}
Recently, Xia et al.~\cite{xia2019hydrogen} predicted that it is possible to obtain high-density hydrogen confining it in an (8,0) single-wall carbon nanotube. Their simulations showed evidence of metallization at moderate pressures realizing a  so-called {\bf physical compression effect}, which, together with chemical precompression, 
extends the possibilities to study materials at reduced pressures.  
Nanoscaffold structures such as nanotubes or fullerenes are ideal systems that could exert further pressure on tiny molecules confined in their interior. A word of caution is, of course, that the stability of nano-scale systems and encapsulation of molecules may be very challenging to achieve experimentally. 

$\bullet$ {\bf Non-hydrostatic pressure:}
Although non-hydrostatic conditions are preferably avoided in high-pressure experiments, if controlled, these effects may also represent a viable route to tune the electronic and vibrational properties of hydrides. However, controlling hydrostatic conditions is very challenging: while it is known that the lack of a suitable pressure-transmitting medium in the region of 100-200\,GPa can lead to conditions of inhomogeneous pressure, it is hard to control these conditions in a reproducible manner. In fact, non-hydrostatic conditions are highly material-, and sample-preparation dependent~\cite{klotz2009hydrostatic}. 

A recent computational work by Liu et al.~\cite{liu2018strain} attempted to examine the effect of non-hydrostatic strain modulation on H$_3$S. This paper pointed out that applied strain can effectively modify the electronic energy levels and the lattice vibrational frequencies, thereby allowing a sensitive control over \tc .  Another attractive possibility to exploit non-hydrostatic conditions is to access metastable phases, which offer a unique possibility to enhance the electronic structure and consequently the superconducting properties of materials under pressure. 
For example, in the recent experimental work on LaH$_x$, non-hydrostaticity seems to play a role in favouring some metastable compositions and structures~\cite{Hemley-LaH10_PRL_2019,Nature_LaH_Eremets_2019}.

$\bullet$ {\bf Metastability:} 
A number of examples have been recently reported, in which metastability adds an extra dimension to the phase diagram of several systems. This was observed, for instance, in elemental phosphorus~\cite{flores_interplay_2017}, 
barium~\cite{Hamlin-Ba} and gallium under pressure~\cite{Campanini-Ga}, tin-nitrides under pressure~\cite{kearney2018pressure},  
PdH~\cite{PdH-HTc-arxiv}, PH$_2$~\cite{flores-sanna_PH3_2016} and 
LaH$_x$~\cite{Hemley-LaH10_PRL_2019,drozdov2018_250}, as well as charge-density-wave superconductors~\cite{MoS2-metastable} and oxides~\cite{WO3-metastable,MoO3-metastable}. These works showed that controlled heating and cooling cycles or pressure-hysteresis cycles permit to stabilize different metastable phases selectively. 

Likely, at least some of the high-\tc\ phases observed in high-pressure hydrides are also metastable at lower pressures. Refining the experimental techniques that permit to control the pressure-hysteresis cycles may hopefully allow, in the near future, to stabilize some of the inaccessible high-\tc\ phases at moderate pressures. 
%%%%%%%%%%%%%%%%%%%%%%%%%%%%%%%%%%%%%%%%%%%%%%%%%%%%%%%%%%%%%%%%%%%%%%%%%%%%%%%%%%%

\subsection{Other (non-hydride) superconductors}\label{Sec:Perspectives_others}

Presently the main candidates  for room- temperature and pressure superconductivity are unconventional superconductors, such as cuprates or iron pnictides and chalcogenides, which currently hold the record for the highest \tc\ at ambient pressure. These systems are still heavily investigated and reserve many surprises. However, a clear advantage of  conventional superconductors is that theoretical and computational  methods can be used to guide the experimental discovery. The same can hardly be said about materials where exotic pairing mechanisms mediate superconductivity, since, in this case, a fully {\it ab initio}  description of the superconducting pairing is missing~\cite{Essenberger_SpinFluctuationsTheory_PRB2014}. 

Below we list the strategies that we consider most promising to search for new superconducting materials, not necessarily restricted to hydrides or conventional superconductivity. Some of these strategies are based on materials which contain  light elements, where the same arguments discussed for hydrides 
(low mass and strong covalent bonding) also hold.

$\bullet$ {\bf Reduced dimensionality:} 
A modern achievement in condensed matter research is the design of materials at the atomic scale, employing techniques such as controlled sputtering deposition~\cite{rucavado2018new}, growth of hetero-structures atom-by-atom, exfoliation and strain engineering with molecular beam epitaxy~\cite{ignatiev2018molecular}.  
These methods allow creating completely artificial materials (FeSe/FeTe on graphene, STO), in which it is possible to induce or tune superconductivity~\cite{PRB_MBE_supra_2011}.  The realization of low-dimensional 1D/2D metallic structures with light covalently-bonded elements (carbon, boron, nitrogen) represents an attractive possibility to discover high-\tc\ superconductors at ambient pressure. Note however that conceptually these approaches are orthogonal to those used at high pressure, where very little can be controlled, and thermodynamic conditions dictate the synthesis. 

One of the most interesting predictions within this class is that of doped graphane~\cite{Giustino_Graphane_PRL2010}, i.e. hydrogenated graphite (see below), indicating that hydrogen can be a fundamental ingredient of superconducting low-dimensional materials. 
Along the same line, the prediction of a superconducting phase in lithium-doped graphene~\cite{Profeta_Graphene_Nat2012} and their subsequent experimentally confirmation~\cite{SC:Ludbrook_Ligraphene_PNAS2015}, further testify the accuracy of first-principles computational methods for superconductivity.

$\bullet$ {\bf Indirect role of hydrogen:} 
Due to its strong ionic potential and high mobility, it is difficult for hydrogen at ambient pressure to be stable in any other chemical configuration than a fully saturated covalent bond. However, due to its small atomic size, hydrogen may be incorporated into existing materials and act indirectly to induce or enhance superconductivity. This is for example the case of palladium hydride, where hydrogen improves the superconducting properties of elemental palladium suppressing spin fluctuations~\cite{boufelfel2019hydrogen}. This approach may not fully benefit from the low-mass/high-phonon frequency argument, but may be essential to achieve high-temperature conventional superconductivity. 
An indirect key role of hydrogen is clearly seen in  high-\tc\ doped graphane; here, hydrogen saturates the $\pi$ bonds of graphene, and, as a result, states with large \ep\ coupling are promoted close to the Fermi level.
 When these states are populated through hole-doping,  superconductivity has been predicted to occur at temperatures as high as 100\,K~\cite{SC:savini_graphane_PRL_2010}. 
Following the same line of thought, and using a \mg\ monolayer as a template material, Bekaert et al.~\cite{MgB2_H_superconductivity} predicted an increased critical temperature in \mg\ from 39\, to 67\,K. 

$\bullet$ {\bf Chemical doping:}
Charge doping through alkali metal incorporation has led to the discovery of a superconducting phase in fullerenes (C$_{60}$)~\cite{C60_superconductivity}, where the current \tc\, record is 40\,K in Cs$_3$C$_{60}$. 
Endohedral doping of the cages is often regarded as the most efficient approach to induce superconductivity, 
but exohedreal doping  could also be an attractive alternative~\cite{de2018stable}. In the same spirit, stable cages of the sodalite-type structure can be doped with elements of the V, VI, and VII columns of the periodic table~\cite{Hapiuk_PRL2012}. 
More recently, theoretical calculations have also predicted superconductivity in doped cages of ZnO~\cite{hapiuk2015superconductivity}. 
Superconductivity has also been observed in intercalated graphite superconductors, notably CaC$_6$ (\tc =11.5 K) and YbC$_6$ (\tc = 11.5 K)~\cite{Weller_CaC6_NAT2005}. Other intercalated systems showing superconductivity are layered binary silicides~\cite{disicilicides_PRB2011}, where it has been shown that, through a high-pressure synthesis procedure, the stabilization of metastable sheets of silicon leads to an enhancement of almost 30\% in \tc~\cite{enhancing_PRL2011}. Estimates of superconductivity in the high-doping limit were given for generic families of intercalated group 4 honeycomb lattices~\cite{Sanna_PRB-2015}. Other examples of superconductivity induced by chemical doping include potassium-doping of polycyclic aromatic hydrocarbons molecular crystals like picene~\cite{picene_kubozono,mine:Subedi_PRB_2011,SC:Casula_picene_PRB_2012}, phenanthrene~\cite{phenanthrene}, dibenzopentacene~\cite{dibenzopentacene} and terphenyl~\cite{terphenyl1,terphenyl2}, which have also been reported to exhibit critical temperature as high as 100\,K~\cite{terphenyl1,terphenyl2}. 

$\bullet$ {\bf Doping by field-effect:} 
Modern techniques based on liquid electrolytes permit to achieve doping levels as high as 10$^{15}$ carriers/cm$^3$, corresponding to tenths of electrons/atom, roughly three orders of magnitude larger than standard solid-state techniques. However, field gating does not act by simply charging the system but, due to the intense fields involved,  can also lead to permanent or long-lived chemical or structural transformations in the region near the surface of the 2D systems. A fascinating speculation is that, since electrolytes usually contain hydrogen or light-elements, field gating may induce a sizable migration of H$^+$ ions onto the gated system, leading to a metallic transition and possibly to superconductivity. Such an insulator-to-metal transition has been unequivocally seen in hydrogen-doped WO$_3$~\cite{WO3-H_synthesis}, where claims of induced superconductivity at 120\,K were reported~\cite{WO3-H_superconductivity}. 

Moreover, the possibility to induce metallic surface states in insulators through an intense electric field and to turn two-dimensional materials into superconductors~\cite{2Dsuper_Iwasa} has opened a new avenue of research. The successful synthesis of a plethora of two-dimensional systems over the last years such as transition metal dichalcogenides~\cite{MoS2-synthesis}, layered transition metal oxides~\cite{MoO3_synthesis,La2CuO4_synthesis}, hexagonal-boron nitride~\cite{BN_synthesis}, topological insulators~\cite{topological_transport}, 2D-cuprates~\cite{cuprate_2D}, ferromagnetic 2D systems~\cite{CrI3_synthesis}, phosphorene~\cite{blackP-_synthesis}, bismuthene\cite{Bismuthene_synthesis}, silicene, germanene~\cite{silicene_synthesis,germanene_synthesis} and borophene~\cite{Borophene_synthesis} has fuelled the field, permitting to investigate doping- and field-induced superconductivity in a large array of different systems. 

\subsection{Theoretical and experimental outlook}\label{Sec:Perspectives_future}

As discussed in Sec.~\ref{sec:StategiesHydrides}, there is still a long path ahead in the study of conventional superconductivity. In particular, concerning hydrides under pressure, we have shown how computational methods can help and guide experimental search to discover novel conventional superconductors. However, there are several theoretical,  computational and experimental challenges left for future developments. Before concluding we would like to list some
of them (not ordered by importance): 

{\it Theoretical Challenges:}

\begin{itemize}
\item[] 1) The residual Coulomb interaction between electrons is often treated within a crude Morel-Anderson approximation,
which is not guaranteed to hold generally. 
\item[] 2) The effect of vertex corrections beyond the Migdal approximation on the electronic self energy 
is usually neglected, but they have been shown to play a role in hydrogen-rich 
materials~\cite{PRB_Sano_Van-Hove_H3S_2016}. 
Generally, they are expected to play an important role for systems where phonon frequencies
are large and the \ep\ interaction is sizable~\cite{Th:pietronero_PRL_1995,Boeri_PRB_2005}. 
\item[] 3) Non-adiabatic effects on phonons have been linked to vibrational anomalies observed by Raman spectroscopy in \mg, 
with no significant implications for superconductivity~\cite{Calandra_AnharmonicAndNonAdiabaticMgB2Raman_PhysicaC2007}. 
Similar effects in hydrides, where the Migdal's approximation 
for the phonon polarization is questionable, are unexplored.    
\item[] 4) Non-linear electron-phonon coupling and its repercussions on \tc, are largely overlooked. 
\item[] 5) Plasmon and polaron effects have been shown to play an essential role for some materials under pressure but are usually disregarded. 
\item[] 6) Currently, a KS-DFT functional explicitly constructed for the extreme compression regime does not exist, 
and virtually all studies conducted so far rely on functionals based on the standard electron gas 
(tuned to reproduce ambient pressure properties). 
\item[] 7) Quantum lattice effects have been conclusively shown to be of tremendous importance, 
specifically at high pressure and for hydrogen-rich materials~\cite{Nature_Errea_2016,work_on_LaHx_2019}. 
Presently, their simulation requires an overhead computational cost, and improvement in computational schemes, algorithms and implementations is strongly desirable.
\end{itemize}

{\it Computational Challenges:} 
Most of the future research effort will undoubtedly be directed towards the automation of procedures 
and the inclusion of computational informatics (machine learning, artificial intelligence, among others) 
into the prediction of crystal structures and new compounds. 
The automatic prediction of phase diagrams is a realistic goal to be reached within the next years. 
In this respect, the computational exploration of binary hydrides is almost completed, 
and ternaries represent the next venue for searching. 
Other aspects that have to be addressed employing computational methods are reaction paths between 
structures and stoichiometries, rare events and nucleation~\cite{Luigi_nucleation_PRL2018}
processes, which are practically unexplored issues in hydrides under pressure. 

{\it Experimental Challenges:} 
The future in high pressure foresees many developments, mainly on two fronts: 
\begin{itemize}
\item[] 1) Focus ion beam techniques will play a fundamental role in DACs, sculpting futuristic micro-anvils 
that will extend the range of pressure well above the 400\,GPa (the ceiling pressure for one-stage DAC). 
\item[] 2) The advancement in new X-ray sources (synchrotron and X-ray free-electron laser) will allow to perform spectroscopic measurements at the sub-micron scale, and the availability of brighter and more intense sources with shorter pulses will enable the study compression volumes of only tens of picolitres. 
Other areas of experimental high pressure research that will increase 
substantially are the study of magnetism and defects in the megabar range. 
\end{itemize}

\section{Conclusions}\label{Sec:conclusions}

In this Review, we offered an up-to-date perspective on the field of conventional superconductivity at high pressures. 
The field has been recently fueled by the experimental discovery of two record-breaking hydrides, 
\sh\ (2015, 200\,K) and \lah\ (2018, 250\,K). 
These materials, which currently hold the record for the highest and second-highest \tc, are different in almost every respect from what would have been considered an ideal superconductor a few years ago. 
They have a rather simple stoichiometry, highly-symmetric structures and are standard metals that obey a conventional mechanism for superconductivity. Given the extremely rapid progress over the last five years, it is a safe bet to predict that hydrides will be the first class of materials where room temperature superconductivity (300\,K) will be achieved in the coming years. 

Hydrides, despite their apparent simplicity when compared to more complex unconventional superconductors, such as cuprates and Fe pnictides and chalcogenides, do possess one aspect which makes them unusual. 
These materials require pressures  more than one million times larger than ambient pressure in order to exist and superconduct. 
Currently, only a handful of groups worldwide can perform the complex characterizations needed to detect superconductivity at these pressures. 
It is important to realize, however, that the fact that the number of groups currently performing these measurements 
is so small does not imply that the necessary set-ups are impossibly expensive or sophisticated.
Counter-intuitively, a HP set-up such as the diamond anvil cell is considerably more accessible than even the cheapest scanning electronic microscope or ultra-high-vacuum deposition chamber. Many groups have access to commercial diamond anvil cells and can routinely reach megabar pressures. The main additional difficulty in the study of high-pressure hydrides resides in the mastering of techniques needed to detect the properties of superconductors under pressure. 
The particular set of skills and instruments have been developed only by a few groups through many years of careful experimentation. \\

Furthermore, the excitement produced by the discovery of warm-superconductors is only one aspect of the even 
more exciting paradigm shift that is taking place in material research. The discovery of high-\tc\ superconductivity in hydrides is, in fact, a spectacular demonstration of the extraordinary symbiosis between experimental research, theoretical methods and computational tools that have taken place over the last years. 

Currently, the main drawback of superconducting hydrides lies in the impossibility for any practical application, 
which limits their near-future interest to the level of basic research. Nevertheless, it is reasonable to expect that this field will increase considerably in size and flourish, possibly delivering futuristic applications. In order to push this field forward, it is highly desirable to determine whether high-temperature conventional superconductivity can be extended to other hydrides which exist in the solid-state and, in the long term, whether the pressure can be drastically decreased. The large portfolio of existing hydrides, which includes energy storage and active materials for lithium‐ion storage, contains many promising materials for achieving this goal. Another route to find better superconductors is to apply the same concepts and methods that led to the discovery of high-\tc\ superconductivity in hydrides to other compounds. \\

In the previous sections, we have reviewed the main aspects of the extraordinary machinery used to discover materials under high pressure, which aided the ground-breaking discoveries we compare it to other interesting of \sh\ and \lah, and elucidated the underlying physics in these materials. We provided a fresh update on experimental techniques and reviewed the key points on the theoretical understanding of superconductors. Other critical active areas of research and future direction were also discussed in this Review. 

Due to the rapid progress of the field, we are confident that some parts of the Review will become obsolete in a short time. 
Interestingly, over the last five years (2014-2019), the goal of superconducting research in hydrides has shifted from the seemingly impossible observation of room-temperature superconductivity, to find ways to achieve the same phenomenon at ambient conditions, both of pressure and temperature. We hope that the methodological framework described and the combined knowledge from experiments, theory and computation may serve as a useful reference and inspire many more exciting discoveries in the years to follow. \\

%% many hours. many many hours of my shitty life in this... I am sad. 4.Dec.2019, JAFL (never again, do the same mistake). 
% > I am happy, 9 gen 2020. it is done. 

{\bf Note added in proof}\\ 
The first version of this Review was completed in May 2019. 
At that time, only \sh\ and \lah\ had been discovered to superconduct at high-\tc.  
A few months after we completed the first version, several important papers have appeared, reporting
the observation of superconductivity in yttrium hydride at 240\,K~\cite{Kong_arxiv_YH9_2019,Troyan_arxiv_YH6_2019}, 
and in thorium hydride at 170\,K~\cite{Semenok_arxiv_ThH10_2019}, as well as the prediction of room temperature superconductivity in the Li$_2$MgH$_{16}$ system~\cite{Ma_RoomT_PRL2019,hot_Ma_view_point_PRL2019}.
Moreover, a thorough theoretical explanation of the open issues in LaH$_{10}$ has also been formulated~\cite{work_on_LaHx_2019}.
For the revised version of this Review, January 2020, 
we updated the relevant references and figures, but not included an in-depth description of the results.\\

{\bf Acknowledgements:}
J.A.F.-L. Acknowledges Stefan Goedecker for willingness to 
support and the NCCR MARVEL funded by the Swiss National Science Foundation. 
J.A.F.-L. and M.E. are thankful to A. Drozdov for valuable discussions. 
Computational resources from project s970 of the 
Swiss National Supercomputing Center (CSCS) in Lugano gratefully acknowledged.  
%----------------------------------------------------------
L.B. acknowledges support from Fondo Ateneo Sapienza 2017-18. 
%----------------------------------------------------------
G.P. acknowledges financial support from the Italian Ministry for Research and Education 
through PRIN-2017 project Tuning and understanding Quantum phases in 2D materials - Quantum 2D 
(IT-MIUR Grant No. 2017Z8TS5B) and CINECA (ISCRA initiative) for computing resources.
%----------------------------------------------------------
R.A. was supported by a Grant-in-Aid for Scientific Research (No.16H06345, 19H05825) 
from Ministry of Education, Culture, Sports, Science and Technology, Japan.
%----------------------------------------------------------- 
M.E. is thankful to the Max Planck community for the invaluable support, 
and U. P\"oschl for the constant encouragement.
%-----------------------------------------------------------
Some of the authors acknowledges the hospitality of the {\it cini-Sardegna} meeting 
where parts of this work were written.

\newpage
\textbf{Appendix}\label{Sec:Appendix}

{\bf Periodic Tables:}

The information shown in the periodic table of superconducting elements (Fig.~\ref{fig:Elemental_supra}) 
has been collected from references
~\cite{buzea2004assembling,Shimizu_EHP2005,SC:HAMLIN_highp_physicaC_2015,shimizu2015superconductivity}. 
%%%%%%%%%%%%%%%%%%%%%%%%%%%%%%%%% 
The following works refer to studies of binary hydrides 
under pressure that was used to construct Fig. ~\ref{fig:Periodic-table_hydrides}.  
Li~\cite{Li_xie2014superconductivity}, K~\cite{K_zhou2012ab}, Fr~\cite{semenok2018distribution}, 
Be~\cite{Be_yu2014exploration}, Mg~\cite{Mg_feng2015compressed}, Ca~\cite{Ca_PNAS_2012superconductive}, 
Sr~\cite{Sr_hooper2014composition,Sr_wang2015structural} 
Ba~\cite{Ba_hooper2013polyhydrides}, 
Ra~\cite{semenok2018distribution}, 
Sc~\cite{Sc_ye2018high}, 
Y~\cite{Kong_arxiv_YH9_2019,Troyan_arxiv_YH6_2019}, 
La~\cite{PNAS_LaHx_2017_Hemley,Clathrate_REHX_PRL_2017}, 
(Ce, Pr, Nd, Ho, Er, Tm, and Lu)~\cite{semenok2018distribution}, 
Ac~\cite{Ac_semenok2018actinium}, Th~\cite{Semenok_arxiv_ThH10_2019,Th_kvashnin2018high}, Pa~\cite{semenok2018distribution,PaH_shit_of-shits_2019}, 
U~\cite{U_kruglov2018_oganov}, (Np, Cm)~\cite{semenok2018distribution}, 
(Ti,Zr,Hf)~\cite{semenok2018distribution}, V~\cite{V_li_2017superconductivity}, 
Nb~\cite{Nb_Hoffman2013_theoretical,Nb_Durajski2014_phonon}, 
Ta~\cite{Ta_zhuang2017pressure}, Cr~\cite{Cr_yu2015pressure}, W~\cite{W_zheng2018structural}, 
Tc~\cite{Tc_li2016crystal}, Ru~\cite{Ru_liu2016stability}, 
Os~\cite{Os_liu2015structures}, (Rh,Ir)~\cite{PRL_kim2011_subproduct},
Pd~\cite{houari_PdH-stability_JAP_2014,PdH_pressure_PRB_hemmes1989}, 
Pt~\cite{Pt_errea2014anharmonic,Pt_zhou2011superconducting,PRL_kim2011_subproduct,PRB_Matsuoka_PtH_30GPa_2019},
Au~\cite{PRL_kim2011_subproduct}, B~\cite{B_abe2011crystalline}, Al~\cite{Al_hou2015high}, 
Ga~\cite{Ga_szczesniak2013superconducting}, In~\cite{In_liu2015pressure}, 
Si~\cite{Eremets_silane_2008,flores_disilane_2012,Si_jin2010superconducting}, 
Ge~\cite{Ge_abe2013quantum,Ge_strobel2010vibrational,Ge_szcze2014thermodynamics,Ge_zhong2013superconductivity,Ge_hou2015ab,Ge_esfahani2017superconductivity,Ge_abe2013quantum,Ge_esfahani2017superconductivity}, 
Sn~\cite{Sn_esfahani2016superconductivity}, 
Pb~\cite{Pb_cheng2015pressure}, 
P~\cite{flores-sanna_PH3_2016,shamp_decomposition_2016,P_liu2016crystal,P_durajski2016quantitative,PH_metastable_Zurek_2017}, 
As~\cite{As_fu2016high}, 
Sb~\cite{Sb_ma2015unexpected}, 
Bi~\cite{Bi_ma2015high}, 
S~\cite{flores-sanna_HSe_2016}, 
Se~\cite{flores-sanna_HSe_2016,Se_zhang2015phase}, 
Te~\cite{Te_hydrides_Ma_2016}, 
Po~\cite{Po_RSC_Tian_2015}, 
Br~\cite{Br_Cl_APL_2010}, I~\cite{I_iodine_doped_2015,I_jpclett_2015_Iodine}, and Xe~\cite{Xe-H_JCP_2015}. 

{\bf Definition of hydrides and superhydrides:} 
Note that the term hydride is an old concept and could have several definitions. 
For example, Gibbs used it to describe a compound in which there is a metal-to-hydrogen bond. 
Along this Review we treated hydride only as a combination of hydrogen and any other element.
Under this definition, hydrides can be generally classified into four categories based on the nature 
of their hydrogen bonding: ionic or saline (i.e., hydrides of alkali and alkaline earth metals), 
covalent or volatile (i.e., hydrides of Group IIIA-VIIA elements), metallic (i.e., hydrides of transition metals), 
or vdW (i.e., hydrides of rare gas elements). 
It is clear that this classification does not offer a complete 
vision of hydrides as the nature of the hydrogen bond can be significantly altered by pressure. 
As such, for simplicity in the construction of periodic tables 
(Fig.~\ref{fig:Periodic-table_hydrides} and Fig.~\ref{fig:Elemental_supra}) 
we omit the naming of compounds accordingly to chemistry conventions. 
For example note that in the periodic table, H$_3$S appears as SH$_3$. 
It also neglects further distinctions such as {\bf superhydrides}, 
which have unusually high hydrogen contents (e.g., LaH$_{10}$). 

{\bf Statistics on publications:} 
In order to track the number of publications on the field 
(see Fig.~\ref{fig:size_of_field}) we used {\sc Google scholar} engine 
together with searches and keywords such as 
--hydrides-- followed by filtering options with --superconducting-- over the selected interval. 
Needless to mention that these numbers are approximate, 
especially on decades when hydrides started to be explored on other domains such as energy materials.   

\newpage

\bibliographystyle{elsarticle-num} 
\bibliography{BIB_master}

\begin{thebibliography}{100}
\expandafter\ifx\csname url\endcsname\relax
  \def\url#1{\texttt{#1}}\fi
\expandafter\ifx\csname urlprefix\endcsname\relax\def\urlprefix{URL }\fi
\expandafter\ifx\csname href\endcsname\relax
  \def\href#1#2{#2} \def\path#1{#1}\fi

\bibitem{DrozdovEremets_Nature2015}
A.~P. Drozdov, M.~I. Eremets, I.~A. Troyan, V.~Ksenofontov, S.~I. Shylin,
  {{Conventional superconductivity at 203 kelvin at high pressures in the
  sulfur hydride system}}, Nature 525 (2015) 2015/08/17/online.

\bibitem{Nature_LaH_Eremets_2019}
A.~P. Drozdov, P.~P. Kong, V.~S. Minkov, S.~P. Besedin, M.~A. Kuzovnikov,
  S.~Mozaffari, L.~Balicas, F.~F. Balakirev, D.~E. Graf, V.~B. Prakapenka,
  E.~Greenberg, D.~A. Knyazev, M.~Tkacz, M.~I. Eremets, Superconductivity at
  250 {K} in lanthanum hydride under high pressures, Nature 569~(7757) (2019)
  528--531.

\bibitem{Hemley-LaH10_PRL_2019}
M.~Somayazulu, M.~Ahart, A.~K. Mishra, Z.~M. Geballe, M.~Baldini, Y.~Meng,
  V.~V. Struzhkin, R.~J. Hemley, {Evidence for Superconductivity above 260 K in
  Lanthanum Superhydride at Megabar Pressures}, Phys. Rev. Lett. 122 (2019)
  027001.

\bibitem{ginzburg1999problems}
V.~L. Ginzburg, {What problems of physics and astrophysics seem now to be
  especially important and interesting (thirty years later, already on the
  verge of {XXI} century)?}, Physics-Uspekhi 42~(4) (1999) 353.

\bibitem{mukhin2003centenary}
K.~N. Mukhin, A.~F. Sustavov, V.~N. Tikhonov, On the centenary of the nobel
  prize: Russian laureates in physics, Physics-Uspekhi 46~(5) (2003) 493.

\bibitem{wigner1935possibility}
E.~Wigner, H.~B. Huntington, On the possibility of a metallic modification of
  hydrogen, The Journal of Chemical Physics 3~(12) (1935) 764--770.

\bibitem{Ashcroft_PRL1968}
N.~Ashcroft, {Metallic Hydrogen: A High-Temperature Superconductor?}, Phys.
  Rev. Lett. 21 (1968) 1748--1749.

\bibitem{Ginzburg_1969}
V.~L. Ginzburg, Superfluidity and superconductivity in the universe, Journal of
  Statistical Physics 1~(1) (1969) 3--24.

\bibitem{Dias_hydrogen_Science2017}
R.~P. Dias, I.~F. Silvera, Observation of the wigner-huntington transition to
  metallic hydrogen, Science.

\bibitem{mcmahon2012properties}
J.~M. McMahon, M.~A. Morales, C.~Pierleoni, D.~M. Ceperley, The properties of
  hydrogen and helium under extreme conditions, Reviews of modern physics
  84~(4) (2012) 1607.

\bibitem{Militzer_JupiterInterior_JGeoResPlanets2016}
B.~Militzer, F.~Soubiran, S.~M. Wahl, W.~Hubbard, {Understanding Jupiter's
  interior}, Journal of Geophysical Research: Planets 121~(9) (2016)
  1552--1572.

\bibitem{Gilman_1971_PRL_LiHF}
J.~J. Gilman, {Lithium Dihydrogen Fluoride --An Approach to Metallic Hydrogen},
  Phys. Rev. Lett. 26 (1971) 546--548.

\bibitem{Ashcroft_PRL2004}
N.~W. Ashcroft, {Hydrogen Dominant Metallic Alloys: High Temperature
  Superconductors?}, Phys. Rev. Lett. 92 (2004) 187002.

\bibitem{earth_pressure-core_2002}
D.~Alf{\`e}, M.~Gillan, G.~Price, Composition and temperature of the earth’s
  core constrained by combining ab initio calculations and seismic data, Earth
  and Planetary Science Letters 195~(1-2) (2002) 91--98.

\bibitem{molodets2005scaling}
A.~Molodets, Scaling law for high pressure isotherms of solids, High Pressure
  Research 25~(4) (2005) 267--276.

\bibitem{loubeyre1996x}
P.~Loubeyre, R.~LeToullec, D.~Hausermann, M.~Hanfland, R.~Hemley, H.~Mao,
  L.~Finger, X-ray diffraction and equation of state of hydrogen at megabar
  pressures, Nature 383~(6602) (1996) 702.

\bibitem{hemley2000effects}
R.~J. Hemley, Effects of high pressure on molecules, Annual Review of Physical
  Chemistry 51~(1) (2000) 763--800.

\bibitem{hermann2017chemical}
A.~Hermann, {Chemical Bonding at High Pressure}, Reviews in Computational
  Chemistry 30 (2017) 1--41.

\bibitem{yoo2013physical}
C.-S. Yoo, Physical and chemical transformations of highly compressed carbon
  dioxide at bond energies, Physical Chemistry Chemical Physics 15~(21) (2013)
  7949--7966.

\bibitem{onnes1913further}
H.~K. Onnes, {Further experiments with liquid helium. H. On the electrical
  resistance of pure metals etc. VII. The potential difference necessary for
  the electric current through mercury below 4◦ 19K}, in: KNAW, Proceedings,
  Vol.~15, 1913, pp. 1406--1430.

\bibitem{Nb3Ge_PRB1965}
B.~T. Matthias, T.~H. Geballe, R.~H. Willens, E.~Corenzwit, G.~W. Hull,
  {Superconductivity of Nb$_{3}$Ge}, Phys. Rev. 139 (1965) A1501--A1503.

\bibitem{Th:Cohen_anderson}
M.~Cohen, P.~Anderson, Superconductivity in $d-$ and $f-$ band Metals, American
  Inst. of Physics, 1972.

\bibitem{SC:akimitsu_mgb2}
J.~Nagamatsu, N.~Nakagawa, T.~Muranaka, Y.~Zenitani, J.~Akimitsu,
  Superconductivity at 39 {K} in magnesium diboride, Nature (London) 410 (2001)
  63.

\bibitem{Schilling_cuprate_nature1993}
A.~Schilling, M.~Cantoni, J.~D. Guo, H.~R. Ott,
  \href{https://doi.org/10.1038/363056a0}{{Superconductivity above 130 K in the
  Hg--Ba--Ca--Cu--O system}}, Nature 363~(6424) (1993) 56--58.
\newblock \href {http://dx.doi.org/10.1038/363056a0}
  {\path{doi:10.1038/363056a0}}.
\newline\urlprefix\url{https://doi.org/10.1038/363056a0}

\bibitem{fesc:kamihara2006}
Y.~Kamihara, H.~Hiramatsu, M.~Hirano, R.~Kawamura, H.~Yanagi, T.~Kamiya,
  H.~Hosono, \href{https://doi.org/10.1021/ja063355c}{{Iron-Based Layered
  Superconductor: {LaOFeP}}}, Journal of the American Chemical Society 128~(31)
  (2006) 10012--10013.
\newblock \href {http://dx.doi.org/10.1021/ja063355c}
  {\path{doi:10.1021/ja063355c}}.
\newline\urlprefix\url{https://doi.org/10.1021/ja063355c}

\bibitem{fesc:kamihara_JACS_2008}
Y.~Kamihara, T.~Watanabe, M.~Hirano, H.~Hosono, {Iron-Based Layered
  Superconductor {LaOFeAs}}, Journal of the American Chemical Society 130~(11)
  (2008) 3296–3297.

\bibitem{matthiasrules}
B.~Matthias, Chapter v superconductivity in the periodic system, in: Progress
  in low temperature physics, Vol.~2, Elsevier, 1957, pp. 138--150.

\bibitem{DFT:Savrasov_PRB_1996}
S.~Y. Savrasov, D.~Y. Savrasov, Electron-phonon interactions and related
  physical properties of metals from linear-response theory, Phys. Rev. B 54
  (1996) 16487.

\bibitem{geballe2006never}
T.~H. Geballe, The never-ending search for high-temperature superconductivity,
  Journal of superconductivity and novel magnetism 19~(3-5) (2006) 261--276.

\bibitem{BCS_1957}
J.~Bardeen, L.~N. Cooper, J.~R. Schrieffer, Theory of superconductivity, Phys.
  Rev. 108 (1957) 1175--1204.

\bibitem{Eliashberg}
G.~Eliashberg, {Eliashberg Equations of Strong Coupling Theory}, Teor. Fiz
  \textbf{38}, 966 (1960)[Sov. Phys. JETP 38.

\bibitem{OGK_SCDFT_PRL1988}
L.~N. Oliveira, E.~K.~U. Gross, W.~Kohn, Density-functional theory for
  superconductors, Phys. Rev. Lett. 60 (1988) 2430--2433.

\bibitem{Lueders_SCDFT_PRB2005}
M.~L\"uders, M.~A.~L. Marques, N.~N. Lathiotakis, A.~Floris, G.~Profeta,
  L.~Fast, A.~Continenza, S.~Massidda, E.~K.~U. Gross, Ab-initio theory of
  superconductivity. i. density functional formalism and approximate
  functionals, Phys. Rev. B 72 (2005) 024545.

\bibitem{Marques_SCDFT_PRB2005}
M.~A.~L. Marques, M.~L\"uders, N.~N. Lathiotakis, G.~Profeta, A.~Floris,
  L.~Fast, A.~Continenza, E.~K.~U. Gross, S.~Massidda, Ab initio theory of
  superconductivity. ii. application to elemental metals, Phys. Rev. B 72
  (2005) 024546.

\bibitem{Margine_anisoEliashberg_PRB2013}
E.~R. Margine, F.~Giustino, {Anisotropic Migdal-Eliashberg theory using Wannier
  functions}, Phys. Rev. B 87 (2013) 024505.

\bibitem{Arita_Nonempirical_AdvMat2017}
R.~Arita, T.~Koretsune, S.~Sakai, R.~Akashi, Y.~Nomura, W.~Sano, Nonempirical
  calculation of superconducting transition temperatures in light-element
  superconductors, Advanced Materials 29~(25) (2017) 1602421.

\bibitem{sanna-flores_2018_Eliashberg}
A.~Sanna, J.~A. Flores-Livas, A.~Davydov, G.~Profeta, K.~Dewhurst, S.~Sharma,
  E.~Gross, {Ab initio Eliashberg Theory: Making Genuine Predictions of
  Superconducting Features}, Journal of the Physical Society of Japan 87~(4)
  (2018) 041012.

\bibitem{Baroni_LinearResponse_PRL1987}
S.~Baroni, P.~Giannozzi, A.~Testa, Green's-function approach to linear response
  in solids, Phys. Rev. Lett. 58 (1987) 1861--1864.

\bibitem{Baroni_DFPT_RMP2001}
S.~Baroni, S.~de~Gironcoli, A.~Dal~Corso, P.~Giannozzi, Phonons and related
  crystal properties from density-functional perturbation theory, Rev. Mod.
  Phys. 73 (2001) 515--562.

\bibitem{DeGironoli_DFPTmetals_PRB1995}
S.~de~Gironcoli, Lattice dynamics of metals from density-functional
  perturbation theory, Phys. Rev. B 51 (1995) 6773--6776.

\bibitem{Savrasov_ep-PRL-1994}
S.~Y. Savrasov, D.~Y. Savrasov, O.~K. Andersen, Linear-response calculations of
  electron-phonon interactions, Phys. Rev. Lett. 72 (1994) 372--375.

\bibitem{Savrasov2_PRB}
S.~Y. Savrasov, D.~Y. Savrasov, Electron-phonon interactions and related
  physical properties of metals from linear-response theory, Phys. Rev. B 54
  (1996) 16487--16501.

\bibitem{Eremets_silane_2008}
M.~I. Eremets, I.~A. Trojan, S.~A. Medvedev, J.~S. Tse, Y.~Yao,
  Superconductivity in hydrogen dominant materials: Silane, Science 319~(5869)
  (2008) 1506--1509.

\bibitem{maddox1988crystals}
J.~Maddox, Crystals from first principles, Nature 335~(6187) (1988) 201.

\bibitem{Bednorz_Muller}
J.~G. Bednorz, K.~A. M\"{u}ller, Possible high {Tc} superconductivity in the
  {B}a{L}a{C}u{O} system, Zeitschrift f\"{u}r Physik B Condensed Matter
  64~(189-193).

\bibitem{Review_Oganov-Pickard_2019}
A.~R. Oganov, C.~J. Pickard, Q.~Zhu, R.~J. Needs, Structure prediction drives
  materials discovery, Nature Reviews Materials 4~(5) (2019) 331--348.

\bibitem{Duan_2019_review}
D.~Duan, H.~Yu, H.~Xie, T.~Cui, Ab initio approach and its impact on
  superconductivity, Journal of Superconductivity and Novel Magnetism 32~(1)
  (2019) 53--60.

\bibitem{wang2018hydrogen}
H.~Wang, X.~Li, G.~Gao, Y.~Li, Y.~Ma, Hydrogen-rich superconductors at high
  pressures, Wiley Interdisciplinary Reviews: Computational Molecular Science
  8~(1) (2018) e1330.

\bibitem{Ma_NatRevMaterials_2017}
L.~Zhang, Y.~Wang, J.~Lv, Y.~Ma, Materials discovery at high pressures, Nature
  Reviews Materials 2~(4) (2017) 17005.

\bibitem{wang2014perspective}
Y.~Wang, Y.~Ma, Perspective: Crystal structure prediction at high pressures,
  The Journal of chemical physics 140~(4) (2014) 040901.

\bibitem{Review_Zurek-2018}
T.~Bi, N.~Zarifi, T.~Terpstra, E.~Zurek,
  \href{http://www.sciencedirect.com/science/article/pii/B9780124095472114350}{The
  search for superconductivity in high pressure hydrides}, in: Reference Module
  in Chemistry, Molecular Sciences and Chemical Engineering, Elsevier, 2019, p.
  050901.
\newblock \href
  {http://dx.doi.org/https://doi.org/10.1016/B978-0-12-409547-2.11435-0}
  {\path{doi:https://doi.org/10.1016/B978-0-12-409547-2.11435-0}}.
\newline\urlprefix\url{http://www.sciencedirect.com/science/article/pii/B9780124095472114350}

\bibitem{kresin_reviewH_RMP2018}
L.~P. Gor'kov, V.~Z. Kresin, {Colloquium: High pressure and road to room
  temperature superconductivity}, Rev. Mod. Phys. 90 (2018) 011001.

\bibitem{Zurek_HiTcSC_Polyhydriedes_JCP2019}
E.~Zurek, T.~Bi, \href{https://doi.org/10.1063/1.5079225}{High-temperature
  superconductivity in alkaline and rare earth polyhydrides at high pressure: A
  theoretical perspective}, The Journal of Chemical Physics 150~(5) (2019)
  050901.
\newblock \href {http://dx.doi.org/10.1063/1.5079225}
  {\path{doi:10.1063/1.5079225}}.
\newline\urlprefix\url{https://doi.org/10.1063/1.5079225}

\bibitem{Pickard_Review_ARCMP2019}
C.~J. Pickard, I.~Errea, M.~I. Eremets, {Superconducting Hydrides Under
  Pressure}, Annual Review of Condensed Matter Physics 11~(1) (2020) null.
\newblock \href {http://dx.doi.org/10.1146/annurev-conmatphys-031218-013413}
  {\path{doi:10.1146/annurev-conmatphys-031218-013413}}.

\bibitem{bridgman1959way}
P.~W. Bridgman, Nobel lecture: General survey of certain results in the field
  of high-pressure physics (1946).

\bibitem{weir1959infrared}
C.~Weir, E.~Lippincott, A.~Van~Valkenburg, E.~Bunting, Infrared studies in the
  1-to 15-micron region to 30,000 atmospheres, J. Res. Natl. Bur. Stand. A 63
  (1959) 55--62.

\bibitem{mao1978high}
H.~Mao, High-pressure physics: sustained static generation of 1.36 to 1.72
  megabars, Science 200~(4346) (1978) 1145--1147.

\bibitem{nellis2017ultracondensed}
W.~J. Nellis, Ultracondensed matter by dynamic compression, Cambridge
  University Press, 2017.

\bibitem{dornheim2018uniform}
T.~Dornheim, S.~Groth, M.~Bonitz, The uniform electron gas at warm dense matter
  conditions, Physics Reports.

\bibitem{SC:HAMLIN_highp_physicaC_2015}
J.~Hamlin, Superconductivity in the metallic elements at high pressures,
  Physica C: Superconductivity and its Applications 514 (2015) 59 -- 76,
  superconducting Materials: Conventional, Unconventional and Undetermined.

\bibitem{buzea2004assembling}
C.~Buzea, K.~Robbie, Assembling the puzzle of superconducting elements: a
  review, Superconductor Science and Technology 18~(1) (2004) R1.

\bibitem{Shimizu_EHP2005}
K.~Shimizu, K.~Amaya, N.~Suzuki, {Pressure-induced Superconductivity in
  Elemental Materials}, Journal of the Physical Society of Japan 74~(5) (2005)
  1345--1357.

\bibitem{shimizu2015superconductivity}
K.~Shimizu, Superconductivity from insulating elements under high pressure,
  Physica C: Superconductivity and its Applications 514 (2015) 46--49.

\bibitem{akahama1995new}
Y.~Akahama, H.~Kawamura, D.~H{\"a}usermann, M.~Hanfland, O.~Shimomura, New
  high-pressure structural transition of oxygen at 96 {GPa} associated with
  metallization in a molecular solid, Physical Review Letters 74~(23) (1995)
  4690.

\bibitem{shimizu1998superconductivity}
K.~Shimizu, K.~Suhara, M.~Ikumo, M.~Eremets, K.~Amaya, Superconductivity in
  oxygen, Nature 393~(6687) (1998) 767.

\bibitem{schilling2006superconductivity}
J.~S. Schilling, Superconductivity in the alkali metals, High Pressure Research
  26~(3) (2006) 145--163.

\bibitem{wittig1970pressure}
J.~Wittig, Pressure-induced superconductivity in cesium and yttrium, Physical
  Review Letters 24~(15) (1970) 812.

\bibitem{neaton1999pairing}
J.~Neaton, N.~Ashcroft, Pairing in dense lithium, Nature 400~(6740) (1999) 141.

\bibitem{shimizu2002superconductivity}
K.~Shimizu, H.~Ishikawa, D.~Takao, T.~Yagi, K.~Amaya, Superconductivity in
  compressed lithium at 20 {K}, Nature 419~(6907) (2002) 597--599.

\bibitem{deemyad2003superconducting}
S.~Deemyad, J.~S. Schilling, Superconducting phase diagram of {Li} metal in
  nearly hydrostatic pressures up to 67 {GPa}, Physical Review Letters 91~(16)
  (2003) 167001.

\bibitem{struzhkin2002superconductivity}
V.~V. Struzhkin, M.~I. Eremets, W.~Gan, H.-k. Mao, R.~J. Hemley,
  Superconductivity in dense lithium, Science 298~(5596) (2002) 1213--1215.

\bibitem{tuoriniemi2007superconductivity}
J.~Tuoriniemi, K.~Juntunen-Nurmilaukas, J.~Uusvuori, E.~Pentti, A.~Salmela,
  A.~Sebedash, Superconductivity in lithium below 0.4 millikelvin at ambient
  pressure, Nature 447~(7141) (2007) 187.

\bibitem{ma_2009_transparent-Na}
Y.~Ma, M.~Eremets, A.~R. Oganov, Y.~Xie, I.~Trojan, S.~Medvedev, A.~O. Lyakhov,
  M.~Valle, V.~Prakapenka, Transparent dense sodium, Nature 458~(7235) (2009)
  182.

\bibitem{lundegaard2009single}
L.~Lundegaard, E.~Gregoryanz, M.~McMahon, C.~Guillaume, I.~Loa, R.~Nelmes,
  Single-crystal studies of incommensurate {Na} to 1.5 {Mbar}, Physical Review
  B 79~(6) (2009) 064105.

\bibitem{gregoryanz2005melting}
E.~Gregoryanz, O.~Degtyareva, M.~Somayazulu, R.~J. Hemley, H.-k. Mao, Melting
  of dense sodium, Physical Review Letters 94~(18) (2005) 185502.

\bibitem{richardson1997effective}
C.~Richardson, N.~Ashcroft, Effective electron-electron interactions and the
  theory of superconductivity, Physical Review B 55~(22) (1997) 15130.

\bibitem{eremets2001superconductivity}
M.~I. Eremets, V.~V. Struzhkin, H.-k. Mao, R.~J. Hemley, Superconductivity in
  boron, Science 293~(5528) (2001) 272--274.

\bibitem{sakata2011superconducting}
M.~Sakata, Y.~Nakamoto, K.~Shimizu, T.~Matsuoka, Y.~Ohishi, Superconducting
  state of ca-{VII} below a critical temperature of 29 {K} at a pressure of 216
  {GPa}, Physical Review B 83~(22) (2011) 220512(R).

\bibitem{Katsuya_PRL-mafia_Ca_2013}
H.~Fujihisa, Y.~Nakamoto, M.~Sakata, K.~Shimizu, T.~Matsuoka, Y.~Ohishi,
  H.~Yamawaki, S.~Takeya, Y.~Gotoh, Ca-{VII}: A chain ordered host-guest
  structure of calcium above 210 {GPa}, Phys. Rev. Lett. 110 (2013) 235501.

\bibitem{hamlin2006superconductivity}
J.~Hamlin, V.~Tissen, J.~Schilling, {Superconductivity at 17 K in yttrium metal
  under nearly hydrostatic pressures up to 89 GPa}, Physical Review B 73~(9)
  (2006) 094522.

\bibitem{kawamura1985anomalous}
H.~Kawamura, I.~Shirotani, K.~Tachikawa, Anomalous superconductivity and
  pressure induced phase transitions in black phosphorus, Solid state
  communications 54~(9) (1985) 775--778.

\bibitem{karuzawa2002pressure}
M.~Karuzawa, M.~Ishizuka, S.~Endo, The pressure effect on the superconducting
  transition temperature of black phosphorus, Journal of Physics: Condensed
  Matter 14~(44) (2002) 10759.

\bibitem{flores_interplay_2017}
J.~A. Flores-Livas, A.~Sanna, A.~P. Drozdov, L.~Boeri, G.~Profeta, M.~Eremets,
  S.~Goedecker, Interplay between structure and superconductivity: Metastable
  phases of phosphorus under pressure, Physical Review Materials 1~(2) (2017)
  024802.

\bibitem{Akahama_hexgonal_P_PRB1999}
Y.~Akahama, M.~Kobayashi, H.~Kawamura, Simple-cubic-simple-hexagonal transition
  in phosphorus under pressure, Phys. Rev. B 59 (1999) 8520--8525.

\bibitem{Phosphorus_IV_PRL2006}
T.~Ishikawa, H.~Nagara, K.~Kusakabe, N.~Suzuki, Determining the structure of
  phosphorus in phase {IV}, Phys. Rev. Lett. 96 (2006) 095502.

\bibitem{Incommesurate_P-IV-PRL2007}
H.~Fujihisa, Y.~Akahama, H.~Kawamura, Y.~Ohishi, Y.~Gotoh, H.~Yamawaki,
  M.~Sakashita, S.~Takeya, K.~Honda, Incommensurate structure of phosphorus
  phase {IV}, Phys. Rev. Lett. 98 (2007) 175501.

\bibitem{Marques_P-PRB-2008}
M.~Marqu\'es, G.~J. Ackland, L.~F. Lundegaard, S.~Falconi, C.~Hejny, M.~I.
  McMahon, J.~Contreras-Garc\'{i}a, M.~Hanfland, Origin of incommensurate
  modulations in the high-pressure phosphorus {IV} phase, Phys. Rev. B 78
  (2008) 054120.

\bibitem{Toledano_a7structuresPRB-2008}
H.~Katzke, P.~Tol\'edano, Displacive mechanisms and order-parameter symmetries
  for the {A7}-incommensurate-bcc sequences of high-pressure reconstructive
  phase transitions in {G}roup {Va} elements, Phys. Rev. B 77 (2008) 024109.

\bibitem{Sugimoto_Psuperlattice-PRB-2012}
T.~Sugimoto, Y.~Akahama, H.~Fujihisa, Y.~Ozawa, H.~Fukui, N.~Hirao, Y.~Ohishi,
  Identification of superlattice structure $ci16$ in the p-{VI} phase of
  phosphorus at 340 {GPa} and room temperature via x-ray diffraction, Phys.
  Rev. B 86 (2012) 024109.

\bibitem{chalcogens_underpressure_2002}
E.~Gregoryanz, V.~V. Struzhkin, R.~J. Hemley, M.~I. Eremets, H.-k. Mao, Y.~A.
  Timofeev, Superconductivity in the chalcogens up to multimegabar pressures,
  Physical Review B 65~(6) (2002) 064504.

\bibitem{Akahama_S_PRB1993}
Y.~Akahama, M.~Kobayashi, H.~Kawamura, {Structural studies of pressure-induced
  phase transitions in selenium up to 150 GPa}, Phys. Rev. B 47 (1993) 20--26.

\bibitem{Akahama_Se_PRB1993}
Y.~Akahama, M.~Kobayashi, H.~Kawamura, {Pressure-induced structural phase
  transition in sulfur at 83 GPa}, Phys. Rev. B 48 (1993) 6862--6864.

\bibitem{struzhkin1997superconductivity}
V.~V. Struzhkin, R.~J. Hemley, H.-k. Mao, Y.~A. Timofeev, Superconductivity at
  10--17 {K} in compressed sulphur, Nature 390~(6658) (1997) 382--384.

\bibitem{Akahama_Se_met_PRB1997}
Y.~Akahama, M.~Kobayashi, H.~Kawamura, {Pressure-induced metallization and
  structural transition of $\alpha$-monoclinic and amorphous Se}, Phys. Rev. B
  56 (1997) 5027--5031.

\bibitem{kometani1997observation}
S.~Kometani, M.~I. Eremets, K.~Shimizu, M.~Kobayashi, K.~Amaya, Observation of
  pressure-induced superconductivity of sulfur, Journal of the Physical Society
  of Japan 66~(9) (1997) 2564--2565.

\bibitem{mao2003bonding}
W.~L. Mao, H.-k. Mao, P.~J. Eng, T.~P. Trainor, M.~Newville, C.-c. Kao, D.~L.
  Heinz, J.~Shu, Y.~Meng, R.~J. Hemley, Bonding changes in compressed superhard
  graphite, Science 302~(5644) (2003) 425--427.

\bibitem{amsler-flores_Zcarbon_2012}
M.~Amsler, J.~A. Flores-Livas, L.~Lehtovaara, F.~Balima, S.~A. Ghasemi,
  D.~Machon, S.~Pailhes, A.~Willand, D.~Caliste, S.~Botti, et~al., Crystal
  structure of cold compressed graphite, Physical Review Letters 108~(6) (2012)
  065501.

\bibitem{Schwarz_RbIV_PRL1999}
U.~Schwarz, A.~Grzechnik, K.~Syassen, I.~Loa, M.~Hanfland, Rubidium-{IV}: A
  high pressure phase with complex crystal structure, Phys. Rev. Lett. 83
  (1999) 4085--4088.

\bibitem{Christensen_BaIV-P_PRB2004}
F.~J.~H. Ehlers, N.~E. Christensen, Phosphorus under pressure: {Ba}-{IV}-type
  structure as a candidate for {P}-{IV}, Phys. Rev. B 69 (2004) 214112.

\bibitem{Hamlin-Ba}
D.~E. Jackson, D.~VanGennep, Y.~K. Vohra, S.~T. Weir, J.~J. Hamlin,
  Superconductivity of barium-vi synthesized via compression at low
  temperatures, Phys. Rev. B 96 (2017) 184514.

\bibitem{Re_under_pressure_Vohra}
Y.~K. Vohra, S.~J. Duclos, A.~L. Ruoff, High-pressure x-ray diffraction studies
  on rhenium up to 216 gpa (2.16 mbar), Phys. Rev. B 36 (1987) 9790--9792.

\bibitem{holzapfel1996physics}
W.~Holzapfel, Physics of solids under strong compression, Reports on Progress
  in Physics 59~(1) (1996) 29.

\bibitem{tonkov2004phase}
E.~Y. Tonkov, E.~Ponyatovsky, Phase transformations of elements under high
  pressure, CRC press, 2004.

\bibitem{jayaraman1983diamond}
A.~Jayaraman, Diamond anvil cell and high-pressure physical investigations,
  Reviews of Modern Physics 55~(1) (1983) 65.

\bibitem{bassett_50th-DAC_2009}
W.~A. Bassett, Diamond anvil cell, 50th birthday, High Pressure Research 29~(2)
  (2009) 163--186.

\bibitem{wang2014review}
X.~Wang, K.~Kamenev, Review of modern instrumentation for magnetic measurements
  at high pressure and low temperature, Low Temperature Physics 40~(8) (2014)
  735--746.

\bibitem{shen2016high}
G.~Shen, H.~K. Mao, High-pressure studies with x-rays using diamond anvil
  cells, Reports on Progress in Physics 80~(1) (2016) 016101.

\bibitem{mcmillan2005pressing}
P.~F. McMillan, {Pressing on: The legacy of Percy W. Bridgman}, Nature
  Materials 4~(10) (2005) 715.

\bibitem{mao2018solids}
H.-K. Mao, X.-J. Chen, Y.~Ding, B.~Li, L.~Wang, Solids, liquids, and gases
  under high pressure, Reviews of Modern Physics 90~(1) (2018) 015007.

\bibitem{goncharov2012raman}
A.~F. Goncharov, Raman spectroscopy at high pressures, International Journal of
  Spectroscopy 2012 (2012) Article ID 617528.

\bibitem{goncharov2009laser}
A.~F. Goncharov, J.~A. Montoya, N.~Subramanian, V.~V. Struzhkin, A.~Kolesnikov,
  M.~Somayazulu, R.~J. Hemley, Laser heating in diamond anvil cells:
  developments in pulsed and continuous techniques, Journal of synchrotron
  radiation 16~(6) (2009) 769--772.

\bibitem{meier2018its}
T.~Meier, At its extremes: {NMR} at giga-pascal pressures, in: Annual Reports
  on NMR Spectroscopy, Vol.~93, Elsevier, 2018, pp. 1--74.

\bibitem{meier2017magnetic}
T.~Meier, N.~Wang, D.~Mager, J.~G. Korvink, S.~Petitgirard, L.~Dubrovinsky,
  Magnetic flux tailoring through lenz lenses for ultrasmall samples: A new
  pathway to high-pressure nuclear magnetic resonance, Science advances 3~(12)
  (2017) 5242.

\bibitem{eremets1996high}
M.~Eremets, High pressure experimental methods, Oxford University Press, 1996.

\bibitem{mozaffari2019superconducting}
S.~Mozaffari, D.~Sun, V.~S. Minkov, A.~P. Drozdov, D.~Knyazev, J.~B. Betts,
  M.~Einaga, K.~Shimizu, M.~I. Eremets, L.~Balicas, F.~F. Balakirev,
  {Superconducting Phase-Diagram of {H}$_3$S under High Magnetic Fields},
  Nature Communications 10~(1) (2019) 2522.

\bibitem{kobayashi2007nonmagnetic}
T.~Kobayashi, H.~Hidaka, H.~Kotegawa, K.~Fujiwara, M.~Eremets, {Nonmagnetic
  indenter-type high-pressure cell for magnetic measurements}, Review of
  scientific instruments 78~(2) (2007) 023909.

\bibitem{meier2018nmr}
T.~Meier, S.~Petitgirard, S.~Khandarkhaeva, L.~Dubrovinsky,
  \href{https://doi.org/10.1038/s41467-018-05164-x}{Observation of nuclear
  quantum effects and hydrogen bond symmetrisation in high pressure ice},
  Nature Communications 9~(1) (2018) 2766.
\newblock \href {http://dx.doi.org/10.1038/s41467-018-05164-x}
  {\path{doi:10.1038/s41467-018-05164-x}}.
\newline\urlprefix\url{https://doi.org/10.1038/s41467-018-05164-x}

\bibitem{boehler2000laser}
R.~Boehler, Laser heating in the diamond cell: techniques and applications,
  Hyperfine Interactions 128~(1-3) (2000) 307--321.

\bibitem{gonchy_laser2010x2}
A.~F. Goncharov, V.~B. Prakapenka, V.~V. Struzhkin, I.~Kantor, M.~L. Rivers,
  D.~A. Dalton, X-ray diffraction in the pulsed laser heated diamond anvil
  cell, Review of Scientific Instruments 81~(11) (2010) 113902.

\bibitem{boehler2004new}
R.~Boehler, K.~De~Hantsetters, New anvil designs in diamond-cells, High
  Pressure Research 24~(3) (2004) 391--396.

\bibitem{dubrovinskaia2016terapascal}
N.~Dubrovinskaia, L.~Dubrovinsky, N.~A. Solopova, A.~Abakumov, S.~Turner,
  M.~Hanfland, E.~Bykova, M.~Bykov, C.~Prescher, V.~B. Prakapenka, et~al.,
  Terapascal static pressure generation with ultrahigh yield strength
  nanodiamond, Science advances 2~(7) (2016) e1600341.

\bibitem{vohra2015high}
Y.~K. Vohra, G.~K. Samudrala, S.~L. Moore, J.~M. Montgomery, G.~M. Tsoi,
  N.~Velisavljevic, High pressure studies using two-stage diamond micro-anvils
  grown by chemical vapor deposition, High Pressure Research 35~(3) (2015)
  282--288.

\bibitem{lobanov2015pressure}
S.~S. Lobanov, V.~B. Prakapenka, C.~Prescher, Z.~Kon{\^o}pkov{\'a}, H.-P.
  Liermann, K.~L. Crispin, C.~Zhang, A.~F. Goncharov, Pressure, stress, and
  strain distribution in the double-stage diamond anvil cell, Journal of
  Applied Physics 118~(3) (2015) 035905.

\bibitem{sakai2015high}
T.~Sakai, T.~Yagi, H.~Ohfuji, T.~Irifune, Y.~Ohishi, N.~Hirao, Y.~Suzuki,
  Y.~Kuroda, T.~Asakawa, T.~Kanemura, High-pressure generation using double
  stage micro-paired diamond anvils shaped by focused ion beam, Review of
  Scientific Instruments 86~(3) (2015) 033905.

\bibitem{mcmahon2018diamond}
M.~McMahon, Diamond sculpting pushes extremes, Nature Materials 17 (2018)
  858--859.

\bibitem{jenei2018single}
Z.~Jenei, E.~O'Bannon, S.~Weir, H.~Cynn, M.~Lipp, W.~Evans, Single crystal
  toroidal diamond anvils for high pressure experiments beyond 5 megabar,
  Nature Communications 9~(1) (2018) 3563.

\bibitem{dewaele2018toroidal}
A.~Dewaele, P.~Loubeyre, F.~Occelli, O.~Marie, M.~Mezouar, Toroidal diamond
  anvil cell for detailed measurements under extreme static pressures, Nature
  Communications 9~(1) (2018) 2913.

\bibitem{o2018contributed}
E.~F. O'Bannon~III, Z.~Jenei, H.~Cynn, M.~J. Lipp, J.~R. Jeffries, Contributed
  review: Culet diameter and the achievable pressure of a diamond anvil cell:
  Implications for the upper pressure limit of a diamond anvil cell, Review of
  Scientific Instruments 89~(11) (2018) 111501.

\bibitem{mao1981electrical}
H.~Mao, P.~Bell, Electrical resistivity measurements of conductors in the
  diamond-window, high-pressure cell, Review of Scientific Instruments 52~(4)
  (1981) 615--616.

\bibitem{eremets2001semiconducting}
M.~I. Eremets, R.~J. Hemley, H.-k. Mao, E.~Gregoryanz, Semiconducting
  non-molecular nitrogen up to 240 {GPa} and its low-pressure stability, Nature
  411~(6834) (2001) 170.

\bibitem{rotundu2013high}
C.~Rotundu, T.~{\'C}uk, R.~Greene, Z.-X. Shen, R.~J. Hemley, V.~Struzhkin,
  High-pressure resistivity technique for quasi-hydrostatic compression
  experiments, Review of Scientific Instruments 84~(6) (2013) 063903.

\bibitem{das1992review}
K.~Das, V.~Venkatesan, K.~Miyata, D.~Dreifus, J.~Glass, A review of the
  electrical characteristics of metal contacts on diamond, Thin Solid Films
  212~(1-2) (1992) 19--24.

\bibitem{evans2009diamond}
D.~Evans, O.~Roberts, G.~Williams, A.~Vearey-Roberts, F.~Bain, S.~Evans,
  D.~Langstaff, D.~Twitchen, Diamond--metal contacts: interface barriers and
  real-time characterization, Journal of Physics: Condensed Matter 21~(36)
  (2009) 364223.

\bibitem{werner1995effect}
M.~Werner, O.~Dorsch, H.-U. Baerwind, E.~Obermeier, C.~Johnston, P.~R. Chalker,
  S.~Romani, {The effect of metallization on the ohmic contact resistivity to
  heavily B-doped polycrystalline diamond films}, IEEE Transactions on Electron
  Devices 42~(7) (1995) 1344--1351.

\bibitem{forman1972pressure}
R.~A. Forman, G.~J. Piermarini, J.~D. Barnett, S.~Block, Pressure measurement
  made by the utilization of ruby sharp-line luminescence, Science 176~(4032)
  (1972) 284--285.

\bibitem{flores2012raman}
J.~A. Flores-Livas, L.~Lehtovaara, M.~Amsler, S.~Goedecker, S.~Pailhes,
  S.~Botti, A.~San~Miguel, M.~A. Marques, {Raman activity of $sp^3$ carbon
  allotropes under pressure: A density functional theory study}, Physical
  Review B 85~(15) (2012) 155428.

\bibitem{hanfland1985raman}
M.~Hanfland, K.~Syassen, A {Raman} study of diamond anvils under stress,
  Journal of applied physics 57~(8) (1985) 2752--2756.

\bibitem{eremets2003megabar}
M.~Eremets, Megabar high-pressure cells for {Raman} measurements, Journal of
  {Raman} Spectroscopy 34~(7-8) (2003) 515--518.

\bibitem{akahama2006pressure}
Y.~Akahama, H.~Kawamura, Pressure calibration of diamond anvil {Raman} gauge to
  310 {GPa}, Journal of Applied Physics 100~(4) (2006) 043516.

\bibitem{akahama2010pressure}
Y.~Akahama, H.~Kawamura, Pressure calibration of diamond anvil {Raman} gauge to
  410 {GPa}, in: Journal of Physics: Conference Series, Vol. 215, IOP
  Publishing, 2010, p. 012195.

\bibitem{akahama2007diamond}
Y.~Akahama, H.~Kawamura, {Diamond anvil Raman gauge in multimegabar pressure
  range}, High Pressure Research 27~(4) (2007) 473--482.

\bibitem{li2014metallization}
Y.~Li, J.~Hao, H.~Liu, Y.~Li, Y.~Ma, The metallization and superconductivity of
  dense hydrogen sulfide, The Journal of chemical physics 140~(17) (2014)
  174712.

\bibitem{mao1994ultrahigh}
H.-k. Mao, R.~J. Hemley, Ultrahigh-pressure transitions in solid hydrogen,
  Reviews of modern physics 66~(2) (1994) 671.

\bibitem{LeToullec2002}
P.~Loubeyre, F.~Occelli, R.~LeToullec, {Optical studies of solid hydrogen to
  320\,GPa and evidence for black hydrogen}, Nature 416~(6681) (2002) 13--17.

\bibitem{Eremets_NatMat2011}
M.~I. Eremets, I.~A. Troyan, Conductive dense hydrogen, Nature Materials 10
  (2011) 927--931.

\bibitem{eremets2016low}
M.~Eremets, I.~Troyan, A.~Drozdov, {Low temperature phase diagram of hydrogen
  at pressures up to 380 {GPa}. A possible metallic phase at 360 {GPa} and 200
  {K}}, arXiv preprint arXiv:1601.04479.

\bibitem{mcmahon_high_2011}
J.~M. McMahon, D.~M. Ceperley, High-temperature superconductivity in atomic
  metallic hydrogen, Phys. Rev. B 84 (2011) 144515.

\bibitem{cazorla2017simulation}
C.~Cazorla, J.~Boronat, Simulation and understanding of atomic and molecular
  quantum crystals, Reviews of Modern Physics 89~(3) (2017) 035003.

\bibitem{carbotte2018detecting}
J.~Carbotte, E.~Nicol, T.~Timusk, Detecting superconductivity in the high
  pressure hydrides and metallic hydrogen from optical properties, Physical
  Review Letters 121~(4) (2018) 047002.

\bibitem{borinaga2018strong}
M.~Borinaga, J.~Iba{\~n}ez-Azpiroz, A.~Bergara, I.~Errea, Strong
  electron-phonon and band structure effects in the optical properties of high
  pressure metallic hydrogen, Physical Review Letters 120~(5) (2018) 057402.

\bibitem{monserrat2018structure}
B.~Monserrat, N.~D. Drummond, P.~Dalladay-Simpson, R.~T. Howie, P.~L.
  R{\'\i}os, E.~Gregoryanz, C.~J. Pickard, R.~J. Needs, Structure and
  metallicity of phase {V} of hydrogen, Physical Review Letters 120~(25) (2018)
  255701.

\bibitem{Hemley_PRL2012}
C.-S. Zha, Z.~Liu, R.~Hemley, {Synchrotron Infrared Measurements of Dense
  Hydrogen to 360 {GPa}}, Phys. Rev. Lett. 108 (2012) 146402.

\bibitem{HRussell_hydrogenJACS2014}
I.~I. Naumov, R.~J. Hemley, {Aromaticity, Closed-Shell Effects, and
  Metallization of Hydrogen}, Accounts of Chemical Research 47~(12) (2014)
  3551--3559.

\bibitem{Salamat-Silvera_2016}
M.~Zaghoo, A.~Salamat, I.~F. Silvera, Evidence of a first-order phase
  transition to metallic hydrogen, Physical Review B 93~(15) (2016) 155128.

\bibitem{eremets2017molecular}
M.~I. Eremets, A.~P. Drozdov, P.~P. Kong, H.~Wang, Semimetallic molecular
  hydrogen at pressure above 350 {GPa}, Nature Physics.

\bibitem{azadi2017role}
S.~Azadi, G.~J. Ackland, The role of van der {W}aals and exchange interactions
  in high-pressure solid hydrogen, Physical Chemistry Chemical Physics 19~(32)
  (2017) 21829--21839.

\bibitem{magduau2017simple}
I.~B. Magd{\u{a}}u, M.~Marqu{\'e}s, B.~Borgulya, G.~J. Ackland, Simple
  thermodynamic model for the hydrogen phase diagram, Physical Review B 95~(9)
  (2017) 094107.

\bibitem{liu2017comment}
X.-D. Liu, P.~Dalladay-Simpson, R.~T. Howie, B.~Li, E.~Gregoryanz, { Comment on
  {\it Observation of the Wigner-Huntington transition to metallic hydrogen}},
  Science 357~(6353) (2017) 2286.

\bibitem{goncharov2017comment}
A.~F. Goncharov, V.~V. Struzhkin, { Comment on {\it Observation of the
  Wigner-Huntington transition to metallic hydrogen}}, Science 357~(6353)
  (2017) eaam9736.

\bibitem{zha2017melting}
C.-s. Zha, H.~Liu, S.~T. John, R.~J. Hemley, Melting and high {P}- {T}
  transitions of hydrogen up to 300 {GPa}, Physical Review Letters 119~(7)
  (2017) 075302.

\bibitem{zaghoo2017conductivity}
M.~Zaghoo, I.~F. Silvera, Conductivity and dissociation in liquid metallic
  hydrogen and implications for planetary interiors, Proceedings of the
  National Academy of Sciences (2017) 201707918.

\bibitem{rillo2018coupled}
G.~Rillo, M.~A. Morales, D.~M. Ceperley, C.~Pierleoni, Coupled electron-ion
  monte carlo simulation of hydrogen molecular crystals, The Journal of
  chemical physics 148~(10) (2018) 102314.

\bibitem{Eremets_Science2008}
M.~I. Eremets, I.~A. Trojan, S.~A. Medvedev, J.~S. Tse, Y.~Yao,
  Superconductivity in hydrogen dominant materials: Silane, Science 319~(5869)
  (2008) 1506--1509.

\bibitem{Chen_disilane_PNAS2008}
X.-J. Chen, V.~V. Struzhkin, Y.~Song, A.~F. Goncharov, M.~Ahart, Z.~Liu, H.-k.
  Mao, R.~J. Hemley, Pressure-induced metallization of silane, Proceedings of
  the National Academy of Sciences 105~(1) (2008) 20--23.

\bibitem{Wang_PNAS2009}
S.~Wang, H.-k. Mao, X.-J. Chen, W.~L. Mao, High pressure chemistry in the
  {H}$_2$-{SiH}$_4$ system, Proceedings of the National Academy of Sciences
  106~(35) (2009) 14763--14767.

\bibitem{Hanfland_PRL2011}
M.~Hanfland, J.~E. Proctor, C.~L. Guillaume, O.~Degtyareva, E.~Gregoryanz,
  High-pressure synthesis, amorphization, and decomposition of silane, Phys.
  Rev. Lett. 106~(9) (2011) 095503.

\bibitem{PtH_PRL2011}
D.~Y. Kim, R.~H. Scheicher, C.~J. Pickard, R.~J. Needs, R.~Ahuja, Predicted
  formation of superconducting platinum-hydride crystals under pressure in the
  presence of molecular hydrogen, Phys. Rev. Lett. 107 (2011) 117002.

\bibitem{Pt-H_2019experimental}
G.~Liu, Z.~Yu, S.~Li, H.~Wang, The experimental compression behavior of
  platinum hydride to 128 {GPa}, Materials Letters 249 (2019) 84.

\bibitem{flores_disilane_2012}
J.~A. Flores-Livas, M.~Amsler, T.~J. Lenosky, L.~Lehtovaara, S.~Botti, M.~A.~L.
  Marques, S.~Goedecker, {High-Pressure Structures of Disilane and Their
  Superconducting Properties}, Phys. Rev. Lett. 108 (2012) 117004.

\bibitem{hirsch2009bcs}
J.~E. Hirsch, {BCS theory of superconductivity: it is time to question its
  validity}, Physica Scripta 80~(3) (2009) 035702.

\bibitem{Duan_SciRep2014}
D.~Duan, Y.~Liu, F.~Tian, D.~Li, X.~Huang, Z.~Zhao, H.~Yu, B.~Liu, W.~Tian,
  T.~Cui, {Pressure-induced metallization of dense ({H}$_2${S})$_2${H}$_2$ with
  high-{T}c superconductivity}, Sci. Rep. 4.

\bibitem{drozdov2014conventional}
A.~Drozdov, M.~Eremets, I.~Troyan, Conventional superconductivity at 190 {K} at
  high pressures, arXiv preprint arXiv:1412.0460.

\bibitem{shimizu1995pressure}
H.~Shimizu, H.~Yamaguchi, S.~Sasaki, A.~Honda, S.~Endo, M.~Kobayashi,
  Pressure-temperature phase diagram of solid hydrogen sulfide determined by
  {Raman} spectroscopy, Physical Review B 51~(14) (1995) 9391.

\bibitem{sakashita1997pressure}
M.~Sakashita, H.~Yamawaki, H.~Fujihisa, K.~Aoki, S.~Sasaki, H.~Shimizu,
  {Pressure-induced molecular dissociation and metallization in hydrogen-bonded
  {H}$_2${S} solid}, Physical Review Letters 79~(6) (1997) 1082.

\bibitem{fujihisa2004molecular}
H.~Fujihisa, H.~Yamawaki, M.~Sakashita, A.~Nakayama, T.~Yamada, K.~Aoki,
  {Molecular dissociation and two low-temperature high-pressure phases of
  {H}$_2${S}}, Physical Review B 69~(21) (2004) 214102.

\bibitem{Drozdov_thesis_2016}
A.~Drozdov, {Superconductivity in hydrogen-rich materials at high pressures},
  Ph.D. thesis, Johannes Gutenberg-Universitat Mainz (2016).

\bibitem{Einaga_H3S-crystal_NatPhys-2016}
M.~Einaga, M.~Sakata, T.~Ishikawa, K.~Shimizu, M.~I. Eremets, A.~P. Drozdov,
  I.~A. Troyan, N.~Hirao, Y.~Ohishi, Crystal structure of the superconducting
  phase of sulfur hydride, Nature Physics.

\bibitem{flores-sanna_PH3_2016}
J.~A. Flores-Livas, M.~Amsler, C.~Heil, A.~Sanna, L.~Boeri, G.~Profeta,
  C.~Wolverton, S.~Goedecker, E.~K.~U. Gross, Superconductivity in metastable
  phases of phosphorus-hydride compounds under high pressure, Phys. Rev. B 93
  (2016) 020508.

\bibitem{Drozdov_ph3_arxiv2015}
A.~Drozdov, M.~Eremets, I.~Troyan, {Superconductivity above 100 K in PH$_3$ at
  high pressures}, arXiv preprint arXiv:1508.06224.

\bibitem{Ivan_Science-H2S-2016}
I.~Troyan, A.~Gavriliuk, R.~R{\"u}ffer, A.~Chumakov, A.~Mironovich,
  I.~Lyubutin, D.~Perekalin, A.~P. Drozdov, M.~I. Eremets, Observation of
  superconductivity in hydrogen sulfide from nuclear resonant scattering,
  Science 351~(6279) (2016) 1303--1306.

\bibitem{Errea_anha-SH3_PRL2015}
I.~Errea, M.~Calandra, C.~J. Pickard, J.~Nelson, R.~J. Needs, Y.~Li, H.~Liu,
  Y.~Zhang, Y.~Ma, F.~Mauri, High-pressure hydrogen sulfide from first
  principles: A strongly anharmonic phonon-mediated superconductor, Phys. Rev.
  Lett. 114 (2015) 157004.

\bibitem{Heil-Boeri_PRB2015}
C.~Heil, L.~Boeri, Influence of bonding on superconductivity in high-pressure
  hydrides, Phys. Rev. B 92 (2015) 060508.

\bibitem{flores-sanna_HSe_2016}
A.~J. Flores-Livas, A.~Sanna, E.~Gross, High temperature superconductivity in
  sulfur and selenium hydrides at high pressure, Eur. Phys. J. B 89~(3) (2016)
  1--6.

\bibitem{flores_accelerated_2017}
J.~A. Flores-Livas, A.~Sanna, S.~Goedecker, Accelerated materials design
  approaches based on structural classification: application to low enthalpy
  high pressure phases of {SH}$_3$ and {SeH}$_3$, Novel Superconducting
  Materials 3~(1) (2017) 6--13.

\bibitem{akashi_mangeli-phases}
R.~Akashi, W.~Sano, R.~Arita, S.~Tsuneyuki, Possible ``magn\'eli'' phases and
  self-alloying in the superconducting sulfur hydride, Phys. Rev. Lett. 117
  (2016) 075503.

\bibitem{Ca_PNAS_2012superconductive}
H.~Wang, S.~T. John, K.~Tanaka, T.~Iitaka, Y.~Ma, Superconductive sodalite-like
  clathrate calcium hydride at high pressures, Proceedings of the National
  Academy of Sciences 109~(17) (2012) 6463--6466.

\bibitem{Clathrate_REHX_PRL_2017}
F.~Peng, Y.~Sun, C.~J. Pickard, R.~J. Needs, Q.~Wu, Y.~Ma, {Hydrogen Clathrate
  Structures in Rare Earth Hydrides at High Pressures: Possible Route to
  Room-Temperature Superconductivity}, Phys. Rev. Lett. 119 (2017) 107001.

\bibitem{PNAS_LaHx_2017_Hemley}
H.~Liu, I.~I. Naumov, R.~Hoffmann, N.~Ashcroft, R.~J. Hemley, {Potential
  high-Tc superconducting lanthanum and yttrium hydrides at high pressure},
  Proceedings of the National Academy of Sciences 114~(27) (2017) 6990--6995.

\bibitem{Li_PNAS2010}
Y.~Li, G.~Gao, Y.~Xie, Y.~Ma, T.~Cui, G.~Zou, {Superconductivity at 100 K in
  dense SiH$_4$(H$_2$)$_2$ predicted by first principles}, Proceedings of the
  National Academy of Sciences 107~(36) (2010) 15708--15711.

\bibitem{work_on_LaHx_2019}
I.~Errea, F.~Belli, L.~Monacelli, A.~Sanna, T.~Koretsune, T.~Tadano, R.~Bianco,
  M.~Calandra, R.~Arita, F.~Mauri, J.~A. Flores-Livas, Quantum crystal
  structure in the 250 {K} superconducting lanthanum hydride (2019).
\newblock \href {http://arxiv.org/abs/1907.11916} {\path{arXiv:1907.11916}}.

\bibitem{drozdov2018_215}
A.~Drozdov, V.~Minkov, S.~Besedin, P.~Kong, M.~Kuzovnikov, D.~Knyazev,
  M.~Eremets, Superconductivity at 215 k in lanthanum hydride at high
  pressures, arXiv preprint arXiv:1808.07039.

\bibitem{drozdov2018_250}
A.~Drozdov, P.~Kong, V.~Minkov, S.~Besedin, M.~Kuzovnikov, S.~Mozaffari,
  L.~Balicas, F.~Balakirev, D.~Graf, V.~Prakapenka, et~al., {Superconductivity
  at 250 K in lanthanum hydride under high pressures}, arXiv preprint
  arXiv:1812.01561.

\bibitem{flores_thesis_2012}
J.~Flores~Livas, {Computational and experimental studies of $sp^3$-materials at
  high pressure}, Ph.D. thesis, University of Lyon 1 (2012).

\bibitem{grochala2007chemical}
W.~Grochala, R.~Hoffmann, J.~Feng, N.~W. Ashcroft, The chemical imagination at
  work in very tight places, Angewandte Chemie International Edition 46~(20)
  (2007) 3620--3642.

\bibitem{hot_Ma_view_point_PRL2019}
J.~A. Flores-Livas, R.~Arita, {A Prediction for ``Hot" Superconductivity},
  Physics viewpoint 12 (2019) 96.

\bibitem{nguyen2006high}
J.~H. Nguyen, D.~Orlikowski, F.~H. Streitz, J.~A. Moriarty, N.~C. Holmes,
  High-pressure tailored compression: Controlled thermodynamic paths, Journal
  of applied physics 100~(2) (2006) 023508.

\bibitem{degtyareva_formation_2009}
O.~Degtyareva, J.~E. Proctor, C.~L. Guillaume, E.~Gregoryanz, M.~Hanfland,
  Formation of transition metal hydrides at high pressures, Solid State
  Communications 149~(39-40) (2009) 1583--1586.

\bibitem{Strobel_PRL2011}
T.~Strobel, P.~Ganesh, M.~Somayazulu, P.~Kent, R.~Hemley, {Novel Cooperative
  Interactions and Structural Ordering in
  ${\mathrm{H}}_{2}\mathrm{S}\mathrm{\text{-}}{\mathrm{H}}_{2}$}, Phys. Rev.
  Lett. 107 (2011) 255503.

\bibitem{levitas2018high}
V.~I. Levitas, High pressure phase transformations revisited, Journal of
  Physics: Condensed Matter 30~(16) (2018) 163001.

\bibitem{geballe2018synthesis}
Z.~M. Geballe, H.~Liu, A.~K. Mishra, M.~Ahart, M.~Somayazulu, Y.~Meng,
  M.~Baldini, R.~J. Hemley, Synthesis and stability of lanthanum superhydrides,
  Angewandte Chemie International Edition 57~(3) (2018) 688--692.

\bibitem{Maramatsu-Hemley_2015}
T.~Muramatsu, W.~K. Wanene, M.~Somayazulu, E.~Vinitsky, D.~Chandra, T.~A.
  Strobel, V.~V. Struzhkin, R.~J. Hemley, Metallization and superconductivity
  in the hydrogen-rich ionic salt bareh9, The Journal of Physical Chemistry C
  119~(32) (2015) 18007--18013.

\bibitem{Semenok_arxiv_ThH10_2019}
D.~V. Semenok, A.~G. Kvashnin, A.~G. Ivanova, V.~Svitlyk, V.~Y. Fominski, A.~V.
  Sadakov, O.~A. Sobolevskiy, V.~M. Pudalov, I.~A. Troyan, A.~R. Oganov,
  {Superconductivity at 161 K in Thorium Hydride ThH$_{10}$: Synthesis and
  Properties} (2019).
\newblock \href {http://arxiv.org/abs/1902.10206} {\path{arXiv:1902.10206}}.

\bibitem{Troyan_arxiv_YH6_2019}
I.~A. Troyan, D.~V. Semenok, A.~G. Kvashnin, A.~G. Ivanova, V.~B. Prakapenka,
  E.~Greenberg, A.~G. Gavriliuk, I.~S. Lyubutin, V.~V. Struzhkin, A.~R. Oganov,
  {Synthesis and Superconductivity of Yttrium Hexahydride Im$\bar3$m-YH$_6$}
  (2019).
\newblock \href {http://arxiv.org/abs/1908.01534} {\path{arXiv:1908.01534}}.

\bibitem{Kong_arxiv_YH9_2019}
P.~P. Kong, V.~S. Minkov, M.~A. Kuzovnikov, S.~P. Besedin, A.~P. Drozdov,
  S.~Mozaffari, L.~Balicas, F.~F. Balakirev, V.~B. Prakapenka, E.~Greenberg,
  D.~A. Knyazev, M.~I. Eremets, {Superconductivity up to 243 K in yttrium
  hydrides under high pressure} (2019).
\newblock \href {http://arxiv.org/abs/1909.10482} {\path{arXiv:1909.10482}}.

\bibitem{Carbotte_RMP1990}
J.~P. Carbotte, Properties of boson-exchange superconductors, Rev. Mod. Phys.
  62 (1990) 1027--1157.

\bibitem{born1927quantentheorie}
M.~Born, R.~Oppenheimer, Zur quantentheorie der molekeln, Annalen der physik
  389~(20) (1927) 457--484.

\bibitem{born1954dynamical}
M.~Born, K.~Huang, Dynamical theory of crystal lattices, Oxford University
  Press, Oxford, 1954.

\bibitem{DreizlerGross_DFT1990}
R.~M. Dreizler, E.~Gross, {D}ensity {F}unctional {T}heory, Springer-Verlag
  Berlin Heidelberg, 1990.

\bibitem{KohnSham_PR1965}
W.~Kohn, L.~J. Sham, Self-consistent equations including exchange and
  correlation effects, Phys. Rev. 140 (1965) A1133--A1138.

\bibitem{PBE_PRL1996}
J.~P. Perdew, K.~Burke, M.~Ernzerhof, Generalized gradient approximation made
  simple, Phys. Rev. Lett. 77 (1996) 3865--3868.

\bibitem{PerdewZunger_LDA_PRB1981}
J.~P. Perdew, A.~Zunger, Self-interaction correction to density-functional
  approximations for many-electron systems, Phys. Rev. B 23 (1981) 5048--5079.

\bibitem{SunRuzsinskyPerdew_SCAN_PRL2015}
J.~Sun, A.~Ruzsinszky, J.~P. Perdew, Strongly constrained and appropriately
  normed semilocal density functional, Phys. Rev. Lett. 115 (2015) 036402.

\bibitem{Zein-1984}
N.~Zein, Density functional calculations of elastic moduli and phonon spectra
  of crystals, Sov. Phys. Solid State 26 (19884) 1825.

\bibitem{MariniPonceGonze_MBPTandElectronPhonon_PRB2015}
A.~Marini, S.~Ponc\'e, X.~Gonze, Many-body perturbation theory approach to the
  electron-phonon interaction with density-functional theory as a starting
  point, Phys. Rev. B 91 (2015) 224310.

\bibitem{Born1955}
M.~Born, D.~J. Hooton, Statistische dynamik mehrfach periodischer systeme,
  Zeitschrift fur Physik 142~(2) (1955) 201--218.

\bibitem{Ashcroft_book_yes}
N.~W. Ashcroft, N.~D. Mermin, {Solid State Physics, Cornell University} (1976).

\bibitem{ziman2001electrons}
J.~M. Ziman, Electrons and phonons: the theory of transport phenomena in
  solids, Oxford university press, 2001.

\bibitem{maradudin1962scattering}
A.~Maradudin, A.~Fein, Scattering of neutrons by an anharmonic crystal,
  Physical Review 128~(6) (1962) 2589.

\bibitem{koehler1966theory}
T.~R. Koehler, Theory of the self-consistent harmonic approximation with
  application to solid neon, Physical Review Letters 17~(2) (1966) 89.

\bibitem{wang1990tight}
C.~Wang, C.~Chan, K.~Ho, Tight-binding molecular-dynamics study of phonon
  anharmonic effects in silicon and diamond, Physical Review B 42~(17) (1990)
  11276.

\bibitem{hellman2011lattice}
O.~Hellman, I.~Abrikosov, S.~Simak, Lattice dynamics of anharmonic solids from
  first principles, Physical Review B 84~(18) (2011) 180301.

\bibitem{hellman2013_PRB-1}
O.~Hellman, P.~Steneteg, I.~A. Abrikosov, S.~I. Simak, Temperature dependent
  effective potential method for accurate free energy calculations of solids,
  Physical Review B 87~(10) (2013) 104111.

\bibitem{hellman2013_PRB-2}
O.~Hellman, I.~A. Abrikosov, Temperature-dependent effective third-order
  interatomic force constants from first principles, Physical Review B 88~(14)
  (2013) 144301.

\bibitem{ceperley1995path}
D.~M. Ceperley, Path integrals in the theory of condensed helium, Reviews of
  Modern Physics 67~(2) (1995) 279.

\bibitem{bowman1978self}
J.~M. Bowman, Self-consistent field energies and wavefunctions for coupled
  oscillators, The Journal of Chemical Physics 68~(2) (1978) 608--610.

\bibitem{monserrat2013anharmonic}
B.~Monserrat, N.~Drummond, R.~Needs, Anharmonic vibrational properties in
  periodic systems: energy, electron-phonon coupling, and stress, Physical
  Review B 87~(14) (2013) 144302.

\bibitem{hooton1955lviii}
D.~Hooton, Some vibrational properties of solid helium, The London, Edinburgh,
  and Dublin Philosophical Magazine and Journal of Science 46~(376) (1955)
  485--498.

\bibitem{hooton1958_book}
D.~Hooton, The use of a model in anharmonic lattice dynamics, Philosophical
  Magazine 3~(25) (1958) 49--54.

\bibitem{Werthamer_SCHA_PRB1970}
N.~R. Werthamer, Self-consistent phonon formulation of anharmonic lattice
  dynamics, Phys. Rev. B 1 (1970) 572--581.

\bibitem{CochranCowley_phononsInPerfectCrystals_Book1967}
R.~A. Cochran, W.and~Cowley, Phonons in Perfect Crystals, Springer Berlin
  Heidelberg, Berlin, Heidelberg, 1967.

\bibitem{Cardona_RamanAnharmonic_PRB1984}
J.~Men\'endez, M.~Cardona, Temperature dependence of the first-order {Raman}
  scattering by phonons in {Si}, {Ge}, and
  $\ensuremath{\alpha}\ensuremath{-}\mathrm{S}\mathrm{n}$: Anharmonic effects,
  Phys. Rev. B 29 (1984) 2051--2059.

\bibitem{souvatzis2008entropy}
P.~Souvatzis, O.~Eriksson, M.~Katsnelson, S.~Rudin, Entropy driven
  stabilization of energetically unstable crystal structures explained from
  first principles theory, Physical Review Letters 100~(9) (2008) 095901.

\bibitem{SCHA-method_PRB_Errea_2014}
I.~Errea, M.~Calandra, F.~Mauri, Anharmonic free energies and phonon
  dispersions from the stochastic self-consistent harmonic approximation:
  Application to platinum and palladium hydrides, Phys. Rev. B 89 (2014)
  064302.

\bibitem{Terumasa_PRB_2015}
T.~Tadano, S.~Tsuneyuki, {Self-consistent phonon calculations of lattice
  dynamical properties in cubic SrTiO$_{3}$ with first-principles anharmonic
  force constants}, Phys. Rev. B 92 (2015) 054301.

\bibitem{MaradudinFein_ScatternigAnharmonicCrystal_PR1962}
A.~A. Maradudin, A.~E. Fein, Scattering of neutrons by an anharmonic crystal,
  Phys. Rev. 128 (1962) 2589--2608.

\bibitem{LazzeriCalandraMauri_AnharmonicRamanMgB2_PRB2003}
M.~Lazzeri, M.~Calandra, F.~Mauri, Anharmonic phonon frequency shift in
  ${\mathrm{mgb}}_{2}$, Phys. Rev. B 68 (2003) 220509.

\bibitem{Errea_PdH_PRL2013}
I.~Errea, M.~Calandra, F.~Mauri, First-principles theory of anharmonicity and
  the inverse isotope effect in superconducting palladium-hydride compounds,
  Phys. Rev. Lett. 111 (2013) 177002.

\bibitem{Bianco_PRB2017}
R.~Bianco, I.~Errea, L.~Paulatto, M.~Calandra, F.~Mauri, Second-order
  structural phase transitions, free energy curvature, and
  temperature-dependent anharmonic phonons in the self-consistent harmonic
  approximation: Theory and stochastic implementation, Phys. Rev. B 96 (2017)
  014111.
\newblock \href {http://dx.doi.org/10.1103/PhysRevB.96.014111}
  {\path{doi:10.1103/PhysRevB.96.014111}}.

\bibitem{Monacelli_PRB2018}
L.~Monacelli, I.~Errea, M.~Calandra, F.~Mauri, Pressure and stress tensor of
  complex anharmonic crystals within the stochastic self-consistent harmonic
  approximation, Phys. Rev. B 98 (2018) 024106.
\newblock \href {http://dx.doi.org/10.1103/PhysRevB.98.024106}
  {\path{doi:10.1103/PhysRevB.98.024106}}.

\bibitem{Terumasa_PRL_2018}
T.~Tadano, S.~Tsuneyuki, Quartic anharmonicity of rattlers and its effect on
  lattice thermal conductivity of clathrates from first principles, Physical
  Review Letters 120~(10) (2018) 105901.

\bibitem{CSLD_2014_ozolins}
F.~Zhou, W.~Nielson, Y.~Xia, V.~Ozoli{\c{n}}{\v{s}}, et~al., Lattice
  anharmonicity and thermal conductivity from compressive sensing of
  first-principles calculations, Physical Review Letters 113~(18) (2014)
  185501.

\bibitem{Hedin_GW_PR1965}
L.~Hedin, {New Method for Calculating the One-Particle Green's Function with
  Application to the Electron-Gas Problem}, Phys. Rev. 139 (1965) A796--A823.

\bibitem{AryasetiawanGunnarsson_GWMethod_RepProgPhys1998}
F.~Aryasetiawan, O.~Gunnarsson, The {GW} method, Reports on Progress in Physics
  61~(3) (1998) 237.

\bibitem{VonsovskySuperconductivityTransitionMetals}
S.~Vonsovsky, Y.~Izyumov, E.~Kurmaev, E.~Brandt, A.~Zavarnitsyn,
  Superconductivity of Transition Metals: Their Alloys and Compounds, Springer
  Series in Solid-State Sciences Series, Springer London, Limited, 1982.

\bibitem{Gross_TDDFT_PRL1984}
E.~Runge, E.~K.~U. Gross, Density-functional theory for time-dependent systems,
  Phys. Rev. Lett. 52 (1984) 997--1000.

\bibitem{MorelAnderson_1962}
P.~Morel, P.~W. Anderson, Calculation of the superconducting state parameters
  with retarded electron-phonon interaction, Phys. Rev. 125 (1962) 1263--1271.

\bibitem{Akashi_PlasmonSCDFT_PRL2013}
R.~Akashi, R.~Arita, Development of density-functional theory for a
  plasmon-assisted superconducting state: Application to lithium under high
  pressures, Phys. Rev. Lett. 111 (2013) 057006.

\bibitem{Stewart_UncnventionalSuperconductivity_Review2017}
G.~R. Stewart, Unconventional superconductivity, Advances in Physics 66~(2)
  (2017) 75--196.

\bibitem{Scalapino_UconventionalPairing_RevModPhys2012}
D.~J. Scalapino, A common thread: The pairing interaction for unconventional
  superconductors, Rev. Mod. Phys. 84 (2012) 1383--1417.

\bibitem{Essenberger_SpinFluctuationsTheory_PRB2014}
F.~Essenberger, A.~Sanna, A.~Linscheid, F.~Tandetzky, G.~Profeta, P.~Cudazzo,
  E.~K.~U. Gross, Superconducting pairing mediated by spin fluctuations from
  first principles, Phys. Rev. B 90 (2014) 214504.

\bibitem{Gorkov_EnergySpectrumSC_JETP1958}
L.~P. Gor'kov, On the energy spectrum of superconductors, Sov. Phys. JETP 7~(3)
  (1958) 505.

\bibitem{AllenMitrovic1983}
P.~B. Allen, B.~Mitrovi{\' c}, Theory of Superconducting Tc, Vol.~37 of Solid
  State Physics, Academic Press, 1983.

\bibitem{Abrikosov_QuantumFieldTheory_Book1975}
A.~A. Abrikosov, I.~Dzyaloshinskii, L.~P. Gor'kov,
  \href{https://cds.cern.ch/record/107441}{{Methods of quantum field theory in
  statistical physics}}, Dover, New York, NY, 1975.
\newline\urlprefix\url{https://cds.cern.ch/record/107441}

\bibitem{FetterWalecka_QuantumTheoryOfManyPatricleSystems_Book1971}
A.~Fetter, J.~D. Walecka, Quantum Theory of Many-Particle Systems, Dover, New
  York, 1971, 2003.

\bibitem{DeGennes_SupercondMetalsAlloys_Book1966}
P.~G. de~Gennes, Superconductivity of Metals and Alloys, Westview Press, 1966.

\bibitem{Migdal1958}
A.~B. Migdal, {Interaction between electrons and lattice vibrations in a normal
  metal}, Sov. Phys. JETP 7~(6) (1958) 996--1001.

\bibitem{Tsuei_3crystalexperimentYBCO_PRL1994}
C.~C. Tsuei, J.~R. Kirtley, C.~C. Chi, L.~S. Yu-Jahnes, A.~Gupta, T.~Shaw,
  J.~Z. Sun, M.~B. Ketchen, {Pairing Symmetry and Flux Quantization in a
  Tricrystal Superconducting Ring of {YBa}$_2${Cu}$_3${O}$_{7-\delta}$}, Phys.
  Rev. Lett. 73 (1994) 593--596.

\bibitem{Ummarino_Eliashberg2013}
G.~A.~C. Ummarino, Eliashberg theory, in: {Pavarini, Eva and Koch, Erik and
  Schollw\"ock, Ulrich} (Ed.), Emergent Phenomena in Correlated Matter, {Verlag
  des Forschungszentrum J\"ulich}, {J\"ulich}, 2013, Ch.~13, p. 395.

\bibitem{PRB_Sano_Van-Hove_H3S_2016}
W.~Sano, T.~Koretsune, T.~Tadano, R.~Akashi, R.~Arita, Effect of van hove
  singularities on high-${T}_{\mathrm{c}}$ superconductivity in {H}$_3${S},
  Phys. Rev. B 93 (2016) 094525.

\bibitem{ScalapinoSchriefferWilkins_StrongCouplingSC_PR1966}
D.~J. Scalapino, J.~R. Schrieffer, J.~W. Wilkins, Strong-coupling
  superconductivity. i, Phys. Rev. 148 (1966) 263--279.

\bibitem{McMillanTC_PR_1968}
W.~L. McMillan, {Transition Temperature of Strong-Coupled Superconductors},
  Phys. Rev. 167 (1968) 331--344.

\bibitem{dynes1972mcmillan}
R.~Dynes, {McMillan's equation and the Tc of superconductors}, Solid State
  Communications 10~(7) (1972) 615--618.

\bibitem{AllenDynes_PRB1975}
P.~B. Allen, R.~C. Dynes, Transition temperature of strong-coupled
  superconductors reanalyzed, Phys. Rev. B 12 (1975) 905--922.

\bibitem{PhDKurth}
S.~{Kurth Ph.D. thesis}, Exchange-Correlation Functionals for Inhomogeneous
  Superconductors, Bayerische Julius-Maximilians Universit\"at W\"urzburg,
  1995.

\bibitem{KreibichGross_MulticomponentDFT_PRL2001}
T.~Kreibich, E.~K.~U. Gross, Multicomponent density-functional theory for
  electrons and nuclei, Phys. Rev. Lett. 86 (2001) 2984--2987.

\bibitem{flores-Sanna_honeycombs_2015}
J.~A. Flores-Livas, A.~Sanna, Superconductivity in intercalated {group-IV}
  honeycomb structures, Phys. Rev. B 91 (2015) 054508.

\bibitem{sanna2018superconductivity}
A.~Sanna, A.~Davydov, J.~K. Dewhurst, S.~Sharma, J.~A. Flores-Livas,
  Superconductivity in hydrogenated carbon nanostructures, The European
  Physical Journal B 91~(8) (2018) 177.

\bibitem{HohenbergKohn_DFT_PR1964}
P.~Hohenberg, W.~Kohn, Inhomogeneous electron gas, Phys. Rev. 136 (1964)
  B864--B871.

\bibitem{schmidt2019representability}
J.~Schmidt, C.~L. Benavides-Riveros, M.~A. Marques, Representability problem of
  density functional theory for superconductors, Physical Review B 99~(2)
  (2019) 024502.

\bibitem{RDM_SCDFT_schmidt2019}
J.~Schmidt, C.~L. Benavides-Riveros, M.~A. Marques, Reduced density matrix
  functional theory for superconductors, arXiv preprint arXiv:1903.01516.

\bibitem{Mermin_ThermalElectronGas_PR65}
N.~D. Mermin, Thermal properties of the inhomogeneous electron gas, Phys. Rev.
  137 (1965) A1441--A1443.

\bibitem{Bogoljubov_NewMethodSC_FdP1958}
N.~N. Bogoljubov, V.~V. Tolmachov, D.~V. Širkov, A new method in the theory of
  superconductivity, Fortschritte der Physik 6~(11-12) (1958) 605--682.

\bibitem{Giustino_elecronphonon_RMP2017}
F.~Giustino, Electron-phonon interactions from first principles, Rev. Mod.
  Phys. 89 (2017) 015003.

\bibitem{PhDLueders}
M.~{L\"uders Ph.D. thesis}, Density Functional Theory for Superconductors, A
  first Principles Approach to the SC Phase, Bayerische Julius-Maximilians
  Universit\"at W\"urzburg, 1998.

\bibitem{PhDMarques}
M.~{Marques Ph.D. thesis}, Density Functional Theory for Superconductors,
  Exchange and Correlation Potentials for Inhomogeneous Systems, Bayerische
  Julius-Maximilians Universit\"at W\"urzburg, 2000.

\bibitem{Sanna_Migdal}
A.~Sanna, C.~Pellegrini, E.~K.~U. Gross, to be published (2020).

\bibitem{Akashi_SCDFTplasmons_JPSJ2014}
R.~Akashi, R.~Arita, {Density Functional Theory for Plasmon-Assisted
  Superconductivity}, Journal of the Physical Society of Japan 83~(6) (2014)
  061016.

\bibitem{Floris_Pb_PRB2007}
A.~Floris, A.~Sanna, S.~Massidda, E.~K.~U. Gross, {Two-band superconductivity
  in Pb from ab initio calculations}, Phys. Rev. B 75 (2007) 054508.

\bibitem{Cudazzo_PRL2008}
P.~Cudazzo, G.~Profeta, A.~Sanna, A.~Floris, A.~Continenza, S.~Massidda,
  E.~K.~U. Gross, {\textit{Ab Initio} Description of High-Temperature
  Superconductivity in Dense Molecular Hydrogen}, Phys. Rev. Lett. 100 (2008)
  257001.

\bibitem{Massidda_SUST_CoulombSCDFT_2009}
S.~Massidda, F.~Bernardini, C.~Bersier, A.~Continenza, P.~Cudazzo, A.~Floris,
  H.~Glawe, M.~Monni, S.~Pittalis, G.~Profeta, A.~Sanna, S.~Sharma, E.~K.~U.
  Gross, The role of coulomb interaction in the superconducting properties of
  {CaC}$_6$ and {H} under pressure, Superconductor Science and Technology
  22~(3) (2009) 034006.

\bibitem{Essenberger_FeSe_PRB2016}
F.~Essenberger, A.~Sanna, P.~Buczek, A.~Ernst, L.~Sandratskii, E.~K.~U. Gross,
  Ab initio, Phys. Rev. B 94 (2016) 014503.

\bibitem{Takada_plasmonicSC_JPSJ1978}
Y.~Takada, Plasmon mechanism of superconductivity in two- and three-dimensional
  electron systems, Journal of the Physical Society of Japan 45~(3) (1978)
  786--794.

\bibitem{radhakrishnan1965superconductivity}
V.~Radhakrishnan, Superconductivity in transition elements, Physics Letters 16
  (1965) 247--248.

\bibitem{Rietschel1983}
H.~Rietschel, L.~J. Sham, {Role of electron Coulomb interaction in
  superconductivity}, Physical Review B 28~(9) (1983) 5100--5108.

\bibitem{frohlich1968superconductivity}
H.~Fr{\"o}hlich, Superconductivity in metals with incomplete inner shells,
  Journal of Physics C: Solid State Physics 1~(2) (1968) 544.

\bibitem{Akashi_PRB2015}
R.~Akashi, M.~Kawamura, S.~Tsuneyuki, Y.~Nomura, R.~Arita, First-principles
  study of the pressure and crystal-structure dependences of the
  superconducting transition temperature in compressed sulfur hydrides, Phys.
  Rev. B 91 (2015) 224513.

\bibitem{Schon-Jensen_1996_angewandte}
J.~C. Sch{\"o}n, M.~Jansen, First step towards planning of syntheses in
  solid-state chemistry: determination of promising structure candidates by
  global optimization, Angewandte Chemie International Edition in English
  35~(12) (1996) 1286--1304.

\bibitem{stillinger1999exponential}
F.~H. Stillinger, Exponential multiplicity of inherent structures, Physical
  Review E 59~(1) (1999) 48.

\bibitem{Wales_PES_2003_book}
D.~Wales, Energy landscapes: Applications to clusters, biomolecules and
  glasses, Cambridge University Press, 2003.

\bibitem{pickard_PRB-2019_hyperspatial}
C.~J. Pickard, Hyperspatial optimization of structures, Physical Review B
  99~(5) (2019) 054102.

\bibitem{original-idea_AIRSS_2006}
C.~J. Pickard, R.~J. Needs, High-pressure phases of silane, Phys. Rev. Lett. 97
  (2006) 045504.

\bibitem{AIRSS_2011_Pickard-Needs}
C.~J. Pickard, R.~Needs, Ab initio random structure searching, Journal of
  Physics: Condensed Matter 23~(5) (2011) 053201.

\bibitem{pickard_NatPhys_2007_structure-H}
C.~J. Pickard, R.~J. Needs, Structure of phase iii of solid hydrogen, Nature
  Physics 3~(7) (2007) 473.

\bibitem{pickard_NatMat_2010_Al}
C.~J. Pickard, R.~Needs, Aluminium at terapascal pressures, Nature Materials
  9~(8) (2010) 624.

\bibitem{pickard_NatMat_2008_amonia}
C.~J. Pickard, R.~Needs, Highly compressed ammonia forms an ionic crystal,
  Nature Materials 7~(10) (2008) 775.

\bibitem{Needs_APL_2016_perspective-review}
R.~J. Needs, C.~J. Pickard, Perspective: Role of structure prediction in
  materials discovery and design, APL Materials 4~(5) (2016) 053210.

\bibitem{CASTEP_2002}
M.~Segall, P.~J. Lindan, M.~a. Probert, C.~J. Pickard, P.~J. Hasnip, S.~Clark,
  M.~Payne, First-principles simulation: ideas, illustrations and the castep
  code, Journal of Physics: Condensed Matter 14~(11) (2002) 2717.

\bibitem{eberhart_PSO_1995}
R.~Eberhart, J.~Kennedy, A new optimizer using particle swarm theory, in: Micro
  Machine and Human Science, 1995. MHS'95., Proceedings of the Sixth
  International Symposium on, IEEE, 1995, pp. 39--43.

\bibitem{wang_CALYPSS-2012_CPC}
Y.~Wang, J.~Lv, L.~Zhu, Y.~Ma, Calypso: A method for crystal structure
  prediction, Computer Physics Communications 183~(10) (2012) 2063--2070.

\bibitem{wang_CALYPSO-2010_PRB}
Y.~Wang, J.~Lv, L.~Zhu, Y.~Ma, Crystal structure prediction via particle-swarm
  optimization, Physical Review B 82~(9) (2010) 094116.

\bibitem{lv2012particle}
J.~Lv, Y.~Wang, L.~Zhu, Y.~Ma, Particle-swarm structure prediction on clusters,
  The Journal of Chemical Physics 137~(8) (2012) 084104.

\bibitem{zhang2013first}
X.~Zhang, Y.~Wang, J.~Lv, C.~Zhu, Q.~Li, M.~Zhang, Q.~Li, Y.~Ma,
  First-principles structural design of superhard materials, The Journal of
  chemical physics 138~(11) (2013) 114101.

\bibitem{lu2014self}
S.~Lu, Y.~Wang, H.~Liu, M.-s. Miao, Y.~Ma, Self-assembled ultrathin nanotubes
  on diamond (100) surface, Nature Communications 5 (2014) 3666.

\bibitem{gao_XRD-searches-assisted_2017}
P.~Gao, Q.~Tong, J.~Lv, Y.~Wang, Y.~Ma, X-ray diffraction data-assisted
  structure searches, Computer Physics Communications 213 (2017) 40--45.

\bibitem{shi2016investigation}
J.~Shi, W.~Cui, J.~A. Flores-Livas, A.~San-Miguel, S.~Botti, M.~A. Marques,
  Investigation of new phases in the {Ba--Si} phase diagram under high pressure
  using ab initio structural search, Physical Chemistry Chemical Physics
  18~(11) (2016) 8108--8114.

\bibitem{Ma_CALYPSO_database_2017}
C.~Su, J.~Lv, Q.~Li, H.~Wang, L.~Zhang, Y.~Wang, Y.~Ma, Construction of crystal
  structure prototype database: methods and applications, Journal of Physics:
  Condensed Matter 29~(16) (2017) 165901.

\bibitem{Machine-larning_pot_CALYPSO_2018}
Q.~Tong, L.~Xue, J.~Lv, Y.~Wang, Y.~Ma, Accelerating calypso structure
  prediction by data-driven learning of potential energy surface, Faraday
  Discussions.

\bibitem{PNAS_Oxygen-chain_pressure_Ma}
L.~Zhu, Z.~Wang, Y.~Wang, G.~Zou, H.-k. Mao, Y.~Ma, Spiral chain o4 form of
  dense oxygen, Proceedings of the National Academy of Sciences 109~(3) (2012)
  751--753.

\bibitem{amonia_PNAS_2017_Hermann}
V.~N. Robinson, Y.~Wang, Y.~Ma, A.~Hermann, Stabilization of ammonia-rich
  hydrate inside icy planets, Proceedings of the National Academy of Sciences
  114~(34) (2017) 9003--9008.

\bibitem{clathrate_CaH6_PNAS_Ma}
H.~Wang, S.~T. John, K.~Tanaka, T.~Iitaka, Y.~Ma, Superconductive sodalite-like
  clathrate calcium hydride at high pressures, Proceedings of the National
  Academy of Sciences 109~(17) (2012) 6463--6466.

\bibitem{superconductivity_ScHx_2018}
X.~Ye, N.~Zarifi, E.~Zurek, R.~Hoffmann, N.~W. Ashcroft, High hydrides of
  scandium under pressure: Potential superconductors, The Journal of Physical
  Chemistry C 122~(11) (2018) 6298--6309.

\bibitem{nonmetallic_FeH6_JPCC_2018}
S.~Zhang, J.~Lin, Y.~Wang, G.~Yang, A.~Bergara, Y.~Ma, Nonmetallic feh6 under
  high pressure, The Journal of Physical Chemistry C.

\bibitem{Glass_2006_uspex_original}
C.~W. Glass, A.~R. Oganov, N.~Hansen, Uspex—evolutionary crystal structure
  prediction, Computer physics communications 175~(11-12) (2006) 713--720.

\bibitem{oganov_boook-2011}
A.~R. Oganov, Modern methods of crystal structure prediction, John Wiley \&
  Sons, 2011.

\bibitem{oganov2006crystal}
A.~R. Oganov, C.~W. Glass, Crystal structure prediction using ab initio
  evolutionary techniques: Principles and applications, The Journal of chemical
  physics 124~(24) (2006) 244704.

\bibitem{oganov2008evolutionary}
A.~R. Organov, C.~W. Glass, Evolutionary crystal structure prediction as a tool
  in materials design, Journal of Physics: Condensed Matter 20~(6) (2008)
  064210.

\bibitem{kruglov2017refined}
I.~Kruglov, R.~Akashi, S.~Yoshikawa, A.~R. Oganov, M.~M.~D. Esfahani, {Refined
  phase diagram of the HS system with high-Tc superconductivity}, Physical
  Review B 96~(22) (2017) 220101.

\bibitem{ma2008high}
Y.~Ma, A.~R. Oganov, Y.~Xie, High-pressure structures of lithium, potassium,
  and rubidium predicted by an {\it ab initio} evolutionary algorithm, Physical
  Review B 78~(1) (2008) 014102.

\bibitem{zaleski2011high}
P.~Zaleski-Ejgierd, R.~Hoffmann, N.~Ashcroft, {High pressure stabilization and
  emergent forms of PbH$_4$}, Physical Review Letters 107~(3) (2011) 037002.

\bibitem{lyakhov2013new}
A.~O. Lyakhov, A.~R. Oganov, H.~T. Stokes, Q.~Zhu, {New developments in
  evolutionary structure prediction algorithm USPEX}, Computer Physics
  Communications 184~(4) (2013) 1172--1182.

\bibitem{oganov_boook-2018}
A.~Oganov, G.~Saleh, A.~Kvashnin, Computational Materials Discovery, Royal
  Society of Chemistry, 2018.

\bibitem{martinez2009novel}
M.~Martinez-Canales, A.~R. Oganov, Y.~Ma, Y.~Yan, A.~O. Lyakhov, A.~Bergara,
  Novel structures and superconductivity of silane under pressure, Physical
  Review Letters 102~(8) (2009) 087005.

\bibitem{duan2015pressure}
D.~Duan, X.~Huang, F.~Tian, D.~Li, H.~Yu, Y.~Liu, Y.~Ma, B.~Liu, T.~Cui,
  Pressure-induced decomposition of solid hydrogen sulfide, Physical Review B
  91~(18) (2015) 180502.

\bibitem{Twisted_paper_LaH10_2018_Oganov}
I.~A. Kruglov, e.~a. Semenok, {Superconductivity in LaH$_{10}$: a new twist of
  the story}, arXiv preprint arXiv:1810.01113.

\bibitem{lonie_2012_Xtalcomp}
D.~C. Lonie, E.~Zurek, {Identifying duplicate crystal structures: XtalComp, an
  open-source solution}, Computer Physics Communications 183~(3) (2012)
  690--697.

\bibitem{lonie_2011_xtalopt}
D.~C. Lonie, E.~Zurek, {XtalOpt: An open-source evolutionary algorithm for
  crystal structure prediction}, Computer Physics Communications 182~(2) (2011)
  372--387.

\bibitem{avery2017xtalopt}
P.~Avery, Z.~Falls, E.~Zurek, {XtalOpt Version r10: An open--source
  evolutionary algorithm for crystal structure prediction}, Computer Physics
  Communications 217 (2017) 210--211.

\bibitem{Zurek_FeH5_JPCC_2018}
N.~Zarifi, T.~Bi, H.~Liu, E.~Zurek, {Crystal Structures and Properties of Iron
  Hydrides at High Pressure}, The Journal of Physical Chemistry C 122~(42)
  (2018) 24262--24269.

\bibitem{Zurek_CaH2_mix_JPCC_2018}
A.~K. Mishra, T.~Muramatsu, H.~Liu, Z.~M. Geballe, M.~Somayazulu, M.~Ahart,
  M.~Baldini, Y.~Meng, E.~Zurek, R.~J. Hemley, {New Calcium Hydrides with Mixed
  Atomic and Molecular Hydrogen}, The Journal of Physical Chemistry C 122~(34)
  (2018) 19370--19378.

\bibitem{abinit_2009}
X.~Gonze, B.~Amadon, P.-M. Anglade, J.-M. Beuken, F.~Bottin, P.~Boulanger,
  F.~Bruneval, D.~Caliste, R.~Caracas, M.~C{\^o}t{\'e}, et~al., Abinit:
  First-principles approach to material and nanosystem properties, Computer
  Physics Communications 180~(12) (2009) 2582--2615.

\bibitem{Goedecker_mhm_2004}
S.~Goedecker, {Minima hopping: An efficient search method for the global
  minimum of the potential energy surface of complex molecular systems}, The
  Journal of Chemical Physics 120~(21) (2004) 9911.

\bibitem{Goedecker_mhm_2005}
S.~Goedecker, W.~Hellmann, T.~Lenosky, {Global Minimum Determination of the
  {Born-Oppenheimer} Surface within Density Functional Theory}, Phys. Rev.
  Lett. 95~(5) (2005) 055501.

\bibitem{FIRE_algorithm_2006}
E.~Bitzek, P.~Koskinen, F.~G\"ahler, M.~Moseler, P.~Gumbsch, Structural
  relaxation made simple, Phys. Rev. Lett. 97 (2006) 170201.

\bibitem{Amsler_mhm_2010}
M.~Amsler, S.~Goedecker, Crystal structure prediction using the minima hopping
  method, The Journal of Chemical Physics 133~(22) (2010) 224104.

\bibitem{Amsler_Thesis_2012}
M.~Amsler~K, Crystal structure prediction based on density functional theory,
  Ph.D. thesis, University of Basel (2012).

\bibitem{jensen2017introduction}
F.~Jensen, Introduction to computational chemistry, John wiley \& sons, 2017.

\bibitem{Amsler_MHM_handobook}
M.~Amsler, Minima hopping method for predicting complex structures and chemical
  reaction pathways, Handbook of Materials Modeling: Applications: Current and
  Emerging Materials (2018) 1--20.

\bibitem{Amsler_FeBi_HP_ChemScien_2017}
M.~Amsler, S.~S. Naghavi, C.~Wolverton, Prediction of superconducting
  iron--bismuth intermetallic compounds at high pressure, Chemical science
  8~(3) (2017) 2226--2234.

\bibitem{Clarke_ChemMat_CuBi_2017}
S.~M. Clarke, M.~Amsler, J.~P. Walsh, T.~Yu, Y.~Wang, Y.~Meng, S.~D. Jacobsen,
  C.~Wolverton, D.~E. Freedman, Creating binary cu--bi compounds via
  high-pressure synthesis: a combined experimental and theoretical study,
  Chemistry of Materials 29~(12) (2017) 5276--5285.

\bibitem{Amsler_Ternary_HSeS_PRB2019}
M.~Amsler, {Thermodynamics and superconductivity of S$_x$Se$_{1-x}$H$_3$},
  Physical Review B 99~(6) (2019) 060102.

\bibitem{Amsler_ternary_mixed_ML_2019}
M.~Amsler, L.~Ward, V.~I. Hegde, M.~G. Goesten, X.~Yi, C.~Wolverton, Ternary
  mixed-anion semiconductors with tunable band gaps from machine-learning and
  crystal structure prediction, arXiv preprint arXiv:1812.02708.

\bibitem{Heusler_HT_ChmMat_Wolverton_2018}
J.~He, S.~S. Naghavi, V.~I. Hegde, M.~Amsler, C.~Wolverton, Designing and
  discovering a new family of semiconducting quaternary heusler compounds based
  on the 18-electron rule, Chemistry of Materials 30~(15) (2018) 4978--4985.

\bibitem{PRB_Cohen_oldschool_Si3N4}
A.~Y. Liu, M.~L. Cohen, {Structural properties and electronic structure of
  low-compressibility materials:
  \ensuremath{\beta}-${\mathrm{Si}}_{3}$${\mathrm{N}}_{4}$ and hypothetical
  \ensuremath{\beta}-${\mathrm{C}}_{3}$${\mathrm{N}}_{4}$}, Phys. Rev. B 41
  (1990) 10727--10734.

\bibitem{pannetier_simulated_ann-1990_nature}
J.~Pannetier, J.~Bassas-Alsina, J.~Rodriguez-Carvajal, V.~Caignaert, Prediction
  of crystal structures from crystal chemistry rules by simulated annealing,
  Nature 346~(6282) (1990) 343.

\bibitem{PRL_SiO-Tsuneyuki_1988}
S.~Tsuneyuki, M.~Tsukada, H.~Aoki, Y.~Matsui, {First-Principles Interatomic
  Potential of Silica Applied to Molecular Dynamics}, Phys. Rev. Lett. 61
  (1988) 869--872.

\bibitem{kirkpatrick1983optimization}
S.~Kirkpatrick, C.~D. Gelatt, M.~P. Vecchi, Optimization by simulated
  annealing, science 220~(4598) (1983) 671--680.

\bibitem{vcerny1985thermodynamical}
V.~{\v{C}}ern{\`y}, {Thermodynamical approach to the traveling salesman
  problem: An efficient simulation algorithm}, Journal of optimization theory
  and applications 45~(1) (1985) 41--51.

\bibitem{van1987simulated}
P.~Van~Laarhoven, {Simulated Annealing/PJM van Laarhoven, EHL Aarts}, Theory
  and Applications.--Dordrecht.--Springer.

\bibitem{doll2007global}
K.~Doll, J.~Sch{\"o}n, M.~Jansen, Global exploration of the energy landscape of
  solids on the ab initio level, Physical Chemistry Chemical Physics 9~(46)
  (2007) 6128--6133.

\bibitem{schon2010predicting}
J.~C. Sch{\"o}n, K.~Doll, M.~Jansen, Predicting solid compounds via global
  exploration of the energy landscape of solids on the ab initio level without
  recourse to experimental information, physica status solidi (b) 247~(1)
  (2010) 23--39.

\bibitem{zagorac2011ab}
D.~Zagorac, K.~Doll, J.~Sch{\"o}n, M.~Jansen, Ab initio structure prediction
  for lead sulfide at standard and elevated pressures, Physical Review B 84~(4)
  (2011) 045206.

\bibitem{curtarolo2013high}
S.~Curtarolo, G.~L. Hart, M.~B. Nardelli, N.~Mingo, S.~Sanvito, O.~Levy, The
  high-throughput highway to computational materials design, Nature Materials
  12~(3) (2013) 191.

\bibitem{hattrick_APL_2016_datamining}
J.~R. Hattrick-Simpers, J.~M. Gregoire, A.~G. Kusne, Perspective:
  Composition--structure--property mapping in high-throughput experiments:
  Turning data into knowledge, APL Materials 4~(5) (2016) 053211.

\bibitem{glawe_2016_pettifor_data-minig}
H.~Glawe, A.~Sanna, E.~Gross, M.~A. Marques, The optimal one dimensional
  periodic table: a modified pettifor chemical scale from data mining, New
  Journal of Physics 18~(9) (2016) 093011.

\bibitem{PYMATGEN_2013-paper}
S.~P. Ong, W.~D. Richards, A.~Jain, G.~Hautier, M.~Kocher, S.~Cholia,
  D.~Gunter, V.~L. Chevrier, K.~A. Persson, G.~Ceder, {Python Materials
  Genomics (pymatgen): A robust, open-source python library for materials
  analysis}, Computational Materials Science 68 (2013) 314--319.

\bibitem{NatMat_Ceder_datamining_2006}
C.~C. Fischer, K.~J. Tibbetts, D.~Morgan, G.~Ceder, Predicting crystal
  structure by merging data mining with quantum mechanics, Nature Materials
  5~(8) (2006) 641.

\bibitem{Amsler_PRX_2018}
M.~Amsler, V.~I. Hegde, S.~D. Jacobsen, C.~Wolverton, Exploring the
  high-pressure materials genome, Phys. Rev. X 8 (2018) 041021.

\bibitem{Ceder_thermoscale_2016_SciAdv}
W.~Sun, S.~T. Dacek, S.~P. Ong, G.~Hautier, A.~Jain, W.~D. Richards, A.~C.
  Gamst, K.~A. Persson, G.~Ceder, The thermodynamic scale of inorganic
  crystalline metastability, Science advances 2~(11) (2016) e1600225.

\bibitem{aykol_2018thermodynamic}
M.~Aykol, S.~S. Dwaraknath, W.~Sun, K.~A. Persson, Thermodynamic limit for
  synthesis of metastable inorganic materials, Science advances 4~(4) (2018)
  eaaq0148.

\bibitem{mueller2016machine}
T.~Mueller, A.~G. Kusne, R.~Ramprasad, Machine learning in materials science:
  Recent progress and emerging applications, Reviews in Computational Chemistry
  29 (2016) 186--273.

\bibitem{ward2017atomistic}
L.~Ward, C.~Wolverton, {Atomistic calculations and materials informatics: A
  review}, Current Opinion in Solid State and Materials Science 21~(3) (2017)
  167--176.

\bibitem{Togo_book_2018}
A.~Seko, A.~Togo, I.~Tanaka, Descriptors for machine learning of materials
  data, in: Nanoinformatics, Springer, Singapore, 2018, pp. 3--23.

\bibitem{deep_learning_2015}
Y.~LeCun, Y.~Bengio, G.~Hinton, Deep learning, nature 521~(7553) (2015) 436.

\bibitem{xie2018crystal}
T.~Xie, J.~C. Grossman, Crystal graph convolutional neural networks for an
  accurate and interpretable prediction of material properties, Physical Review
  Letters 120~(14) (2018) 145301.

\bibitem{deep-learning_ryan2018crystal}
K.~Ryan, J.~Lengyel, M.~Shatruk, Crystal structure prediction via deep
  learning, Journal of the American Chemical Society 140~(32) (2018)
  10158--10168.

\bibitem{behler2016perspective}
J.~Behler, {Perspective: Machine learning potentials for atomistic
  simulations}, The Journal of chemical physics 145~(17) (2016) 170901.

\bibitem{podryabinkin2019accelerating}
E.~V. Podryabinkin, E.~V. Tikhonov, A.~V. Shapeev, A.~R. Oganov, Accelerating
  crystal structure prediction by machine-learning interatomic potentials with
  active learning, Physical Review B 99~(6) (2019) 064114.

\bibitem{faber2015crystal}
F.~Faber, A.~Lindmaa, O.~A. von Lilienfeld, R.~Armiento, Crystal structure
  representations for machine learning models of formation energies,
  International Journal of Quantum Chemistry 115~(16) (2015) 1094--1101.

\bibitem{ye2018deep}
W.~Ye, C.~Chen, Z.~Wang, I.-H. Chu, S.~P. Ong, Deep neural networks for
  accurate predictions of crystal stability, Nature Communications 9~(1) (2018)
  3800.

\bibitem{curtarolo_universal}
O.~Isayev, C.~Oses, C.~Toher, E.~Gossett, S.~Curtarolo, A.~Tropsha, {Universal
  fragment descriptors for predicting properties of inorganic crystals}, Nature
  Communications 8 (2017) 15679.

\bibitem{Olle_klintenberg2013possible}
M.~Klintenberg, O.~Eriksson, Possible high-temperature superconductors
  predicted from electronic structure and data-filtering algorithms,
  Computational materials science 67 (2013) 282--286.

\bibitem{owolabi2015estimation}
T.~O. Owolabi, K.~O. Akande, S.~O. Olatunji, Estimation of superconducting
  transition temperature {T}$_c$ for superconductors of the doped {MgB}$_2$
  system from the crystal lattice parameters using support vector regression,
  Journal of Superconductivity and Novel Magnetism 28~(1) (2015) 75--81.

\bibitem{norman2016materials}
M.~Norman, Materials design for new superconductors, Reports on Progress in
  Physics 79~(7) (2016) 074502.

\bibitem{ML_superconducting_curatolos_2018}
V.~Stanev, C.~Oses, A.~G. Kusne, E.~Rodriguez, J.~Paglione, S.~Curtarolo,
  I.~Takeuchi, Machine learning modeling of superconducting critical
  temperature, npj Computational Materials 4~(1) (2018) 29.

\bibitem{mehta2019high}
P.~Mehta, M.~Bukov, C.-H. Wang, A.~G. Day, C.~Richardson, C.~K. Fisher, D.~J.
  Schwab, A high-bias, low-variance introduction to machine learning for
  physicists, Physics Reports.

\bibitem{DFT_Delta-test_Science_2016}
K.~Lejaeghere, G.~Bihlmayer, T.~Bj{\"o}rkman, P.~Blaha, S.~Bl{\"u}gel, V.~Blum,
  D.~Caliste, I.~E. Castelli, S.~J. Clark, A.~Dal~Corso, et~al.,
  Reproducibility in density functional theory calculations of solids, Science
  351~(6280) (2016) aad3000.

\bibitem{Reference_wiki_dft-codes}
{List of quantum chemistry and solid-state physics software},
  \url{https://en.wikipedia.org/wiki/List_of_quantum_chemistry_and_solid-state_physics_software},
  accessed: 2019-02-21.

\bibitem{gale2003general}
J.~D. Gale, A.~L. Rohl, {The general utility lattice program (GULP)}, Molecular
  Simulation 29~(5) (2003) 291--341.

\bibitem{GULP}
J.~D. Gale, \href{http://dx.doi.org/10.1039/A606455H}{Gulp: A computer program
  for the symmetry-adapted simulation of solids}, J. Chem. Soc.{,} Faraday
  Trans. 93 (1997) 629--637.
\newblock \href {http://dx.doi.org/10.1039/A606455H}
  {\path{doi:10.1039/A606455H}}.
\newline\urlprefix\url{http://dx.doi.org/10.1039/A606455H}

\bibitem{LAMMPS}
S.~Plimpton, Fast parallel algorithms for short-range molecular dynamics,
  Journal of Computational Physics 117~(1) (1995) 1 -- 19.

\bibitem{TBDFT}
M.~Elstner, D.~Porezag, G.~Jungnickel, J.~Elsner, M.~Haugk, T.~Frauenheim,
  S.~Suhai, G.~Seifert, Self-consistent-charge density-functional tight-binding
  method for simulations of complex materials properties, Phys. Rev. B 58
  (1998) 7260--7268.

\bibitem{wavelets_2008daubechies}
L.~Genovese, A.~Neelov, S.~Goedecker, T.~Deutsch, S.~A. Ghasemi, A.~Willand,
  D.~Caliste, O.~Zilberberg, M.~Rayson, A.~Bergman, et~al., Daubechies wavelets
  as a basis set for density functional pseudopotential calculations, The
  Journal of chemical physics 129~(1) (2008) 014109.

\bibitem{Andersen1975_Linearmethods}
O.~K. Andersen, Linear methods in band theory, Phys. Rev. B 12 (1975)
  3060--3083.

\bibitem{Gulans_muHa-prec_solids_2018}
A.~Gulans, A.~Kozhevnikov, C.~Draxl, Microhartree precision in density
  functional theory calculations, Phys. Rev. B 97 (2018) 161105.

\bibitem{Elephant_paper_2017}
S.~R. Jensen, S.~Saha, J.~A. Flores-Livas, W.~Huhn, V.~Blum, S.~Goedecker,
  L.~Frediani, The elephant in the room of density functional theory
  calculations, The journal of physical chemistry letters 8~(7) (2017)
  1449--1457.

\bibitem{pseudodojo_2018}
M.~Van~Setten, M.~Giantomassi, E.~Bousquet, M.~J. Verstraete, D.~R. Hamann,
  X.~Gonze, G.-M. Rignanese, The pseudodojo: Training and grading a 85 element
  optimized norm-conserving pseudopotential table, Computer Physics
  Communications 226 (2018) 39--54.

\bibitem{SSPS_marvel_2018}
G.~Prandini, A.~Marrazzo, I.~E. Castelli, N.~Mounet, N.~Marzari, Precision and
  efficiency in solid-state pseudopotential calculations, npj Computational
  Materials 4~(1) (2018) 72.

\bibitem{Monkhorst_pack}
H.~J. Monkhorst, J.~D. Pack, Special points for brillouin-zone integrations,
  Phys. Rev. B 13 (1976) 5188--5192.

\bibitem{abinit_2016}
X.~Gonze, F.~Jollet, F.~A. Araujo, D.~Adams, B.~Amadon, T.~Applencourt,
  C.~Audouze, J.-M. Beuken, J.~Bieder, A.~Bokhanchuk, et~al., Recent
  developments in the abinit software package, Computer Physics Communications
  205 (2016) 106--131.

\bibitem{Quantumespresso_2017}
P.~Giannozzi, O.~Andreussi, T.~Brumme, O.~Bunau, M.~B. Nardelli, M.~Calandra,
  R.~Car, C.~Cavazzoni, D.~Ceresoli, M.~Cococcioni, et~al., Advanced
  capabilities for materials modelling with quantum espresso, Journal of
  Physics: Condensed Matter 29~(46) (2017) 465901.

\bibitem{VASP_Kresse}
G.~Kresse, J.~Furthm\"{u}ller, Efficiency of ab-initio total energy
  calculations for metals and semiconductors using a plane-wave basis set,
  Comput. Mat. Sci. 6 (1996) 15--50.

\bibitem{GPU_accelerated-paper_2018}
L.~E. Ratcliff, A.~Degomme, J.~A. Flores-Livas, S.~Goedecker, L.~Genovese,
  {Affordable and accurate large-scale hybrid-functional calculations on
  GPU-accelerated supercomputers}, Journal of Physics: Condensed Matter 30~(9)
  (2018) 095901.

\bibitem{butler2018machine}
K.~T. Butler, D.~W. Davies, H.~Cartwright, O.~Isayev, A.~Walsh, Machine
  learning for molecular and materials science, Nature 559~(7715) (2018) 547.

\bibitem{Philip_Ball_Note_Nature-Materials}
P.~Ball, Material witness: Metallic hydrogen in the spotlight, Nature Materials
  16~(3) (2017) 288.

\bibitem{Th:Hopfield_PR_1969}
J.~Hopfield, Angular momentum and transition-metal superconductivity, Phys.
  Rev. 186 (1969) 443.

\bibitem{mueller_metal-H_ed1968}
W.~M. Mueller, J.~P. Blackledge, G.~G. Libowitz, Metal hydrides, Academic
  Press, N. Y., 1968.

\bibitem{Maksimov_1975}
E.~G. Maksimov, O.~A. Pankratov, Hydrogen in metals, Soviet Physics Uspekhi
  18~(7) (1975) 481--495.

\bibitem{Satterthwaite_ThH_1970}
C.~B. Satterthwaite, I.~L. Toepke, {Superconductivity of Hydrides and
  Deuterides of Thorium}, Phys. Rev. Lett. 25 (1970) 741--743.

\bibitem{Th4H15_pressure_supra1974}
M.~Dietrich, W.~Gey, H.~Rietschel, C.~Satterthwaite, Pressure dependence of the
  superconducting transition temperature of {Th}$_4${H}$_15$, Solid State
  Communications 15~(5) (1974) 941--943.

\bibitem{PdH1}
T.~Skoskiewicz, Superconductivity in the palladium-hydrogen and
  palladium-nickel-hydrogen systems, physica status solidi (a) 11~(2) (1972)
  K123--K126.

\bibitem{PdH2}
B.~Stritzker, W.~Buckel, Superconductivity in the palladium-hydrogen and the
  palladium-deuterium systems, Zeitschrift f{\"u}r Physik A Hadrons and nuclei
  257~(1) (1972) 1--8.

\bibitem{oesterreicher1976studies}
H.~Oesterreicher, J.~Clinton, H.~Bittner, Studies of hydride formation and
  superconductivity in hydrides of {Th} and {Pd} compounds, Journal of Solid
  State Chemistry France 16 (1976) 209--210.

\bibitem{PdCuH_Stritzker1974_16K}
B.~Stritzker, High superconducting transition temperatures in the
  palladium-noble metal-hydrogen system, Zeitschrift f{\"u}r Physik 268~(2)
  (1974) 261--264.

\bibitem{HfVH_PRL_1976}
P.~Duffer, D.~M. Gualtieri, V.~U.~S. Rao, {Pronounced Isotope Effect in the
  Superconductivity of ${\mathrm{HfV}}_{2}$ Containing Hydrogen (Deuterium)},
  Phys. Rev. Lett. 37 (1976) 1410--1413.

\bibitem{NbHx_Welter1977}
J.~M. Welter, F.~J. Johnen, Superconducting transition temperature and low
  temperature resistivity in the niobium-hydrogen system, Zeitschrift f{\"u}r
  Physik B Condensed Matter 27~(3) (1977) 227--232.

\bibitem{ponyatovskii1985new}
E.~Ponyatovskii, I.~Bashkin, {New phase transitions in hydrides of the IA,
  {III}-{A}, and {IV}-{A} group metals}, Zeitschrift f{\"u}r Physikalische
  Chemie 146~(2) (1985) 137--157.

\bibitem{tonkov1998compounds}
E.~Y. Tonkov, Compounds and alloys under high pressure: a handbook, CRC Press,
  1998.

\bibitem{mazin2015superconductivity}
I.~I. Mazin, Superconductivity: Extraordinarily conventional, Nature 525~(7567)
  (2015) 40--41.

\bibitem{zunger2019beware}
A.~Zunger, Beware of plausible predictions of fantasy materials (2019).

\bibitem{H2O_terapascal_PRL}
B.~Militzer, H.~F. Wilson, {New Phases of Water Ice Predicted at Megabar
  Pressures}, Phys. Rev. Lett. 105 (2010) 195701.

\bibitem{PhysRevB.84.220104}
J.~M. McMahon, Ground-state structures of ice at high pressures from {\it ab
  initio} random structure searching, Phys. Rev. B 84 (2011) 220104.

\bibitem{PhysRevB.87.024112}
S.~Zhang, H.~F. Wilson, K.~P. Driver, B.~Militzer, {H$_{4}$O and other
  hydrogen-oxygen compounds at giant-planet core pressures}, Phys. Rev. B 87
  (2013) 024112.

\bibitem{PhysRevLett.110.245701}
C.~J. Pickard, M.~Martinez-Canales, R.~J. Needs, {Decomposition and Terapascal
  Phases of Water Ice}, Phys. Rev. Lett. 110 (2013) 245701.

\bibitem{RevModPhys.84.885}
T.~Bartels-Rausch, V.~Bergeron, J.~H.~E. Cartwright, R.~Escribano, J.~L.
  Finney, H.~Grothe, P.~J. Guti\'errez, J.~Haapala, W.~F. Kuhs, J.~B.~C.
  Pettersson, S.~D. Price, C.~I. Sainz-D\'{\i}az, D.~J. Stokes, G.~Strazzulla,
  E.~S. Thomson, H.~Trinks, N.~Uras-Aytemiz, {Ice structures, patterns, and
  processes: A view across the icefields}, Rev. Mod. Phys. 84 (2012) 885--944.

\bibitem{donnerer2013high}
C.~Donnerer, T.~Scheler, E.~Gregoryanz, High-pressure synthesis of noble metal
  hydrides, The Journal of chemical physics 138~(13) (2013) 134507.

\bibitem{Cu-H_binns2019structural}
J.~Binns, M.~Pe{\~n}a-Alvarez, M.-E. Donnelly, E.~Gregoryanz, R.~T. Howie,
  P.~Dalladay-Simpson, {Structural Studies on the {Cu-H} System under
  Compression}, Engineering.

\bibitem{antonov2002phase}
V.~Antonov, Phase transformations, crystal and magnetic structures of
  high-pressure hydrides of d-metals, Journal of alloys and compounds 330
  (2002) 110--116.

\bibitem{eremets2003exploring}
M.~I. Eremets, V.~V. Struzhkin, H.-k. Mao, R.~J. Hemley, Exploring
  superconductivity in low-z materials at megabar pressures, Physica B:
  Condensed Matter 329 (2003) 1312--1316.

\bibitem{He-H_Gonchy_shit}
Y.~Wang, X.~Zhang, S.~Jiang, Z.~M. Geballe, T.~Pakornchote, M.~Somayazulu,
  V.~B. Prakapenka, E.~Greenberg, A.~F. Goncharov, Helium-hydrogen
  immiscibility at high pressures, The Journal of Chemical Physics 150~(11)
  (2019) 114504.

\bibitem{H:Pepin_PNAS_Li_2015}
C.~P{\'e}pin, P.~Loubeyre, F.~Occelli, P.~Dumas, Synthesis of lithium
  polyhydrides above 130 {GPa} at 300 {K}, Proceedings of the National Academy
  of Sciences 112~(25) (2015) 7673--7676.

\bibitem{H:Struzhkin_Na_Natcomm_2016}
V.~V. Struzhkin, D.~Y. Kim, E.~Stavrou, T.~Muramatsu, H.-k. Mao, C.~J. Pickard,
  R.~J. Needs, V.~B. Prakapenka, A.~F. Goncharov, Synthesis of sodium
  polyhydrides at high pressures, Nature Communications 7 (2016) 12267,
  article.

\bibitem{ponyatovsky1992pressure}
E.~Ponyatovsky, O.~Barkalov, Pressure—induced amorphous phases, Materials
  Science Reports 8~(4) (1992) 147--191.

\bibitem{goncharov1996compression}
A.~Gonchirov, V.~Struzhkin, M.~Somayazulu, R.~Hemley, H.~Mao, Compression of
  ice to 210 gigapascals: Infrared evidence for a symmetric hydrogen-bonded
  phase, Science 273~(5272) (1996) 218--220.

\bibitem{AlH_experiment_PRL_2008}
I.~Goncharenko, M.~I. Eremets, M.~Hanfland, J.~S. Tse, M.~Amboage, Y.~Yao,
  I.~A. Trojan, {Pressure-Induced Hydrogen-Dominant Metallic State in Aluminum
  Hydride}, Phys. Rev. Lett. 100 (2008) 045504.

\bibitem{PhysRevLett.113.265504}
C.~M. P\'epin, A.~Dewaele, G.~Geneste, P.~Loubeyre, M.~Mezouar, {New Iron
  Hydrides under High Pressure}, Phys. Rev. Lett. 113 (2014) 265504.

\bibitem{H:Pepin_FEH_2017}
C.~M. P{\'e}pin, G.~Geneste, A.~Dewaele, M.~Mezouar, P.~Loubeyre, Synthesis of
  feh5: A layered structure with atomic hydrogen slabs, Science 357~(6349)
  (2017) 382--385.

\bibitem{H:Heil_FEH_2018}
L.~B. Christoph~Heil, Giovanni B.~Bachelet, No superconductivity in iron
  polyhydrides at high pressures, arXiv preprint, arXiv:1804.03572.

\bibitem{FeH5_meier2019pressure}
T.~Meier, F.~Trybel, S.~Khandarkhaeva, G.~Steinle-Neumann, S.~Chariton,
  T.~Fedotenko, S.~Petitgirard, M.~Hanfland, K.~Glazyrin, N.~Dubrovinskaia,
  et~al., Pressure induced hydrogen-hydrogen interaction in metallic {FeH}
  revealed by {NMR}, arXiv preprint arXiv:1902.03182.

\bibitem{PhysRevB.97.214510}
C.~Heil, G.~B. Bachelet, L.~Boeri, Absence of superconductivity in iron
  polyhydrides at high pressures, Phys. Rev. B 97 (2018) 214510.

\bibitem{Co-H_wang2018high}
M.~Wang, J.~Binns, M.-E. Donnelly, M.~Pe{\~n}a-Alvarez, P.~Dalladay-Simpson,
  R.~T. Howie, High pressure synthesis and stability of cobalt hydrides, The
  Journal of chemical physics 148~(14) (2018) 144310.

\bibitem{NiH-hydrides_under_pressure_exp}
J.~Binns, M.-E. Donnelly, M.~Wang, A.~Hermann, E.~Gregoryanz,
  P.~Dalladay-Simpson, R.~T. Howie, {Synthesis of Ni$_{2}$H$_{3}$ at high
  temperatures and pressures}, Phys. Rev. B 98 (2018) 140101.

\bibitem{Ni-H_PRM_2018}
J.~Ying, H.~Liu, E.~Greenberg, V.~B. Prakapenka, V.~V. Struzhkin, Synthesis of
  new nickel hydrides at high pressure, Phys. Rev. Materials 2 (2018) 085409.

\bibitem{PRL_SiH4}
X.-J. Chen, J.-L. Wang, V.~V. Struzhkin, H.-k. Mao, R.~J. Hemley, H.-Q. Lin,
  {Superconducting Behavior in Compressed Solid ${\mathrm{SiH}}_{4}$ with a
  Layered Structure}, Phys. Rev. Lett. 101 (2008) 077002.

\bibitem{PhysRevB.83.144102}
T.~A. Strobel, A.~F. Goncharov, C.~T. Seagle, Z.~Liu, M.~Somayazulu, V.~V.
  Struzhkin, R.~J. Hemley, High-pressure study of silane to 150 {GPa}, Phys.
  Rev. B 83 (2011) 144102.

\bibitem{Gonchy_PRB_SeH3-2018}
X.~Zhang, W.~Xu, Y.~Wang, S.~Jiang, F.~A. Gorelli, E.~Greenberg, V.~B.
  Prakapenka, A.~F. Goncharov, Synthesis and properties of selenium trihydride
  at high pressures, Phys. Rev. B 97 (2018) 064107.

\bibitem{Nb-H_experiments_2017}
G.~Liu, S.~Besedin, A.~Irodova, H.~Liu, G.~Gao, M.~Eremets, X.~Wang, Y.~Ma,
  {Nb}-{H} system at high pressures and temperatures, Phys. Rev. B 95 (2017)
  104110.

\bibitem{Rh_H_exp_li2011rhodium}
B.~Li, Y.~Ding, D.~Y. Kim, R.~Ahuja, G.~Zou, H.-K. Mao, Rhodium dihydride
  ({RhH}$_2$) with high volumetric hydrogen density, Proceedings of the
  National Academy of Sciences 108~(46) (2011) 18618--18621.

\bibitem{I-H_experiments_2018_gregoryanz}
J.~Binns, P.~Dalladay-Simpson, M.~Wang, G.~J. Ackland, E.~Gregoryanz, R.~T.
  Howie, {Formation of H$_{2}$-rich iodine-hydrogen compounds at high
  pressure}, Phys. Rev. B 97 (2018) 024111.

\bibitem{Ta-H_PRB_exp}
M.~A. Kuzovnikov, M.~Tkacz, H.~Meng, D.~I. Kapustin, V.~I. Kulakov,
  High-pressure synthesis of tantalum dihydride, Phys. Rev. B 96 (2017) 134120.

\bibitem{Ce-H_salke2018synthesis}
N.~P. Salke, M.~Esfahani, Y.~Zhang, I.~A. Kruglov, J.~Zhou, Y.~Wang,
  E.~Greenberg, V.~B. Prakapenka, A.~R. Oganov, J.-F. Lin, {Synthesis of
  clathrate cerium superhydride CeH$_9$ at 80 GPa with anomalously short H--H
  distance}, arXiv preprint arXiv:1805.02060.

\bibitem{Eu-H_Matsuoka_2011}
T.~Matsuoka, H.~Fujihisa, N.~Hirao, Y.~Ohishi, T.~Mitsui, R.~Masuda, M.~Seto,
  Y.~Yoda, K.~Shimizu, A.~Machida, K.~Aoki, Structural and valence changes of
  europium hydride induced by application of high-pressure ${\mathrm{h}}_{2}$,
  Phys. Rev. Lett. 107 (2011) 025501.

\bibitem{La-Ni_hydrides_1976}
H.~Oesterreicher, J.~Clinton, H.~Bittner, Hydrides of {La}-{Ni} compounds,
  Materials Research Bulletin 11~(10) (1976) 1241--1247.

\bibitem{REM-H_MENG201729344}
H.~Meng, M.~Kuzovnikov, M.~Tkacz, Phase stability of some rare earth
  trihydrides under high pressure, International Journal of Hydrogen Energy
  42~(49) (2017) 29344 -- 29349.

\bibitem{Papaconstantopoulos_2019}
P.-H. Chang, S.~Silayi, D.~Papaconstantopoulos, M.~Mehl, {Pressure-induced high
  temperature superconductivity in {H}$_3${X} ({X}= {As}, {Se}, {Br}, {Sb},
  {Te} and {I})}, arXiv preprint arXiv:1903.03255.

\bibitem{sennikov1994weak}
P.~Sennikov, {Weak hydrogen-bonding by second-row (PH$_3$, H$_2$S) and
  third-row ({AsH}$_3$, {H}$_2${S}e) hydrides}, The Journal of Physical
  Chemistry 98~(19) (1994) 4973--4981.

\bibitem{Pickett_2019_PRB}
Y.~Quan, S.~S. Ghosh, W.~E. Pickett,
  \href{https://link.aps.org/doi/10.1103/PhysRevB.100.184505}{Compressed
  hydrides as metallic hydrogen superconductors}, Phys. Rev. B 100 (2019)
  184505.
\newblock \href {http://dx.doi.org/10.1103/PhysRevB.100.184505}
  {\path{doi:10.1103/PhysRevB.100.184505}}.
\newline\urlprefix\url{https://link.aps.org/doi/10.1103/PhysRevB.100.184505}

\bibitem{PRB_2015_Abe-Ashcroft}
K.~Abe, N.~W. Ashcroft, {Stabilization and highly metallic properties of heavy
  group-V hydrides at high pressures}, Phys. Rev. B 92 (2015) 224109.

\bibitem{gupta1981trends}
M.~Gupta, J.~Burger, Trends in the electron-phonon coupling parameter in some
  metallic hydrides, Physical Review B 24~(12) (1981) 7099.

\bibitem{burger1984electron}
J.~Burger, Electron-phonon coupling and superconductivity in metal-hydrogen
  systems, Journal of the Less Common Metals 101 (1984) 53--67.

\bibitem{semenok2018distribution}
D.~V. Semenok, I.~A. Kruglov, A.~G. Kvashnin, A.~R. Oganov, On distribution of
  superconductivity in metal hydrides, arXiv preprint arXiv:1806.00865.

\bibitem{kim_general_2010}
D.~Y. Kim, R.~H. Scheicher, H.-k. Mao, T.~W. Kang, R.~Ahuja, General trend for
  pressurized superconducting hydrogen-dense materials, PNAS 107~(7) (2010)
  2793--2796.

\bibitem{DFT:COHP_Dronskowski_JPC1993}
R.~Dronskowski, P.~E. Bloechl, Crystal orbital hamilton populations ({COHP}):
  energy-resolved visualization of chemical bonding in solids based on
  density-functional calculations, The Journal of Physical Chemistry 97~(33)
  (1993) 8617--8624.

\bibitem{Th:Berk_Schrieffer_PRL_1966}
N.~F. Berk, J.~R. Schrieffer, Effect of ferromagnetic spin correlations on
  superconductivity, Phys. Rev. Lett. 17 (1966) 433--435.

\bibitem{PdH-neutron}
K.~G. McLennan, E.~M. Gray, J.~F. Dobson, Deuterium occupation of tetrahedral
  sites in palladium, Phys. Rev. B 78 (2008) 014104.

\bibitem{Alavi}
R.~Caputo, A.~Alavi, Where do the {H} atoms reside in {PdH}$_x$ systems?,
  Molecular Physics 101~(11) (2003) 1781--1787.

\bibitem{houari_PdH-stability_JAP_2014}
A.~Houari, S.~F. Matar, V.~Eyert, Electronic structure and crystal phase
  stability of palladium hydrides, Journal of Applied Physics 116~(17) (2014)
  173706.

\bibitem{PdH-300K}
P.~Tripodi, D.~Di~Gioacchino, R.~Borelli, J.~D. Vinko, Possibility of high
  temperature superconducting phases in {PdH}, Physica C: Superconductivity 388
  (2003) 571--572.

\bibitem{PdH-HTc-arxiv}
H.~Syed, T.~Gould, C.~Webb, E.~Gray, Superconductivity in palladium hydride and
  deuteride at 52-61 kelvin, arXiv preprint arXiv:1608.01774.

\bibitem{PdH-tetra}
S.~Ostanin, V.~Borisov, D.~Fedorov, E.~Salamatov, A.~Ernst, I.~Mertig, Role of
  tetrahedrally coordinated dopants in palladium hydrides on their
  superconductivity and inverse isotope effect, Journal of Physics: Condensed
  Matter.

\bibitem{PdH-resistivity}
J.~Burger, S.~Senoussi, B.~Soufaché, Electrical and magnetic properties of
  palladium hydrides compared with those of pure palladium, Journal of the Less
  Common Metals 49 (1976) 213 -- 222, hydrogen in metals.

\bibitem{PdH-photoemission}
D.~E. Eastman, J.~K. Cashion, A.~C. Switendick, Photoemission studies of energy
  levels in the palladium-hydrogen system, Phys. Rev. Lett. 27 (1971) 35--38.

\bibitem{PdH-phonons1}
J.~M. Rowe, J.~J. Rush, H.~G. Smith, M.~Mostoller, H.~E. Flotow, Lattice
  dynamics of a single crystal of {Pd}${\mathrm{d}}_{0.63}$, Phys. Rev. Lett.
  33 (1974) 1297--1300.

\bibitem{PdH-phonons2}
O.~Blaschko, R.~Klemencic, P.~Weinzierl, L.~Pintschovius, {Lattice dynamics of
  $\ensuremath{\beta}$-{Pd}${\mathrm{D}}_{0.78}$ at 85 K and ordering effects
  at 75 K}, Phys. Rev. B 24 (1981) 1552--1555.

\bibitem{burger_1981_PdH_tunneling}
J.~Burger, {Electron-Phonon Coupling and Superconductivity in Palladium
  Hydrides and Deuterides}, in: Metal Hydrides, Springer, 1981, pp. 243--253.

\bibitem{shamp_decomposition_2016}
A.~Shamp, T.~Terpstra, T.~Bi, Z.~Falls, P.~Avery, E.~Zurek, {Decomposition
  Products of Phosphine Under Pressure: PH$_2$ Stable and Superconducting?}, J.
  Am. Chem. Soc 138~(6) (2016) 1884--1892, pMID: 26777416.

\bibitem{Nature_Errea_2016}
I.~Errea, M.~Calandra, C.~J. Pickard, J.~R. Nelson, R.~J. Needs, Y.~Li, H.~Liu,
  Y.~Zhang, Y.~Ma, F.~Mauri, Quantum hydrogen-bond symmetrization in the
  superconducting hydrogen sulfide system, Nature 532~(7597) (2016) 81.

\bibitem{H:Capitani_SH3_nphys_2017}
F.~Capitani, B.~Langerome, J.-B. Brubach, P.~Roy, A.~Drozdov, M.~I. Eremets,
  E.~J. Nicol, J.~P. Carbotte, T.~Timusk, {Spectroscopic evidence of a new
  energy scale for superconductivity in H$_3$S}, Nature Physics 13 (2017) 859
  EP --, article.

\bibitem{H3S-anharm-Bianco}
R.~Bianco, I.~Errea, M.~Calandra, F.~Mauri, High-pressure phase diagram of
  hydrogen and deuterium sulfides from first principles: Structural and
  vibrational properties including quantum and anharmonic effects, Phys. Rev. B
  97 (2018) 214101.

\bibitem{Yanming_JCP2014}
Y.~Wang, Y.~Ma, Perspective: Crystal structure prediction at high pressures,
  The Journal of Chemical Physics 140~(4) (2014) --.

\bibitem{Tse_magneli-non}
Y.~Yao, J.~S. Tse, Superconducting hydrogen sulfide, Chemistry – A European
  Journal 24~(8) (2018) 1769--1778.

\bibitem{HxS_majumdar2019mechanism}
A.~Majumdar, S.~T. John, Y.~Yao, Mechanism for the structural transformation to
  the modulated superconducting phase of compressed hydrogen sulfide,
  Scientific reports 9~(1) (2019) 5023.

\bibitem{van_ovfs-singuluarity}
L.~Van~Hove, {The Occurrence of Singularities in the Elastic Frequency
  Distribution of a Crystal}, Phys. Rev. 89 (1953) 1189--1193.

\bibitem{SH3-vHs-Pickett}
Y.~Quan, W.~E. Pickett, {Van Hove singularities and spectral smearing in
  high-temperature superconducting ${\mathrm{H}}_{3}\mathrm{S}$}, Phys. Rev. B
  93 (2016) 104526.

\bibitem{H3S-vHs-Marsiglio}
T.~X.~R. Souza, F.~Marsiglio, {The possible role of van Hove singularities in
  the high Tc of superconducting H$_3$S}, International Journal of Modern
  Physics B 31~(25) (2017) 1745003.

\bibitem{H3S-vHs-Bianconi}
T.~Jarlborg, A.~Bianconi, {Breakdown of the Migdal approximation at Lifshitz
  transitions with giant zero-point motion in the H$_3$S superconductor},
  Scientific Reports 6 (2016) 24816.

\bibitem{ortenzi_TB_2015}
L.~Ortenzi, E.~Cappelluti, L.~Pietronero, {Band Structure and electron-phonon
  coupling in {H}$_3${S}: a tight-binding model}, ArXiv e-prints.

\bibitem{Liu_MicroscopicMechanismLaH10_PRB2019}
L.~Liu, C.~Wang, S.~Yi, K.~W. Kim, J.~Kim, J.-H. Cho, {Microscopic mechanism of
  room-temperature superconductivity in compressed ${\mathrm{LaH}}_{10}$},
  Phys. Rev. B 99 (2019) 140501.

\bibitem{heil_YH_2019superconductivity}
C.~Heil, S.~Di~Cataldo, G.~B. Bachelet, L.~Boeri, Superconductivity in
  sodalite-like yttrium hydride clathrates, arXiv preprint arXiv:1901.04001.

\bibitem{liu2018dynamics}
H.~Liu, I.~I. Naumov, Z.~M. Geballe, M.~Somayazulu, S.~T. John, R.~J. Hemley,
  Dynamics and superconductivity in compressed lanthanum superhydride, Physical
  Review B 98~(10) (2018) 100102.

\bibitem{dalladay2016evidence}
P.~Dalladay-Simpson, R.~T. Howie, E.~Gregoryanz, Evidence for a new phase of
  dense hydrogen above 325 gigapascals, Nature 529~(7584) (2016) 63--67.

\bibitem{H:Borinaga_PRB_2016}
M.~Borinaga, I.~Errea, M.~Calandra, F.~Mauri, A.~Bergara, Anharmonic effects in
  atomic hydrogen: Superconductivity and lattice dynamical stability, Phys.
  Rev. B 93 (2016) 174308.

\bibitem{FloresLivas_MagneticNitridePerovskites_2019}
J.~A. Flores-Livas, R.~Sarmiento-P{\'e}rez, S.~Botti, S.~Goedecker, M.~A.
  Marques, Rare-earth magnetic nitride perovskites, Journal of Physics:
  Materials 2~(2) (2019) 025003.

\bibitem{filip2018geometric}
M.~R. Filip, F.~Giustino, The geometric blueprint of perovskites, Proceedings
  of the National Academy of Sciences 115~(21) (2018) 5397--5402.

\bibitem{xu2018rationalizing}
Q.~Xu, Z.~Li, M.~Liu, W.-J. Yin, Rationalizing perovskite data for machine
  learning and materials design, The journal of physical chemistry letters
  9~(24) (2018) 6948--6954.

\bibitem{PhysRevMaterials.2.085201}
J.~A. Flores-Livas, D.~Tomerini, M.~Amsler, A.~Boziki, U.~Rothlisberger,
  S.~Goedecker, {Emergence of hidden phases of methylammonium lead iodide
  $({\mathrm{CH}}_{3}{\mathrm{NH}}_{3}{\mathrm{PbI}}_{3})$ upon compression},
  Phys. Rev. Materials 2 (2018) 085201.

\bibitem{gong2017discovery}
C.~Gong, L.~Li, Z.~Li, H.~Ji, A.~Stern, Y.~Xia, T.~Cao, W.~Bao, C.~Wang,
  Y.~Wang, et~al., Discovery of intrinsic ferromagnetism in two-dimensional van
  der waals crystals, Nature 546~(7657) (2017) 265.

\bibitem{novoselov2004electric}
K.~S. Novoselov, A.~K. Geim, S.~V. Morozov, D.~Jiang, Y.~Zhang, S.~V. Dubonos,
  I.~V. Grigorieva, A.~A. Firsov, Electric field effect in atomically thin
  carbon films, science 306~(5696) (2004) 666--669.

\bibitem{mounet2018two}
N.~Mounet, M.~Gibertini, P.~Schwaller, D.~Campi, A.~Merkys, A.~Marrazzo,
  T.~Sohier, I.~E. Castelli, A.~Cepellotti, G.~Pizzi, et~al., Two-dimensional
  materials from high-throughput computational exfoliation of experimentally
  known compounds, Nature nanotechnology 13~(3) (2018) 246.

\bibitem{reymond2012exploring}
J.-L. Reymond, M.~Awale, Exploring chemical space for drug discovery using the
  chemical universe database, ACS chemical neuroscience 3~(9) (2012) 649--657.

\bibitem{Goldschmidt_Krystallochemie1926}
V.~M. Goldschmidt, {{Die Gesetze der Krystallochemie}}, Naturwissenschaften
  14~(21) (1926) 477--485.

\bibitem{H:Heil_PRB_2015}
C.~Heil, L.~Boeri, Influence of bonding on superconductivity in high-pressure
  hydrides, Phys. Rev. B 92 (2015) 060508.

\bibitem{Binbin_PRB_2018_Ternary-wrong}
B.~Liu, W.~Cui, J.~Shi, L.~Zhu, J.~Chen, S.~Lin, R.~Su, J.~Ma, K.~Yang, M.~Xu,
  J.~Hao, A.~P. Durajski, J.~Qi, Y.~Li, Y.~Li, Effect of covalent bonding on
  the superconducting critical temperature of the {H-S-Se} system, Phys. Rev. B
  98 (2018) 174101.

\bibitem{CaYH_12_ternary2019}
X.~Liang, A.~Bergara, L.~Wang, B.~Wen, Z.~Zhao, X.-F. Zhou, J.~He, G.~Gao,
  Y.~Tian, {Potential high-${T}_{c}$ superconductivity in
  ${\mathrm{CaYH}}_{12}$ under pressure}, Phys. Rev. B 99 (2019) 100505.

\bibitem{CaYH_12_clathrate2019}
H.~Xie, D.~Duan, Z.~Shao, H.~Song, Y.~Wang, X.~Xiao, D.~Li, F.~Tian, B.~Liu,
  T.~Cui, {High-temperature superconductivity in ternary clathrate YCaH$_{12}$
  under high pressures}, Journal of Physics: Condensed Matter.

\bibitem{mohtadi_Henergy-material-2017_review}
R.~Mohtadi, S.-i. Orimo, The renaissance of hydrides as energy materials,
  Nature Reviews Materials 2~(3) (2017) 16091.

\bibitem{H:Kokail_LIBH_PRM_2017}
C.~Kokail, W.~von~der Linden, L.~Boeri, Prediction of high-${T}_{c}$
  conventional superconductivity in the ternary lithium borohydride system,
  Phys. Rev. Materials 1 (2017) 074803.

\bibitem{H-O-N_Hermann_2018}
V.~Naden~Robinson, M.~Marqu{\'e}s, Y.~Wang, Y.~Ma, A.~Hermann, Novel phases in
  ammonia-water mixtures under pressure, The Journal of chemical physics
  149~(23) (2018) 234501.

\bibitem{Amonia_Hermann_2017_icy}
V.~N. Robinson, Y.~Wang, Y.~Ma, A.~Hermann, Stabilization of ammonia-rich
  hydrate inside icy planets, Proceedings of the National Academy of Sciences
  114~(34) (2017) 9003--9008.

\bibitem{Shimizu-Pdoping-SH3}
A.~Nakanishi, T.~Ishikawa, K.~Shimizu, First-principles study on
  superconductivity of {P}-and {Cl}-doped {H}$_3${S}, Journal of the Physical
  Society of Japan 87~(12) (2018) 124711.

\bibitem{Yungui_PRB_2016_ternaryCE}
Y.~Ge, F.~Zhang, Y.~Yao, First-principles demonstration of superconductivity at
  280 k in hydrogen sulfide with low phosphorus substitution, Phys. Rev. B 93
  (2016) 224513.

\bibitem{ekimov_superconductivity_2004}
E.~A. Ekimov, V.~A. Sidorov, E.~D. Bauer, N.~N. Mel'nik, N.~J. Curro, J.~D.
  Thompson, S.~M. Stishov, Superconductivity in diamond, Nature 428~(6982)
  (2004) 542--545.

\bibitem{Boeri_BdopedDiamond_PRL2004}
L.~Boeri, J.~Kortus, O.~K. Andersen, {Three-Dimensional
  ${\mathrm{M}\mathrm{g}\mathrm{B}}_{2}$-Type Superconductivity in Hole-Doped
  Diamond}, Phys. Rev. Lett. 93 (2004) 237002.

\bibitem{bustarret_superconductivity_2006}
E.~Bustarret, C.~Marcenat, P.~Achatz, J.~Ka{\v{c}}mar{\v{c}}ik, F.~L{\'e}vy,
  A.~Huxley, L.~Ort{\'e}ga, E.~Bourgeois, X.~Blase, D.~D{\'e}barre, et~al.,
  Superconductivity in doped cubic silicon, Nature 444~(7118) (2006) 465--468.

\bibitem{flores-sanna_ice_2017}
J.~A. Flores-Livas, A.~Sanna, M.~Grauzinyte, A.~Davydov, S.~Goedecker, M.~A.~L.
  Marques, {Emergence of superconductivity in doped H$_2$O ice at high
  pressure}, Scientific Reports 7~(1) (2017) 6825.

\bibitem{Flores_polyethylene}
J.~A. Flores-Livas, M.~Grau{\v{z}}inyt{\.e}, L.~Boeri, G.~Profeta, A.~Sanna,
  Superconductivity in doped polyethylene at high pressure, The European
  Physical Journal B 91~(8) (2018) 176.

\bibitem{willardson_HP-semiconductors-book_1998}
R.~K. Willardson, E.~R. Weber, W.~Paul, T.~Suski, High pressure semiconductor
  physics I, Vol.~54, Academic Press, 1998.

\bibitem{grauzinyte2017computational}
M.~Grau{\v{z}}inyt{\.e}, S.~Goedecker, J.~A. Flores-Livas, Computational
  screening of useful hole--electron dopants in {SnO}$_2$, Chemistry of
  Materials 29~(23) (2017) 10095--10103.

\bibitem{Freysold_PointDefectsInSolids_RMP2014}
C.~Freysoldt, B.~Grabowski, T.~Hickel, J.~Neugebauer, G.~Kresse, A.~Janotti,
  C.~G. Van~de Walle, First-principles calculations for point defects in
  solids, Rev. Mod. Phys. 86 (2014) 253--305.

\bibitem{Bruneval_FormationVolumeChargedDefects_PRB2012}
F.~Bruneval, J.-P. Crocombette, {Ab initio formation volume of charged
  defects}, Phys. Rev. B 86 (2012) 140103.

\bibitem{PCCP_defects_2019}
M.~Grau{\v{z}}inyt{\.e}, S.~Botti, M.~A. Marques, S.~Goedecker, J.~A.
  Flores-Livas, Computational acceleration of prospective dopant discovery in
  cuprous iodide, Physical Chemistry Chemical Physics 21~(35) (2019)
  18839--18849.

\bibitem{xia2019hydrogen}
Y.~Xia, B.~Yang, F.~Jin, Y.~Ma, X.~Liu, M.~Zhao, Hydrogen confined in a single
  wall carbon nanotube becomes a metallic and superconductive nanowire under
  high pressure, Nano letters.

\bibitem{klotz2009hydrostatic}
S.~Klotz, J.~Chervin, P.~Munsch, G.~Le~Marchand, Hydrostatic limits of 11
  pressure transmitting media, Journal of Physics D: Applied Physics 42~(7)
  (2009) 075413.

\bibitem{liu2018strain}
C.~Liu, H.~Zhai, Y.~Sun, W.~Gong, Y.~Yan, Q.~Li, W.~Zheng, {Strain-induced
  modulations of electronic structure and electron-phonon coupling in dense
  H$_3$S}, Physical Chemistry Chemical Physics 20~(8) (2018) 5952--5957.

\bibitem{Campanini-Ga}
D.~Campanini, Z.~Diao, A.~Rydh, {Raising the superconducting ${\mathrm{T}}_{c}$
  of gallium: In situ characterization of the transformation of
  $\ensuremath{\alpha}$-Ga into $\ensuremath{\beta}$-Ga}, Phys. Rev. B 97
  (2018) 184517.

\bibitem{kearney2018pressure}
J.~S. Kearney, M.~Grau{\v{z}}inyt{\.e}, D.~Smith, D.~Sneed, C.~Childs,
  J.~Hinton, C.~Park, J.~S. Smith, E.~Kim, S.~D. Fitch, et~al.,
  {Pressure-Tuneable Visible-Range Band Gap in the Ionic Spinel Tin Nitride},
  Angewandte Chemie International Edition 57~(36) (2018) 11623--11628.

\bibitem{MoS2-metastable}
C.~Shang, Y.~Q. Fang, Q.~Zhang, N.~Z. Wang, Y.~F. Wang, Z.~Liu, B.~Lei, F.~B.
  Meng, L.~K. Ma, T.~Wu, Z.~F. Wang, C.~G. Zeng, F.~Q. Huang, Z.~Sun, X.~H.
  Chen, Superconductivity in the metastable $1t'$ and $1t{''}$ phases of
  $\mathrm{Mo}\mathrm{S}_2$ crystals, Phys. Rev. B 98 (2018) 184513.

\bibitem{WO3-metastable}
A.~Palnichenko, O.~Vyaselev, A.~Mazilkin, I.~Zver‘kova, S.~Khasanov,
  Metastable superconductivity of {W}/{WO}$_3$ interface, Physica C:
  Superconductivity and its Applications 534 (2017) 61 -- 67.

\bibitem{MoO3-metastable}
A.~Palnichenko, I.~Zver‘kova, D.~Shakhrai, O.~Vyaselev, {Metastable
  superconductivity of Mo/MoO$_{3−x}$ interface}, Physica C:
  Superconductivity and its Applications 558 (2019) 25 -- 29.

\bibitem{rucavado2018new}
E.~Rucavado, M.~Graužinytė, J.~A. Flores-Livas, Q.~Jeangros, F.~Landucci,
  Y.~Lee, T.~Koida, S.~Goedecker, A.~Hessler-Wyser, C.~Ballif, et~al., {New
  Route for ``Cold-Passivation" of Defects in Tin-Based Oxides}, The Journal of
  Physical Chemistry C 122~(31) (2018) 17612--17620.

\bibitem{ignatiev2018molecular}
A.~Ignatiev, A.~Freundlich, O.~Pchelyakov, A.~Nikiforov, L.~Sokolov,
  D.~Pridachin, V.~Blinov, {Molecular Beam Epitaxy in the Ultravacuum of Space:
  Present and Near Future}, in: Molecular Beam Epitaxy, Elsevier, 2018, pp.
  741--749.

\bibitem{PRB_MBE_supra_2011}
C.-L. Song, Y.-L. Wang, Y.-P. Jiang, Z.~Li, L.~Wang, K.~He, X.~Chen, X.-C. Ma,
  Q.-K. Xue, Molecular-beam epitaxy and robust superconductivity of
  stoichiometric {FeSe} crystalline films on bilayer graphene, Phys. Rev. B 84
  (2011) 020503.

\bibitem{Giustino_Graphane_PRL2010}
G.~Savini, A.~C. Ferrari, F.~Giustino, {First-Principles Prediction of Doped
  Graphane as a High-Temperature Electron-Phonon Superconductor}, Phys. Rev.
  Lett. 105 (2010) 037002.

\bibitem{Profeta_Graphene_Nat2012}
G.~Profeta, M.~Calandra, F.~Mauri, Phonon-mediated superconductivity in
  graphene by lithium deposition, Nature Phys. 8 (2012) 131--134.

\bibitem{SC:Ludbrook_Ligraphene_PNAS2015}
B.~M. Ludbrook, G.~Levy, P.~Nigge, M.~Zonno, M.~Schneider, D.~J. Dvorak, C.~N.
  Veenstra, S.~Zhdanovich, D.~Wong, P.~Dosanjh, C.~Stra{\ss}er, A.~St{\"o}hr,
  S.~Forti, C.~R. Ast, U.~Starke, A.~Damascelli, {Evidence for
  superconductivity in Li-decorated monolayer graphene}, Proceedings of the
  National Academy of Sciences 112~(38) (2015) 11795--11799.

\bibitem{boufelfel2019hydrogen}
A.~Boufelfel, {Hydrogen Effect on Electron-Phonon Interactions in
  {L}1$_0$-{FePd}}, Journal of Superconductivity and Novel Magnetism (2019)
  1--9.

\bibitem{SC:savini_graphane_PRL_2010}
G.~Savini, A.~C. Ferrari, F.~Giustino, {First-Principles Prediction of Doped
  Graphane as a High-Temperature Electron-Phonon Superconductor}, Phys. Rev.
  Lett. 105 (2010) 037002.

\bibitem{MgB2_H_superconductivity}
J.~Bekaert, M.~Petrov, A.~Aperis, P.~M. Oppeneer, M.~V. Milo\ifmmode
  \check{s}\else \v{s}\fi{}evi\ifmmode~\acute{c}\else \'{c}\fi{},
  {Hydrogen-Induced High-Temperature Superconductivity in Two-Dimensional
  Materials: The Example of Hydrogenated Monolayer ${\mathrm{MgB}}_{2}$}, Phys.
  Rev. Lett. 123 (2019) 077001.
\newblock \href {http://dx.doi.org/10.1103/PhysRevLett.123.077001}
  {\path{doi:10.1103/PhysRevLett.123.077001}}.

\bibitem{C60_superconductivity}
S.~Margadonna, K.~Prassides, Recent advances in fullerene superconductivity,
  Journal of Solid State Chemistry 168~(2) (2002) 639--652.

\bibitem{de2018stable}
D.~S. De, J.~A. Flores-Livas, S.~Saha, L.~Genovese, S.~Goedecker, {Stable
  structures of exohedrally decorated C$_{60}$-fullerenes}, Carbon 129 (2018)
  847--853.

\bibitem{Hapiuk_PRL2012}
D.~Hapiuk, M.~A.~L. Marques, P.~Melinon, J.~A. Flores-Livas, S.~Botti,
  B.~Masenelli, $p$ doping in expanded phases of {ZnO}: An ab initio study,
  Phys. Rev. Lett. 108 (2012) 115903.

\bibitem{hapiuk2015superconductivity}
D.~Hapiuk, M.~Marques, P.~M{\'e}linon, S.~Botti, B.~Masenelli, J.~Flores-Livas,
  Superconductivity in an expanded phase of {ZnO}: an ab initio study, New
  Journal of Physics 17~(4) (2015) 043034.

\bibitem{Weller_CaC6_NAT2005}
T.~E. Weller, M.~Ellerby, S.~S. Saxena, R.~P. Smith, N.~T. Skipper,
  {Superconductivity in the intercalated graphite compounds C$_6$Yb and
  C$_6$Ca}, Nat. Phys. 1~(39--41).

\bibitem{disicilicides_PRB2011}
J.~A. Flores-Livas, R.~Debord, S.~Botti, A.~San~Miguel, S.~Pailh\`es, M.~A.~L.
  Marques, Superconductivity in layered binary silicides: A density functional
  theory study, Phys. Rev. B 84 (2011) 184503.

\bibitem{enhancing_PRL2011}
J.~A. Flores-Livas, R.~Debord, S.~Botti, A.~San~Miguel, M.~A.~L. Marques,
  S.~Pailh\`es, Enhancing the superconducting transition temperature of
  ${\mathrm{basi}}_{2}$ by structural tuning, Phys. Rev. Lett. 106 (2011)
  087002.

\bibitem{Sanna_PRB-2015}
J.~A. Flores-Livas, A.~Sanna, {Superconductivity in intercalated group-IV
  honeycomb structures}, Phys. Rev. B 91 (2015) 054508.

\bibitem{picene_kubozono}
R.~Mitsuhashi, Y.~Suzuki, Y.~Yamanari, H.~Mitamura, T.~Kambe, N.~Ikeda,
  H.~Okamoto, A.~Fujiwara, M.~Yamaji, N.~Kawasaki, Y.~Maniwa, Y.~Kubozono,
  {Superconductivity in alkali-metal-doped picene}, Nature 464 (2010) 76.

\bibitem{mine:Subedi_PRB_2011}
A.~Subedi, L.~Boeri, Vibrational spectrum and electron-phonon coupling of doped
  solid picene from first principles, Phys. Rev. B 84 (2011) 020508.

\bibitem{SC:Casula_picene_PRB_2012}
M.~Casula, M.~Calandra, F.~Mauri, Local and nonlocal electron-phonon couplings
  in {K}$_{3}$ picene and the effect of metallic screening, Phys. Rev. B 86
  (2012) 075445.

\bibitem{phenanthrene}
X.~F. Wang, R.~H. Liu, Z.~Gui, Y.~L. Xie, Y.~J. Yan, J.~J. Ying, X.~G. Luo,
  X.~H. Chen, Superconductivity at 5 k in alkali-metal-doped phenanthrene,
  Nature Communications 2 (2011) 507.

\bibitem{dibenzopentacene}
M.~Xue, T.~Cao, D.~Wang, Y.~Wu, H.~Yang, X.~Dong, J.~He, F.~Li, G.~F. Chen,
  {Superconductivity above 30 K in alkali-metal-doped hydrocarbon}, Scientific
  Reports 2 (2012) 389.

\bibitem{terphenyl1}
R.-S. {Wang}, Y.~{Gao}, Z.-B. {Huang}, X.-J. {Chen}, {{Superconductivity above
  120 kelvin in a chain link molecule}}, arXiv e-prints (2017)
  arXiv:1703.06641.

\bibitem{terphenyl2}
W.~Liu, H.~Lin, R.~Kang, X.~Zhu, Y.~Zhang, S.~Zheng, H.-H. Wen, Magnetization
  of potassium-doped $p$-terphenyl and $p$-quaterphenyl by high-pressure
  synthesis, Phys. Rev. B 96 (2017) 224501.

\bibitem{WO3-H_synthesis}
X.~Leng, J.~Pereiro, J.~Strle, G.~Dubuis, A.~T. Bollinger, A.~Gozar, J.~Wu,
  N.~Litombe, C.~Panagopoulos, D.~Pavuna, I.~Bo{\v z}ovi{\'c}, Insulator to
  metal transition in {WO}$_3$ induced by electrolyte gating, npj Quantum
  Materials 2~(1) (2017) 35.

\bibitem{WO3-H_superconductivity}
S.~Reich, G.~Leitus, R.~Popovitz-Biro, A.~Goldbourt, S.~Vega, {A possible 2D
  H$_x${WO}$_3$ superconductor with a T$_c$ of 120 K}, Journal of
  Superconductivity and Novel Magnetism 22 (2009) 343--346.

\bibitem{2Dsuper_Iwasa}
Y.~Saito, T.~Nojima, Y.~Iwasa, Gate-induced superconductivity in
  two-dimensional atomic crystals, Superconductor Science and Technology 29~(9)
  (2016) 093001.

\bibitem{MoS2-synthesis}
K.~F. Mak, C.~Lee, J.~Hone, J.~Shan, T.~F. Heinz, {Atomically Thin
  $\mathrm{MoS}_{2}$: A New Direct-Gap Semiconductor}, Phys. Rev. Lett. 105
  (2010) 136805.

\bibitem{MoO3_synthesis}
A.~J. Molina-Mendoza, J.~L. Lado, J.~O. Island, M.~A. Ni{\~n}o, L.~Aballe,
  M.~Foerster, F.~Y. Bruno, A.~L{\'o}pez-Moreno, L.~Vaquero-Garzon, H.~S.~J.
  van~der Zant, G.~Rubio-Bollinger, N.~Agra{\"\i}t, E.~M. P{\'e}rez,
  J.~Fern{\'a}ndez-Rossier, A.~Castellanos-Gomez, {Centimeter-Scale Synthesis
  of Ultrathin Layered MoO$_3$ by van der Waals Epitaxy}, Chemistry of
  Materials 28~(11) (2016) 4042--4051.

\bibitem{La2CuO4_synthesis}
M.~P.~M. Dean, R.~S. Springell, C.~Monney, K.~J. Zhou, J.~Pereiro, I.~Bo{\v
  z}ovi{\'c}, B.~Dalla~Piazza, H.~M. R{\o}nnow, E.~Morenzoni, J.~van~den Brink,
  T.~Schmitt, J.~P. Hill, Spin excitations in a single {La}$_2${CuO}$_4$ layer,
  Nature Materials 11 (2012) 850 EP --.

\bibitem{BN_synthesis}
D.~Golberg, Y.~Bando, Y.~Huang, T.~Terao, M.~Mitome, C.~Tang, C.~Zhi, Boron
  nitride nanotubes and nanosheets, ACS Nano 4~(6) (2010) 2979--2993.

\bibitem{topological_transport}
H.~Tang, D.~Liang, R.~L.~J. Qiu, X.~P.~A. Gao, {Two-Dimensional
  Transport-Induced Linear Magneto-Resistance in Topological Insulator
  Bi$_2$Se$_3$ Nanoribbons}, ACS Nano 5~(9) (2011) 7510--7516.

\bibitem{cuprate_2D}
D.~Jiang, T.~Hu, L.~You, Q.~Li, A.~Li, H.~Wang, G.~Mu, Z.~Chen, H.~Zhang,
  G.~Yu, J.~Zhu, Q.~Sun, C.~Lin, H.~Xiao, X.~Xie, M.~Jiang, {High-Tc
  superconductivity in ultrathin Bi$_2$Sr$_2$CaCu$_2$O$_{8+x}$ down to
  half-unit-cell thickness by protection with graphene}, Nature Communications
  5 (2014) 5708.

\bibitem{CrI3_synthesis}
B.~Huang, G.~Clark, E.~Navarro-Moratalla, D.~R. Klein, R.~Cheng, K.~L. Seyler,
  D.~Zhong, E.~Schmidgall, M.~A. McGuire, D.~H. Cobden, W.~Yao, D.~Xiao,
  P.~Jarillo-Herrero, X.~Xu, Layer-dependent ferromagnetism in a van der waals
  crystal down to the monolayer limit, Nature 546 (2017) 270 EP --.

\bibitem{blackP-_synthesis}
L.~Li, Y.~Yu, G.~J. Ye, Q.~Ge, X.~Ou, H.~Wu, D.~Feng, X.~H. Chen, Y.~Zhang,
  Black phosphorus field-effect transistors, Nature Nanotechnology 9 (2014) 372
  EP --.

\bibitem{Bismuthene_synthesis}
F.~Reis, G.~Li, L.~Dudy, M.~Bauernfeind, S.~Glass, W.~Hanke, R.~Thomale,
  J.~Sch{\"a}fer, R.~Claessen, Bismuthene on a sic substrate: A candidate for a
  high-temperature quantum spin hall material, Science 357~(6348) (2017)
  287--290.

\bibitem{silicene_synthesis}
B.~Aufray, A.~Kara, S.~Vizzini, H.~Oughaddou, C.~Léandri, B.~Ealet, G.~Le~Lay,
  {Graphene-like silicon nanoribbons on Ag(110): A possible formation of
  silicene}, Applied Physics Letters 96~(18) (2010) 183102.

\bibitem{germanene_synthesis}
J.~Yuhara, H.~Shimazu, K.~Ito, A.~Ohta, M.~Araidai, M.~Kurosawa, M.~Nakatake,
  G.~Le~Lay, Germanene epitaxial growth by segregation through ag(111) thin
  films on ge(111), ACS Nano 12~(11) (2018) 11632--11637.

\bibitem{Borophene_synthesis}
Z.~Zhang, Y.~Yang, G.~Gao, B.~I. Yakobson, Two-dimensional boron monolayers
  mediated by metal substrates, Angewandte Chemie International Edition 54~(44)
  (2015) 13022--13026.

\bibitem{Th:pietronero_PRL_1995}
C.~Grimaldi, L.~Pietronero, S.~Straessler, Nonadiabatic superconductivity:
  electron-phonon interaction beyond migdal's theorem, Phys. Rev. Lett. 75
  (1995) 1158.

\bibitem{Boeri_PRB_2005}
L.~Boeri, E.~Cappelluti, L.~Pietronero, {Small Fermi energy, zero-point
  fluctuations, and nonadiabaticity in MgB$_{2}$}, Phys. Rev. B 71 (2005)
  012501.

\bibitem{Calandra_AnharmonicAndNonAdiabaticMgB2Raman_PhysicaC2007}
M.~Calandra, M.~Lazzeri, F.~Mauri,
  \href{http://www.sciencedirect.com/science/article/pii/S0921453407000135}{{{Anharmonic
  and non-adiabatic effects in MgB$_2$: Implications for the isotope effect and
  interpretation of Raman spectra}}}, Physica C: Superconductivity 456~(1)
  (2007) 38 -- 44, recent Advances in MgB2 Research.
\newblock \href {http://dx.doi.org/https://doi.org/10.1016/j.physc.2007.01.021}
  {\path{doi:https://doi.org/10.1016/j.physc.2007.01.021}}.
\newline\urlprefix\url{http://www.sciencedirect.com/science/article/pii/S0921453407000135}

\bibitem{Luigi_nucleation_PRL2018}
L.~Bonati, M.~Parrinello, Silicon liquid structure and crystal nucleation from
  ab initio deep metadynamics, Phys. Rev. Lett. 121 (2018) 265701.

\bibitem{Ma_RoomT_PRL2019}
Y.~Sun, J.~Lv, Y.~Xie, H.~Liu, Y.~Ma, {Route to a Superconducting Phase above
  Room Temperature in Electron-Doped Hydride Compounds under High Pressure},
  Phys. Rev. Lett. 123 (2019) 097001.

\bibitem{Li_xie2014superconductivity}
Y.~Xie, Q.~Li, A.~R. Oganov, H.~Wang, Superconductivity of lithium-doped
  hydrogen under high pressure, Acta Crystallographica Section C: Structural
  Chemistry 70~(2) (2014) 104--111.

\bibitem{K_zhou2012ab}
D.~Zhou, X.~Jin, X.~Meng, G.~Bao, Y.~Ma, B.~Liu, T.~Cui, Ab initio study
  revealing a layered structure in hydrogen-rich {KH}$_6$ under high pressure,
  Physical Review B 86~(1) (2012) 014118.

\bibitem{Be_yu2014exploration}
S.~Yu, Q.~Zeng, A.~R. Oganov, C.~Hu, G.~Frapper, L.~Zhang, Exploration of
  stable compounds, crystal structures, and superconductivity in the {Be}-{H}
  system, AIP Advances 4~(10) (2014) 107118.

\bibitem{Mg_feng2015compressed}
X.~Feng, J.~Zhang, G.~Gao, H.~Liu, H.~Wang, Compressed sodalite-like {MgH}$_6$
  as a potential high-temperature superconductor, RSC Advances 5~(73) (2015)
  59292--59296.

\bibitem{Sr_hooper2014composition}
J.~Hooper, T.~Terpstra, A.~Shamp, E.~Zurek, Composition and constitution of
  compressed strontium polyhydrides, The Journal of Physical Chemistry C
  118~(12) (2014) 6433--6447.

\bibitem{Sr_wang2015structural}
Y.~Wang, H.~Wang, S.~T. John, T.~Iitaka, Y.~Ma, Structural morphologies of
  high-pressure polymorphs of strontium hydrides, Physical Chemistry Chemical
  Physics 17~(29) (2015) 19379--19385.

\bibitem{Ba_hooper2013polyhydrides}
J.~Hooper, B.~Altintas, A.~Shamp, E.~Zurek, Polyhydrides of the alkaline earth
  metals: a look at the extremes under pressure, The Journal of Physical
  Chemistry C 117~(6) (2013) 2982--2992.

\bibitem{Sc_ye2018high}
X.~Ye, N.~Zarifi, E.~Zurek, R.~Hoffmann, N.~W. Ashcroft, High hydrides of
  scandium under pressure: potential superconductors, The Journal of Physical
  Chemistry C 122~(11) (2018) 6298--6309.

\bibitem{Ac_semenok2018actinium}
D.~V. Semenok, A.~G. Kvashnin, I.~A. Kruglov, A.~R. Oganov, {Actinium hydrides
  AcH$_{10}$, AcH$_{12}$, and AcH$_{16}$ as high-temperature conventional
  superconductors}, {The Journal of Physical Chemistry Letters} 9~(8) (2018)
  1920--1926.

\bibitem{Th_kvashnin2018high}
A.~G. Kvashnin, D.~V. Semenok, I.~A. Kruglov, I.~A. Wrona, A.~R. Oganov,
  High-temperature superconductivity in a {Th}--{H} system under pressure
  conditions, ACS applied materials \& interfaces 10~(50) (2018) 43809--43816.

\bibitem{PaH_shit_of-shits_2019}
X.~Xiao, D.~Duan, H.~Xie, Z.~Shao, D.~Li, F.~Tian, H.~Song, H.~Yu, K.~Bao,
  T.~Cui, Structure and superconductivity of protactinium hydrides under high
  pressure, Journal of Physics: Condensed Matter.

\bibitem{U_kruglov2018_oganov}
I.~A. Kruglov, A.~G. Kvashnin, A.~F. Goncharov, A.~R. Oganov, S.~S. Lobanov,
  N.~Holtgrewe, S.~Jiang, V.~B. Prakapenka, E.~Greenberg, A.~V. Yanilkin,
  Uranium polyhydrides at moderate pressures: {P}rediction, synthesis, and
  expected superconductivity, Science advances 4~(10) (2018) eaat9776.

\bibitem{V_li_2017superconductivity}
X.~Li, F.~Peng, Superconductivity of pressure-stabilized vanadium hydrides,
  Inorganic chemistry 56~(22) (2017) 13759--13765.

\bibitem{Nb_Hoffman2013_theoretical}
G.~Gao, R.~Hoffmann, N.~W. Ashcroft, H.~Liu, A.~Bergara, Y.~Ma, Theoretical
  study of the ground-state structures and properties of niobium hydrides under
  pressure, Physical Review B 88~(18) (2013) 184104.

\bibitem{Nb_Durajski2014_phonon}
A.~P. Durajski, Phonon-mediated superconductivity in compressed {NbH}$_4$
  compound, The European Physical Journal B 87~(9) (2014) 210.

\bibitem{Ta_zhuang2017pressure}
Q.~Zhuang, X.~Jin, T.~Cui, Y.~Ma, Q.~Lv, Y.~Li, H.~Zhang, X.~Meng, K.~Bao,
  Pressure-stabilized superconductive ionic tantalum hydrides, Inorganic
  chemistry 56~(7) (2017) 3901--3908.

\bibitem{Cr_yu2015pressure}
S.~Yu, X.~Jia, G.~Frapper, D.~Li, A.~R. Oganov, Q.~Zeng, L.~Zhang,
  Pressure-driven formation and stabilization of superconductive chromium
  hydrides, Scientific reports 5 (2015) 17764.

\bibitem{W_zheng2018structural}
S.~Zheng, S.~Zhang, Y.~Sun, J.~Zhang, J.~Lin, G.~Yang, A.~Bergara, {Structural
  and Superconducting Properties of Tungsten Hydrides under High Pressure},
  Frontiers in Physics 6 (2018) 101.

\bibitem{Tc_li2016crystal}
X.~Li, H.~Liu, F.~Peng, Crystal structures and superconductivity of technetium
  hydrides under pressure, Physical Chemistry Chemical Physics 18~(41) (2016)
  28791--28796.

\bibitem{Ru_liu2016stability}
Y.~Liu, D.~Duan, F.~Tian, C.~Wang, Y.~Ma, D.~Li, X.~Huang, B.~Liu, T.~Cui,
  Stability and properties of the {Ru}--{H} system at high pressure, Physical
  Chemistry Chemical Physics 18~(3) (2016) 1516--1520.

\bibitem{Os_liu2015structures}
Y.~Liu, D.~Duan, X.~Huang, F.~Tian, D.~Li, X.~Sha, C.~Wang, H.~Zhang, T.~Yang,
  B.~Liu, et~al., Structures and properties of osmium hydrides under pressure
  from first principle calculation, The Journal of Physical Chemistry C
  119~(28) (2015) 15905--15911.

\bibitem{PRL_kim2011_subproduct}
D.~Y. Kim, R.~H. Scheicher, C.~J. Pickard, R.~Needs, R.~Ahuja, Predicted
  formation of superconducting platinum-hydride crystals under pressure in the
  presence of molecular hydrogen, Physical Review Letters 107~(11) (2011)
  117002.

\bibitem{PdH_pressure_PRB_hemmes1989}
H.~Hemmes, A.~Driessen, R.~Griessen, M.~Gupta, {Isotope effects and pressure
  dependence of the Tc of superconducting stoichiometric {PdH} and {PdD}
  synthesized and measured in a diamond anvil cell}, Physical Review B 39~(7)
  (1989) 4110.

\bibitem{Pt_errea2014anharmonic}
I.~Errea, M.~Calandra, F.~Mauri, Anharmonic free energies and phonon
  dispersions from the stochastic self-consistent harmonic approximation:
  Application to platinum and palladium hydrides, Physical Review B 89~(6)
  (2014) 064302.

\bibitem{Pt_zhou2011superconducting}
X.-F. Zhou, A.~R. Oganov, X.~Dong, L.~Zhang, Y.~Tian, H.-T. Wang, et~al.,
  Superconducting high-pressure phase of platinum hydride from first
  principles, Physical Review B 84~(5) (2011) 054543.

\bibitem{PRB_Matsuoka_PtH_30GPa_2019}
T.~Matsuoka, M.~Hishida, K.~Kuno, N.~Hirao, Y.~Ohishi, S.~Sasaki, K.~Takahama,
  K.~Shimizu, Superconductivity of platinum hydride, Phys. Rev. B 99 (2019)
  144511.

\bibitem{B_abe2011crystalline}
K.~Abe, N.~Ashcroft, Crystalline diborane at high pressures, Physical Review B
  84~(10) (2011) 104118.

\bibitem{Al_hou2015high}
P.~Hou, X.~Zhao, F.~Tian, D.~Li, D.~Duan, Z.~Zhao, B.~Chu, B.~Liu, T.~Cui, High
  pressure structures and superconductivity of {AlH}$_3$({H}$_2$) predicted by
  first principles, RSC Advances 5~(7) (2015) 5096--5101.

\bibitem{Ga_szczesniak2013superconducting}
R.~Szcze{\'s}niak, A.~Durajski, Superconducting state above the boiling point
  of liquid nitrogen in the gah3 compound, Superconductor Science and
  Technology 27~(1) (2013) 015003.

\bibitem{In_liu2015pressure}
Y.~Liu, D.~Duan, F.~Tian, H.~Liu, C.~Wang, X.~Huang, D.~Li, Y.~Ma, B.~Liu,
  T.~Cui, Pressure-induced structures and properties in indium hydrides,
  Inorganic chemistry 54~(20) (2015) 9924--9928.

\bibitem{Si_jin2010superconducting}
X.~Jin, X.~Meng, Z.~He, Y.~Ma, B.~Liu, T.~Cui, G.~Zou, H.-k. Mao,
  Superconducting high-pressure phases of disilane, Proceedings of the National
  Academy of Sciences 107~(22) (2010) 9969--9973.

\bibitem{Ge_abe2013quantum}
K.~Abe, N.~Ashcroft, {Quantum disproportionation: The high hydrides at elevated
  pressures}, Physical Review B 88~(17) (2013) 174110.

\bibitem{Ge_strobel2010vibrational}
T.~A. Strobel, X.-J. Chen, M.~Somayazulu, R.~J. Hemley, Vibrational dynamics,
  intermolecular interactions, and compound formation in {GeH}$_4$--{H}$_2$
  under pressure, The Journal of chemical physics 133~(16) (2010) 164512.

\bibitem{Ge_szcze2014thermodynamics}
R.~Szczesniak, D.~Szczesniak, A.~Durajski, et~al., Thermodynamics of the
  superconducting phase in compressed {GeH}$_4$ ({H}$_2$)$_2$, Solid State
  Communications 184 (2014) 6--11.

\bibitem{Ge_zhong2013superconductivity}
G.~Zhong, C.~Zhang, G.~Wu, J.~Song, Z.~Liu, C.~Yang, {Superconductivity in
  {GeH}$_4$ ({H}$_2$)$_2$ above 220 GPa high-pressure}, Physica B: Condensed
  Matter 410 (2013) 90--92.

\bibitem{Ge_hou2015ab}
P.~Hou, F.~Tian, D.~Li, Z.~Zhao, D.~Duan, H.~Zhang, X.~Sha, B.~Liu, T.~Cui, {Ab
  initio study of germanium-hydride compounds under high pressure}, RSC
  Advances 5~(25) (2015) 19432--19438.

\bibitem{Ge_esfahani2017superconductivity}
M.~M.~D. Esfahani, A.~R. Oganov, H.~Niu, J.~Zhang, Superconductivity and
  unexpected chemistry of germanium hydrides under pressure, Physical Review B
  95~(13) (2017) 134506.

\bibitem{Sn_esfahani2016superconductivity}
M.~M.~D. Esfahani, Z.~Wang, A.~R. Oganov, H.~Dong, Q.~Zhu, S.~Wang, M.~S.
  Rakitin, X.-F. Zhou, {Superconductivity of novel tin hydrides
  ({Sn}$_n${H}$_m$) under pressure}, Scientific reports 6 (2016) 22873.

\bibitem{Pb_cheng2015pressure}
Y.~Cheng, C.~Zhang, T.~Wang, G.~Zhong, C.~Yang, X.-J. Chen, H.-Q. Lin,
  {Pressure-induced superconductivity in H$_2$-containing hydride
  {PbH}$_4$({H}$_2$)$_2$}, Scientific reports 5 (2015) 16475.

\bibitem{P_liu2016crystal}
H.~Liu, Y.~Li, G.~Gao, J.~S. Tse, I.~I. Naumov, {Crystal structure and
  superconductivity of PH$_3$ at high pressures}, The Journal of Physical
  Chemistry C 120~(6) (2016) 3458--3461.

\bibitem{P_durajski2016quantitative}
A.~P. Durajski, {Q}uantitative analysis of nonadiabatic effects in dense
  {H}$_2${S} and {PH}$_3$ superconductors, Scientific reports 6 (2016) 38570.

\bibitem{PH_metastable_Zurek_2017}
T.~Bi, D.~P. Miller, A.~Shamp, E.~Zurek, Superconducting phases of phosphorus
  hydride under pressure: stabilization by mobile molecular hydrogen,
  Angewandte Chemie 129~(34) (2017) 10326--10329.

\bibitem{As_fu2016high}
Y.~Fu, X.~Du, L.~Zhang, F.~Peng, M.~Zhang, C.~J. Pickard, R.~J. Needs, D.~J.
  Singh, W.~Zheng, Y.~Ma, High-pressure phase stability and superconductivity
  of pnictogen hydrides and chemical trends for compressed hydrides, Chemistry
  of Materials 28~(6) (2016) 1746--1755.

\bibitem{Sb_ma2015unexpected}
Y.~Ma, D.~Duan, D.~Li, Y.~Liu, F.~Tian, X.~Huang, Z.~Zhao, H.~Yu, B.~Liu,
  T.~Cui, {The unexpected binding and superconductivity in SbH$_4$ at high
  pressure}, arXiv preprint arXiv:1506.03889.

\bibitem{Bi_ma2015high}
Y.~Ma, D.~Duan, D.~Li, Y.~Liu, F.~Tian, H.~Yu, C.~Xu, Z.~Shao, B.~Liu, T.~Cui,
  High-pressure structures and superconductivity of bismuth hydrides, arXiv
  preprint arXiv:1511.05291.

\bibitem{Se_zhang2015phase}
S.~Zhang, Y.~Wang, J.~Zhang, H.~Liu, X.~Zhong, H.-F. Song, G.~Yang, L.~Zhang,
  Y.~Ma, Phase diagram and high-temperature superconductivity of compressed
  selenium hydrides, Scientific reports 5 (2015) 15433.

\bibitem{Te_hydrides_Ma_2016}
X.~Zhong, H.~Wang, J.~Zhang, H.~Liu, S.~Zhang, H.-F. Song, G.~Yang, L.~Zhang,
  Y.~Ma, {Tellurium Hydrides at High Pressures: High-Temperature
  Superconductors}, Phys. Rev. Lett. 116 (2016) 057002.

\bibitem{Po_RSC_Tian_2015}
Y.~Liu, D.~Duan, F.~Tian, C.~Wang, G.~Wu, Y.~Ma, H.~Yu, D.~Li, B.~Liu, T.~Cui,
  {{Prediction of stoichiometric PoH$_n$ compounds: Crystal structures and
  properties}}, RSC Adv. 5 (2015) 103445--103450.
\newblock \href {http://dx.doi.org/10.1039/C5RA19223D}
  {\path{doi:10.1039/C5RA19223D}}.

\bibitem{Br_Cl_APL_2010}
D.~Duan, F.~Tian, Z.~He, X.~Meng, L.~Wang, C.~Chen, X.~Zhao, B.~Liu, T.~Cui,
  {Hydrogen bond symmetrization and superconducting phase of HBr and HCl under
  high pressure: An ab initio study}, The Journal of Chemical Physics 133~(7)
  (2010) 074509.

\bibitem{I_iodine_doped_2015}
D.~Duan, F.~Tian, Y.~Liu, X.~Huang, D.~Li, H.~Yu, Y.~Ma, B.~Liu, T.~Cui,
  {{Enhancement of Tc in the atomic phase of iodine-doped hydrogen at high
  pressures}}, Phys. Chem. Chem. Phys. 17 (2015) 32335--32340.
\newblock \href {http://dx.doi.org/10.1039/C5CP05218A}
  {\path{doi:10.1039/C5CP05218A}}.

\bibitem{I_jpclett_2015_Iodine}
A.~Shamp, E.~Zurek, {Superconducting High-Pressure Phases Composed of Hydrogen
  and Iodine}, The Journal of Physical Chemistry Letters 6~(20) (2015)
  4067--4072, pMID: 26722778.

\bibitem{Xe-H_JCP_2015}
X.~Yan, Y.~Chen, X.~Kuang, S.~Xiang, Structure, stability, and
  superconductivity of new {Xe}–{H} compounds under high pressure, The
  Journal of Chemical Physics 143~(12) (2015) 124310.

\end{thebibliography}
 
\end{document}